\documentclass[11pt,a4paper]{report}
\usepackage[latin1]{inputenc}
\usepackage{amssymb}
\usepackage{amsmath}
\usepackage{amsthm}
\usepackage{amsfonts}
\usepackage[pdftex]{graphicx}
\usepackage[usenames,dvipsnames,table]{xcolor}
\usepackage{setspace}\usepackage{subfig,cancel,setspace,bm,mathrsfs,subfig,makeidx,slashed}
\usepackage[left=30mm,right=30mm,top=20mm,bottom=30mm]{geometry}
\usepackage[toc,page]{appendix}
\usepackage{float,makecell}

\pdfoutput=1

\usepackage{latexsym}
\usepackage{bbm}
\usepackage{multirow}
\usepackage{wrapfig}
\usepackage{braket}

\numberwithin{equation}{section}

\usepackage{hyperref}
\hypersetup{
colorlinks=true,
linktoc=page,
citecolor=BlueViolet,
linkcolor=BlueViolet,
urlcolor=BlueViolet} 

\def\D{\Delta}

\newcommand{\black}{\mathord{\parbox[c]{1em}{\includegraphics[width=0.115\textwidth]{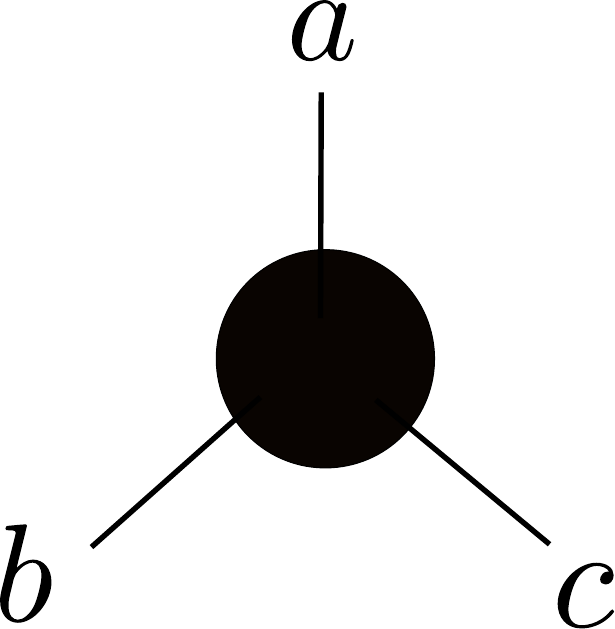}}}}
\newcommand{\white}{\mathord{\parbox[c]{1em}{\includegraphics[width=0.11\textwidth]{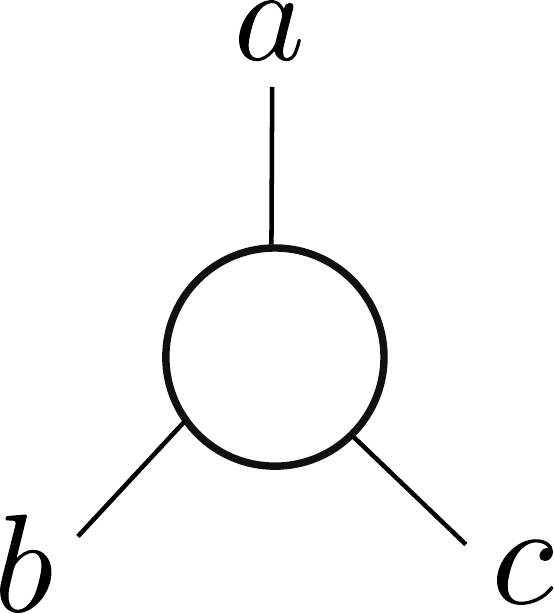}}}}

\newcommand*{\matminus}{%
  \leavevmode
  \hphantom{0}%
  \llap{%
    \settowidth{\dimen0 }{$0$}%
    \resizebox{1.1\dimen0 }{\height}{$-$}%
  }%
}

\newcommand{\sref}[1]{\S\ref{#1}}
\newcommand{\fref}[1]{Figure \ref{#1}}

\newcommand{\pl}{Pl\"ucker }

\newcommand{\be}{\begin{equation}}
\newcommand{\ee}{\end{equation}}

\def\beqa{\begin{eqnarray}}
\def\eeqa{\end{eqnarray}}
\def\beq{\begin{equation}}
\def\eeq{\end{equation}}

\def\one{\mbox{1 \kern-.59em {\rm l}}}

\def\uno{\mbox{1 \kern-.59em {\rm l}}}

\newcommand{\barray}{\begin{eqnarray}}
\newcommand{\earray}{\end{eqnarray}}

\newcommand{\HRule}{\rule{\linewidth}{0.4mm}}
\newcommand{\comments}[1]{}
\newcommand{\la}{\langle}
\newcommand{\ra}{\rangle}
\newcommand{\tl}{\widetilde\lambda}

\newcommand{\A}{\mathcal{A}}
\newcommand{\F}{\mathcal{F}}
\newcommand{\T}{\mathcal{T}}

\newcommand{\N}{\mathcal{N}}

\newcommand{\Tr}{\text{Tr}}
\newcommand{\MHVb}{\overline{\text{MHV}}}
\renewcommand{\b}[1]{\braket{#1}}

\renewcommand{\O}{\mathcal{O}}

\def\cA{{\cal A}} \def\cB{{\cal B}} \def\cC{{\cal C}}
  \def\cF{{\cal F}}
\def\cG{{\cal G}}  
  \def\cL{{\cal L}}
 \def\cN{{\cal N}} \def\cO{{\cal O}}
  \def\cR{{\cal R}}
\def\cS{{\cal S}} \def\cT{{\cal T}}

\def\drawbox#1#2{\hrule height#2pt 
        \hbox{\vrule width#2pt height#1pt \kern#1pt \vrule width#2pt}
              \hrule height#2pt}

\def\Asym#1#2{\vcenter{\vbox{\drawbox{#1}{#2}
              \kern-#2pt       
              \drawbox{#1}{#2}}}}

\def\lan{\langle}
\def\ran{\rangle}

\def\eps{\epsilon}
\def\Li{{\rm Li}_2}

\spacing{1.5}

\begin{document}

\pagenumbering{arabic}

\begin{titlepage}
\begin{center}

\begin{center}
\def\svgwidth{6cm}
\includegraphics[width=0.5\linewidth]{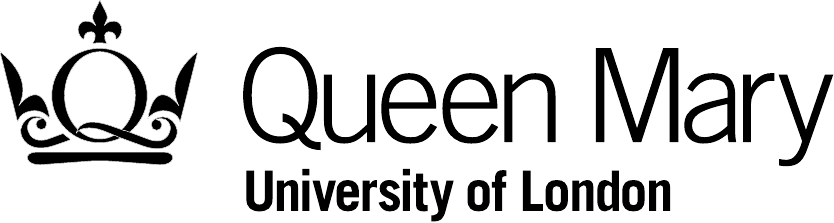}
\end{center}
\vspace{15pt}

\textsc{\Large Thesis submitted for the degree of\\[10pt] Doctor of Philosophy}\\[0.5cm]

\HRule \\[0.4cm]
{ \LARGE \bfseries On-shell methods for off-shell quantities\\
 in $\N=4$ Super Yang-Mills:\\[10pt]}
{\Large \bfseries From scattering amplitudes to form factors\\ and the dilatation operator \\[0.4cm] }

\HRule \\[1.5cm]

\begin{minipage}{0.4\textwidth}
\begin{flushleft} \large
\emph{Author:}\\
Brenda Corr\^{e}a de\\
Andrade Penante
\end{flushleft}
\end{minipage}
\begin{minipage}{0.4\textwidth}
\begin{flushright} \large
\emph{Supervisors:} \\
Prof. Gabriele Travaglini\\
Prof. Bill Spence
\end{flushright}
\end{minipage}

\vspace{30pt}
{\large April 2016}

\vfill

\def\svgwidth{6cm}
\includegraphics[width=0.4\linewidth]{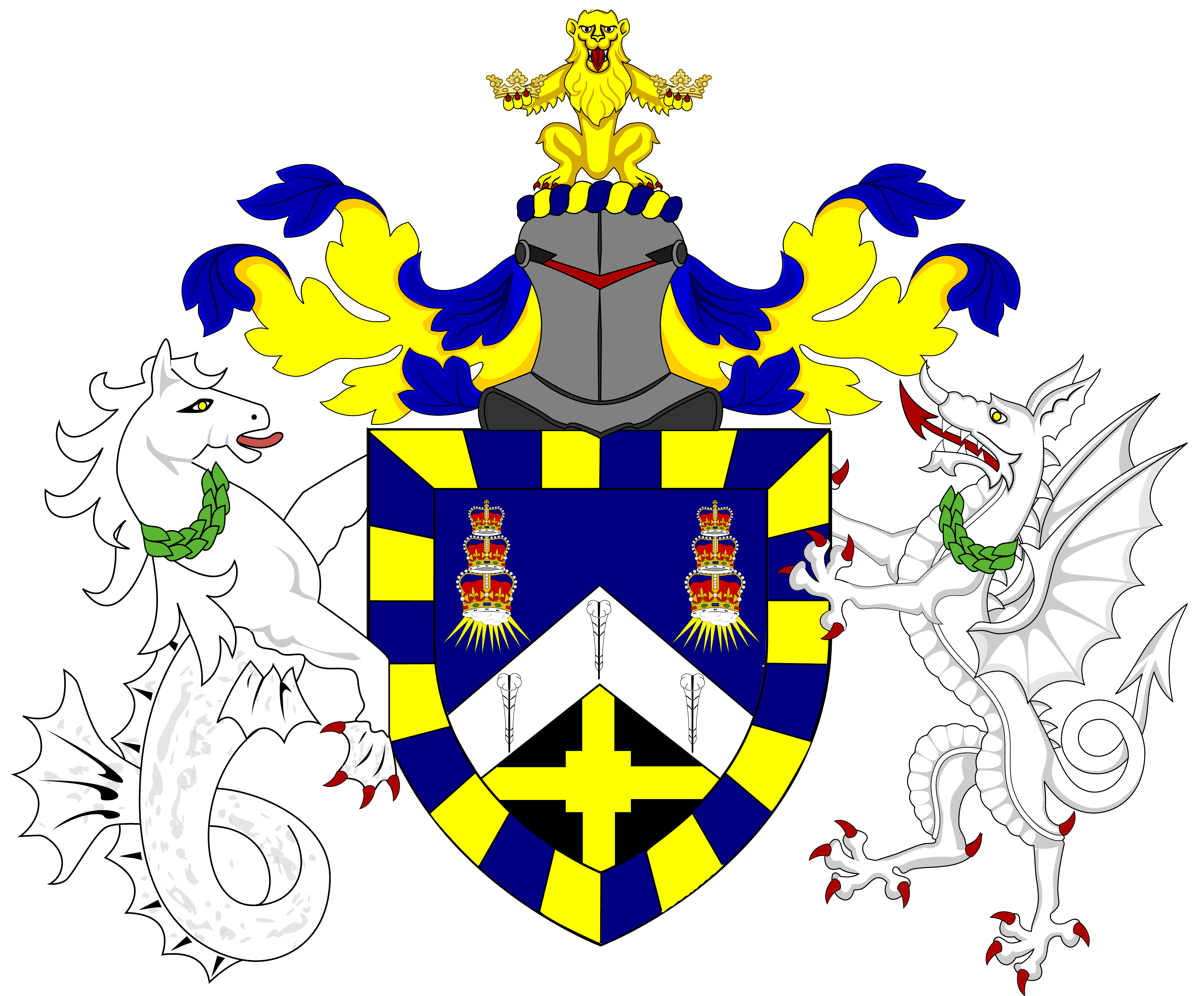}
\end{center}
\end{titlepage} 
\abstract{
Planar maximally supersymmetric Yang-Mills theory ($\N=4$ SYM) is a special quantum field theory. A few of its remarkable features are conformal symmetry at the quantum level, evidence of integrability and, moreover, it is a prime example of the AdS/CFT duality. Triggered by Witten's twistor string theory \cite{Witten:GaugeAsStringInTwistor2003}, the past 15 years have witnessed enormous progress in reformulating this theory to make as many of these special features manifest, from the choice of convenient variables to recursion relations that allowed new mathematical structures to appear, like the Grassmannian \cite{ArkaniHamed:2009dn}. These methods are collectively referred to as on-shell methods. The ultimate hope is that, by understanding $\N=4$ SYM in depth, one can learn about other, more realistic quantum field theories. The overarching theme of this thesis is the investigation of how on-shell methods can aid the computation of quantities other than scattering amplitudes. In this spirit we study form factors and correlation functions, said to be partially and completely off-shell quantities, respectively. More explicitly, we compute form factors of half-BPS operators up to two loops, and study the dilatation operator in the $SO(6)$ and $SU(2|3)$ sectors using techniques originally designed for amplitudes. A second part of the work is dedicated to the study of scattering amplitudes beyond the planar limit, an area of research which is still in its infancy, and not much is known about which special features of the planar theory survive in the non-planar regime. In this context, we generalise some aspects of the on-shell diagram formulation of Arkani-Hamed et al. \cite{ArkaniHamed:2012nw} to take into account non-planar corrections.
} 
\vspace{0.5cm}

\noindent This thesis is based on \cite{Penante:2014sza,Brandhuber:2014ica,Brandhuber:2014pta,Brandhuber:2015boa,Franco:2015rma} and has considerable overlap with these papers.
\newpage


\tableofcontents

\chapter[Introduction]{Introduction}

A generic quantum field theory is completely specified by the knowledge of all its \emph{correlation functions}, the key objects that encode how excitations propagate in spacetime. Correlation functions of local gauge invariant operators $\O_i$ are defined as the following vacuum expectation values\footnote{Time ordering is implicit.},
\begin{align}
\begin{split}
C_{\O_{1},\dots,\O_{n}}(x_1,\dots,x_n)\,&\equiv\,\b{\O_1(x_1)\cdots\O_n(x_n)}\ .
\end{split}
\end{align}
In theories with a Lagrangian description in terms of fundamental fields $\Psi$, the correlators can be written inside the path integral as
\begin{align}
\label{eq:correlator-definition}
C_{\O_{1},\dots,\O_{n}}(x_1,\dots,x_n)
\,&=\,\int\!\! d[\Psi] \O_{1}(x_1)\cdots\O_n(x_n)e^{-S_{\text{Euc}}[\Psi]}\ ,
\end{align}
where $d[\Psi]$ corresponds to the integration over all possible field configurations and $S_{\rm Euc}$ is the Euclidean action (obtained by Wick rotation $t\mapsto -it$),
\begin{align}
S_{\rm Euc}[\Psi]\,\equiv\, \int\!\! d^4 x \, \mathcal{L}[\Psi(x)] \ .
\end{align}
The exact functional form for all correlators is in general not known and is available only for very simple models. If a theory is weakly coupled, it is possible to decompose $\mathcal{L}$ into a free piece $\mathcal{L}_{\rm free}$ and an interaction term $\mathcal{L}_{\rm int}$ which comes multiplied by a small parameter $g$,
\begin{equation}
\mathcal{L}\,=\, \mathcal{L}_{\rm free}+g\mathcal{L}_{\rm int}\,,\quad g\ll 1\ .
\end{equation}
In this situation one can expand the exponential in \eqref{eq:correlator-definition} and thus all correlation functions \eqref{eq:correlator-definition} become a series in $g$.

From the correlation functions it is possible to extract observable quantities that relate theory predictions to measurable cross sections. This is done through the Lehmann-Symanzik-Zimmermann (LSZ) reduction prescription \cite{Lehmann:1954rq}; it amounts to Fourier transforming the correlator of fundamental fields to momentum space and requiring that the fields are momentum eigenstates, i.e. plane waves. This procedure leads to \emph{scattering amplitudes}, which are then used to calculate cross sections of physical processes. More precisely, the cross sections are obtained from amplitudes by taking its modulus squared, integrating over the phase space of the outgoing particles and performing an average over the quantum numbers of the initial particles and a sum over the quantum numbers of the outgoing particles.


In momentum space all momenta entering the scattering amplitudes must satisfy the on-shell condition $p_i^2=m_i^2$ whereas for the correlators they are unconstrained. For this reason, correlation functions are said to be \emph{off-shell} quantities whereas scattering amplitudes are said to be \emph{on-shell}.

Scattering amplitudes are formally defined as the overlap between an incoming state of $n_i$ particles and an outgoing state of $n_f$ particles. They are the elements of the \emph{$S$-matrix},
\begin{equation}
S_{if}\,\equiv\,\b{1,\dots,n_f|1,\dots,n_i}\ .
\end{equation}
Here $|1,\dots,n\rangle$ stands for an $n$-particle momentum eigenstate and similarly for $\langle 1,\dots,n|$. The incoming and outgoing states are free and the elements of the $S$-matrix account for the interactions at finite time.

Interpolating between these completely on-shell and off-shell quantities lie \emph{form factors}, defined as the expectation value of a gauge invariant local operator $\O(x)$ computed between the vacuum and an $n$-particle on-shell state $\langle 1, \ldots , n |$. Conventionally in the definition of a form factor the spacetime dependence of the operator is also Fourier transformed to momentum space,
\begin{align}
\label{eq:Fourier-FF}
\begin{split}
&\int d^4x\, e^{-iq\cdot x}\la 1\ldots n|\mathcal{O}(x)|0\rangle\,=\,\int d^4x\, e^{-iq \cdot x}\la 1\ldots n|e^{iP\cdot x}\mathcal{O}(0)e^{-iP\cdot x}|0\rangle\\
=&\,\int d^4x\, e^{-i(q-\sum_{i=1}^n p_i)\cdot x}\la 1\ldots n|\mathcal{O}(0)|0\rangle\,=\,\delta^{(4)}\Big(q-\sum_{i=1}^n p_i\Big)\la 1\ldots n|\mathcal{O}(0)|0\rangle\ ,
\end{split}
\end{align}
where we used that $\la 1\ldots n|$ is an eigenstate of the momentum operator $P$ with eigenvalue $\sum_i p_i$ and that the vacuum is translation invariant. The overall delta-function in \eqref{eq:Fourier-FF} is a consequence of translation symmetry. Thus, the quantitiy to consider is
\begin{equation}
\label{eq:FF-definition}
F_\O(1, \ldots, n;q) \, \equiv \,   \langle 1, \ldots , n |\cO (0)  |0\rangle\ ,
\end{equation}
which is a function of a set of on-shell momenta $p_1^2=\dots=p_n^2=0$ as well as one off-shell momentum $q^2\neq 0$ associated with the operator.

\begin{figure}[htb]
\centering
\includegraphics[width=0.41\textwidth]{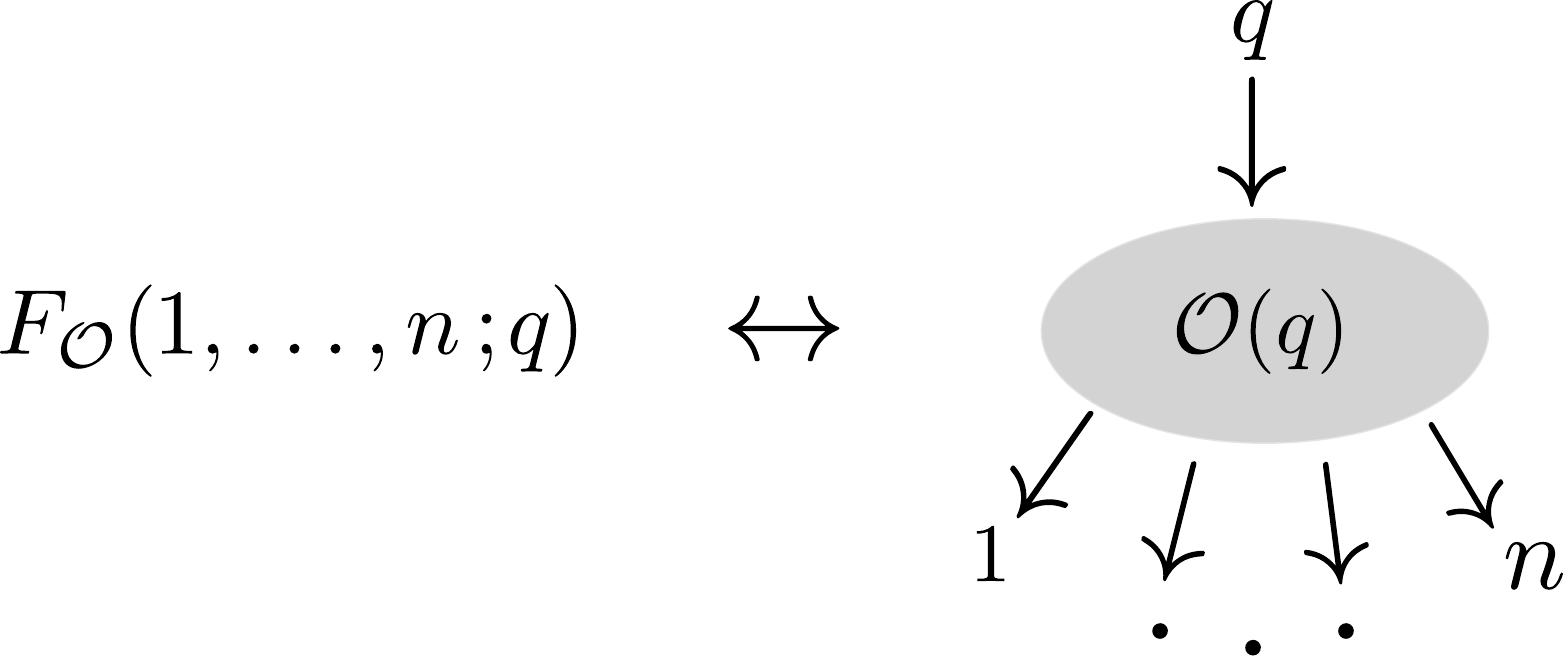}
\caption{\it A form factor models an effective interaction producing a specified $|\text{out}\ra$ state starting from the vacuum.}
\label{fig:FF_figure}
\end{figure}

Form factors can be used to model interactions where the detailed physical process is not fully known and $\O$ stands for an effective interaction, as the one shown in \fref{fig:FF_figure}.
\\

\begin{figure}[h]
\centering
\includegraphics[width=0.8\linewidth]{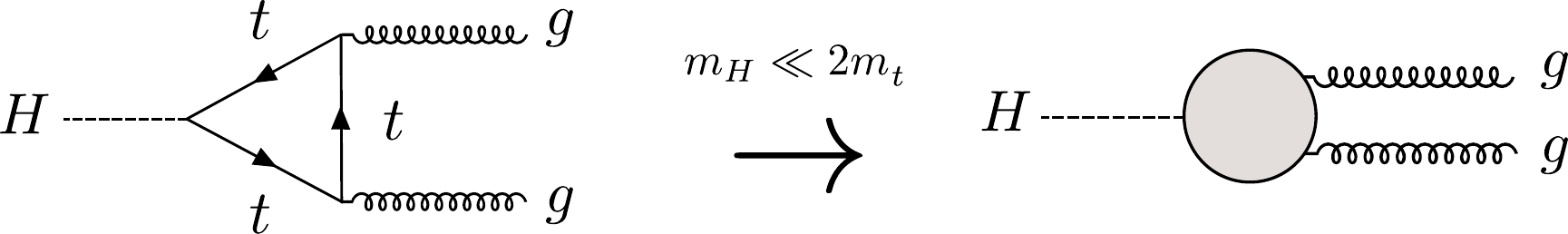}
\caption{\it A Higgs particle decaying into two gluons. In the limit where  $m_H \ll 2 m_t$ this process is approximated by a form factor.}
\label{fig:higgs-gluon}
\end{figure}

Form factors appear in various contexts, an interesting one is the decay of a Higgs boson into gluons. This process is mediated by a fermion loop, and the leading contribution is from a top quark running the loop. In the limit where $m_H \ll 2 m_t$\footnote{Using the values $m_H\sim 126\,\text{GeV}$ and $m_t\sim 173\,\text{GeV}$ the ratio $m_H/2m_t \sim 0.36$.} the mass of the top can be sent to infinity, giving rise to an effective vertex $H \,\Tr \, F_{\rm SD}^2$, where $F_{\rm SD}$ is the self-dual part of the field strength \cite{Wilczek:1977zn, Shifman:1978zn} (see also \cite{Dixon:2004za} for a recent discussion). This is shown in \fref{fig:higgs-gluon} underneath, notice that the Higgs particle can be produced as an intermediate state and thus does not need to satisfy $p_H^2\,=\,m_H^2$.

The quantum field theory we will consider is maximally supersymmetric ($\N=4$) Yang-Mills (SYM) with gauge group $SU(N)$ \cite{Brink:1976bc}, which can been obtained by dimensional reduction of ten-dimensional $\N=1$ SYM down to $d=4$. This theory has been extensively studied in the past decades and displays very special properties like quantum conformality \cite{Green:1982sw}, integrability in the planar limit (also called large $N$ or 't Hooft limit \cite{'tHooft:1973jz}) and it is the most well understood example of the AdS/CFT correspondence \cite{Maldacena:1997re}, under which it is dual to type IIB string theory. Due to these properties, this theory is commonly used to develop new ideas and mathematical techniques that can in principle revolutionise the current understanding of quantum field theory and gravity. These achievements were triggered by a duality between amplitudes in $\N=4$ SYM and an instanton expansion in a particular twistor string theory, found by Witten in  \cite{Witten:GaugeAsStringInTwistor2003}.

Scattering amplitudes at weak coupling are given as a perturbative expansion around a free theory, and to each order in perturbation theory there is a set of Feynman diagrams that formally encode the mathematical expressions that sum to the amplitude. Each diagram looks like a sequence of local interactions in spacetime, where physical particles exchange virtual particles and sometimes particles with unfixed momentum can run in loops. Feynman diagrams are therefore easy to picture, and they make each interaction manifestly local and unitary. There are, however, many drawbacks also in the package~---~manifest locality and unitarity come at the expense of a large amount of off-shell information associated to virtual particles and gauge redundancies. All these unnecessary ingredients obscure an underlying simplicity of the amplitudes that manifests itself as a high degree of cancellations, at least for $\N=4$ SYM. The prime examples are the so-called Parke-Taylor amplitudes \cite{ParkeTaylor:1986}: consider amplitudes with $n$ outgoing gluons, $k$ of which of helicity $-1$ and $n-k$ of helicity $+1$. For $k=0, 1$ these amplitudes are zero, and for $k=2$ they are given by a single-term expression. In terms of Feynman diagrams one may have to, for instance, show a cancellation between 10 million terms for 10 gluons!
\begin{figure}[h]
\centering
\includegraphics[width=\linewidth]{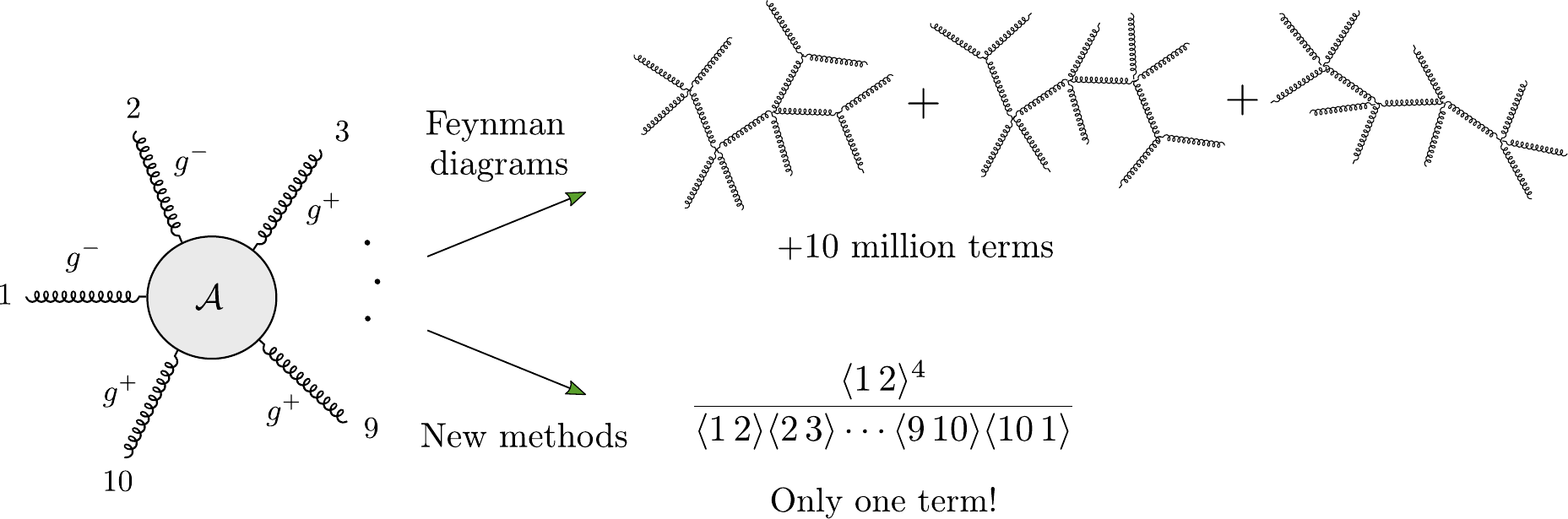}
\caption{\it The simplicity of the tree-level scattering of 10 gluons is not manifest from the Feynman diagram perspective.}
\label{fig:old-new-methods}
\end{figure}

The progress made in the last two decades is much related to reformulating $\N=4$ SYM in a different way in order to expose the underlying structures responsible for the simplicity of the final amplitudes. In this way, one can say that the study of mathematical properties of scattering amplitudes has become an area of research in its own right, and a very active one indeed. Furthermore, it does not concern only $\N=4$ SYM; much has also been learned about theories with fewer supersymmetries, gravity, and theories in dimensions different than four.

For massless theories such as $\N=4$ SYM, there is a variety of methods that simplify the calculation of on-shell quantities enormously, both at tree and loop level. These techniques are collectively referred to as \emph{on-shell methods}, some of which are reviewed in the following chapter (for a comprehensive review we indicate \cite{Elvang:2013cua} and its rich bibliography). 

Although the simplicity of $\N=4$ SYM is remarkably seen by studying its scattering amplitudes, it does not stop there and also features in the study of off-shell quantities, and even in other theories. For instance, the anomalous magnetic moment of the electron in Quantum Electrodynamics (QED) is given by the form factor $F_{J^\mu_{\rm EM}}(e^+,e^-;q)$ where $J^\mu_{\rm EM}$ is the electromagnetic current. This form factor was computed at three loops in \cite{Cvitanovic:1974um,Laporta:1996mq} and, while there were $\sim 70$ Feynman diagrams to be summed, each of which with a value which oscillated between $\pm 10$ and 100, they combined to a result of $O(1)$ (times $(\alpha/2\pi)^3$, where $\alpha\sim 1/137$ is the fine structure constant). Cvitanovic later found that if one first organises the terms in gauge invariant subsets then each combination has a value of $O(1)$. These enormous cancellations suggest that a better approach is in order. Another example of simplicties of off-shell quantities are the supersymmetric form factors computed in \cite{Brandhuber:2011tv}; their expressions closely resemble that of the Parke-Taylor amplitudes mentioned earlier.

The main theme of this thesis is the study of on-shell methods in $\N=4$ SYM. Inspired by the simplicities mentioned above, in Chapters \ref{ch:formfactors} and \ref{ch:dilatation} we investigate how on-shell methods can be used to unravel simple structures for off-shell quantities. The second interesting question we investigate in Chapter \ref{ch:onshelldiagrams} is how to move beyond the well understood planar limit of $\N=4$ SYM.


The on-shell methods that will be used throughout this thesis, as well as some other useful concepts, will be reviewed in Chapter \ref{ch:Review}. In Chapter \ref{ch:formfactors} we apply the above methods to compute sypersymmetric form factors of a particular kind of operators, called half-BPS operators, up to two loops and find once again a remarkable simplicity in the results.


In Chapter \ref{ch:dilatation} we move on to the study of the one-loop dilatation operator in $\N=4$ SYM. This operator accounts for the renormalisation of the scaling dimension of composite operators (the quantum corrections are called \emph{anomalous dimensions}).
The study of $\N=4$ SYM has led to the discovery of integrability in the planar limit,  providing the tools to compute the anomalous dimensions of local operators for any value of the coupling.
It is  widely expected that the integrability of the planar anomalous dimension problem and  the hidden structures and symmetries of scattering amplitudes  are related in some interesting way. For this reason, we investigate the application of on-shell methods to the dilatation operator. Similar ideas were also applied in \cite{Koster:2014fva,Wilhelm:2014qua,Nandan:2014oga} and the the interplay between the integrability of the spectral problem and scattering amplitudes has started to be established in the opposite direction too, see for instance the spectral parameter deformation introduced in \cite{Ferro:2012xw,Ferro:2013dga}.

For a single trace local composite operator $\O_L(x)\,=\,\Tr(\Psi_1\Psi_2\cdots\Psi_L)(x)$ where $\Psi_{i},\, i\,=\,1,\dots, L$ are fundamental fields, one way to compute its anomalous dimension is by studying the following $(L+1)$-point correlation function,
\begin{align}
\label{eq:correlator-ff}
\mathcal{C}^{\text{1-loop}}_{\O_L, \Psi_1, \dots, \Psi_L}(x,x_1,x_2,\dots,x_L)\,=\,\b{\Tr(\Psi_1\Psi_2\cdots\Psi_L)(x)\Psi_1(x_1)\Psi_2(x_2)\cdots\Psi_L(x_L)}^{\text{one-loop}}\ .
\end{align}
Since every $\Psi_i$ is a fundamental field, the Fourier transform of \eqref{eq:correlator-ff} is a form factor.
The complete one-loop dilatation operator is known \cite{Beisert:2003jj,Beisert:2003yb} and was reproduced from a form factor perspective in \cite{Wilhelm:2014qua}\footnote{We also indicate \cite{Wilhelm:2016izi} for many applications.}. The approach taken here is different, we consider instead the two-point function of an operator $\O(x)$ with its conjugate $\bar{\O}(y)$ at one loop,
\begin{align}
\mathcal{C}_{\O,\bar{\O}}^{\text{1-loop}}(x,y)\,=\,\b{\O(x)\bar{\O}(y)}^{\text{one-loop}}\ .
\end{align}
Working also in momentum space allows us to use two different methods originally designed for scattering amplitudes~---~MHV rules \cite{Cachazo:2004kj} and generalised unitarity \cite{Bern:1994zx,Bern:1994cg,Bern:1997sc,Britto:2004nc}~---~for the computation of the dilatation operator in two sectors (called $SO(6)$ and $SU(2|3)$, as will be reviewed in the corresponding chapter). As we will see, the calculation becomes very transparent and simple, involving only one single-scale integral.

$\N=4$ SYM with gauge group $SU(N)$ has been extensively studied in the large $N$ limit. The idea of a planar limit was introduced by 't Hooft in the 70's and relies in exchanging the expansion in the Yang-Mills coupling constant $g_{\rm YM}$ for $1/N$ and $\lambda= N g_{\rm YM}^2$ \cite{'tHooft:1973jz}. The latter is called the 't Hooft coupling and is held fixed (and small) as $g_{\rm YM}\rightarrow 0$ and $N\rightarrow\infty$. In this formulation, scattering amplitudes which are of leading order in $1/N$ can be drawn on a plane whereas corrections can only be drawn on surfaces of higher genus, a property which naturally fits with the genus expansion in the dual string theory picture.

Planar $\N=4$ SYM is, however, not the full theory and it is important to investigate quantities which are subleading in $1/N$ and in particular which features of the planar theory survive in the non-planar corrections. In this spirit, Chapter \ref{ch:onshelldiagrams} is dedicated to the generalisation of an on-shell formulation of planar scattering amplitudes in $\N=4$ SYM, introduced by Arkani-Hamed et. al. in \cite{ArkaniHamed:2012nw}, beyond the planar limit. In this formulation, all off-shell information commonly associated to virtual particles are encoded in internal variables that parametrise an auxiliary space~---~the \emph{Grassmannian} $Gr_{k,n}$, which is the space of $k$-dimensional planes in $\mathbb{C}^n$. For the sake of clarity we postpone a brief review of this method to Chapter \ref{ch:onshelldiagrams}, followed by a generalisation of the formulation away from the planar limit.

Finally, Chapter \ref{ch:conclusions} contains concluding remarks of the work presented throughout the thesis and some future research directions.

\label{ch:Introduction}

\chapter{Review}
\label{ch:Review}
\section{On-shell methods for scattering amplitudes}

The purpose of this section is to give an introduction to the first two manipulations one performs on scattering amplitudes to expose some of the simplicities mentioned in Chapter \ref{ch:Introduction}: \emph{colour decomposition} and the \emph{spinor-helicity formalism}. Part of it will be based on \cite{Review}.

\subsection{Colour decomposition} 
\label{sec:colordecomp}

In general, scattering amplitudes in gauge theory are functions of the momenta, wavefunctions and colour charges of the external states, as well as the coupling constant(s). As a first simplification, it is useful to separate the dependence on the colour charges from the kinematics.

\begin{figure}[h]
\centering
\includegraphics[width=0.8\linewidth]{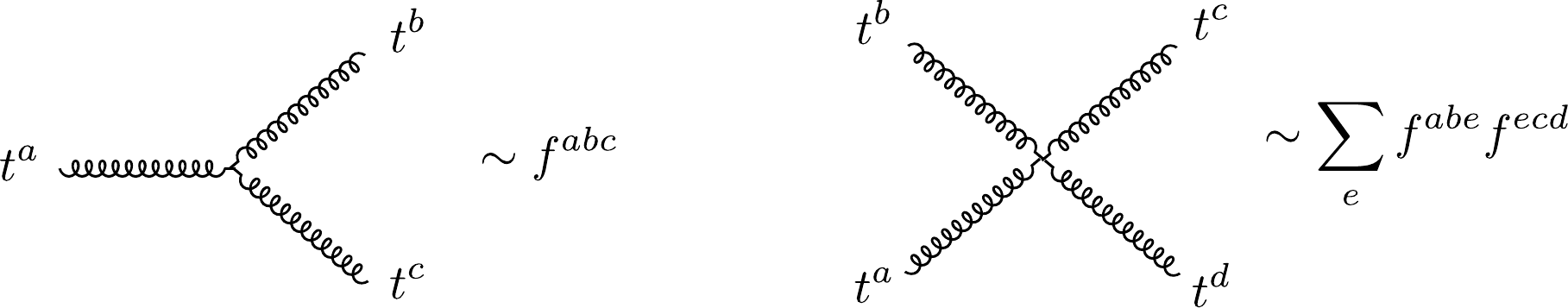}
\caption{\it Colour structure corresponding to the three- and four-gluon interaction vertices.}
\label{fig:gluon-vertices-colour}
\end{figure}

\begin{figure}[h]
\centering
\includegraphics[width=0.8\linewidth]{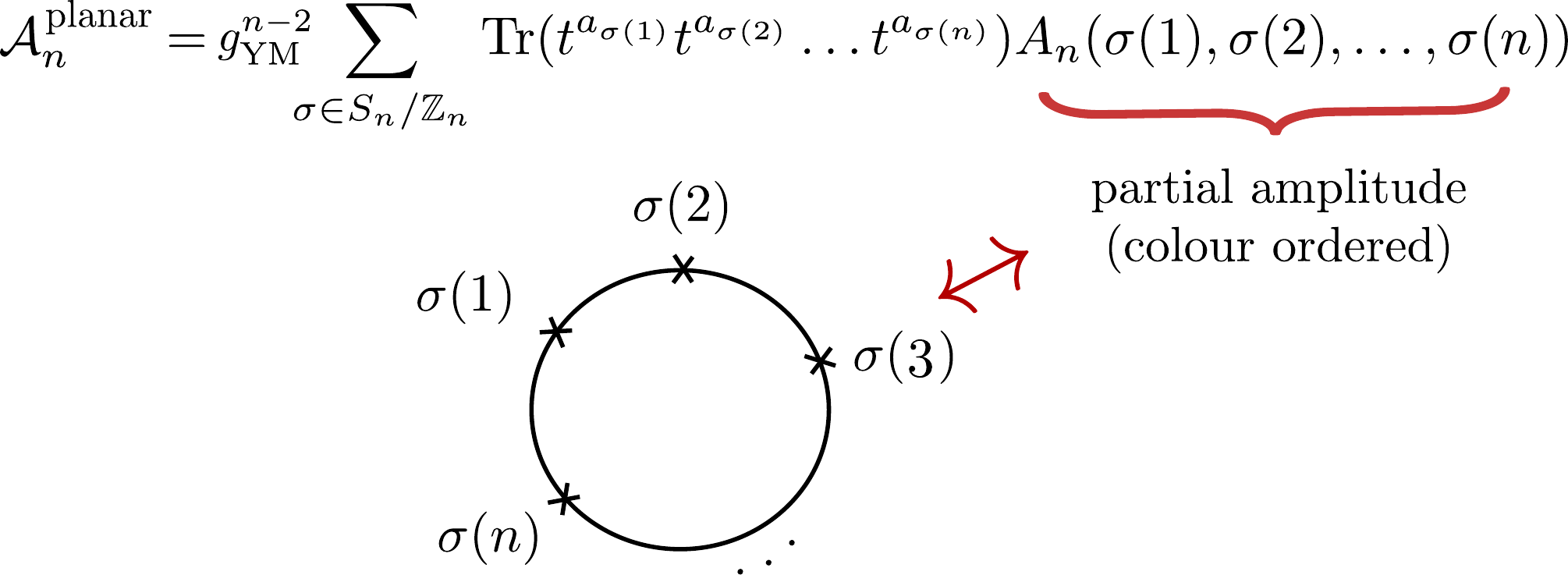}
\caption{\textit{Single-trace amplitudes can be drawn on a disk and as such are referred to as \emph{planar}.}}
\label{fig:amp-disk}
\end{figure}

The dependence on the gauge group appears in the interaction terms in the Lagrangian in terms of structure constants of the colour algebra. The colour structure of the three- and four-gluon vertices are shown in \fref{fig:gluon-vertices-colour}.
\\
In order to absorb factors of $2$ it is useful to redefine the generators of the fundamental representation of $SU(N)$, $t^a$ with $a=1,\dots,N^2-1$, such that
\begin{align}
\label{eq:normalisation-t}
\Tr(t^a t^b)\,=\,\delta^{ab}\qquad \Rightarrow \qquad t^a \rightarrow \sqrt{2}\, t^a\ .
\end{align}
The structure constants must also be redefined as $f^{abc}\rightarrow \sqrt{2}\, f^{abc}$ so that the commutation relation of the Lie algebra,
\begin{equation}
\label{eq:lie-algebra}
[t^a,t^b]\,=\sum_{c=1}^{N^2-1} f^{abc}t^c\ ,
\end{equation}
remains valid. Using \eqref{eq:normalisation-t} and \eqref{eq:lie-algebra}, the structure constants can be written in terms of $t^a$ as
\begin{equation}
\label{eq:struc-trace}
f^{abc}=\text{Tr}(t^a[t^b,t^c])\ .
\end{equation}
Doing so, after representing each structure constant as in \eqref{eq:struc-trace} one may use use the completeness relation 
\begin{equation}
\sum_{a=1}^{N^2-1}(t^a)_{i}^{\phantom{i}j}(t^a)_{k}^{\phantom{k}l}=\delta_{i}^{\phantom{i}l}\delta_{k}^{\phantom{k}j}-\frac{1}{N}\delta_{i}^{\phantom{i}j}\delta_{k}^{\phantom{k}l}
\end{equation}
to merge traces. This can be done easily for large $N$\footnote{In this limit there is no distinction between $SU(N)$ and $U(N)$ and the extra $U(1)$ gives rise to the \emph{$U(1)$ decoupling identities}.}, where the $1/N$ term above drops out and all amplitudes are proportional to a single trace over the $n$ generators associated to each of the $n$ particles,
\begin{align} 
\label{eq:colordecomp}
\mathcal{A}^{\rm planar}_n=g_{\text{YM}}^{n-2}\sum_{\sigma\in S_n/\mathbb{Z}_n}\mathrm{Tr}(t^{a_{\sigma(1)}}t^{a_{\sigma(2)}}\dots t^{a_{\sigma(n)}})A_n(\sigma(1),\sigma(2),\dots, \sigma(n))\ ,
\end{align}
where $g_{\rm YM}$ is the coupling constant. The object $A_n$ on the right-hand side is called  \textit{partial amplitude}; it is a function of the kinematics only and the order of its arguments (particle labels) follows that of the generators in the trace that multiplies it. Due to this natural ordering, it is possible to draw planar partial amplitudes on a disk, as shown in Figure \ref{fig:amp-disk}.
Accounting for $1/N$ corrections amounts to considering multiple traces in \eqref{eq:colordecomp}, thus the non-planar partial amplitudes can be drawn on surfaces with more than one boundary. Those will be further explored in Chapter \ref{ch:onshelldiagrams}. In the next sections, however, we will be strictly considering planar amplitudes (and form factors\footnote{The discussion of planarity in the context of form factors is a bit more subtle, see Chapter \ref{ch:formfactors}, in particular \sref{sec:colour-planar}.}).

\subsection{Spinor-helicity formalism}
\label{sec:SH_Formalism}

After colour decomposition, the next step is to write the partial amplitudes in convenient variables. Here convenient means ``making as many symmetries manifest as one can''. This is the subject of this subsection and it goes by the name of \emph{spinor-helicity formalism}.

The aim here is to make manifest the on-shell condition $p_i^2=0$\footnote{It is also possible to make manifest momentum conservation condition $\sum_{i=1}^n p_i=0$ by means of \emph{momentum twistors}, however this will not be relevant for the work presented here.}. For massless particles, this is easily achieved with the observation that the contraction between the momentum four-vector $p_i^\mu$ with the Pauli matrices $\sigma^\mu$ gives a matrix of less than maximal rank,
\begin{equation}
\label{eq:p-matrix}
 p_{\alpha\dot{\alpha}}\,=\,p_\mu\sigma^\mu_{\alpha\dot{\alpha}}\,=\, \begin{pmatrix}
p^0+p^3 & p^1-ip^2\\
p^1+ip^2 & p^0-p^3
\end{pmatrix}_{\alpha\dot{\alpha}}\,, \quad \det(p)\,=\,(p^0)^2-(p^1)^2-(p^2)^2-(p^3)^2\,=\,0\ ,
\end{equation}
where the four Pauli matrices are
\begin{equation}
\sigma^0=\begin{pmatrix}
1 & 0 \\
0 & 1
\end{pmatrix},\quad \sigma^1=\begin{pmatrix}
0 & 1 \\
1 & 0
\end{pmatrix},\quad \sigma^2=\begin{pmatrix}
0 & \matminus i \\
i & 0
\end{pmatrix},\quad \sigma^3=\begin{pmatrix}
1 & 0 \\
0 & \matminus 1
\end{pmatrix}.
\end{equation}
The vanishing determinant on \eqref{eq:p-matrix} implies that the matrix $p$ factorises as a product of two spinors of opposite chirality,
\begin{equation}
\label{eq:P-spinor}
 p_{\alpha\dot{\alpha}}=\lambda_{\alpha} \widetilde{\lambda}_{\dot{\alpha}}\ .
\end{equation}
The Lorentz group acts on the spinors as $SU(2)_L\times SU(2)_R$. The indices $\alpha,\,\dot{\alpha}=1,2$ transform in the fundamental representation of one of the $SU(2)_{L,R}$, respectively, and are a singlet under the other $SU(2)$. Thus the irreducible representations of the Lorentz group can be characterised by a pair of integers or half-integers $(p,q)$  \cite{Witten:GaugeAsStringInTwistor2003}. The spinors $\lambda,\,,\widetilde{\lambda}$ and the four-vector $p$ are in the representations shown in Table \ref{tab:Representations} below.
\begin{table}[htb]
\centering
\begin{tabular}{cl}
$(1/2,0)$ &  \quad Weyl spinor of negative chirality $\lambda_\alpha$,\\
$(0,1/2)$ & \quad Weyl spinor of positive chirality $\widetilde{\lambda}_{\dot{\alpha}}$,\\
$(1/2,1/2)$ & \quad Four-vector $p_{\alpha\dot{\alpha}}$,
\end{tabular}
\caption{\textit{Representations of the Lorentz group decomposed into representations of $SU(2)_L\times SU(2)_R$.}}
\label{tab:Representations}
\end{table}\\
A massless four-vector has only three independent components. In terms of the spinors, this is a consequence of the following rescaling redundancy,
\begin{equation}
\label{eq:little-group}
 (\lambda,\widetilde\lambda)\,\mapsto\,(t\lambda,t^{-1}\widetilde{\lambda}),\qquad t\neq 0\ .
\end{equation}
This rescaling has a physical meaning, it corresponds to the action of the subset of Lorentz transformations which leave the momentum unchanged, or what is called the \emph{little group}.
As we will see, when doing such a rescaling the amplitude picks up a phase that depends on the helicity $h_i$ of the corresponding particle with momentum $p_i$,
\begin{equation}
\label{eq:little-group-amp}
 (\lambda^i,\widetilde\lambda^i)\rightarrow(t\lambda^i,t^{-1}\widetilde{\lambda}^i) \quad \Rightarrow \quad \mathcal{A}\rightarrow t^{-2h_i}\mathcal{A}\ .
\end{equation}
Lorentz invariant quantities are constructed contracting the spinors with the $SU(2)$ invariant tensors $\epsilon_{\alpha\beta}$ and $\epsilon_{\dot{\alpha}\dot{\beta}}$,
\begin{equation}
\epsilon_{\alpha\beta}=\epsilon_{\dot{\alpha}\dot{\beta}}= \begin{pmatrix}
0 & 1 \\ \matminus 1 & 0
\end{pmatrix}\ .
\end{equation}
This gives rise to the following spinor products\footnote{All spinor conventions used throughout this thesis are presented in Appendix \ref{app:spinor-conventions}.}
\begin{align}
\label{eq:spinor-brackets}
\begin{split}
 \braket{ij}&\,\equiv\,\braket{\lambda^{i}\lambda^{j}}\,=\, \epsilon_{\alpha\beta}\,\lambda^{i\alpha}\lambda^{j \beta}, \qquad \braket{ij}\,=\,-\braket{ji},\\
 [ij]&\,\equiv\,[\widetilde{\lambda}^{i}\,\widetilde{\lambda}^{j}]\,=\,\epsilon^{\dot{\alpha}\dot{\beta}}\, \widetilde{\lambda}^{i}_{\dot{\alpha}}\tl^j_{\dot{\beta}}, \qquad \;\;\; [ij]\,=\,-\,[ji]\ .
 \end{split}
\end{align}
The tensors $\epsilon_{\alpha\beta},\,\epsilon_{\dot{\alpha}\dot{\beta}}$ and their inverses $\epsilon^{\alpha\beta},\,\epsilon^{\dot{\alpha}\dot{\beta}}$ are also used to raise and lower the $\alpha,\,\dot\alpha$ indices.

The scalar product of two momenta $p_i=\lambda^i\widetilde{\lambda}^i$ and $p_j=\lambda^j\widetilde{\lambda}^j$ in this language is simply given by the products of angular and square brackets,
\begin{align}
2\,(p_i\cdot p_j)\,=\,\braket{ij}[ji]\ .
\end{align}
In a massless theory this is equivalent to the Mandelstam variable $s_{ij}=(p_i+p_j)^2$.
Notice that $\braket{ij}=0$ if $\lambda^i \propto \lambda^j $ and equivalently for the square brackets. Physically, vanishing of either bracket means that the two momenta $p_{i}$ and $p_{j}$ are collinear.

In the amplitudes literature, often the momenta are taken to be complex, and thus the Lorentz group is $SL(2,\mathbb{C})\times SL(2,\mathbb{C})$. In this case $\lambda$ and $\widetilde{\lambda}$ are independent spinors with complex components. The requirement that the momenta are real imposes constraints or relations between the $\lambda$'s and $\widetilde{\lambda}$'s. The relations depend on the signature of space-time and are listed below:
\begin{table}[h]
\centering
\begin{tabular}{ll}
$(+ + - -)$ &  $\lambda$ and $\widetilde{\lambda}$ are real and independent,\\
$(+ - - -)$ & $\lambda$ and $\widetilde{\lambda}$ are complex and $\widetilde{\lambda}=\pm\bar{\lambda}.$
\end{tabular}
\end{table}\\
It is also important to mention that the spinors $\lambda$, $\widetilde\lambda$ satisfy the \emph{Schouten identities}:
\begin{align}
\label{eq:schouten}
\begin{split}
 \braket{ij}\braket{kl}\,+\braket{ik}\braket{lj}+\braket{il}\braket{jk} &=0\ , \\
 [ij]\;[kl]\;+\;[ik]\;[lj]\;+\;[il]\;[jk]&=0\ ,
\end{split}
\end{align}
which are very important for simplifying computations.

The last elements present in the partial amplitude which still remain to be written in terms of the spinor variables are the polarisation vectors/spinors. For a given helicity, they can be read off the plane wave solutions of the equations of motion of the free theory. Here it is useful to investigate fermions and gauge bosons separately.

\subsubsection*{Fermions -- helicity $ \boldsymbol{h}\mathbf{=\pm 1/2}$}

The Dirac equation in the massless case decouples into two Weyl equations, so the four-component Dirac spinor can be written as a direct sum of two two-component Weyl spinors of opposite chirality which satisfy 
\begin{center}
\begin{tabular}{ll}
$i\sigma^\mu_{\alpha\dot\alpha}p_\mu\psi^{\dot\alpha}\,=\,ip_{\alpha\dot\alpha}\psi^{\dot\alpha}\,=\,0$ & $\  \  \ $ (positive chirality),\\
$i\overline{\sigma}^\mu_{\alpha\dot\alpha}p_\mu\psi^{\alpha}\,=\,i\overline p_{\alpha\dot\alpha}\bar{\psi}^{\alpha}\,=\,0$ & $ \  \  \ $ (negative chirality),
\end{tabular}
\end{center}
where $ \sigma^\mu=(\uno,\vec{\sigma})$ and $ \overline{\sigma}^\mu=(\uno,-\vec{\sigma}) $.
The plane wave solutions are
\begin{center}
\begin{tabular}{ll}
$\psi^{\dot\alpha}\,=\,c\widetilde \lambda^{\dot\alpha}e^{ix_{\alpha\dot\alpha}\lambda^{\alpha}\widetilde{\lambda}^{\dot\alpha}}$ & $ \  \  \ $ (positive chirality),\\
$\bar{\psi}^{\alpha}\,=\,c'\lambda^{\alpha}e^{ix_{\alpha\dot\alpha}\lambda^{\alpha}\widetilde{\lambda}^{\dot\alpha}}$ & $\  \  \ $ (negative chirality),
\end{tabular}
\end{center}
where $c,\,c'$ are non-zero constants. Thus, the polarisation spinors are just
\begin{equation}
\label{eq:polarisation-fermions}
\epsilon^{(-1/2)}_\alpha\,=\,\lambda_\alpha\, , \qquad \epsilon^{(+1/2)}_{\dot\alpha}\,=\,\tl_{\dot\alpha}\ .
\end{equation}
\subsubsection*{Gauge bosons -- helicity $ \boldsymbol{h}\mathbf{=\pm 1}$}
For helicities $\pm 1$, the corresponding polarisation vectors $(\epsilon^{(\pm)})^{\mu}$ satisfy $p_\mu(\epsilon^{(\pm)})^{\mu}=0$.
Thus, they can be written, up to a gauge transformation, as 
\begin{align}
\label{eq:polarisation-gluons}
\epsilon^{(-)}_{\alpha\dot{\alpha}}\,=\,\frac{\lambda_\alpha\widetilde{\mu}_{\dot{\alpha}}}{[\widetilde{\lambda}\widetilde{\mu}]}\, ,\qquad
 \epsilon^{(+)}_{\alpha\dot{\alpha}}\,=\,\frac{\mu_\alpha\widetilde{\lambda}_{\dot{\alpha}}}{\langle\lambda\mu\rangle}\ . 
\end{align}
where $\mu$ and $\widetilde{\mu}$ are reference spinors that are linearly independent of $\lambda$ and $\widetilde{\lambda}$. Note that a redefinition of the reference spinors
\begin{align}
\widetilde{\mu}\rightarrow a\, \widetilde{\mu} + b\, \tl\,, \qquad\qquad \mu \rightarrow c \,\mu + d\, \lambda\ ,
\end{align} would only change $\epsilon^{(+)}$ and $\epsilon^{(-)}$ by a shift proportional to $p_{\alpha\dot{\alpha}}$, thus the independence of $\epsilon^{(-)}$ and $\epsilon^{(+)}$ under rescaling of the reference spinors ensures the independence of the choice of $\mu$ and $\widetilde{\mu}$ up to a gauge transformation.\\ \newline
In summary, the polarisation vectors/spinors for each helicity are shown in Table \ref{tab:polarisation}.
\begin{table}[h]
\centering
\begin{tabular}{cc}
Helicity &  Polarisation vector/spinor \\[10pt]
$+1/2$ &  $ \epsilon^{(+1/2)}_{\dot\alpha}\,=\,\tl^{\dot\alpha}$ \\[10pt]
$-1/2$ & $\epsilon^{(-1/2)}_{\alpha}\,=\,\lambda^{\alpha}$ \\[5pt]
$+1$ & $\epsilon^{(+)}_{\alpha\dot{\alpha}}\,=\,\dfrac{\mu_\alpha\widetilde{\lambda}_{\dot{\alpha}}}{\langle\lambda\mu\rangle}$\\[10pt]
$-1$ & $\epsilon^{(-)}_{\alpha\dot{\alpha}}\,=\,\dfrac{\lambda_\alpha\widetilde{\mu}_{\dot{\alpha}}}{[\widetilde{\lambda}\widetilde{\mu}]}$ 
\end{tabular}
\caption{\textit{Polarisation vectors/spinors for given helicities.}}
\label{tab:polarisation}
\end{table}\\
As expected, under the little group rescaling the gluon polarisation vectors scale according to \eqref{eq:little-group-amp},
\begin{equation}
(\lambda,\widetilde\lambda)\rightarrow(t\lambda,t^{-1}\widetilde{\lambda})\qquad\Rightarrow 
\qquad(\epsilon^{(-)},\epsilon^{(+)})\rightarrow(t^2\epsilon^{(-)},t^{-2}\epsilon^{(+)})\ .
\end{equation} 
The partial amplitudes $A_n$ inherit the scaling properties of the polarisation vectors. This can be easily seen in the simple expressions for the scattering amplitudes of gluons of which two have negative helicity and all remaining gluons have positive helicity. These are called the \emph{Maximally Helicity Violating}, or simply MHV amplitudes, and are given by the Parke-Taylor formula \cite{ParkeTaylor:1986,Mangano:1987xk}
\begin{align} 
\label{eq:MHV}
 A_n^{\text{MHV}}(1^+,\dots,i^-,\dots,j^-,\dots,n^+)&\,=\,\delta^{(4)}\Big(\sum_{a=1}^n \lambda^a\tl^a\Big)\frac{\braket{ij}^4}{\braket{12}\braket{23}\cdots\braket{n1}}\ .
\end{align}
The overall delta-function is common to every amplitude and imposes momentum conservation. The parity conjugate of the MHV amplitude (obtained by reversing all helicities) is called \emph{anti-MHV}, or $\MHVb$ amplitude. Is is the same as \eqref{eq:MHV} with angular brackets replaced by square brackets,
\begin{align} 
\label{eq:MHVb}
A_n^{\MHVb}(1^-,\dots,i^+,\dots,j^+,\ldots,n^-)&\,=\,\delta^{(4)}\Big(\sum\limits_{a=1}^n \lambda^a\tl^a\Big)\frac{[ij]^4}{[12][23]\cdots[n1]}\ .
\end{align}
Note that, as mentioned before, these one-term expressions for the MHV and $\MHVb$ amplitudes arise as a sum of numerous diagrams in the Feynman expansion.

\subsection{Supersymmetry} 
\label{sec:SUSY}

The discussion above regarded amplitudes involving only gluons.
Since $\N=4$ is the maximal number of supersymmetry (SUSY) generators in four dimensions, all helicity states are related to each other via SUSY transformations. The aim of this subsection is to introduce how supersymmetry is made manifest in the context of the spinor-helicity formalism.

The $\N=4$ SUSY algebra is generated by the supercharges $Q_{\alpha}^{A}$ and $\bar{Q}_{\dot{\alpha} A}$ (in addition to the Poincar\'{e} generators, as will be explained in detail in \sref{sec:N=4 algebra}), where $A=1,\dots 4 $ is a fundamental (upper) or anti-fundamental (lower) $SU(4)$ $R$-symmetry index. The states with maximum/minimum helicity are the gluons $g^+$ and $g^-$; these define the ground states of the supercharges
\begin{equation}
Q_\alpha^A |g^+\ra\,=\,0,\qquad\qquad \bar{Q}_{\dot\alpha A} |g^-\ra\,=\,0
\end{equation}  on which the $ \bar{Q}/Q$ generators act as raising/lowering operators to generate the complete tower of helicity states. Explicitly:
\begin{equation}
Q_{\alpha}^{A}|g^-\ra\,=\,\lambda_\alpha|\bar{\psi}\ra^A,\qquad\qquad     \bar{Q}_{\dot\alpha A}|g^+\ra\,=\,\tl_{\dot\alpha}|\psi\ra_A\ .
\end{equation}
Further action of the SUSY generators and the use of the superalgebra commutation and anti-commutation relations generates the field content of $\N=4$ SYM, shown in Table \ref{tab:field-content}.
\begin{table}[htb]
\centering
\begin{tabular}{c|p{5cm} c|c}
Multiplicity & \multicolumn{2}{|c|}{Field}  & $SU(2)_{L}\times SU(2)_{R} $ \\
\hline
2 & gluons  & $ g^-,\; g^+ $ & $(1/2,1/2)$ \\
4 & chiral fermions & $ \psi_{\dot\alpha A} $ & $(1/2,0)$ \\
4 & antichiral fermions  & $ \bar\psi_{\alpha}^{A} $ & $(0,1/2)$\\
6 & (real) scalars &  $ \phi_{AB} = -\phi_{BA} $ & $(0,0)$
\end{tabular}
\caption{\it Field content of $\N=4$ SYM.}
\label{tab:field-content}
\end{table}\\
The only representation of this superalgebra is an on-shell vector supermultiplet which comprises all the states above and transforms in the adjoint of the gauge group. A convenient way to write the supermultiplet is by means of the Nair representation \cite{Nair:1988bq} which uses an auxiliary fermionic variable $\eta^A $ of helicity $h=-1/2$ to write a superfield (or, more precisely, a super-creation operator) in which each component multiplies a different combination of $\{\eta^A \}$:
\begin{equation} \label{eq:supermultiplet}
\ket{\Phi}=\ket{g^+}+\eta^A\ket{\psi_A}+\frac{1}{2!}\eta^A \eta^B\ket{\phi_{AB}} + \frac{1}{3!}\eta^A\eta^B\eta^C \ket{\psi_{ABC}}+\eta^1\eta^2 \eta^3\eta^4\ket{g^-}\ ,
\end{equation}
where we used $|\psi_{ABC}\rangle\,\equiv\,\epsilon_{ABCD}|\bar{\psi}^{D}\rangle$.

The on-shell chiral superspace is obtained by augmenting the space-time coordinates by a set of four extra fermionic coordinates $\{\theta^\alpha_A\}$. The state $\ket{\Phi}$ is an eigenstate of the $Q$ generators: $
Q_{\alpha}^{A}\ket{\Phi}=\lambda_\alpha\eta^A\ket{\Phi}$ whose eigenvalue $q_{\alpha}^{A}=\lambda_\alpha\eta^A$ is the \emph{super-momentum} carried by the state $\ket{\Phi}$ in the $\theta^\alpha_A$ fermionic direction. \\[0.3cm]
In this basis, the full tree amplitude can be expanded as
$ A_n=\sum_{k=2}^{n-2}A_n^k$
where $A_n^k$ has fixed fermionic degree $4k$ and comprises the purely gluonic amplitude as well as the complete family of amplitudes with fermions and scalars related to the gluonic one by SUSY transformations. For a given $k$-sector, the helicities of the scattering particles sum to $n-2k$.
In particular, the lowest Grassmann weight $k=2$ is the MHV amplitude \eqref{eq:MHV} recast in the supersymmetric form:
\begin{equation}\label{eq:MHV_susy}
 A_n^{\textrm{MHV}}\,=\,\frac{\delta^{(0|8)} \Big(\sum\limits_{i=1}^n\lambda^i\eta^i\Big) \delta^{(4)}\Big(\sum\limits_{i=1}^n\lambda^i\widetilde{\lambda}^{i}\Big)}{\braket{12}\braket{23}\dots\braket{n1}}\ .
\end{equation}
The notation $\delta^{(a|b)}$ stands for a combination of $a$ bosonic constraints and $b$ fermionic. We will often denote $\delta^{(0|b)}\equiv \delta^{(b)}$ as the context will be sufficient to specify its bosonic of fermionic nature.
The additional fermionic delta-functions impose supermomentum conservation. Amplitudes including fermions, gluons or scalars are obtained by integrating the final expression with the correct power of $\eta$'s. To show this explicitly, recall from Grassmann calculus
\begin{equation*}
(\eta^{A})^2=0\,,\qquad \int d\eta^A\,1=0\,, \qquad\text{and}\qquad\int d\eta^A\,\eta^A=1\ .
\end{equation*}
Hence the integration of the superamplitude over $\eta$'s with an adequate measure for each particle selects a specific state from \eqref{eq:supermultiplet} as follows:
\begin{align}
\label{eq:etaintegration}
\begin{split}
\int d^4\eta\,1 &\quad\longleftrightarrow\quad \ket{
g^-}\ , \\
\epsilon_{ABCD}\int d^4\eta\,\eta^A &\quad\longleftrightarrow \quad \ket{\psi}_{BCD}\ ,  \\
\frac{1}{2!}\epsilon_{ABCD}\int d^4\eta\,\eta^A\eta^B &\quad\longleftrightarrow\quad \ket{\phi}_{CD}\ , \\
\frac{1}{3!}\epsilon_{ABCD} \int d^4\eta\,\eta^A\eta^B\eta^C  &\quad\longleftrightarrow \quad\ket{\psi}_{D}\ , \\
\frac{1}{4!}\epsilon_{ABCD}\int d^4\eta\,\eta^A\eta^B\eta^C \eta^D  &\quad\longleftrightarrow\quad\ket{g^+}\ ,
\end{split}
\end{align}
where\begin{equation}
\int d^4\eta\,\equiv\,\int d\eta^1d\eta^2d\eta^3d \eta^4\ .
\end{equation}

\section{Tree-level recursion relations}
\label{sec:Recursion-Relations}

As mentioned in the previous section, MHV and $\MHVb$ amplitudes are the simplest non-zero amplitudes. The next level in complexity is the helicity configuration consisting of three gluons with distinct helicity. Those are referred to as \emph{Next to Maximally Helicity Violating}, \emph{Next to Next to Maximally Helicity Violating} and so forth, in short N$^{k-2}$MHV amplitudes. 

In this section, we review two methods which are used to compute N$^{k-2}$MHV tree amplitudes in terms of simpler amplitudes, while a discussion regarding loop amplitudes is presented in \sref{sec:Integrands}. The methods are:
\begin{enumerate}
\item \emph{Britto-Cachazo-Feng-Witten (BCFW) recursion relation} \cite{Britto:2004ap,Britto:2005}: Generic amplitudes $A_n^k$ are written in terms of three-particle building blocks $A_3^2$ (MHV) and $A_3^1$ ($\MHVb$).
\item \emph{MHV diagrams} \cite{Cachazo:2004kj}: Generic amplitudes $A_n^k$ are expanded in terms of only MHV building blocks (but not necessarily with only three particles) $A_n^2$.
\end{enumerate}

\subsection{BCFW recursion relation}
\label{sec:BCFW}
The BCFW recursion relation relies on the analytic properties of amplitudes (the location of singularities) and allows one to ultimately express any tree level amplitude in terms of amplitudes with only three particles. The kinematics of a three-particle scattering is quite special for a massless theory, so let us start by exploring it. Using the spinor algebra from \sref{sec:SH_Formalism}, momentum conservation yields
\begin{align}
\label{eq:3-pt-constraints}
\begin{split}
(p_1+p_2)^2=p_3^2=0 \qquad &\Rightarrow \qquad \la 12 \ra [21] =0\ , \\
(p_2+p_3)^2=p_1^2=0 \qquad &\Rightarrow \qquad \la 23 \ra [32] =0\ , \\
(p_1+p_3)^2=p_2^2=0 \qquad &\Rightarrow \qquad \la 13 \ra [31] =0\ .
\end{split}
\end{align}
Since $\braket{a b}=\overline{[ab]}$ for real momenta in Minkowski signature, the only solution is the trivial $p_1 \parallel p_2 \parallel p_3$, so there is no scattering. This provides motivation to consider complex momenta instead, so that angular and square brackets are independent and there exist non-trivial solutions to \eqref{eq:3-pt-constraints}. Of course ultimately one is interested in real momenta, but considering complex momenta for the intermediate steps is a very useful mathematical tool which is crucial for the BCFW recursion relation.

A general tree-level amplitude is a product of propagators $\sim 1/P^2$ and factors associated with the interaction vertices. For planar amplitudes, as a consequence of locality, $P$ can only be a sum of momenta which are adjacent in colour space, so the physical poles are or the form $ \dfrac{1}{(p_i+p_{i+1}+p_{i+2}+\dots)^2}$.

For complex momenta the amplitude is a meromorphic function with poles associated to kinematical configurations where some internal propagator becomes null, $P^2=0$. The main idea behind the BCFW recursion relation is to use this knowledge to determine the amplitude as a function of its singularities.

\begin{figure}[htb]
\centering
\includegraphics[width=0.4\textwidth]{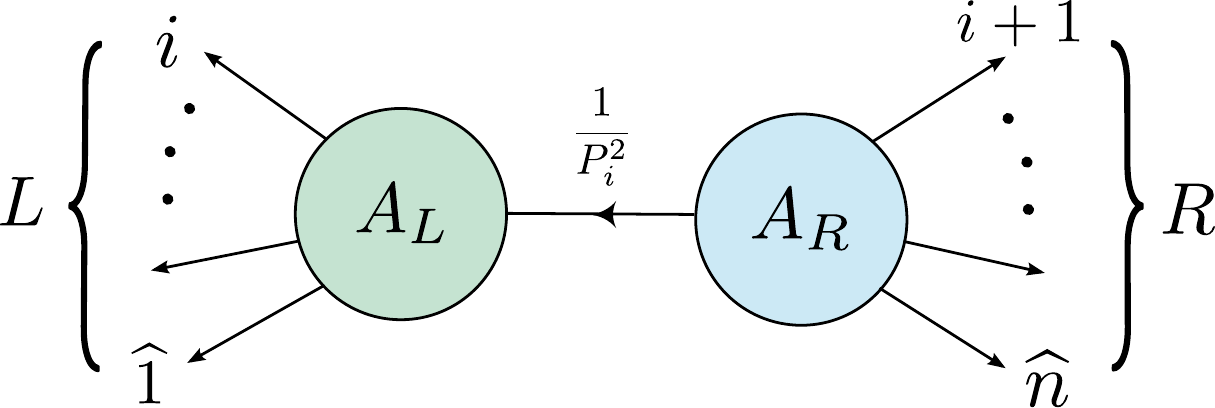}
\caption{\textit{BCFW factorisation of an amplitude as two simpler amplitudes connected by a propagator.}}
\label{fig:BCFW_factorisation}
\end{figure}

The method goes as follows. Given an amplitude $A^k_n$, the \emph{BCFW-shift} is defined as a deformation of two adjacent momenta by a complex parameter $z$. Without loss of generality, using cyclicity one can choose the shifted momenta to be that of particles $1$ and $n$:
\begin{align} \label{eq:BCFW_shift}
\begin{split}
 \lambda^1 & \,\rightarrow\,\widehat\lambda^{1}(z)\,=\,\lambda^1+z\lambda^n  \qquad \leftrightarrow \qquad p_1 \,\rightarrow\, \widehat p_{1}(z)\,=\,p_1+z\, \xi\ , \\
 \tl^n & \,\rightarrow\, \widehat\tl^{n}(z)\,=\,\tl^n-z\tl^1 \qquad \leftrightarrow \qquad p_n \,\rightarrow\,  \widehat p_{n}(z)\,=\,p_n-z\,\xi \ ,
 \end{split}
\end{align}
where $\xi=\tl^1 \lambda^n$. For the \emph{super} BCFW-shift, the fermionic variable $\eta$ receives a shift analogous to $\tl$,
\begin{equation}
\eta^n \, \rightarrow \,\widehat\eta^{n}(z)=\eta^n-z\,\eta^1\ .
\end{equation}
All other particles are left invariant under the BCFW shift. Note that the shifted momenta $\widehat{p}_1$ and $\widehat{p}_n$ still represent massless particles  (after all they are still written as a product of two spinors) and the original momentum conservation equation still holds (because $\widehat{p}_1+\widehat{p}_n = p_1+p_n$). 
Under the BCFW shift the amplitude becomes a holomorphic function of $z$, $A\rightarrow \widehat{A}(z)$. The object of interest is the physical (undeformed) amplitude $A(0)$, which can also be written as a contour integral 
\begin{align} 
\label{eq:contour-integral}
 \widehat{A}(0)&=\frac{1}{2\pi i}\oint_{z=0}\frac{dz}{z}\widehat A(z)\ .
\end{align}
By means of the Cauchy's residue theorem, the amplitude can also be expanded as a sum of residues at the poles corresponding to $z_*\neq 0$,
\begin{align} 
\label{eq:CauchyBCFW}
 \widehat{A}(0)\,=\,-\sum\limits_i\frac{1}{z_{i*}}\text{Res}\left[\widehat A(z_{i*})\right]\ .
\end{align}
The poles for finite $z_*$ occur when some internal propagator with a dependence on $z$ becomes on-shell, that is\footnote{It is also possible that \eqref{eq:CauchyBCFW} has a pole for $z\rightarrow\infty$. This pole does not have the physical interpretation of a factorisation and should be investigated separately. This discussion will play a role in Chapter \ref{ch:formfactors}, where this issue will be discussed in more detail. Poles for $z\rightarrow\infty$ were studied by \cite{Jin:2014qya,Jin:2015pua,Huang:2016bmv}.}
\begin{equation}
(p_i+p_{i+1}+\dots+z_*\xi)^2=0\ .
\end{equation}
This internal propagator generically corresponds to a virtual particle, however for $z=z_*$ it becomes physical. For this reason the interpretation of each residue in \eqref{eq:CauchyBCFW} is a factorisation of the original amplitude into two smaller amplitudes $A_L$ and $A_R$. This is shown in Figure \ref{fig:BCFW_factorisation}.
\\
Notice that the shifted particles $\widehat{1}$ and $\widehat{n}$ are necessarily on different sets, otherwise the internal propagator would not depend on $z$. Considering a generic channel $L=\{\widehat{1},\dots,i\},\,R=\{i+1,\dots,\widehat{n}\}$, the momentum flowing in the intermediate propagator is $\widehat{P}_i(z)=P_i+z\xi$, where $P_i=\sum_{a=1}^i p_a$. The solution $z_{i*}$ is
\begin{equation}
\widehat{P}_i^2=0\quad\Rightarrow\quad z_{i*}=-\frac{P^2}{2 (\xi\cdot P)}\ .
\end{equation}
Plugging this in \eqref{eq:CauchyBCFW} results finally on the \emph{BCFW recursion relation}:
\begin{align}
A^k_n&=\sum_{i,h_i}A_L\left(\widehat{1}(z_{i*}),2\dots,i,-\widehat{P}_i(z_{i*})\right)\frac{1}{P_i^2}A_R\left(\widehat{P}_i(z_{i*}),i+1,\dots,\widehat{n}(z_{i*})\right)
\end{align}
where $A_L\equiv A^{k_L}_{n_L} $ and $A_R\equiv A^{k_R}_{n_R}$ are tree-level amplitudes with smaller $n,\,k$, more precisely $$ k_L+k_R=k+1,\qquad n_L+n_R=n+2  $$ and the sum accounts for all factorisation channels, that is, all possible ways of defining the sets $L,\,R$ as well as the sum over all helicities of the internal on-shell particle.

This recursion relation can be used iteratively for $A_L$ and $A_R$. This allows one to ultimately write any tree-level amplitude in terms of three-particle MHV and $\MHVb$ building blocks. This will be essential for the on-shell diagram construction in Chapter \ref{ch:onshelldiagrams}.

\subsection{MHV diagrams}
\label{sec:MHV-diagrams}

The second tree level recursion relation which can also be used to compute any N$^{k-2}$MHV amplitude was proposed by Cachazo, Svr\v{c}ek and Witten (CSW) in \cite{Cachazo:2004kj}. It amounts to decomposing any amplitude as vertices which are \emph{off-shell continuations} of MHV amplitudes. MHV superamplitudes are given in \eqref{eq:MHV_susy} and are completely independent of the anti-holomorphic spinor variables $\tl$. For this reason, the off-shell continuation is defined by associating to each off-shell momentum $L_{\alpha\dot{\alpha}}$ a holomorphic spinor 
\begin{equation}
\label{eq:off-shell-spinor}
L_{\alpha\dot{\alpha}}\quad\leftrightarrow\quad \ell_\alpha\,\equiv\, L_{\alpha\dot{\alpha}}\xi^{\dot{\alpha}}\ ,
\end{equation}
where $\xi^{\dot{\alpha}}$ is a reference spinor.
Using \eqref{eq:off-shell-spinor}, the generic definition of an off-shell MHV vertex is
\begin{align}
\label{eq:MHV-off-shell}
\begin{split}
 A_n^{\textrm{MHV}}\,&=\,\frac{\delta^{(4)}\Big(\sum\limits_{i=1}^n\lambda^i\widetilde{\lambda}^{i}\Big)\delta^{(0|8)} \Big(\sum\limits_{i=1}^n\lambda^i\eta^i\Big) }{\braket{12}\braket{23}\cdots\braket{n1}}\\
\xrightarrow{\rm off-shell}\quad 
V_{n}^{\rm MHV}\, &= \,
\frac{\delta^{(4)} \Big(\sum\limits_{i=1}^n L_i \Big) \delta^{(0|8)} \Big(\sum\limits_{i=1}^n \ell_i \eta_i  \Big) }{ \lan 12\ran \lan 23\ran \cdots \lan n 1\ran} \ .
\end{split}
\end{align}

Gauge invariance demands that the final result is independent of the choice of reference spinor.

The CSW method was shown to arise in many frameworks. In \cite{Risager:2005vk} gluon amplitudes were obtained by a generalised BCFW shift, where for N$^{k-2}$MHV amplitudes, the momenta of the $k$ gluons with negative helicity were shifted, as opposed to just two. In \cite{ArkaniHamed:2009sx} the expansion is a consequence of a residue theorem in the Grassmannian formulation (see Chapter \ref{ch:onshelldiagrams} for an exposition of this formulation). In \cite{Mansfield:2005yd} it was shown to arise as a change of variables in the $\N=4$ SYM Lagrangian in the lightcone gauge; in this description, the action is mapped to a free theory plus an infinite set of interaction vertices, each corresponding to an off-shell MHV amplitude. Lastly, the expansion into MHV vertices was shown in  \cite{Boels:2007qn} to be the Feynman diagrams arising from the action in twistor space \cite{Boels:2006ir,Boels:2007qn,Adamo:2011cb}. The MHV diagram method was used to successfully compute loop amplitudes in \cite{Brandhuber:2005kd,Brandhuber:2011ke} and will be applied to the computation of the one-loop dilatation operator Chapter \ref{ch:dilatation}.

\section{Loop techniques}
\label{sec:loops}


Tree level amplitudes are simple objects, they are just rational functions of the external momenta whose singularities are well understood. The nature of the singularities in massless theories is threefold:
\begin{itemize}
\item Factorisation -- Internal propagator goes on-shell,
\item Collinear -- The momenta of two or more particles become parallel to each other,
\item ``Soft'' -- The momentum of an external particle becomes small.
\end{itemize}
The three types of singularities outlined above occur for particular values of the external momenta. When loop momenta enter the game, the singularity structure of the amplitudes becomes much more involved. Typically the result of loop integrals involve multivalued functions with branch cuts. Moreover, the integrals are hard to carry out, and they are often divergent. The divergences can be of the infrared (IR) kind~---~for small values of loop momenta~---~or ultraviolet (UV)~---~for large values of loop momenta.

$\N=4$ SYM is an especially simple theory due to the fact that it is conformal at the quantum level (i.e. the $\beta$-function,  the equation that governs how the coupling constant $g_{\rm YM}$ varies with energy scale, is zero to all orders in perturbation theory \cite{Brink:1982wv}). Physically this means that there is no inherent length scale, so any short distance phenomenon can be ``zoomed out''. As a result there are no UV divergences present in this theory. IR divergences, on the other hand, have a physical meaning; they tell us that for a massless theory one cannot distinguish a state of one particle from a state where this particle emits one or many particles with soft, undetectable momenta or if the measured particle is indeed one particle or many  with collinear momenta. The observable physical quantities, like cross sections, are however finite. At a given order in perturbation theory, the IR divergent part of a loop integral precisely cancels against soft singularities of the phase space integral of a lower-loop amplitude involving extra undetectable particles.

Indeed one may raise the question that for a conformal theory there is no notion of the asymptotic states that enter in the definition of the $S$-matrix (since in the absence of a length scale it makes no sense to define particles ``at infinity''). However, due to IR divergences one is forced to regulate the integrals to make sense of them (for instance the threshold of the detector), and this often involves introducing a scale which breaks conformal symmetry\footnote{In \cite{Ferro:2013dga,Ferro:2012xw} the authors propose a regularisation procedure in terms of \emph{spectral parameters} which preserves all symmetries of planar amplitudes. However, a proof that the extra parameters actually serve as regulators, i.e. they disappear for physical observables, is still lacking.}. There are many ways to regulate integrals, for a comprehensive presentation of many methods we indicate \cite{Smirnov:2004ym}. The regularisation procedure that will be used throughout the following chapters is \emph{dimensional regularisation}, which consists in evaluating integrals in dimension $d=4-2\epsilon $, where $\eps$ is an infinitesimal parameter,  as opposed to $d=4$. In this framework the result of the integral is a Laurent series in $\epsilon$. The IR divergent terms appear with negative powers of $\epsilon$ and these must cancel for well defined observables, allowing one to finally take the physical limit $\epsilon\rightarrow 0$.

One of the consequences of IR divergences is that the result of loop integrals display fewer symmetries than the tree level amplitudes. For this reason, it is common to study the \emph{loop integrand} itself~\footnote{The loop integrand is only well defined in the planar limit, this is discussed in detail in \sref{sec:Integrands}.}, which prior to integration preserves the symmetries of the tree-level amplitudes and are just rational functions with poles involving both external particles $p_i$ and loop momenta $\ell_j$.

Of course one is ultimately interested in the results of the integrals themselves. To this end there exist a rich collection of techniques~---~integration by parts (IBP) relations \cite{Chetyrkin:1981qh}~---~which allows the representation of a family of integrals in terms of a finite basis called \emph{master integrals}, differential equations \cite{Kotikov:1990kg,Henn:2013pwa}\footnote{For an overview of the method and the most modern formulation, see the review \cite{Henn:2014qga}.}, bootstrap approaches \cite{Dixon:2011pw}, and so forth. In the work presented here we are in the fortunate scenario where it is not necessary to evaluate any new integral and we review below the techniques which will be used in the forthcoming chapters. In \sref{sec:Integrands} we discuss a particular method used to construct integrands and in \sref{sec:Integrals}, after giving an overview of loop amplitudes, we present a particular tool which has revolutionised the way one can deal with the special class of functions that result from loop integrations~---~the \emph{symbol} of transcendental functions~---~which will play a central role in two-loop form factor computation presented in \sref{sec:ff-two-loops}.


\subsection{Integrands}
\label{sec:Integrands}

As mentioned before, loop integrals in general involve a complicated combination of multivalued functions with branch cuts and discontinuities. The integrand of a scattering amplitude at a given order in perturbation theory is, in analogy with tree-level amplitudes, a rational function of external and loop momenta that, after integration, reproduces all branch cuts and discontinuities of the loop integral, plus potential rational terms. At the level of the integrand, however, the singularities are simply poles for which propagators involving one or more loop momenta go on shell. 

This is the main idea behind what is called the \emph{generalised unitarity} method for constructing loop integrands \cite{GeneralizedUnitarity,Bern:1994zx,Bern:1994cg,Bern:1994cg,Bern:1995db,Bern:1996fj,Bern:1996je,Bern:2004cz,Britto:2004nc,Cachazo:2008vp}. But before embarking on this, one needs to first investigate if the integrand is a well defined notion to begin with.

As it turns out, in the planar limit the answer is yes, and for non-planar corrections the answer is, at least until this day, not yet.

\begin{figure}[htb]
\centering
\includegraphics[width=0.7\textwidth]{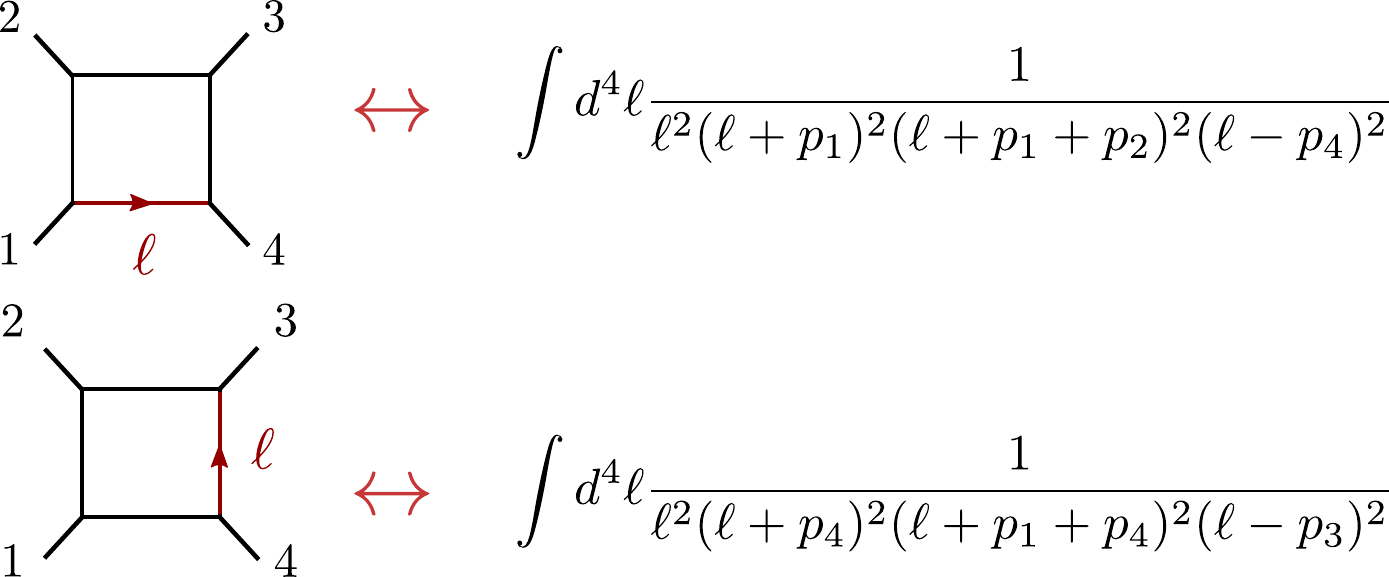}
\caption{\textit{A one-loop four-particle example illustrating how a redefinition of the loop integration variables change the location of the poles of the integrand. These ambiguities is remedied with the use of region momenta which makes the integrand canonically defined.}}
\label{fig:planar-ambiguity}
\end{figure}

A generic loop integral is a sum of many terms. The idea of a well defined loop integrand relies on the possibility of canonically defining loop integration variables which are consistent between all terms. For each integral entering the sum, the loop momenta are dummy integration variables and as such can be redefined as one pleases, however at the level of the integrand a redefinition of the loop variables changes the locations of the poles. In order to combine all functions into a single integrand, one has to find a way to canonically define what is meant by the loop integration variables. This difficulty is illustrated in Figure \ref{fig:planar-ambiguity}.

\begin{figure}[h]
\centering
\includegraphics[width=0.7\linewidth]{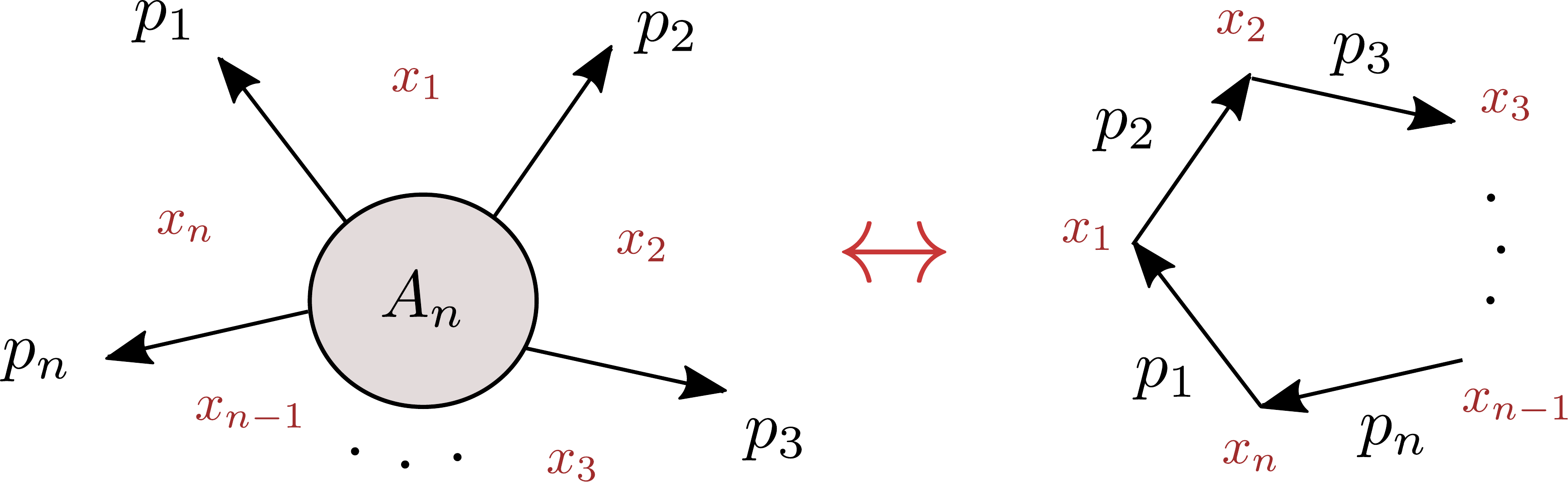}
\caption{\textit{Dual coordinates where $p_i\,\,=\,x_i-x_{i-1}$. Momentum conservation implies that in dual space the amplitudes as supported on a closed polygon with null edges.}}
\label{fig:dual-coords-polygon}
\end{figure}

The solution to this issue for planar integrands comes from the natural ordering of the external states. If instead one assigns variables labelling regions between the momenta, the integration variables are uniquely defined as the variables associated with bounded regions. The map between the standard momenta and the so-called \emph{dual variables} or \emph{region momenta} is:
\begin{align}
\label{eq:dual-variables}
p_i\,&=\,x_i-x_{i+1}\,\equiv\,x_{i\,i+1}\ .
\end{align}
For superamplitudes, one defines the analogous dual supermomentum by
\begin{align}
\label{eq:dual-variables-fermions}
\lambda^i\eta^A_i\,&=\,\theta^A_i-\theta^A_{i+1}\,\equiv\,\theta^A_{i\,i+1}\ ,\quad  A\,=\,1,2,3,4\ ,
\end{align}
where $\theta^A_i $ are $\N=4$ on-shell superspace coordinates. This map is illustrated in Figure \ref{fig:dual-coords-polygon}, where it is also clear that the external momenta form a closed polygon with null edges in the dual space.

\begin{figure}[h]
\centering
\includegraphics[width=0.8\linewidth]{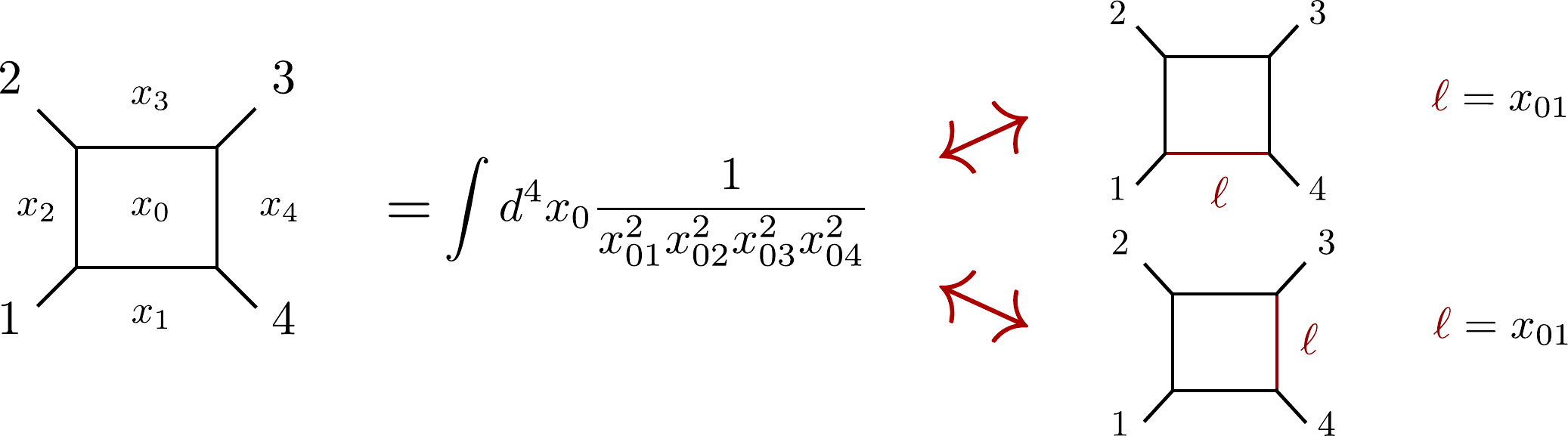}\caption{\textit{The canonically defined planar loop integrand in terms of dual variables $x$. Loop integration are over variables associated to internal faces, symmetrised for $L>1$.}}
\label{fig:planar-integrand-canonical}
\end{figure}

In dual variables, both integrals from Figure \ref{fig:planar-ambiguity} are identical
and given by
\begin{align}
\label{eq:planar-integrand}
I_4^{1\text{-loop}}\,=\,\int\!d^4{x_0} \frac{1}{x_{01}^2x_{02}^2x_{03}^2x_{04}^2}\ .
\end{align}
This integrand is the same for identification of the variable $\ell$ with any edge of the square, that is, $\ell\,=\,x_{i0}\text{ for }i=1,\dots, 4$, as shown in Figure \ref{fig:planar-integrand-canonical}. At higher loop order, the unique integrand is obtained by symmetrising over all possible labellings of internal faces.

The use of the dual coordinates unravels a remarkable duality within $\N=4$ SYM, that between MHV scattering amplitudes and Wilson loops evaluated on the corresponding null polygon in $x$-space \cite{Alday:2007hr,Drummond:2007aua,Brandhuber:2007yx}\footnote{This duality was extended to relate N$^{k-2}$MHV amplitudes and supersymmetric Wilson loops in \cite{CaronHuot:2010ek,Mason:2010yk,Eden:2011yp}}. Polygonal Wilson loops are invariant under ordinary superconformal symmetry in position space. In the context of amplitudes, this corresponds to a hidden symmetry of the planar sector, called dual superconformal symmetry \cite{Drummond:2008vq}. In the amplitude description, this dual symmetry is broken by IR divergences, whereas in the Wilson loop description it is broken by UV divergences associated to the cusps.

Both the BCFW and the MHV-diagram expansions presented in \sref{sec:Recursion-Relations} were originally proposed for tree-level amplitudes, but afterwards extend to construct loop integrands too. For the BCFW recursion relation, the idea behind the loop generalisation is to take into account, in addition to factorisation-like singularities, singularities for which propagators involving loop momenta become on-shell. The latter can be obtained from a lower loop amplitude with two extra particles in the so-called \emph{forward limit}. This allows one to recursively construct the planar loop integrand \cite{ArkaniHamed:2010kv}. We will not expand on the loop BCFW recursion relation since it will not be relevant for the future chapters. The loop MHV-rules, however, will be further discussed and applied in the context of the dilatation operator in \sref{sec:MHV-diagrams}.

\subsubsection{Generalised unitarity}

Recall that generic loop integrals can be a combination of multivalued functions and rational terms. In supersymmetric theories the rational terms are absent. For this reason, the integrand can be found by considering a set of standard integrals that span all possible physical branch cuts the amplitude may have. The idea behind the generalised unitarity method is to write the amplitude as a sum of basis integrals and compute the coefficients of the integrals by matching the singularities of the amplitude with that of the integrals. The name stems from the standard \emph{unitarity cuts}, which makes use of the the unitarity of the $S$-matrix ($S \cdot S^\dagger = \uno$) to represent the discontinuity of the imaginary part of a loop amplitude as a sum over two separate lower-loop factors with two propagators set on-shell. One can compute discontinuities across different unitarity cuts successively, and the name generalised unitarity refers to the situation where any number of propagators can be cut by effectively replacing\footnote{Throughout explicit calculations the factors of $i$ will often be omitted and reinstated at the end.}
\begin{equation}
\frac{i}{P^2}\,\mapsto\,\delta^+(P^2)\,\equiv\,\delta(P^2)\Theta(p^0)\ ,
\end{equation}
where the Heaviside function $\Theta(p^0)$ ensures that the physical state has positive energy. 

\begin{figure}[h]
\centering
\includegraphics[width=0.7\linewidth]{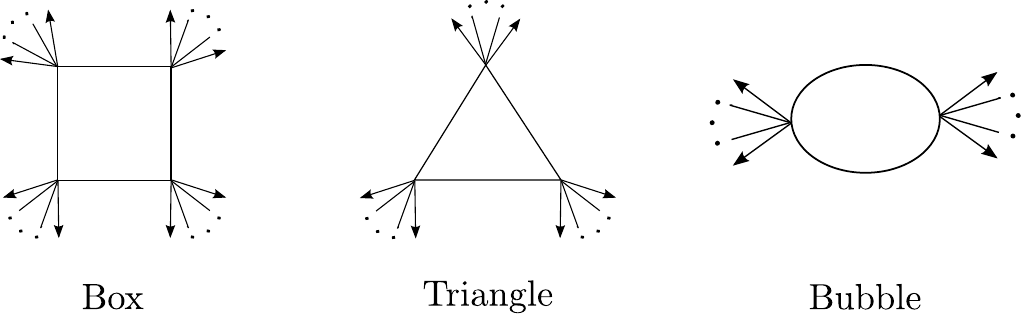}
\caption{\it One-loop integral basis for amplitudes in massless theories without rational terms. Each integral is a sum of Feynman diagrams that share the same propagators.}
\label{fig:one-loop-integrals}
\end{figure}

Disregarding rational terms (i.e. focusing on what is called the \emph{cut-constructible} part of the integrals), one can determine one-loop amplitudes by cutting up to four propagators. A basis of integrals at one loop is formed of scalar boxes, triangles and bubbles and tadpoles \cite{Passarino:1978jh}. For massless theories, the tadpoles do not contribute and thus will be dropped. The box, triangle and bubble integrals are shown in Figure \ref{fig:one-loop-integrals} and their dependence on the dimensional-regularisation parameter $\epsilon$, which are relevant for the future chapters, are written in Appendix \ref{app:scalar-integrals}.

Thus, at one loop, the cut-constructible part of an amplitude in a generic massless theory can be expanded as (this discussion simplifies considerably in the case of $\N=4$ SYM, see below)
\begin{equation}
\label{eq:one-loop-expansion}
\A^{1\text{-loop}}\,=\, \sum_i c_{\text{Box}_i} \text{Box}_i +\sum_i c_{\text{Tri}_i} \text{Tri}_i + \sum_i c_{\text{Bub}_i} \text{Bub}_i\ ,
\end{equation}
where the coefficients $c_{\text{Box}_i},\,c_{\text{Tri}_i},\,c_{\text{Bub}_i}$ are rational functions of the kinematic variables and the sums run over all possible ways of distributing the external momenta on the corners of the integrals. The coefficients of each integral can then be found by matching the discontinuities of the functions of either side of \eqref{eq:one-loop-expansion} in the following way. The coefficients of the boxes are determined by cutting four propagators (also called a quadruple cut or a four-particle cut) as they are the only functions that become singular in this situation. Subsequently, one can determine the coefficients of the triangles by matching the singularities under triple cuts (notice the boxes also become singular, but their coefficients are already fixed). In the same way, two-particle cuts determine finally the coefficient of the bubble integrals, and thus the full cut-constructible part of the loop amplitude.

The situation where as many propagators as possible are cut (the same number of integration variables, $d\times L$) is called a \emph{maximal cut}. The values of the integrand evaluated on solutions of such cuts are called \emph{leading singularities}. Leading singularities are rational functions which correspond to discontinuities of the integral across maximal cuts.

For amplitudes in $\N=4$ SYM this discussion simplifies further \cite{Bern:1994zx,Bern:1994cg,Bern:1993kr}. Bubbles are UV divergent and therefore absent. Moreover, the planar $\N=4$ SYM integrand preserves the dual conformal symmetry present in the tree-level amplitude, this allows one to also eliminate the triangles from the basis above, leaving only boxes. In contrast with amplitudes, for form factors dual conformal symmetry is relaxed and one has to keep the triangles and bubbles in the basis of integrals. For the protected operators studied in Chapter \ref{ch:formfactors} the bubbles are still unnecessary, as will be explicitly show in \sref{sec:ff-one-loop}. In the study of the dilatation operator in Chapter \ref{ch:dilatation}, one is interested in precisely the opposite~---~UV-divergent integrals~---~and thus the bubble will become relevant.

The problem of finding an integral basis that span all cuts at two loops and higher is not solved in general. This problem at the level of the planar integrand in $\N=4$ SYM is solved, in the sense that the integrand satisfies the all-loop BCFW recursion relation \cite{ArkaniHamed:2010kv}. In \cite{ArkaniHamed:2010gh} the authors represent the integrand as a linear combination of functions which are dual conformal invariant and normalised to have unit leading singularity. This is equivalent to the statement that the leading singularities of the planar integrand are enough to determine the full integrand. This topic will be revisited in Chapter \ref{ch:onshelldiagrams} when we discuss non-planar leading singularities.

The generalised unitarity method will be heavily used in the context of loop form factors in Chapter \ref{ch:formfactors} and correlation functions in Chapter \ref{ch:dilatation}.

\subsection{Integrals}
\label{sec:Integrals}

Scattering amplitudes at loop level can be very difficult to compute, but explicit calculations have shown that, to some degree, the special properties of $\N=4$ SYM lead to some structure at the level of the integrated expressions too. To make the treatment of loop amplitudes clearer, it is customary to study instead of the loop amplitude itself, the helicity-independent function obtained by dividing it by the corresponding tree-level amplitude. This is called the \emph{ratio function},
\begin{align}
\label{eq:ratio-function}
M_n^{k(L)}\,\equiv\, \frac{A_n^{k(L)}}{A_n^{k(0)}}\ .
\end{align}
The first hint of an underlying structure in the context of loop amplitudes was the finding of Anastasiou, Bern, Dixon and Kosower (ABDK) \cite{Anastasiou:2003kj}. They observed that the four-particle, two-loop ratio function $M_4^{2(2)}$ could be expressed in terms of the one-loop result. An iterative process was further shown to hold at three loops by Bern, Dixon and Smirnov (BDS) \cite{Bern:2005iz}, which led them to conjecture that the fully resummed MHV ratio function (denoted by $\mathcal{M}_n^{\rm MHV}$) could be obtained from $M_n^{2(1)}$ via an exponential relation called the \emph{BDS/ABDK ansatz},
\begin{align}
\label{eq:BDS-ABDK-ansatz}
\mathcal{M}_n^{\rm MHV}\,=\,\exp \Big[\sum_{L=1}^\infty a^L\big(f^{(L)}(\eps)M_n^{2(1)}(L\eps) + C^{(L)}+\O(\eps)\big)\Big]\ .
\end{align}
The ingredients entering the formula are explained below.
\begin{itemize}
\item $a$ is a convenient function of the 't Hooft coupling $\lambda$ given by
\begin{align}
\label{eq:a}
a\,\equiv\, \frac{\lambda e^{-\epsilon \gamma_{\rm E}}}{(4 \pi )^{2-\epsilon }} \ ,
\end{align}
where $\gamma_{\rm E}\,\approx\,0.577$ is the Euler-Mascheroni constant, often grouped together with the coupling constant to absorb extra factors that arise from loop integrations.
\item $f^{(L)}(\eps)$ is a polynomial of degree two in $\eps$,
\begin{align}
\label{eq:f}
f^{(L)}(\eps)\,\equiv\, f_0^{(L)}+ f_1^{(L)}\eps + f_2^{(L)} \eps^2\ ,
\end{align}
where $f_0^{(L)}$ is called the $L$-loop \emph{cusp anomalous dimension}\footnote{The name stems from the Wilson loop picture where the divergences are of the UV kind, associated with the cusps. The cusp anomalous dimension $\gamma_\text{K} $ \cite{Korchemskaya:1992je} appears in the anomalous Ward identity of the dual special conformal generator acting on the finite part of the Wilson loop and it is predicted for any value of $\lambda$ \cite{Beisert:2006ez}. At $L$ loops, the relation between $f_0^{(L)}$ and $\gamma^{(L)}_\text{K}$ is $f_0^{(L)}\,=\,\gamma^{(L)}_\text{K}/4$.}, and $f_1^{(L)}$ is called the \emph{collinear anomalous dimension}.
\item $C^{(L)}$ is a constant which is independent of $n$ and $\epsilon $\ .
\end{itemize} 

For a while the hope was that \eqref{eq:BDS-ABDK-ansatz} was in fact the final answer to the all-loop MHV amplitudes, with results verified numerically up to five particles at two loops \cite{Bern:2006vw,Cachazo:2006tj}. This would mean that one would only ever have to calculate one-loop integrals, which are comparatively an easy task.  However, before anyone had a chance to prove \eqref{eq:BDS-ABDK-ansatz}, some disagreement was found starting at six particles\footnote{The existence of a deviation from the BDS/ABDK ansatz was first indicated by Alday and Maldacena in \cite{Alday:2007he} from computations at strong coupling. They also constructed the BDS/ABDK ansatz using AdS/CFT in \cite{Alday:2007hr}.}
~---~while \eqref{eq:BDS-ABDK-ansatz} reproduces correctly the IR divergent part of $M_6^{2(2)}$, there is a finite correction which is a function of dual conformal cross ratios \cite{Bern:2008ap}. Indeed, any finite correction to the BDS/ABDK ansatz must be dual conformal invariant and, as such, a function of dual conformal cross ratios. For $n<6$ there are no possible cross-ratios (the number of dual conformal cross ratios in an $n$-particle scattering is $3n-15$, thus non-zero only for $n\geq 6$). 
An interesting quantity to consider is therefore the mismatch between the $L$-loop ratio function and the prediction given by the BDS/ABDK ansatz~---~the \emph{remainder function} \cite{Bern:2008ap, Drummond:2008aq}. The BDS/ABDK ansatz captures all IR divergent terms of the amplitude, thus the remainder is a finite function of dual conformal cross ratios. At two loops, the remainder is simply the difference between the two-loop ration function and the result predicted by the BDS/ABDK ansatz \eqref{eq:BDS-ABDK-ansatz},
\begin{align}
\label{eq:2-loop-remainder}
\mathcal{R}^{2(2)}_n\,=\,M_n^{2(2)}-\big[\tfrac{1}{2}(M_{n}^{(1)}(\epsilon))^2\,+\, f^{(2)}(\eps)M_n^{2(1)}(2\eps) + C^{(2)}\big]\ .
\end{align}
At two loops \cite{Anastasiou:2003kj},
\begin{equation}
\label{eq:f-amp-2-loops}
f^{(2)}(\eps)\,=\,-2(\zeta_2+\epsilon\zeta_3+\eps^2\zeta_4)\,,\quad C^{(2)}\,=\,-\zeta_2^2\ ,
\end{equation}
where $\zeta_n$ is the Riemann zeta function.
\subsubsection{Transcendental functions and symbols}
\label{sec:Trans-func-symbols}

The remainder function itself can be still extremely complicated, as will become clear shortly. It is widely believed, however, that the $L$-loop remainder function in $\N=4$ SYM is a \emph{transcendental function} of weight (or depth) $2L$, that is, a linear combination of iterated integrals that involves $2L$ ``steps''. The formal definition of transcendental function of degree $m$ (also called a \emph{pure function}), $F^{(m)}$, is in terms of its differential,
\begin{align}
\label{eq:pure-function-differential}
dF^{(m)}\,=\,\sum_{i}F_{i}^{(m-1)}d\log{f}_i\ ,
\end{align}
where $f_i$ is an algebraic function and transcendentality zero functions are constants. A simple example of a weight $m$ transcendental function of one variable $x$ is the classical polylogarithm $\text{Li}_m(x)$, which is recursively defined as
\begin{align}
\label{eq:polylog}
\text{Li}_m(x)\,\equiv\,\int_0^x\, \frac{dt}{t}\,\text{Li}_{m-1}(t)\,,\qquad \text{Li}_1(x)\,\equiv -\log
(1-x)\,=\,\int_0^x\frac{dt}{1-t}\ .
\end{align}
Another notation for $\text{Li}_m(x)$ is
\begin{align}
\text{Li}_m(x)\,\equiv\,-\int_0^x d\log(1-t)  \circ \underbrace{\log(t) \circ d\log(t)\circ \cdots  \circ d\log(t)}_{m-1\text{ times}}\ ,
\end{align}
where the outermost terms are meant to be integrated first. A more general kind of iterated integrals are the Goncharov polylogarithms, also recursively defined as
\begin{align}
\label{eq:Goncharov}
\begin{split}
&G(\{a_1,a_2\dots,a_m\};x)\,\equiv\,\int_0^x \frac{dt}{t-a_1} G(\{a_2,\dots,a_m\},t)\,,\\
&G(\{a\},x)\,\equiv\,\int_0^x\frac{dt}{t-a}\,,\quad a\neq 0\,,\qquad G(\{\vec{0}_n\},x)\,\equiv\,\frac{1}{n!}\log^n(x)  \ .
\end{split}
\end{align}
So the classical polylogarithms \eqref{eq:polylog} are special cases of the Goncharov polylogarithms \eqref{eq:Goncharov} for $\{a_1,a_2,\dots,a_m\}=\{0,0,\dots,1\}$.
For an extensive explanation of the properties of transcendental functions and their appearance in various contexts in Physics, we indicate the reader the lecture notes \cite{lec_vergu}.

The combination of transcendental functions that result from integrals at loop orders higher than one can be extremely complicated. For instance, in \cite{DelDuca:2010zg}, Del Duca, Duhr and Smirnov (DDS) computed analytically the remainder function for a six-sided null Wilson Loop (which is dual to an MHV ratio function \cite{Alday:2007hr,Drummond:2007aua,Brandhuber:2007yx}\footnote{This correspondence was later generalised to relate N$^{k-2}$MHV ration functions and supersymmetric Wilson Loops \cite{Mason:2010yk,CaronHuot:2010ek,Eden:2011yp,Eden:2011ku}.}).  Their result is very famous for being (besides very laborious) a 17-page long combination of transcendentality four functions involving many Goncharov polylogarithms. 

Initially it was certainly not expected that this result could be simplified to something simple, but fortunately this is not true. They key point behind the simplification of that beast is the fact that polylogarithms satisfy very complicated relations. For transcendentality one, the relation between logarithms is rather simple,
\begin{equation}
\label{eq:log-identities}
\log(xy)\,=\, \log(x)+\log(y)\ .
\end{equation}
Dilogarithms $\text{Li}_2(x)$ satisfy the so-called five-term identity,
\begin{align}
\label{eq:5-term-identity}
\begin{split}
&\sum_{n=1}^5\big[\text{Li}_2(a_n)+\log(a_{n-1})\log(a_n)\big]\,=\,\frac{\pi^2}{6}\ ,\\
a_1\,=\,x\,,\quad & a_2\,=\,\frac{1-x}{1-xy}\,,\quad a_3\,=\,\frac{1-y}{1-xy}\,,\quad a_4\,=\,y\,,\quad a_5\,=\,1-xy\ .
\end{split}
\end{align}
Clearly \eqref{eq:5-term-identity} is already much more complicated than \eqref{eq:log-identities} and functions of higher transcendentality satisfy very intricate identities that easily get out of hand. It is perhaps important to mention that often factors of $\pi$ appear in relations between transcendental functions as they are associated to discontinuities across branch cuts, the simplest example being
\begin{align}
\log\big(e^{i(\theta+2\pi)}\big)\,=\,\log\big(e^{i\theta}\big) + 2\pi i\ .
\end{align}
To bypass the complication arising from relations like \eqref{eq:5-term-identity} it is very helpful to use the notion of the \emph{symbol} of a transcendental function \cite{Goncharov:2010jf,Duhr:2011zq}. By definition, the symbol of a generic iterated integral of transcendentality $m$ is via the recursion (recall \eqref{eq:pure-function-differential})
\begin{equation}
\label{eq:symbol-recursion}
dF^{(m)}\,=\,\sum_{i}F_{i}^{(m-1)}d\log{f}_i\quad\Rightarrow\quad\mathcal{S}[F^{(m)}]\,\equiv\,\sum_{i}\mathcal{S}[F_{i}^{(m-1)}]\otimes f_i\ .
\end{equation}
One property of the symbol of a transcendental function that is extremely desirable for loop integrals is that is makes manifest the location of its branch cuts, and the discontinuities associated to it. From the definition \eqref{eq:symbol-recursion} one can infer that the function $F^{(m)}$ has branch cuts for $f_i=0$ and, moreover, $F_{i}^{(m-1)}$ are the corresponding discontinuities. This is the heart of the idea behind the bootstrap approaches, where the location of the branch cuts in various kinematic limits, together with integrability data \cite{Basso:2013vsa,Basso:2013aha,Basso:2014koa}, act as physical input to constrain the symbol \cite{Dixon:2011pw,Dixon:2014xca,Dixon:2014iba,Dixon:2014voa}.

The application of \eqref{eq:symbol-recursion} $m$ times culminates in an $m$-fold tensor product. This can be seen easily for a function of a single variable,
\begin{align}
F^{(m)}(x)\,=\,\int_0^x d\log(f_1(t))\circ d\log(f_2(t))\circ \cdots \circ d\log(f_m(t))\ ,
\end{align}
where $f_1\,,\dots,f_m$ are algebraic functions. Its symbol is the $m$-fold tensor product of the arguments of the logarithms in the integrals evaluated at the endpoint of the integration,
\begin{align}
\mathcal{S}[F^{(m)}(x)]\,=\,f_1(x)\otimes f_2(x)\otimes \cdots\otimes f_m(x)\ .
\end{align}
As a consequence of \eqref{eq:log-identities} and $\log(x^n)= n\log(x)$, the symbols obey
\begin{align}
\label{eq:symbols-properties}
\begin{split}
\cdots\otimes x\,y \otimes \cdots\,&=\, \cdots\otimes x \otimes\cdots + \cdots\otimes y \otimes\cdots\ ,\\\
\cdots\otimes x^n \otimes \cdots\,&=\, n\, (\cdots\otimes x \otimes \cdots)\ ,\\
\cdots\otimes (\text{constant})\, x \otimes \cdots \,&=\, \cdots\otimes x \otimes \cdots\ .
\end{split}
\end{align}
The last property follows from $d\log c = 0$ for any constant $c$. Table \ref{tab:symbols-examples} contains some instructive examples of symbols of the functions mentioned earlier.
\begin{table}[h]
\centering
\begin{tabular}{cc}
Function & Symbol \\[5pt]
$\log(x)$ & $x$ \\[5pt]
$\log(x)\log(y)$ & $x\otimes y + y\otimes x$ \\[5pt]
$\text{Li}_n(x)$ &  	$-(1-x)\otimes \underbrace{x \otimes \cdots \otimes x}_{n-1\text{ times}}$
\end{tabular}
\caption{\it Examples of symbols of simple transcendental functions.}
\label{tab:symbols-examples}
\end{table}
The main advantage of using the symbols is that every relation satisfied by transcendental functions turns into an algebraic relation satisfied by the symbols. For example, \eqref{eq:5-term-identity} with $y=0$ (thus only $a_1=x$ and $a_2=1-x$ are not constants) reads
\begin{align}
\label{eq:five-term-example}
\text{Li}_2(x) +\text{Li}_2(1-x)+ \log(x)\log(1-x)\,=\,\frac{\pi^2}{6}\ .
\end{align}
This is easily seen considering the symbol of the above expression (see Table \ref{tab:symbols-examples}),
\begin{align}
\label{eq:five-term-symbol}
-(1-x)\otimes x - x\otimes(1-x) + x\otimes(1-x) + (1-x)\otimes x \,=\,0\ .
\end{align}
Clearly the information about constants which are powers of $\pi$ are lost after taking the symbol (c.f. \eqref{eq:symbols-properties}), as can be seen from going from \eqref{eq:five-term-example} to \eqref{eq:five-term-symbol}. In other words, the symbol loses information about which Riemann sheet the multivalued functions are evaluated on. Terms containing powers of $\pi$ times lower degree functions are referred to as \emph{beyond the symbol} and can, for instance, be determined numerically demanding agreement between the functions before and after simplification. This will be used in \sref{sec:ff-two-loops}

The power of the symbols was first demonstrated by Goncharov, Spradlin, Vergu and Volovich (GSVV) in \cite{Goncharov:2010jf}. There the authors simplified the DDS result for the ratio function of the six-sided two-loop MHV Wilson loop from the 17-page long linear combination of classical and generalised polylogarithms to an expression that fits within a line! Moreover the expression involved only classical polylogarithms \eqref{eq:polylog}, all the more complicated functions cancelled out. The strategy there was to compute the symbol of the DDS expression, which turned out to be very simple, and then reconstruct a simple function that reproduced the same symbol.
The procedure of recovering a function from its symbol is not completely straightforward. In particular, a generic linear combination of tensor products does not necessarily originates from a function, this is only the case if the symbol obeys the \emph{integrability conditions},
\begin{align}
\begin{split}
\mathcal{S}[F^{(m)}]\,&=\,\sum_{i_1,\dots,i_m}f_{i_1}\otimes\cdots\otimes f_{i_m}\, ,\\
d^2F^{(m)}\,&=\,0\quad\Rightarrow \quad  \sum_{r=1}^{m-1}\sum_{i_1,\dots,i_m} f_{i_1}\otimes\cdots\otimes f_{i_r}\wedge f_{i_{r+1}}\otimes\cdots\otimes f_{i_m}\,=\,0\ ,
\end{split}
\end{align}
where $\wedge$ stands for the usual wedge product,
\begin{align}
\label{eq:wedge}
\cdots \otimes x \wedge y  \otimes \cdots = \cdots \otimes x \otimes y  \otimes \cdots - \cdots \otimes y \otimes x  \otimes \cdots\ .
\end{align}
The integrability conditions assure that the iterated integrals do not change if the integration path is slightly deformed keeping the endpoints fixed, which is of course required since the functions depend on the endpoints of integration only. This is also called homotopy invariance. In the case of the GSVV symbol, they observed that it also satisfied the so-called \emph{Goncharov condition} \cite{gonch, Goncharov:2010jf}, described as follows. Since the two-loop remainder function is of transcendentality four, one can denote its symbol schematically by $\mathcal{S}_{abcd}$ where the subscripts stand for the letters which form the symbol keeping the order of the arguments. Then the Goncharov condition reads
\begin{align}
\label{eq:gonch-cond}
\cS_{abcd} - \cS_{bacd} - \cS_{abdc}+ \cS_{badc} - (a \leftrightarrow c\, ,\,  b \leftrightarrow d)\ = \ 0
\ .
\end{align}
When a symbol obeys this criterion, it means that it can be integrated to a combination of classical polylogarithms only\footnote{There exists a conjecture by Goncharov that all weight four functions can be written in a basis formed by classical polylogarithms plus the function $\text{Li}_{2,2}(x,y)=\sum_{a_1>a_2\geq 1}\frac{x^{a_1}y^{a_2}}{a_1^2a_2^g2}=-\int_0^1\frac{x dt}{1-xt}\log{t}\text{Li}_2(xyt)$. Goncharov's condition assures that the function $\text{Li}_{2,2}$ is absent.}. Also, investigating symmetry properties of the symbol with respect to permutations of its arguments it is possible to find the precise combination of classical polylogarithms. The same notions will appear in explicit calculations of form factor remainders in \sref{sec:ff-two-loops}.

When trying to recover a function from its symbol, it is useful to use the notion of the \emph{coproduct} introduced in \cite{Golden:2014xqa}. The idea is, instead of tackling the complete symbol at once, to identify which parts of it correspond to functions of highest degree possible (same as the number of entries in the symbol) and which are products of functions with lower transcendentality. For instance, in \cite{Golden:2014xqa} the authors define a projector $\rho$ which acts on an $m$-fold tensor product and gives a non-zero result only if the function cannot be written as a product of simpler functions\footnote{The idea behind it is that the symbol of products of functions are given in terms of a \emph{shuffle product} and the projector $\rho$ is defined such that it  annihilates any shuffle product. For details, see \cite{Golden:2014xqa}.}. Its action is defined via the recursion
\begin{align}
\label{eq:projector-rho}
\begin{split}
\rho(a_1\otimes\cdots\otimes a_m)\,&\equiv\, \frac{m-1}{m}\big[\rho(a_1\otimes\cdots\otimes a_{m-1})\otimes a_m -\rho(a_2\otimes\cdots\otimes a_m)\times a_1 \big]\,,\\
\rho(a_i)\,&\equiv\,a_i\ .
\end{split}
\end{align}
For a detailed definition of the coproduct we indicate the original work of \cite{Golden:2014xqa}, and also the explicit form factor example considered in \sref{sec:ff-two-loops}.

The notions mentioned above can also be formulated in the context of form factors. In particular, a remainder function was defined in \cite{Brandhuber:2012vm} and computed for the two-loop form factor of the chiral part of the stress tensor multiplet. In \sref{sec:ff-two-loops}, we will compute the remainder function of an infinite class of operators called half-BPS operators (see \sref{sec:BPS} for more details). There the use of symbols will be extremely fruitful, and will allow substantial simplification, similar to that of GSVV, of the form factor remainders.

\label{sec:on-shell-methods}

\section{The $\N=4$ superconformal algebra}
\label{sec:N=4 algebra}

In this section, we will present general aspects of the $\N=4$  superconformal algebra in four dimensions that will be relevant for the discussions on form factors and the dilatation operator.
The conventions are taken from \cite{Dolan:2002zh}.

A superconformal algebra is a combination of the regular conformal algebra with the (Poincar\'{e}) SUSY algebra whose closure require the addition of extra generators called superconformal charges. Let us do it step by step. The Poincar\'{e} algebra is generated by spacetime translations ($P_\mu$) and Lorentz transformations (rotations and boosts, $M_{\mu\nu}=-M_{\nu\mu}\,,\;\mu,\nu=0,1,2,3 $). They satisfy the following commutation relations:
\begin{align}
\label{eq:poincare}
\begin{split}
[M_{\mu\nu},P_{\rho}]\,&=\,-i\,(\eta_{\mu\rho}P_{\nu}-\eta_{\nu\rho}P_{\mu})\ ,\\
[M_{\mu\nu},M_{\rho\sigma}]\,&=\,-i(\eta_{\mu\sigma}M_{\nu\rho}+\eta_{\nu\rho}M_{\mu\sigma}-\eta_{\nu\sigma}M_{\mu\rho}-\eta_{\mu\rho}M_{\nu\sigma})\ ,
\end{split}
\end{align}
where $\eta_{\mu\nu}=\text{diag}(1,-1,-1,-1)$ is the Minkowski metric.
The conformal algebra is generated by augmenting \eqref{eq:poincare} with special conformal transformations $K_\mu$ (also called conformal boosts) and spacetime dilatations $D$. The additional commutation relations are the following,
\begin{align}
\label{eq:conformal-algebra}
\begin{split}
[D,P_\mu]\,=\,iP_\mu\,,\quad &[D,M_{\mu\nu}]\,=\,0\,,\quad [D,K_\mu]\,=\,-iK_\mu\,,\\
[M_{\mu\nu},K_{\rho}]\,&=\,-i\,(\eta_{\mu\rho}K_{\nu}-\eta_{\nu\rho}K_{\mu})\,,\\
[P_\mu,K_\nu]\,&=\,2i\,(M_{\mu\nu}+\eta_{\mu\nu}D)\ .
\end{split}
\end{align}

The action of the dilatation operator on a local scalar operator $\O(x)$ is given by
\begin{equation}
[D,\mathcal{O}(x)]\,=\,i\left(\Delta+x\frac{\partial}{\partial x}\right)\O(x)\ ,
\end{equation}
where $\Delta$, the eigenvalue of $D$ acting on $\O(0)$, is the \emph{conformal dimension} of $\O(x)$. The \emph{bare dimension} $\Delta_0$ of a composite operator is simply the sum of the dimensions of its fundamental constituent fields. For instance in $d=4$ the dimension of the fundamental fields can be read off from the Langrangian density by requiring that all kinetic terms have mass dimension four. Denoting the dimension of a generic field $\Psi$ by $[\Psi]$, scalar fields, fermions and the field strength have dimensions, respectively,
\begin{equation}
[\phi]\,=\,1\,,\quad [\psi]\,=\,[\bar\psi]\,=\,3/2\,,\quad [F_{\mu\nu}]\,=\,2\ .
\end{equation}
In interacting theories,  $\Delta$ gets quantum corrections called \emph{anomalous dimensions}. This topic will be explained in detail in \sref{sec:dilatation}.

Due to the commutation relations between $D$ and the other conformal generators (first line of \eqref{eq:conformal-algebra}), it follows that
\begin{align}
\label{eq:raising-lowering}
[D,[K_\mu,\O(0)]]\,=\,i(\Delta-1)\O(0)\,,\quad [D,[P_\mu,\O(0)]]\,=\,i(\Delta+1)\O(0)\ ,
\end{align}
and thus $P_\mu/K_{\mu}$ act as raising/lowering operators for the conformal dimension, respectively. Together they generate a representation of the conformal group whose highest weight state is called a \emph{conformal primary} operator $\widetilde{\O}(x)$. When evaluated at the origin $x=0$, it is annihilated by $K_\mu$,
\begin{equation}
\label{eq:conf-primary}
[K_\mu,\widetilde{\O}(0)]\,=\,0 \ .
\end{equation} 
The action of a sequence of $P_\mu$ generates an infinite tower of \emph{descendant} operators which are obtained from $\widetilde{\O}$ by taking derivatives, i.e. $\mathcal{O}_{\mu_1\mu_2\cdots\mu_n}\,=\,\partial_{\mu_1}\partial_{\mu_2}\cdots\partial_{\mu_n}\widetilde{\O}$.

In a superconformal theory, in addition to \eqref{eq:conformal-algebra} there are \emph{supercharges}, which are fermionic generators $Q^A_{\alpha},\,\bar{Q}_{\dot{\alpha}A}$ and $S^{\alpha}_{A},\,\bar{S}^{\dot{\alpha} A}$, where $A\,=\,1,\dots \N$ classifies the number of supersymmetries. From now on we will use $\N=4$ which is the relevant case for the remaining chapters.
The $Q,\bar{Q}$ generators together with \eqref{eq:poincare} form a closed algebra which is called Poincar\'{e} supersymmetry. To begin with, it is helpful to write the generators in terms of spinor indices, in the same way as in \sref{sec:SH_Formalism}. Using the Pauli matrices $\sigma^{\mu}_{\alpha\dot\alpha}$ and $(\bar{\sigma}^{\mu})^{\dot\alpha \alpha}= \epsilon^{\alpha\beta}\sigma^{\mu}_{\beta\dot\beta}\epsilon^{\dot\alpha\dot\beta}$, the generators of translations, conformal boosts and Lorentz transformations are represented as
\begin{align}
\begin{split}
P_{\alpha\dot\alpha}\,&=\,\sigma^{\mu}_{\alpha\dot\alpha}P_\mu\,,\qquad\qquad\qquad\; \bar{K}^{\dot\alpha \alpha}\,=\,(\bar{\sigma}^{\mu})^{\dot\alpha \alpha}K_\mu\ ,\\
M_{\alpha}^{\phantom{\alpha}\beta}\,&=\,-\frac{i}{4} \sigma_{\alpha\dot\alpha}^{\mu}(\bar{\sigma}^{\nu})^{\dot\alpha\beta} M_{\mu\nu}\,,\quad \bar{M}_{\phantom{\beta}\dot\alpha}^{\dot\beta}\,=\,-\frac{i}{4}(\bar{\sigma}^{\mu})^{\dot\beta \alpha}\sigma_{\alpha\dot\alpha}^{\nu}M_{\mu\nu}
\end{split}
\end{align}

\noindent The additional non-zero (anti-)commutation relations are
\begin{align}
\label{eq:Poincare-susy}
\begin{split}
&\{Q_{\alpha}^{A},\bar{Q}_{\dot\alpha B}\}\,=\,2 P_{\alpha\dot\alpha}\delta^A_{\phantom{A}B} \,,\\
&[M_{\alpha}^{\phantom{\alpha}\beta},Q_{\rho}^{A}]\,=\,\delta_{\rho}^{\phantom{\alpha}\beta} Q_{\alpha}^{A}-\frac{1}{2}\delta_{\alpha}^{\phantom{\alpha}\beta} Q_{\rho}^{A}\,,\quad [M^{\dot\alpha}_{\phantom{\alpha}\dot\beta},\bar{Q}_{\dot\rho A}]\,=\,-\delta^{\alpha}_{\phantom{\beta}\rho}\bar{Q}_{\dot\beta A}+\frac{1}{2}\delta^{\alpha}_{\phantom{\beta}\beta}\bar{Q}_{\dot\rho A}\ .
\end{split}
\end{align}
The commutators between the supercharges and the momentum operator vanish as a consequence of the independence of $Q/\bar{Q}$ on the spacetime coordinates (they are global). 

Finally, the superconformal algebra is the conjunction of \eqref{eq:conformal-algebra} and \eqref{eq:poincare}. Closure of the algebra demands the existence of a second set of supercharges~-- called \emph{superconformal charges}~--  $S_A^{\alpha}/\bar{S}^{\dot\alpha A}$ which are obtained by the action of $K_\mu$ on the supercharges $Q/\bar{Q}$,
\begin{align}
[K^{\mu},Q^A_{\alpha}]\,=\,-\sigma^{\mu}_{\alpha\dot\alpha} \bar{S}^{\dot\alpha A}\,,\quad [K^{\mu},\bar{Q}_{\dot\alpha A}]\,=\,S^{\alpha}_{A}\sigma^{\mu}_{\alpha\dot\alpha}\,,\quad  \{\bar{S}^{\dot\alpha A},S^{\alpha}_{B}\}
\,=\,2\delta^A_{\phantom{A}B} \bar{K}^{\dot\alpha \alpha}\,,
\end{align}
as well as the $SU(4)\cong SO(6)$ $R$-symmetry generators $R_{AB},\,A,B=1,\dots,4$. The commutation relations between the dilatation operator and the supercharges $Q,\bar{Q}/S,\bar{S}$ reveal their scaling dimensions to be $+1/2$ and $-1/2$, respectively, 
\begin{align}
\label{eq:dilatation-supercharges}
\begin{split}
&[D,Q_{\alpha}^{A}]\,=\,\frac{i}{2} Q_{\alpha}^{A}\,,\quad [D,\bar{Q}_{\dot\alpha A}]\,=\,\frac{i}{2} \bar{Q}_{\dot\alpha A}\,,\quad [D,S^{\alpha}_{A}]\,=\,-\frac{i}{2} S^{\alpha}_{A}\,,\quad [D,\bar{S}^{\dot\alpha A }]\,=\, -\frac{i}{2} \bar{S}^{\dot\alpha A}\ .
\end{split}
\end{align}
A particular anti-commutation relation that is crucial for the discussion in \sref{sec:BPS} is that between $Q$ and $S$,
\begin{align}
\begin{split}
&\{Q_{\alpha}^{A},S^{\beta}_B\}\,=\,4\big[ \delta^{A}_{\phantom{A}B} (M_\alpha^{\phantom{\alpha}\beta}-\frac{i}{2}\delta_\alpha^{\phantom{\alpha}\beta}D)-\delta_\alpha^{\phantom{\alpha}\beta} R^{A}_{\phantom{A}B}\big]\ .
\end{split}
\end{align}
The symmetry group of $\N=4$ SYM is $PSU(2,2|4)$ whose maximal bosonic subgroup is the Lorentz $SU(2)_L \times SU(2)_R $ times the $R$-symmetry group $SU(4)$.

\section{Half-BPS operators}
\label{sec:BPS}

\emph{Superconformal primary operators} are defined as the operators with lowest conformal dimension. Since according to \eqref{eq:dilatation-supercharges} the superconformal charges lower the dimension by half a unit, superconformal primary operators obey, in addition to \eqref{eq:conf-primary},
\begin{equation}
\label{eq:superconformal-primary}
[S^{\alpha}_A,\widetilde{O}(0)]\,=\,0\,,\quad [\bar{S}^{\dot\alpha A},\widetilde{O}(0)]\,=\,0\ .
\end{equation}
A special situation occurs when a superconformal primary operator is annihilated by one or more extra SUSY generators. For instance,  for some $Q_{\alpha}^{A}$ it obeys
\begin{equation}
[Q_{\alpha}^{A},\widetilde{O}(0)]\,=\,0\ .
\end{equation}
In the following chapters we will be interested in scalar operators. In this case it follows from \eqref{eq:superconformal-primary} and \eqref{eq:superconformal-primary} that
\begin{align}
\label{eq:BPS}
\begin{split}
&[\{Q_{\alpha}^{A},S^{\beta}_B\},\widetilde{O}(0)]\,=\,0\\\Rightarrow \quad
&4\,\delta^{A}_{\phantom{A}B} \underbrace{[M_\alpha^{\phantom{\alpha}\beta},\widetilde{\O}(0)]}_{=\,0\text{ for scalar }\widetilde{O}} - 2i\, \delta^{A}_{\phantom{A}B}\, \delta_\alpha^{\phantom{\alpha}\beta}[D,\widetilde{\O}(0)] - 4\, \delta_\alpha^{\phantom{\alpha}\beta}[ R^{A}_{\phantom{A}B},\widetilde{\O}(0)]\,=\,0\ .
\end{split}
\end{align}
Therefore the conformal dimension and $R$-charge of $\widetilde{\O}$ are related. A remarkable consequence of this relation is that the conformal dimensions of these operators, called \emph{BPS} operators or \emph{chiral primary operators} (CPO), do not receive quantum corrections (and thus the operators are said to be \emph{protected}). This is the case because the $R$-charges are integers while the anomalous dimensions are smooth functions of the coupling constant. Thus, for \eqref{eq:BPS} to hold, $\Delta=\Delta_0$ for any value of the coupling constant. BPS operators are said to give rise to \emph{short representations} since their multiplets are constrained by additional SUSY generators (even though the representations are still infinite-dimensional). 

In Chapter \ref{ch:formfactors} we will consider form factors of half-BPS operators, that is, operators which preserve half of the SUSY generators. One example is the scalar bilinear half-BPS operator in $\N=4$ SYM defined as
\begin{equation}
\label{eq:bilinear-BPS}
\O_{ABCD} \equiv
{\rm Tr} (\phi_{AB} \phi_{CD})  \, - \,  (1/12) \,  \eps_{ABCD} {\rm Tr} ( \bar\phi^{LM} \phi_{LM} )\ ,
\end{equation}
where
\begin{equation}
\label{eq:phi-bar}
\bar \phi^{AB} \equiv \tfrac{1}{2} \eps^{ABCD} \phi_{CD}\ .
\end{equation} 
This operator  belongs to the  $\mathbf{20}^\prime$ representation of the  $SU(4)$  $R$-symmetry group.

\section{The dilatation operator}
\label{sec:dilatation}

Conformal field theories (CFTs) have, by definition, no mass spectrum. The usual way one thinks of states in euclidean CFTs is through the map between states and local operators inserted at the origin,
\begin{equation}
\label{eq:op-state-correspondence}
|\O\rangle\,=\,\lim_{x\rightarrow 0}\O(x)|0\rangle\ .
\end{equation}
This correspondence is inherent to CFTs because it relies on a map between $\mathbb{R}^d$ and the cylinder $S^{d-1}\times \mathbb{R}_{\rm time}$, under which the origin of $\mathbb{R}^d$ is mapped to the far past in the cylinder. In this correspondence, the time evolution in the cylinder corresponds to the \emph{dilatation operator} on $\mathbb{R}^d$, that is, the generator of rescaling of spacial coordinates,
\begin{equation}
\label{eq:dilatation}
x^\mu\, \rightarrow\, \lambda x^\mu\ .
\end{equation}
For this reason, the analogous notion of a mass spectrum in a CFT is the conformal dimension of local operators, which dictates how they transform under a dilatation. A scalar local operator with dimension $\Delta$, denoted by $\O_{\Delta}(x) $, transforms under \eqref{eq:dilatation} like
\begin{equation}
\O_{\Delta}(x)\,\rightarrow\, \lambda^{-\Delta}\O_{\Delta}(\lambda x) \ .
\end{equation}
The conformal dimension of a scalar operator $\O_{\Delta}$ can be read off from the two point function between itself and its conjugate, which is fixed by conformal symmetry to be\footnote{In general, two-point functions of different operators with definite anomalous dimension $\O_{\Delta_1}(x)$ and $\O_{\Delta_2}(y)$ are given by $\langle \mathcal{O}_{\Delta_1}(x)\mathcal{O}_{\Delta_2}(y)\rangle\, =\, \dfrac{\delta_{\Delta_1\Delta_2}}{|x-y|^{2\Delta_1}}$.}
\begin{align}
\label{eq:two-pt-function}
\langle \mathcal{O}_{\Delta}(x)\bar{\mathcal{O}}_{\Delta}(y)\rangle\, =\, \frac{1}{|x-y|^{2\Delta}} \ .
\end{align}
For a free theory $\Delta$ coincides with the bare dimension $\Delta_0$. However, for interacting theories the scaling dimension gets renormalised. This happens because the two-point functions suffer from UV divergences arising from the integration over the interaction points. For small values of the coupling constant, the first correction is a small perturbation of the bare dimension,
\begin{equation}
\label{eq:anomalous dimension}
\Delta\,=\,\Delta_0+\gamma\,,\qquad \gamma\ll\Delta_0\ .
\end{equation}
The factor $\gamma$ is called the \emph{one-loop anomalous dimension}. In this case, \eqref{eq:two-pt-function} can be expanded as
\begin{align}
\label{eq:two-point-function-divengent}
 \langle \mathcal {O}_{\Delta}(x)\bar{\mathcal{O}}_{\Delta}(y)\rangle\, =\, \frac{1}{|x-y|^{2\Delta}} \,=\, \frac{1}{|x-y|^{2\Delta_0}} \big[1-\gamma\,\log(|x-y|^2\Lambda^2)\,+\,\dots \big]\ ,
\end{align}
where $\Lambda$ is the UV cutoff scale. When computing two-point functions in interacting theories, one generally finds that the UV divergences are not always proportional to the initial tree-level correlator, but instead receive contributions of tree-level two-point functions of different operators. This is referred to as the \emph{mixing problem} and as a consequence one should indeed compute a \emph{matrix} of anomalous dimensions. For this reason, the
dilatation operator is represented as an expansion in the 't Hooft coupling $\lambda$ as
\begin{align}
\label{eq:dilatation-op}
D\,=\,\sum_{n=0}^\infty \lambda^n D^{(2n)}\ ,
\end{align}
where the eigenvalues of $D^{(0)}$ are the bare dimensions of operators, the eigenvalues of $D^{(2)}$ are the one-loop anomalous dimensions and so forth. Normally in the literature $D^{(2)}$ is represented by the letter $\Gamma$.

Therefore, to be precise, \eqref{eq:two-point-function-divengent} is only valid for operators said to have \emph{definite anomalous dimension}, and the one-loop anomalous dimension $\gamma$ entering a ``diagonal'' two-point function is the corresponding eigenvalue of the matrix $\Gamma$. The operators with definite anomalous dimension are linear combinations of single trace operators that diagonalise $\Gamma$. So the idea behind the solution to the spectral problem is to, at one loop,
\begin{enumerate}
\item Find the matrix of anomalous dimensions $\Gamma$, also called the \emph{one-loop dilatation operator},
\item Find the eigenvalues of $\Gamma$, that is, the spectrum of anomalous dimensions.
\item Find the eigenvectors of $\Gamma$, that is, the operators with definite anomalous dimension.
\end{enumerate}
The solution to the mixing problem is in general very hard. Fortunately there are some cases where a set of operators only mix among themselves at a given order in perturbation theory. These are called \emph{closed sectors}. Table \ref{tab:closed-sectors} shows two closed sectors that will be studied later in Chapter \ref{ch:dilatation}: $SO(6)$ and $SU(2|3)$ at one loop. They consist of composite local operators formed of a particular set of fundamental fields, or letters.
\begin{table}[h]
\centering
\begin{tabular}{c|c}
Sector & Letters \\[3pt]
\hline
$SO(6)$  & $\qquad\phi_{AB}\,,\;\; A,B\,=\,1,2,3,4$\\[5pt]
$SU(2|3)$  & $\big\{\psi_{1\,\alpha}, \phi_{1A}\big\}\,,\;A=2,3,4$   
\end{tabular}
\caption{\textit{Two closed sectors of the mixing problem at one loop that will be studied in Chapter \ref{ch:dilatation}.}}
\label{tab:closed-sectors}
\end{table}

The solution to the spectral problem was revolutionised by Minahan and Zarembo (MZ) in \cite{Minahan:2002ve} where they showed that the one-loop dilatation operator in the $SO(6)$ sector is equivalent to the Hamiltonian of a spin chain with nearest-neighbour interactions and, moreover, this Hamiltonian is \emph{integrable}. In this picture, single trace operators are mapped to a periodic spin chain where each site carries an $SO(6)$ vector index. For illustrative purposes, we briefly present the main results of MZ.

\noindent Generic operators in the $SO(6)$ sectors are of the form
\begin{align}
\label{eq:SO(6)-operators}
\O_{A_1 B_1, A_2 B_2, \ldots, A_L B_L} (x) \ \equiv \ \Tr \big( \phi_{A_1 B_1} (x) \cdots  \phi_{A_L B_L} (x) \big) \ . 
\end{align}
According to \eqref{eq:two-point-function-divengent}, to obtain the one-loop dilatation operator one must investigate the UV divergent part of the two-point function (suppressing indices),
\begin{align}
 \langle \mathcal {O}(x_1)\bar{\mathcal{O}}(x_2)\rangle\Big|_{\rm UV}^{\text{one-loop}} \ .
\end{align}
In the planar limit and at one loop, only interactions between  scalar fields which are adjacent in colour space are relevant, and thus one can equivalent study the two-point function
\begin{align}
\label{eq:only-two-scalars}
\left\langle (\phi^a_{AB} \phi^b_{CD})(x_1)(\phi^c_{A^\prime B^\prime} \phi^d_{C^\prime D^\prime})(x_2)  \right\rangle\Big|_{\rm UV}^{\text{one-loop}} \ ,
\end{align}
where $a,b,c,d$ are $SU(N)$ colour labels (note that only the full operator is gauge invariant) and we used \eqref{eq:phi-bar}. An equivalent statement is that the dilatation operator can be expanded as a sum of operators acting on two adjacent sites at a time,
\begin{align}
\Gamma \,=\,  \frac{\lambda}{8\pi^2} \sum_{i=1}^L \Gamma_{i\, i+1}\,,\qquad \Gamma_{L\,L+1}\equiv \Gamma_{L\,1}\ ,
\end{align}
and thus it is enough to study, at one loop, only a two-site operator $\Gamma_{i\,i+1}$. 

\begin{table}[h]
\centering
\begin{tabular}{ccccl}

$\mathord{\parbox[c]{1em}{\includegraphics[width=3cm]{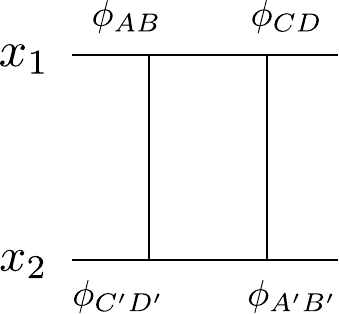}}}$ & \hspace{4cm}  & $\epsilon_{ABC'D'}\epsilon_{CDA'B'}$ & \hspace{1cm} & Identity ($\uno$) \\[2cm]
$\mathord{\parbox[c]{1em}{\includegraphics[width=3cm]{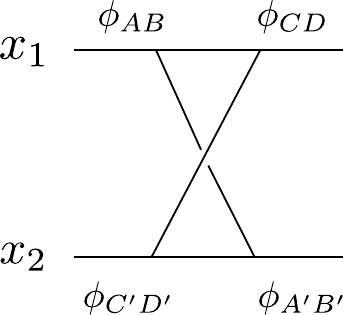}}}$ & \hspace{4cm}
& $\epsilon_{ABA'B'}\epsilon_{CDC'D'}$ & \hspace{1cm} & Permutation ($\mathbb{P}$) \\[2cm]
$\mathord{\parbox[c]{1em}{\includegraphics[width=3cm]{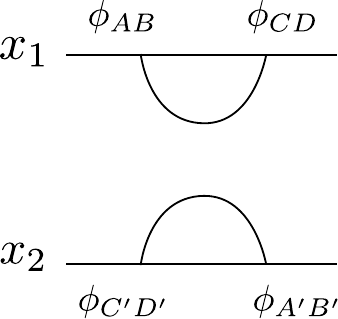}}}$ & \hspace{4cm} & $\epsilon_{ABCD}\epsilon_{A'B'C'D'}$ & \hspace{1cm} & Trace ($\Tr$) 
\end{tabular}
\caption{\textit{The three possible ways to contract the $R$-symmetry indices of the two-point function \eqref{eq:only-two-scalars}. For planar contractions at tree level only the identity is allowed, whereas at one loop all three structures are present.}}
\label{tab:R-contractions}
\end{table}

\begin{figure}[h]
\centering
\includegraphics[width=0.8\linewidth]{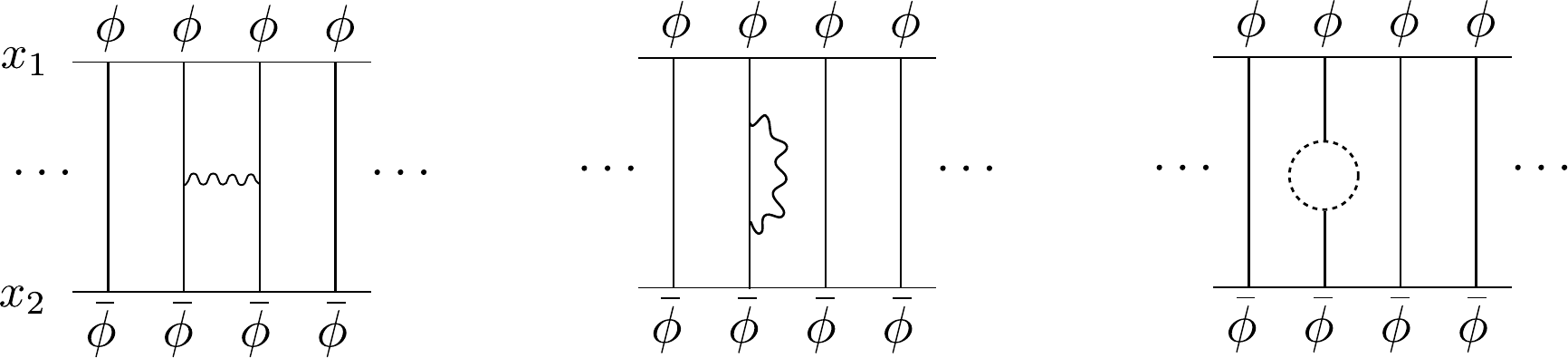}
\caption{\textit{Feynman diagrams that contribute to the identity part of the dilatation operator}~\cite{Minahan:2002ve}.}
\label{fig:one-loop-identity}
\end{figure}

There are three possible ways to contract the $R$-symmetry indices of \eqref{eq:only-two-scalars}, shown in Table \ref{tab:R-contractions}. At tree level, the only planar contraction is the identity, whereas at one loop also the permutation and trace structures contribute to the dilatation operator.
Considering all possible interactions of the theory, MZ observed that self-energy diagrams and terms where the scalars exchange a gluon (shown in Figure \ref{fig:one-loop-identity}) contribute only to the identity part and can be fixed by imposing that $\gamma=0$ for $R$-symmetry assignments corresponding to a protected operator.

\begin{figure}[h]
\centering
\includegraphics[width=0.15\linewidth]{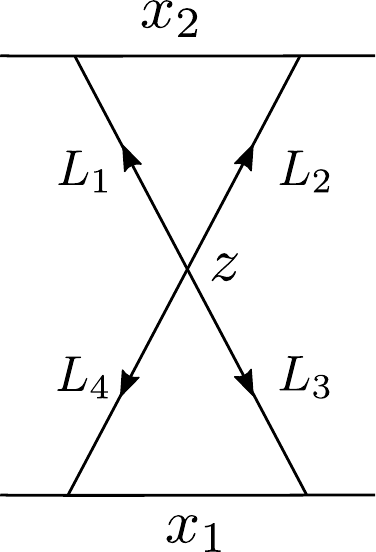}
\caption{\it The particular one-loop integral in configuration space contributing to the dilatation operator.}
\label{fig:MZintegral}
\end{figure}

The only interaction that contributes to the permutation and trace structures at one loop comes from the term involving four scalars in the Lagrangian of $\N=4$ SYM. This term is of the form
\begin{equation}
V_{\text{scalar}} \,\sim\, g^2 \text{Tr}([\phi_{AB},\phi_{CD}][\bar{\phi}^{AB},\bar{\phi}^{CD}])\ .
\end{equation}
So the only integral to consider corresponds to the interaction between four scalar fields, depicted in Figure \ref{fig:MZintegral}.
It is given by
\begin{align}
\label{eq:X-position}
I(x_{12})\,=\,\int\!\!d^d z \ \D^2 (x_1 - z) \, \D^2(x_2-z)\, , 
\end{align}
where $x_{12}\equiv x_1 - x_2$ and  
\begin{align}
\D  (x)  \ \equiv \ 
- {\pi^{2 - {d\over 2}} \over 4 \pi^2}
\Gamma \Big(\frac{d}{2} - 1 \Big)
 {1\over (x^2)^{{d\over 2} - 1}}
\ , 
\end{align}
is the (Euclidean) scalar propagator in $d$ dimensions. 

Note that $I(x_{12})$  has UV divergences arising from the regions $z\rightarrow x_1$ and $z\rightarrow x_{2}$.
The result for the one loop dilatation operator found by MZ is
\begin{align}
\label{eq:one-loop-SO(6)}
 \Gamma\,=\,  \frac{\lambda}{8\pi^2} \sum_{n=1}^L  (\uno - \mathbb{P}+\tfrac{1}{2} \text{Tr})_{n\,n+1}\ .
\end{align}

The discovery of this underlying spin chain introduced a completely new perspective to the spectral problem, and techniques used in the context of integrable systems~---~the various kinds of Bethe ans\"{a}tze~---~could now be applied for $\N=4$ SYM. This illustrates how special $\N=4$ SYM is; integrability~---~factorisation of the $S$-matrix into a sequence of $2\rightarrow 2$ scattering processes~---~is usually thought of as a phenomenon intrinsic to two-dimensional systems, and is unlikely to feature in a four-dimensional theory. There is, however, a hidden two-dimensionality in $\N=4$ SYM which can be thought of as a spin chain \cite{Beisert:2003yb}, or indeed the two-dimensional worldsheet of the dual string theory picture. Integrability in the context of the AdS/CFT duality has been largely studied and a detailed review is contained in \cite{Beisert:2010jr}.

Since the discovery of MZ, the dilatation operator has been extensively studied, and it is known completely at one loop \cite{Beisert:2003jj,Beisert:2003yb}.
At higher loop order, the $SU(2|3)$ sector remains closed at, but the $SO(6)$ sector does not. Direct perturbative calculations at higher loops~---~without the assumption of integrability~---~have been performed only up to two   \cite{Eden:2005bt,Belitsky:2005bu,Georgiou:2011xj}, three  
\cite{Beisert:2003ys, Eden:2005ta,Sieg:2010tz} and four loops \cite{Beisert:2007hz}. 


The aim of the work presented in Chapter \ref{ch:dilatation} is to establish a connection between the on-shell methods presented in \sref{sec:on-shell-methods} and the dilatation operator. Inspired by \cite{Koster:2014fva}, where the one-loop dilatation operator in the $SO(6)$ sector \eqref{eq:one-loop-SO(6)} was rederived in twistor space, we do the same using MHV rules in \sref{sec:MHV-diagrams} and, subsequently, using generalised unitarity~---~thus only on-shell information~---~in the $SO(6)$ and $SU(2|3)$ sectors.

\chapter[Form factors of half-BPS operators]{Form factors of half-BPS operators}
\label{ch:formfactors}

\section{Introduction}

Recently, there has been a resurgence of interest in the study of form factors in $\N=4$ SYM. One reason behind this is that, as discussed in Chapter \ref{ch:Introduction}, form factors interpolate between fully on-shell quantities, i.e.~scattering amplitudes, and correlation functions, which are  off shell. Indeed, recalling the definition presented in Chapter \ref{ch:Introduction}, a form factor is obtained by taking a gauge-invariant, local operator $\O (x)$ in the theory, applying it to the vacuum $|0\rangle$, and considering the overlap with a multi-particle state $\langle 1, \ldots , n |$, as in \eqref{eq:FF-definition},
\begin{equation}
F_\O(1, \ldots, n;q) \, \equiv \,   \langle 1, \ldots , n |\cO (0)  |0\rangle\ ,
\end{equation}

Once we fix a certain operator, one can study how the form factor changes as we vary the state. In a pioneering paper \cite{vanNeerven:1985ja} almost thirty years ago, van Neerven considered the simplest form factor of  the operator $\Tr (\phi_{12}^2) $, namely the two-point (also called Sudakov) form factor, deriving its expression at one and two loops. Operators of the kind $\Tr(\phi_{12}^k)$ are called \emph{half-BPS} operators, reviewed in \sref{sec:BPS}. These operators are special, and in particular have their scaling dimension protected from quantum corrections.

More recently, the computation of form factors at strong coupling was considered in \cite{Alday:2007he,Maldacena:2010kp}, and at weak coupling in a number of papers  in $\N=4$ SYM \cite{Brandhuber:2010ad,Bork:2010wf,Brandhuber:2011tv,Bork:2011cj,Broedel:2012rc,Henn:2011by,Gehrmann:2011xn,Brandhuber:2012vm,Bork:2012tt,Engelund:2012re,Boels:2012ew,Bork:2014eqa,Wilhelm:2014qua,Nandan:2014oga,Loebbert:2015ova,Bork:2015fla,Frassek:2015rka,Boels:2015yna,Huang:2016bmv,Koster:2016ebi,Koster:2016loo} and also in ABJM theory\footnote{Aharony-Bergman-Jafferis-Maldacena (ABJM) theories are three-dimensional $\N=6$ Chern-Simmons theories constructed in \cite{Aharony:2008ug}. They display many special features analogous to $\N=4$ SYM, for instance a 't Hooft limit as well as Yangian symmetry in the planar limit.} \cite{Brandhuber:2013gda,Young:2013hda,Bianchi:2013pfa}. In particular, in \cite{Brandhuber:2010ad} it was pointed out that on-shell methods can successfully be applied to the computation of such quantities, and the expression for the infinite sequence of MHV form factors of the simplest dimension-two, scalar half-BPS operators was computed. Perhaps unsurprisingly, this computation revealed the remarkable simplicity of this quantity~---~for instance, the form factor of two scalars and $n-2$ positive-helicity gluons is very reminiscent of the  Parke-Taylor MHV amplitude \eqref{eq:MHV},  
\begin{equation}
\lan  g^+(p_1) \cdots  \phi_{12} (p_i) \cdots    
 \phi_{12} (p_j) \cdots g^+ (p_n) | \O (0) | 0 \ran \, = \, {\lan ij\ran^2 \over \lan 12\ran \cdots \lan n1\ran }
\ , 
\end{equation}
where $\O \equiv \Tr  (\phi_{12}^2)$.    These form factors maintain this simplicity also at one loop~---~they are proportional to their tree-level expression, multiplied by a sum of one-mass triangles and two-mass easy box functions\footnote{See Appendix \ref{app:scalar-integrals} for the definition of these integral functions.}. Other common features between form factors and amplitudes include the presence of a version of colour-kinematics duality \cite{Boels:2012ew} similar to that of BCJ \cite{Bern:2008qj}, and the possibility of computing form factors at strong coupling using Y-systems \cite{Maldacena:2010kp,Gao:2013dza} which extend those of the  amplitudes  \cite{Alday:2010vh}.
 A second motivation to study form factors is therefore to explore to what extent their  simplicity is preserved as we vary the choice of the operator and of the external state. 

There are interesting distinctive features of form factors as compared to scattering amplitudes. One of them is the presence of non-planar integral topologies in their perturbative expansion. Indeed, the presence of a colour-singlet   operator introduces an element of non-planarity in the computation even when we consider   external states that are colour ordered, as  is usual in scattering amplitudes. Specifically, the external leg carrying  the momentum of the operator does not participate in the colour ordering, and hence non-planar integrals are expected to appear  at loop level. Even the simple  two-loop Sudakov form factor of  \cite{vanNeerven:1985ja} is expressed in terms of a planar as well as a non-planar two-loop triangle integral. In general, non-planar contributions for single trace form factors of $\Tr(\phi_{12}^k)$ arise at $k^{\rm th}$ loop order.

One may wonder if higher-loop corrections can spoil the simple structures observed at tree level and one loop. There is a number of examples which indicate that, fortunately, this is not the case. For instance, in \cite{Gehrmann:2011xn} the three-loop corrections to the Sudakov form factor were computed and found to be given by a maximally transcendental expression. 
%
Exponentiation of the infrared divergences leads one to define a finite remainder function in the same spirit of the BDS remainder function \eqref{eq:2-loop-remainder} \cite{Bern:2008ap,Drummond:2008aq}. Using the concept of the symbol of a transcendental function  \cite{Goncharov:2010jf} as well as various physical constraints, it was found that the form factor remainder is given by a remarkably simple, two-line expression written in terms of classical polylogarithms only. Moreover, the remainder function  was found to be closely related to the analytic expression of the MHV amplitude six-point remainder at two-loops found in \cite{Goncharov:2010jf}. 

Similarly to the miraculous simplifications which occur in going from the result of an explicit calculation \cite{DelDuca:2010zg} to the expression of  \cite{Goncharov:2010jf}, the (complicated) two-loop planar and non-planar functions found in \cite{Brandhuber:2012vm} combined into a maximally transcendental, compact result. Surprising agreement was furthermore found between this form factor and the maximally transcendental part of certain very different quantities, namely  the Higgs plus three-gluon amplitudes in QCD computed in \cite{Gehrmann:2011aa}. A hint of a possible connection between such unrelated quantities (and a further reason to study half-BPS form factors in $\N=4$ SYM) is that the top component of the stress-tensor multiplet operator (of which $\Tr \, (\phi_{12}^2)$ is the lowest component) is the on-shell Lagrangian of the theory, which contains the term  $\Tr \, F_{\rm SD}^2$, where $F_{\rm SD}$ is the self-dual part of the field strength. In turn, it is known that Higgs plus multi-gluon amplitudes in the large top mass limit can be obtained from  an effective interaction of the form $H \,\Tr \, F_{\rm SD}^2$, shown in \fref{fig:higgs-gluon}. 

Incidentally, we note that form factors can be used to compute correlation functions using generalised unitarity as in \cite{Engelund:2012re,Engelund:2015cfa}. They also appear in the intermediate sums defining total cross sections, or the event shapes considered in \cite{Belitsky:2013ofa,Belitsky:2013bja,Belitsky:2013xxa}, and in the computation of the dilatation operator in \cite{Wilhelm:2014qua,Loebbert:2015ova}.

In this chapter we concentrate on the calculation of form factors of half-BPS operators in $\cN=4$ (SYM).
In particular we look at operators of the form $\cO_k \equiv \Tr (\phi_{12}^k)$, 
with $k>2$,  and their superpartners,  which can be packaged into a single superfield $\cT_k$.  
Here $\phi_{AB}=-\phi_{BA}$ denotes the three complex scalar fields of the theory, satisfying the reality condition $\bar{\phi}^{AB} =(1/2) \eps^{ABCD} \phi_{CD}$, where  $A, \ldots , D$ are $SU(4)$ {\it R}-symmetry indices.

Sudakov form factors of $\cO_2$ (the lowest component of $\cT_2$) have been constructed
up to four loops \cite{vanNeerven:1985ja, Gehrmann:2011xn, Boels:2012ew}, 
while in \cite{Brandhuber:2010ad, Brandhuber:2012vm} form factors of $\cO_2$ with more than two external on-shell states were computed.
Later,   the supersymmetric form factors of $\cT_2$  were presented in \cite{Brandhuber:2011tv,Bork:2011cj} using  harmonic and Nair's on-shell superspace \cite{Nair:1988bq}, extending the results obtained for the bosonic operator $\cO_2$.

The superfield $\cT_k$ is a generalisation of the stress-tensor multiplet $\cT_2$. For $k >2$   it is dual to massive Kaluza-Klein modes of the $\text{AdS}_5 \times S^5$ compactification of type IIB supergravity\footnote{Their four-point functions were studied in \cite{Arutyunov:2002fh}.}, while  for $k = 2$  it is dual to the massless graviton multiplet.

In this chapter we study form factors of $\cO_k$ and super form factors of $\cT_k$ with $k>2$, quoting the results of \cite{Penante:2014sza} and \cite{Brandhuber:2014ica}. For our purposes we find  it convenient to introduce a more concise notation,
\begin{align*}
\F_{k,n}\ &\leftrightarrow\ \F_{\T_k,n}\qquad\, \text{Supersymmetric form factor of } \T_k \ , \\
F_{k,n}\ &\leftrightarrow\ F_{\O_k,n}\qquad\text{Form factor of } \Tr\big[(\phi_{12})^k\big]\ .
\end{align*}
Notice that in order to have  a non-vanishing result for the form factor $F_{k,k}$, all external states must be equal to $\phi_{12}$.

In \sref{sec:ff-tree} we present MHV form factors $\F_{k,n}$ of $\T_k$ with $n$ external legs at tree level and in \sref{sec:ff-one-loop} at one loop.
In \sref{sec:ff-two-loops} we will focus on the special class of  form factors $\cF_{k,k}$ which we call ``minimal" because they have the same number of on-shell legs fundamental fields in $\T_k$.
In the case $k\!=\!n\!=\!2$, called Sudakov, the result has trivial kinematic dependence dictated by dimensional analysis and Lorentz invariance. The minimal form factors $\F_{k,k}$ are close cousins of the Sudakov form factors (and for this reason sometimes we refer to them as Sudakov as well, in a slight abuse of nomenclature) and hence it is natural to expect that their kinematic dependence will be simpler, albeit non-trivial, compared to the general case with $n>k$. Indeed, we will be able to present very compact, analytic expression for arbitrary $n=k$ written in terms of simple, universal building blocks.

\section{Tree level}
\label{sec:ff-tree}

So far, most of the available results are concerned with bilinear half-BPS operators.%
\footnote{With the exception of \cite{Bork:2010wf}, where form factors of operators of the form $\Tr \, (\phi^n)$ were considered with an external state containing the same number $n$ of particles as of fields in the operator.} In this section we will focus on form factors of operators of the form $\Tr \, (\phi^k)$ with an $n$-point external state, for arbitrary $k$ and $n$.
In fact, there is no reason to limit our study to scalar operators, as one can supersymmetrise the scalar operators in a similar fashion as  is done in the case of the stress-tensor multiplet operator.  Thus, the  operator we consider is  
\beq
\label{tk}
\cT_k\  \equiv  \ \Tr[ (W_{++})^k]
\ , 
\eeq
where $W_{++}$  is a particular projection  
of the chiral vector multiplet superfield  $W_{AB}(x, \theta)$ of $\N=4$ SYM,  introduced   in the next section.    
For $k=2$ this is the chiral part of  the stress-tensor multiplet operator.  
$\cT_k$  is a half-BPS operator, and its lowest component is simply  the scalar  operator $\Tr [(\phi_{++})^k]$. 

In \sref{sec:Ward} we review a convenient formalism to study these operators, namely  harmonic superspace  \cite{Galperin:1984av,Galperin:2001uw}.  
We will then consider form factors of the  chiral part of the operators $\cT_k$, which preserve half of the supersymmetries off shell \cite{Eden:2011yp, Eden:2011ku}.  External states will be described  naturally  with the  supersymmetric formalism of   Nair \cite{Nair:1988bq}. One can then write down very simple Ward identities, similar to those considered in \cite{Brandhuber:2011tv} for the case of the stress-tensor multiplet operator, which we can then solve finding constraints on the  expressions for the form factors.

In \sref{sec:susyff} we consider the simplest supersymmetric form factors, namely those of $\cT_3$. Using  BCFW recursion relations \cite{Britto:2004ap,Britto:2005} (in the supersymmetric version of  \cite{ArkaniHamed:2008gz,Brandhuber:2008pf}) we will find a compact expression for the  $n$-point form factor of this operator. Interestingly, the standard recursion relation with adjacent shifts contains a boundary term, hence we are led to use a recursion relation with next-to-adjacent shifts.   

The presence of boundary terms in the adjacent-shift recursion relations for the form factor of $\cT_3$ motivates us in \sref{sec:adjBCFWshift} to study their structure for the case of the form factor of $\cT_k$ for general $k$. This will lead us to propose a new supersymmetric recursion relation for the MHV form factors of $\cT_k$, which involves form factors with different operators, namely $\cT_k$ and $\cT_{k-1}$. We also look at a simple generalisation of this recursion to the case of NMHV form factors. Based on some experimentation for lower values of $k$, we propose a general solution for all $n$-point MHV form factors of $\cT_k$ for arbitrary $k$ and $n$. We also check that our proposed solution satisfies the required cyclic symmetry. 

\sref{sec:MHVrules} briefly shows that MHV diagrams \cite{Cachazo:2004kj} can be extended  to compute form factors of the half-BPS operators considered in this chapter, as a simple extension of the work of  \cite{Brandhuber:2011tv} where MHV rules for the stress-tensor multiplet operator were found. 
We present two examples in detail, namely the calculation of a four-point NMHV form factor using bosonic as well as supersymmetric MHV rules.


\subsection{Super form factors  of $\cT_k$ and Ward identities}
\label{sec:Ward}

In this section we will study the  supersymmetric form factors of the operators $\cT_k$ introduced in \eqref{tk}, 
which generalise  those of the stress-tensor multiplet operator studied in  \cite{Brandhuber:2011tv}.

We begin our discussion  by recalling   that  the states in  the $\N=4$ multiplet can be efficiently  described using the formalism introduced by  Nair \cite{Nair:1988bq}, where all helicity states are packaged into the super-wavefunction \eqref{eq:supermultiplet}.

The  supersymmetric operator we wish to consider is a  generalisation of the chiral part of the stress-tensor multiplet operator $\T_2$. 
It is defined as 
\begin{equation}
\label{T}
\T_k (x, \theta_{+}) \equiv  {\Tr} \big[\big(W_{++} (x, \theta_{+})\big)^k\big]\ , 
\end{equation}
where $W_{++}$ is a particular projection of the chiral vector multiplet superfield  $W_{AB}(x, \theta)$, defined as follows.%
\footnote{We follow closely the notation and conventions of 
\cite{Eden:2011yp,Eden:2011ku}, see also \cite{Brandhuber:2011tv}.} 
We introduce the harmonic projections of  the chiral superspace coordinates  $\theta_A^{\alpha}$  and supersymmetry charges $Q_{\alpha}^A$
as 
\begin{align}
\theta^{\alpha}_{\pm a} \,\equiv \,  \theta_A^{\alpha} u^A_{\pm a}\, \qquad  \ Q^{\pm a}_{\alpha} \,\equiv \, \bar{u}_A^{\pm a} Q^A_\alpha \, .
\end{align}
Here  $a\,=\,1,2$ is an $SU(2)$ index, and the
harmonic $SU(4)$  $u$ and $\bar{u}$ variables are normalised as in Section 3 of
\cite{Eden:2011yp}\footnote{The only difference is that all upper/lower indices are swapped, this is to keep the notation consistent with that of Chapter \ref{ch:Review}}.  Then 
\be
W_{+a+b}\, \equiv  \, u^A_{+a} u^B_{+b}  W_{AB}   \, = \, \eps_{ab} \, W_{++} 
\ . 
\eeq
In particular, the chiral part of the stress-tensor multiplet operator is simply  
\beq
\label{cpst}
\T_2 (x, \theta_{+})\, \equiv  \, \Tr( W_{++}W_{++}) (x, \theta_{+})
\, = \, \Tr ( \phi_{++}\phi_{++})\, + \, \cdots \, + \,  {1\over 3} (\theta_{+})^4 \cL\ . 
\eeq
Note that the $(\theta_+)^0$ component is the scalar operator 
$\Tr (\phi_{++} \phi_{++})$, whereas the $(\theta_+)^4$ component is the chiral on-shell Lagrangian  denoted by $\cL$. In complete analogy to \eqref{cpst}, we have 
\beq
\T_k (x, \theta_{+}) \, = \, \Tr \big[( \phi_{++})^k\big] \, + \, \cdots 
\ . 
\eeq
Ward identities associated to supersymmetry can be used to constrain the expression of the super form factor. This was done in \cite{Brandhuber:2011tv} and we briefly review here this procedure.  
We consider a symmetry generator $s$ that annihilates the vacuum. It then follows that 
\beq
\lan 0 | [s \, ,   \Phi(1) \cdots \Phi (n)\, \O \,] | 0  \ran  \ = \ 0  
\ ,
\eeq
or
\beq
\label{WI}
 \lan 0 |  \Phi(1) \cdots \Phi (n)\,  [s \, , \, \O] \,| 0  \ran \, + \, \sum_{i=1}^n  \lan 0 | \,  \Phi(1) \cdots [s \, , \, \Phi(i) ] \cdots \Phi (n)\, \O | 0  \ran \ = \ 0
\ , 
\eeq
where   $\lan 0 | \Phi(1) \cdots \Phi(n)  $ is the superstate $\lan1 \cdots n |$. In this notation, a form factor is simply $\lan 0 | \Phi(1) \cdots \Phi (n)\,\O \, | 0  \ran$ or, more compactly,  $\lan 1\cdots n | \, \O \, | 0  \ran$.
We are  interested in the action of the supersymmetry charges $Q^{\pm}$, which  are realised  on the half-BPS operators $\cT_k$ as  
\beq
\label{two} [Q^{-}  \, , \, \cT_k (x, \theta_{+} ) ] \ = \ 0
 \ , \qquad
 [Q^{+}  \, , \, \cT_k (x, \theta_{+} ) ] \ = \  i {\partial \over \partial \theta_{+} } \cT_k (x, \theta_{+} )  \ .
 \eeq
The first relation is a simple consequence of the fact that $\cT_k (x, \theta_{+} )$ is independent of $\theta_{-}$, while the second shows that $Q^{+}$ can be used to relate the various components in the supermultiplet described by $\cT_k (x, \theta_{+} )$.\\[10pt]
\noindent We now introduce the object we will compute, i.e.~the (super) Fourier transform of the form factor, 
\beq
\cF_{k,n} (1, \ldots, n;  q, \gamma^{+} )
\ \equiv  \ \int\!\!d^4x\, d^4 \theta_{+} \
e^{-(iq\cdot x + i \theta_+\cdot \gamma^+) } \, \langle  \, 1 \cdots n\,  | \cT_k(x, \theta_+)
\,  | 0   \ran\ \ ,
\eeq
where $\theta_+\cdot \gamma^+ =\theta^\alpha_{+a} \gamma^{+a}_{\alpha}$. 

\noindent The Ward identities \eqref{WI} for $Q^{+}$ and $Q^{-}$ then give
\beqa
\label{WIqpm}
\begin{split}
 \big( \sum_{i=1}^n \lambda^i \eta^{-,i} \big) \ \cF_{k} (1, \ldots, n;q, \gamma^{+}) & = & 0\ ,
 \\
\big( \sum_{i=1}^n \lambda^i \eta^{+, i} \,  -  \, \gamma^{+} \big)
\ \cF_{k} (1, \ldots, n; q, \gamma^{+}) & = & 0 \, ,
\end{split}
\eeqa
where 
\beq
\eta^{\pm a, i} \, \equiv  \, \bar{u}^{\pm a}_A \eta^{A, i}\ .
\eeq
Momentum conservation follows from the Ward identity for the momentum generator, 
\beq
\big(q - \sum_{i=1}^n p_i\big) \cF_{k} (1, \ldots, n ;q, \gamma^{+})\  = \ 0
\ .
\eeq
Hence, the Ward identities require that 
\beq
\cF_{k,n} (1, \ldots, n; q, \gamma^{+}) \ \propto \    \delta^{(4)}\big(q-\sum\limits_{i=1}^n\lambda^i\tl^i\big) \delta^{(4)}\big(\gamma^{+}-\sum\limits_{i=1}^n\lambda^i\eta^{+,i}\big) \delta^{(4)}\big(\sum\limits_{i=1}^n\lambda^i\eta^{-,i}\big)\, . 
\eeq
It was shown in   \cite{Brandhuber:2011tv} that the supersymmetric MHV form factor of the the stress-tensor multiplet operator $\T_2$ is simply obtained by multiplying the required delta functions by  a Parke-Taylor denominator:  
\begin{equation} 
\label{eq:stresstensorFF}
\F^{\text{MHV}}_{2,n}(1,\ldots,n;q,\gamma^{+}) =\frac{  \delta^{(4)}\big(q-\sum\limits_{i=1}^n\lambda^i\tl^i\big) \delta^{(4)}\big(\gamma^{+}-\sum\limits_{i=1}^n\lambda^i\eta^{+,i}\big) \delta^{(4)}\big(\sum\limits_{i=1}^n\lambda^i\eta^{-,i}\big)}{\la 12\ra\la 23\ra\cdots\la n1\ra}\ . 
\end{equation}
One of the goals of this work is to determine the form factors of the more general operators $\T_k$ for any $k$ and for a generic number $n$ of external particles.

\subsection{The  super form factor $\F^{\text{MHV}}_{3,n}$}
\label{sec:susyff}

In this section we will study the  form factors of the chiral operator $\cT_3$, where $\cT_k$ is defined in \eqref{T}. In particular we will consider the form factor with the simplest helicity assignment, namely MHV,%
\footnote{Note that in general, the MHV form factor of $\cT_k$  will have  fermionic degree $8+ 2(k-2)$.}
and will show that it is given by the compact expression
\begin{equation}
\label{eq:FF3MHVsusy-simple}
\F^{\text{MHV}}_{3,n}(1,\dots,n;q,\gamma^{+}) \ =\ \F^{\text{MHV}}_{2,n}(1,\ldots,n;q,\gamma^{+})
\  
\Big(\sum\limits_{i < j=1}^{n} \b{i\,j}\eta^{-, i} \cdot\eta^{-,j} \Big) \ , 
\end{equation}
where we have introduced the shorthand notation
\begin{equation}
\eta^{-, i}\cdot\eta^{-,j}\,=\,\eta^{-, j}\cdot\eta^{-,i}\, \equiv\,  \frac{1}{2}\, \eta^{-a, i}\eta^{-b, j}\, \epsilon_{ab}\,,\qquad  \eta^{-, i}\cdot\eta^{-,i}= \eta^{-1, i}\eta^{-2, i}\equiv(\eta^{-, i})^2\ .
\end{equation}
Interestingly, this form factor can be written as a product of the stress-tensor MHV form factor \eqref{eq:stresstensorFF} with an additional term which compensates for the different $R$-charge of the operator $\T_3$. Indeed, 
it is immediate to see that, for  $\F^{\text{MHV}}_{3,n}$ to be non-vanishing for an external state containing three scalars and an arbitrary number of  positive-helicity gluons, the form factor  must have a fermionic degree which exceeds that of  $\F^{\text{MHV}}_{2,n}$ by two units.

We also show an equivalent  expression for the super form factor  $\F^{\text{MHV}}_{3,n}$ given by the following formula, 
\begin{align}
\label{eq:FF3MHVsusy}
\begin{split}
\F^{\text{MHV}}_{3,n}(1,\dots,n;q,\gamma^{+}) \ =\ \F^{\text{MHV}}_{2,n}(1,\ldots,n;q,\gamma^{+})
\, \Big(\sum\limits_{i\leq j=1}^{n-2}(2-\delta_{ij})\dfrac{\b{n\,i}\b{j\,n-1}}{\b{n-1\,n}}\eta^{-, i}\cdot\eta^{-, j}\Big)\, .
\end{split}
\end{align}
Although \eqref{eq:FF3MHVsusy} looks slightly more complicated than \eqref{eq:FF3MHVsusy-simple}, this expression will  prove more convenient for later generalisations to higher $k$ and applications to loop computations. \\[12pt]
To prove the equivalence of \eqref{eq:FF3MHVsusy-simple} and \eqref{eq:FF3MHVsusy}, consider the  expression
\begin{equation}
\label{foll}
\sum\limits_{i < j=1}^{n} \b{i\,j}\eta^{-, i}\cdot\eta^{-,j} + \sum_{i,j=1}^n
\frac{\b{n\,i}\b{j\,n-1}}{\b{n-1\,n}} \eta^{-, i}\cdot\eta^{-, j} \ .  
\end{equation}
The second term on the right-hand side of  \eqref{foll} is in fact zero due to supermomentum conservation in the $Q^{-}$ direction, as can be seen by rewriting it as
\begin{align}
\sum_{i,j=1}^n
\frac{\b{n\,i}\b{j\,n-1}}{\b{n-1\,n}} \eta^{-, i}\cdot\eta^{-, j}\,=\,\frac{1}{\b{n-1\,n}} \Big(\sum_{i=1}^n \b{n\,i} \eta^{-, i}\Big)\cdot \Big(\sum_{j=1}^n \b{j\,n-1} \eta^{-, j}\Big)\ .
\end{align}
Splitting the sum in \eqref{foll} over all $i,j$ in that term into the cases $i=j,\, i<j$ and $j<i$, it is straightforward to show that \eqref{eq:FF3MHVsusy-simple} and \eqref{eq:FF3MHVsusy} are equal. Explicitly we have
\begin{align}
\begin{split}
&\sum_{i<j=1}^n \frac{\b{i\,j} \b{n-1 n} +\b{n\,i}\b{j\,n-1}+\b{n\,j}\b{i\,n-1}}{\b{n-1\,n}}\eta^{-, i}\cdot\eta^{-, j}+\sum_{i=1}^n \frac{\b{n\,i}\b{i\,n-1}}{\b{n-1\,n}} (\eta^{-, i})^2\\ 
\,=\, &\sum_{i<j=1}^n 2\frac{\b{n\,i}\b{j\,n-1}}{\b{n-1\,n}}\eta^{-, i}\cdot\eta^{-, j}+\sum_{i=1}^n \frac{\b{n\,i}\b{i\,n-1}}{\b{n-1\,n}} (\eta^{-, i})^2\ .
\end{split}
\end{align}
We also comment that it is straightforward to show that the expression \eqref{eq:FF3MHVsusy-simple} is cyclically invariant. Defining
\begin{align}
V(1,2,\dots,n)\,\equiv\,\sum_{i < j=1}^{n} \b{i\,j}\eta^{-, i} \cdot\eta^{-, j}\ ,
\end{align}
then isolating the terms with label `1',
\begin{align}
\sum_{i < j=1}^{n} \b{i\,j}\eta^{-, i} \cdot\eta^{-, j}\,=\, \underbrace{\sum_{j=1}^n  \b{1\,j}\eta^{-, 1} \cdot\eta^{-, j}}_{=\,0\text{ due to }\sum\limits_{j=1}^n\lambda^j\eta^{-,j}\,=\,0} + \sum_{i < j=2}^{n} \b{i\,j}\eta^{-, i} \cdot\eta^{-, j} \,=\, V(2,3,\dots,n)\ ,
\end{align}
Using the same argument it follows that $V(2,3,\dots,n)=V(2,3,\dots,n,1)$, so $V(1,2,\dots,n,)= V(2,3,\dots,n,1)$ as required.

For the case of three external legs, the form factor $\F_{3,3}$ is simply equal to one, or  $(\eta^{-,1})^2(\eta^{-,2})^2(\eta^{-,3})^2$  
in the supersymmetric language. Indeed, it is easy to check that  \eqref{eq:FF3MHVsusy-simple} evaluated for $n=3$  reproduces this result. 
Having  established the correctness of $\F_{3,3}$ for three external legs, we will prove the validity of  \eqref{eq:FF3MHVsusy-simple} for all $n$ by induction using the  BCFW recursion relation. 

A caveat is in order here: for adjacent BCFW shifts,   \eqref{eq:FF3MHVsusy-simple} 
has  a residue at $z \rightarrow \infty$. The physical interpretation of   this behaviour is interesting and will be discussed in  \sref{sec:adjBCFWshift}.  On the other hand, $\F_{3,n}$ has a good large-$z$ behaviour under \textit{next-to-adjacent} shifts, which we will use in the next section to prove  \eqref{eq:FF3MHVsusy-simple} for generic $n$.


 \subsubsection{Proof for general $n$ from recursion relations with non-adjacent shifts}
\label{nta} 
 
\begin{figure}[htb]
\centering
\includegraphics[width=0.75\textwidth]{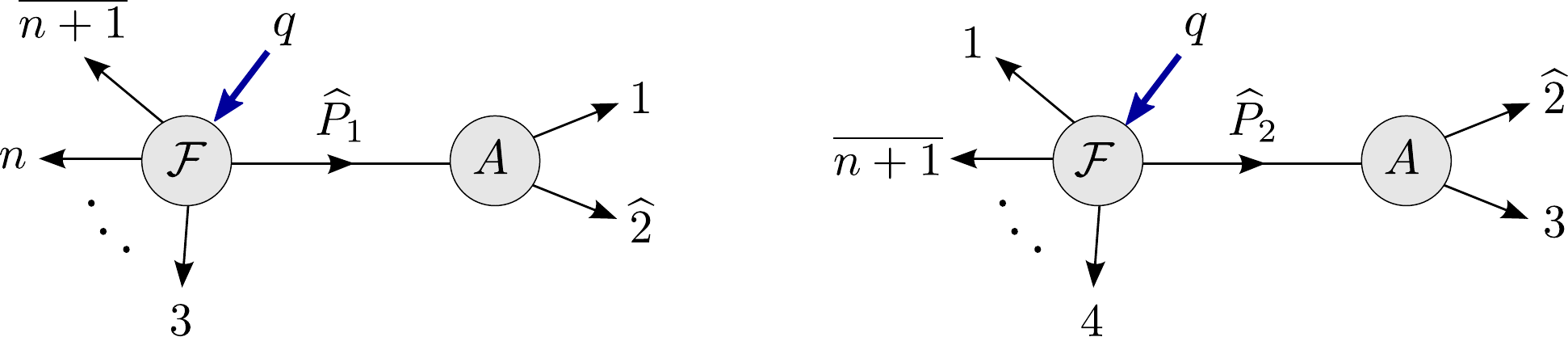}
\caption{\it The two BCFW recursive diagrams contributing to $\F^{\text{\rm MHV}}_{3,n+1}$ under a next-to-adjacent  shift $(\widehat{2}, \overline{n+1})$. The amplitude on the right is $\overline{\rm MHV}$.}
\label{fig:BCFWproof}
\end{figure}

We now move on to proving  \eqref{eq:FF3MHVsusy-simple} using recursion relations. We consider the form factor with $n+1$ external particles under the following next-to-adjacent BCFW shifts, which we denote by $(\hat{2},\overline{n+1})$,
\begin{align}
\begin{split}
\lambda^2&\rightarrow\lambda^2+z\lambda^{n+1}\ ,
 \\
\tl^{n+1}&\rightarrow\tl^{n+1}-z\tl^{2}\ ,
 \\
\eta^{-, n+1}&\rightarrow\eta^{-, n+1}-z\eta^{-, 2}\ .
\end{split}
\end{align} 
Since in the MHV case we only have a three-particle $\MHVb$ amplitude attached to an $n$-particle MHV form factor, there are two diagrams to consider, shown in Figure \ref{fig:BCFWproof}.
\\
These are explicitly given by
\begin{align}
\F^{\text{MHV}}_{3,n}( \widehat{P}_1,3,\dots,n,\overline{n+1}; q,\gamma^{+}) & \dfrac{1}{s_{12}}A^{\MHVb}(-\widehat{P}_1,1,\widehat{2})\ , & \widehat{P}_1&=p_1+\widehat{p}_2 \, , 
\label{eq:diag1}\\
\F^{\text{MHV}}_{3,n}( 1,\widehat{P}_2, 4,\dots, \overline{n,+1}; q,\gamma^{+}) &\dfrac{1}{s_{23}}A^{\MHVb}(-\widehat{P}_2,\widehat{2},3)\ , & \widehat{P}_2&=\widehat{p}_2+p_3 \, , \label{eq:diag2}
\end{align}
where $\F^{\text{MHV}}_{3,n}$ is given in \eqref{eq:FF3MHVsusy-simple} while
\beqa
\label{MHVbsupera}
A^{\MHVb}(1,2,3)& =&\frac{\delta^{(4)}\left(\eta^{1}[23]+\eta^2[31]+\eta^{3}[12 ]\right) }{[12][23][31]}\ . 
\eeqa
It is straightforward to evaluate these two diagrams, and the corresponding results are 
\beqa
\label{eq:diag1-sol}
\text{Diag 1} & = &\F^{\text{MHV}}_{2, n+1} \frac{\b{23}\b{1\,n+1}}{\b{13}\b{2\,n+1}}\left[  \sum\limits_{i>j=4}^{n+1} \b{i\,j} \eta^{-, i} \cdot \eta^{-, j}+  \sum\limits_{j=4}^{n+1}\b{3\,j}\eta^{-,3} \cdot \eta^{-, j} \right.\nonumber \\
 &+&\left. \sum\limits_{j=3}^{n+1}\b{1\,j}\left(\eta^{-, 1}+\frac{\b{2\,n+1}}{1\,n+1} \eta^{-,2}\right)\cdot \eta^{-, j} \right]\ , \cr
 \text{Diag 2} &= &\F^{\text{MHV}}_{2, n+1} \frac{\b{12}\b{3\,n+1}}{\b{13}\b{2\,n+1}} \left[  \sum\limits_{i<j=4}^{n+1}\b{i\,j}\eta^{-, i} \cdot \eta^{-, j}+\b{13} \eta^{-,1} \cdot \left(\eta^{-,3}+\frac{\b{2\,n+1}}{\b{3\,n+1}}\eta^{-,1}\right) \right.
\nonumber \\
  &+ & \left.\sum\limits_{j=4}^{n+1} \b{3\,j}\left(\eta^{-,3}+\frac{\b{2\,n+1}}{\b{3\,n+1}}\eta^{-,1}\right)\cdot \eta^{-,j} \right]  \, .
\eeqa
Summing  these two contributions by collecting coefficients of $\eta^{-,i}\cdot \eta^{-,j}$, we obtain the  expected result  for the  ($n+1$)-particle form factor,
\begin{align}
\label{eq:expected}
\F^{\text{MHV}}_{3,n+1}= \F^{\text{MHV}}_{2, n+1} \sum\limits_{i<j=1}^{n+1} \b{i\,j} \eta^{-, i} \cdot \eta^{-, j} \ .
\end{align}
This completes the proof of our result for $\F_{3, n}$ via the BCFW recursion relation. 

\subsubsection{A few examples of component form factors}

To conclude this section, it is useful to present a couple of examples of component form factors. In particular, we will look at the lowest component of 
$\cT_k$ (i.e. the coefficient of the lowest power of $\theta_+$ in \eqref{cpst}), which is given by the scalar operator\footnote{In  all our computations we will choose the reference directions such that $\phi_{++}\,\equiv\,\phi_{12}$.}
\beq
\label{trfi}
\O_k(x)\, \equiv  \,\Tr \, \big[\phi_{12}(x)^k\big]
\ . 
\eeq
To begin with, we consider  the simple case $k=3$.  
From Feynman diagrams, it is immediate to see that at tree level the  form factor of $\O_3(x)$ is equal to one (apart from a trivial momentum conservation delta function):
\begin{align}
\begin{split}
F_{3,3}(1^{\phi_{12}},2^{\phi_{12}},3^{\phi_{12}};q)\,\equiv &\int\!d^4x \ e^{-i q\cdot x} \ \la 1^{\phi_{12}},2^{\phi_{12}},3^{\phi_{12}} |\Tr\, \big[\big(\phi_{12}(x)\big)^3\big]|0\ra \\
=& \, \delta^{(4)}\big(q-\sum_{i=1}^3 \lambda^i\tl^i\big)\ .
\end{split}
\end{align}
From \eqref{eq:FF3MHVsusy-simple}, 
we can immediately derive the expression for the $n$-point MHV form factor with three scalars and  $n-3$ positive-helicity gluons. This is given by
\begin{equation}
\label{eq:FF3MHVboson}
F_{3,n}^{\text{MHV}}(\{g^+\},a^{\phi_{12}},b^{\phi_{12}},c^{\phi_{12}};q)=\frac{\b{ab}\b{bc}\b{ca}}{\b{12}\b{23}\cdots\b{n1}}\ \delta^{(4)}\big(q-\sum_{i=1}^n \lambda^i\tl^i\big)\ ,
\end{equation}
where the  three scalars $\phi_{12}$ are at positions $a,b,c$. Notice that \eqref{eq:FF3MHVboson} scales as $(\lambda^{i})^0$ for $i \in \{a,b,c\}$ and  $(\lambda^{i})^{-2}$ for $i\notin \{a,b,c\}$ as required.
 
In fact, similar arguments can be used to write down a very concise formula for the MHV form factor of $\O_k$ with $k$ scalars and $n-k$ positive-helicity gluons for general $k$. It contains a ratio of  Parke-Taylor factors, where in the numerator only the (ordered) scalar particle momenta appear, while the denominator is the standard Parke-Taylor expression for $n$ particles,
\begin{equation}
\label{eq:FFmMHVboson}
F_{k,n}^{\text{MHV}}(\{g^+\},i_1^{\phi_{12}},i_2^{\phi_{12}},\dots, i_k^{\phi_{12}};q)=\frac{\b{i_1\,i_2}\b{i_2\,i_3}\cdots \b{i_k\,i_1}}{\b{12}\b{23}\cdots\b{n1}} \ \delta^{(4)}\big(q-\sum_{i=1}^n \lambda^i\tl^i\big)\ .
\end{equation}
The correctness of \eqref{eq:FFmMHVboson}  can easily be shown using BCFW recursion relations \cite{Britto:2004ap,Britto:2005} with adjacent shifts applied to form factors \cite{Brandhuber:2010ad}.
We will not present this proof here, rather we will now consider its supersymmetric generalisation.


\subsection{A new recursion relation and  conjecture for the  MHV  super form factors of  $\cT_k$ }
\label{sec:adjBCFWshift}

In this section we will propose a new recursion relation for the form factors of the half-BPS supersymmetric operators $\cT_k$, shown below in \eqref{eq:recursion}. This recursion relation is quite different from the usual BCFW recursion relation applied to form factors, in the sense that it relates form factors of operators $\cT_k$ with different $k$. In the following we will motivate this recursion relation, whose origin lies in the presence of certain boundary terms in the usual supersymmetric BCFW recursion relation for $\cT_k$ with adjacent shifts. Following this, we will conjecture an expression for the  MHV form factors of the operators $\cT_k$  for general $k$ and show that it satisfies this new recursion relation as well as the cyclicity requirement for some values of $k$ and $n$. 

\subsubsection{A new recursion relation for form factors}

As  observed in \sref{sec:susyff}, the tree-level expression \eqref{eq:FF3MHVsusy-simple}  develops a non-vanishing large-$z$ behaviour under an adjacent BCFW shift. 
In the case of $\T_3$, we can circumvent this problem  by using a next-to-adjacent shift, for which there is no pole at infinity. Indeed, this is the strategy we followed in \sref{nta} in order to determine the form factors of $\T_3$ from recursion relations. The situation is  worse   for the operators $\T_{k}$ with  $k>3$; one can convince oneself that even with non-adjacent  shifts the bad large $z$ behaviour cannot be eliminated.

This feature impels us to look for other means to study form factors of $\T_{k}$ for general $k$. Fortunately,  the exploration of the boundary term for adjacent BCFW shifts brought to our attention an intriguing recursion relation involving the MHV form factors $\F_{k,n}$, $\F_{k-1,n-1}$ and $\F_{k,n-1}$, as we will now  discuss.\\[5pt]
\noindent Considering the $n$-particle form factor $\F^{\text{MHV}}_{k,n}$  shifted according to the  BCFW shifts 
\beqa \label{BCFWshift}
\begin{split}
\lambda^{n} &\rightarrow& \lambda^{n} -z \lambda^{n-1} \ , \\
\tl^{n-1} &\rightarrow &\tl^{n-1} + z \tl^{n} \ ,  \\   
\eta^{n-1} &\rightarrow &\eta^{n-1} + z \eta^{n} \ , 
\end{split}
\eeqa 
the claim is that its residue at $z \rightarrow \infty$, which we denote by $R^{\rm MHV}_{k,n}$,  is given by 
\begin{align} \label{residue}
R^{\rm MHV}_{k,n}  \ = \   (\eta^{-, n})^2 \widetilde{\F}^{\text{MHV}}_{k-1,n-1}(1,\dots,n-1;q,\gamma^{+}) \ .  
\end{align}
In this equation and in the following, $\widetilde{\F}$ is the form factor $\F$ with the momentum and supermomentum conservation delta-functions stripped off. 
For $k=3$,  we can confirm  this by simply using our result for  $\F^{\text{MHV}}_{3,n}$ given in \eqref{eq:FF3MHVsusy-simple}. Performing   the BCFW shift \eqref{BCFWshift} and using supermomentum conservation, we find that the  residue at $z \rightarrow \infty $ is, on the support of the delta- functions,   
\begin{eqnarray}
\begin{split}
R^{\rm MHV}_{3,n}  &= 
\frac {  \sum\limits_{i=1}^{n-2} \b{i\,n-1} \eta^{-, i}\cdot \eta^{-, n}}{\b{12}\b{23}\cdots  \b{n-1 \ 1}} 
\ = \frac{(\eta^{-,n})^2
} {\b{12}\b{23}\cdots  \b{n-1 \, 1}}  \, ,
\end{split}
\end{eqnarray}
which is indeed simply $(\eta^{-,n})^2 \times \widetilde{\F}^{\text{MHV}}_{2,n-1}(1,\dots, n-1;q,\gamma^{+})$. 

Conceptually this result is very interesting since it shows that the form factors of the operator $\T_k$ are related to the form factors of the operator $\T_{k-1}$ in a simple manner. In practice, \eqref{residue}  allows us to determine the $n$-particle form factor $\F^{\text{MHV}}_{k,n}$ from the $(n-1)$-particle form factors $\F^{\text{MHV}}_{k,n-1}$ and $\F^{\text{MHV}}_{k-1,n-1}$ in the following way: 
\begin{align}
\label{eq:recursion}
\begin{split}
\widetilde{\F}^{\text{MHV}}_{k,n}(1, \ldots, n; q,\gamma^{+}) &=   {\b{n-1 \, 1} \over \b{n-1 \, n} \b{n \, 1 } } \widetilde{\F}^{\text{MHV}}_{k,n-1}(1^\prime,2, \ldots, n-2, (n-1)^\prime; q,\gamma^{+})
 \\ \cr
&+ (\eta^{-, n})^2 \ \widetilde{\F}^{\text{MHV}}_{k-1,n-1}(1,\ldots, n-1; q,\gamma^{+}) \ ,  
\end{split}
\end{align}
where we have solved the BCFW diagram in the inverse soft form \cite{ArkaniHamed:2009si, BoucherVeronneau:2011nm, Nandan:2012rk, ArkaniHamed:2012nw}; indeed  the first term in \eqref{eq:recursion} simply adds particle $n$ to the $(n-1)$-particle form factor $\F^{\text{MHV}}_{k,n-1}$ with a soft factor. To maintain momentum conservation, we need to shift the legs adjacent to $n$, i.e. $(n-1)^\prime$ and $1^\prime$, with the corresponding shifted spinors given by
\begin{eqnarray}
\begin{split}
\tl^{(n-1)'} &= &\tl^{n-1} + \frac{ \b{n \, 1} } {\b{ n-1 \, 1} } \tl^{n} \,  , \qquad 
\tl^{1'} = \tl^{1} + \frac{ \b{n \, n-1} } {\b{ 1 \, n-1} } \tl^{n}\, ,  \\ 
\eta^{(n-1)'} &= &\eta^{n-1} + \frac{ \b{n \, 1} } {\b{ n-1 \, 1} } \eta^{n} \, ,  \qquad \
\eta^{1'} = \eta^{1} + \frac{ \b{n \, n-1} } {\b{ 1 \, n-1} } \eta^{n}  \, .
\end{split}
\end{eqnarray}
The second term in \eqref{eq:recursion} is again an $(n-1)$-particle form factor, but now for the operator $\T_{k-1}$. The factor $(\eta^{-, n})^2$ ensures that the fermionic degree of the expression is correct. The recursion relation may be recast into a slightly different form by removing the Parke-Taylor prefactor, 
\begin{align}
\label{recursion2}
\begin{split}
f_{k,n}(1, \ldots, n) &\,= \,  f_{k,n-1}(1',2, \ldots, n-2, (n-1)')
 \\ \cr
&+ (\eta^{-, n})^2 \ f_{k-1,n-1}(1,\ldots, n-1)  { \b{n-1 \, n} \b{n \, 1 } \over \b{n-1 \, 1}  }
\ , 
\end{split}
\end{align}
where we have defined $f_{k,n}(1, \ldots, n)$ from the relation
\begin{equation}
\label{eq:little-f}
\F^{\text{MHV}}_{k,n}(1, \ldots, n; q,\gamma^{+}) \ \equiv  \ \F^{\text{MHV}}_{2,n}(1, \ldots, n; q,\gamma^{+})\  f_{k,n}(1, \ldots, n) \, .
\end{equation}
Given the fact that the form factors of $\T_2$ are  simply given by the Parke-Taylor formula, and the $k$-point form factor of the operator $\T_k$ is just one (or, in a supersymmetric language, $\prod^k_{i=1} (\eta^{-,i})^2$, the recursion relation \eqref{eq:recursion} fully determines all MHV form factors for any operator $\T_k$. Indeed, in the next section  we will propose an explicit solution to the recursion relation for the form factor $\F^{\rm MHV}_{k,n}$. 


\subsubsection{A peek into NMHV form factors of $\T_k$}

Having found a novel recursion relation \eqref{eq:recursion} for MHV super form factors, we would like to study how to generalise it to non-MHV helicity configurations. Non-adjacent shifts also work for non-MHV form factors of $\T_3$,  which in principle fully determines all form factors of this operator. We can use them in order to derive the expression of non-MHV form factors, of which we can then study the large-$z$ behaviour under adjacent shifts. 
\begin{figure}[htb]
\centering
\includegraphics[width=0.85\textwidth]{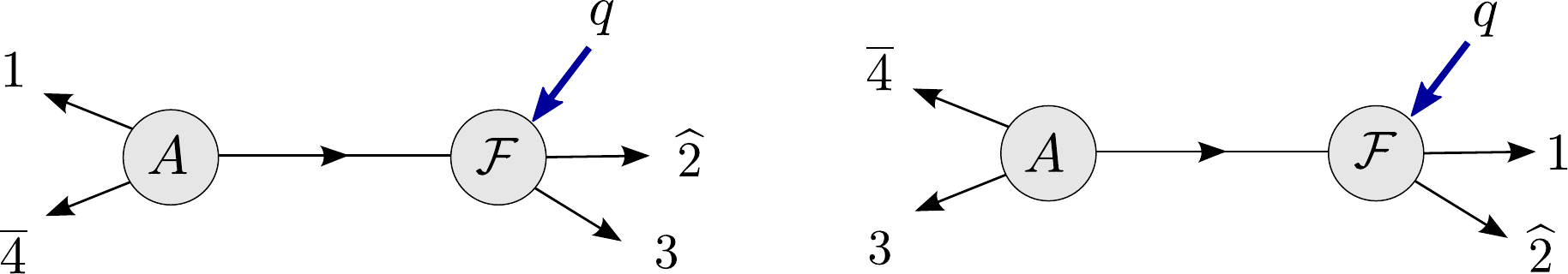}
\caption{\it Recursion relation for the  four-point NMHV form factor of $\T_3$. The amplitude on the left is MHV.}
\label{fig:4ptNMHV}
\end{figure}

The simplest non-MHV form factor is the NMHV four-particle form factor of $\T_3$. From the recursion relation with non-adjacent BCFW shifts on legs $2$ and $4$ given in Figure \ref{fig:4ptNMHV}, we find the following result, 
\begin{align}
\label{sss}
\begin{split}
\F^{\text{NMHV}}_{{3},4}
&=\delta^{(4)}\big(\gamma^{+} - \sum_i \lambda^i \eta^{+, i} \big) \delta^{(4)}
\big(\sum_i \lambda^i \eta^{-, i} \big)
{\delta^{(4)}\big( [23]\eta^4 + [34] \eta^2 + [42] \eta^3 \big)
(\eta^{-,1})^2 \over [23][34][42] s^2_{234} } 
 \\ 
&- (1 \leftrightarrow 3 ) \\  
&=
\delta^{(4)}(\gamma^{+} - \sum_i \lambda^i \eta^{+, i} ) \prod^4_{i=1} (\eta^{-,i})^2
\Big[ 
{\delta^{(2)}( [23]\eta^{+,4} + [34] \eta^{+,2} + [42] \eta^{+,3} ) \over [23][34][42] }  \\ 
&- (1 \leftrightarrow 3 ) 
\Big] \, .
\end{split}
\end{align}
If we expand the fermionic delta function $\delta^{(2)}\big( [23]\eta^{+,4} + [34] \eta^{+,2} + [42] \eta^{+,3} \big)$, we find  non-trivial  agreement with the result \eqref{NMHV4pt1} that we will derive later using MHV rules. 

Having obtained \eqref{sss}, we can find its behaviour under adjacent BCFW shifts, for instance,
\begin{eqnarray}
\begin{split}
\lambda^1 &\rightarrow& \lambda^1 - z\, \lambda^2 \, ,  \\
\tl^2 &\rightarrow& \tl^2  + z\, \tl^1 \, ,  \\
\eta^2 &\rightarrow &\eta^2 + z \,\eta^1 \, ,
\end{split}
\end{eqnarray}
doing so, we find that the residue of $\F^{\text{NMHV}}_{3,4}$ at large $z$ is given by
\begin{eqnarray}
\label{com}
\begin{split}
R^{\text{NMHV}}_{3,4} &= \ 
\delta^{(4)}\big(\gamma^{+} - \sum\limits_{i=1}^4 \lambda^i \eta^{+,i}\big) \delta^{(4)}\big(\sum\limits_{i=1}^4 \lambda^i \eta^{-,i}\big)
{z^4\,\delta^{(4)}( [13]\eta^4 + [34] \eta^1 + [41] \eta^3 )  \over z^2 \, [13][34][41] (2\, q \cdot p_{\widehat{1}})^2 }\, (\eta^{-,1})^2\\
&=\ 
 {q^4 \over \langle 2| q |1]^2 } (\eta^{-, 1})^2 \times \widetilde{\F}^{\text{NMHV}}_{{2},3}(3,4,1; q,\gamma^{+})  \, . 
\end{split}
\end{eqnarray}
In the last step we related  the residue of the NMHV form factor of the operator $\T_3$  at infinity with the 
NMHV form factor of $\T_2$, similarly to the case of MHV form factors considered earlier.  
From \eqref{com} we see that the structure of this boundary term is more complicated than in the MHV case. 
It would be interesting to understand  this  boundary term for a general non-MHV form factor.


\subsubsection{Supersymmetric MHV  form factors of $\T_k$}

In this section we will propose a solution to the recursion relation \eqref{eq:recursion} for the form factor $\F^{\text{MHV}}_{k,n}$.  We begin by considering the  case $k=4$. After computing a few simple examples by using the recursion relation \eqref{eq:recursion}, a clear pattern appears for $\F^{\text{MHV}}_{4,n}$, which is given by 
\begin{align}
\label{eq:O4}
\F^{\text{MHV}}_{4,n}\ = \ \F^{\text{MHV}}_{2,n} \sum\limits_{1\leq i\leq j}^{n-3} \sum\limits_{j< k\leq l}^{n-2} (2-\delta_{ij})(2-\delta_{kl}) \frac{\b{n\,i}\b{j\,k}\b{l\,n-1}}{\b{n-1\,n}}(\eta^{-, i}\cdot \eta^{-,j})(\eta^{-,k}\cdot \eta^{-,l})\, .
\end{align}
This is clearly  a generalisation of the $k=3$ case for $\F^{\text{MHV}}_{3,n}$ considered  in \eqref{eq:FF3MHVsusy}. 

Further generalisation of $\F^{\text{MHV}}_{3,n}$ and $\F^{\text{MHV}}_{4,n}$ leads to a proposal for $\F^{\text{MHV}}_{k,n}$ for arbitrary $k$. 
In general we will have $2(k-2)$ nested sums with  fermionic degree  $2(k-2)$ in  $\eta^{-}$ (besides the delta function of supermomentum conservation).
Our conjecture for $\F^{\text{MHV}}_{k,n}$ is 
\begin{align}
\label{conj}
\begin{split}
\F^{\text{MHV}}_{k,n}\,=\; &\F^{\text{MHV}}_{2,n} 
\sum\limits_{1\leq a_1\leq b_1}^{n-k+1}  \sum\limits_{b_1< a_2 \leq b_2}^{n-k+2} \cdots \!\!
\sum\limits_{b_{k-3} < a_{k-2} \leq b_{k-2}}^{n-2} 
\\ \times &\; C_{a_1,b_1,a_2,b_2, \cdots,a_{k-2},b_{k-2}}
\prod^{k-2}_{\alpha=1 } (\eta^{-, a_{\alpha}}\cdot\eta^{-, b_{\alpha}})
\ ,
\end{split}
\end{align}
where the coefficients  $C_{a_1,b_1,a_2,b_2, \cdots,a_{k-2},b_{k-2}}$ are natural generalisations of the coefficient in \eqref{eq:O4},
\beqa
\begin{split}
 &C_{a_1,b_1,a_2,b_2, \cdots,a_{k-2},b_{k-2}} = \cr
& 
\left( \prod^{k-2}_{\alpha=1 } (2-\delta_{a_{\alpha}b_{\alpha}}) \right)
 \frac {\b{n\,a_{1}}\b{b_{1} \,a_{2}}\cdots \b{b_{k-3}\,a_{k-2}} \b{b_{k-2}\,n-1}}
{\b{n-1 \,n}}  \, .
\end{split}
\eeqa
In the summations in \eqref{conj} we sum over pairs of indices $a_{\alpha}$, $b_{\alpha}$, for $\alpha =1, \ldots ,k-2$. 
We have  compared   \eqref{conj} to the result obtained from  the recursion relation \eqref{eq:recursion} and agreement has been found for all cases we have checked, namely $k\leq 6,\,n\leq 7$.   

We would like to stress that,  unlike the case of the recursion for the form factor $\F_{3,n}$ with  non-adjacent shifts, 
the recursion relation \eqref{eq:recursion} is a conjecture, hence it is important to check the correctness of the resulting $\F_{k,n}$ in \eqref{conj}, obtained from studying  \eqref{eq:recursion}. 
One non-trivial test consists in checking  the cyclicity of the result. 
In Appendix \ref{app:cyclicity} we prove that our result for $\F_{4,n}^{\rm MHV}$ indeed enjoys this symmetry in a very non-trivial way. Unfortunately we have not been able to prove the cyclicity of $\F_{k,n}$ for arbitrary $k$, however we have checked various cases for $k\leq 6$ with \texttt{Mathematica} and found that the required symmetry is indeed present. 
The proof of $\F_{4,n}$ and these checks provide support both to the conjectured recursion relation \eqref{eq:recursion} and  solution \eqref{conj}.


\subsection{MHV rules for $\F_{k,n}$}
\label{sec:MHVrules}

In \cite{Brandhuber:2011tv}, MHV rules for the form factor of the stress-tensor multiplet operator were constructed. Here we show in a number of concrete applications that these MHV rules can directly be extended to the form factors of the operators $\cT_k$ with $k>2$. In this approach, the usual MHV vertices of \cite{Cachazo:2004kj} are augmented by a new set of vertices obtained by continuing off-shell the holomorphic form factor expression for  $\cF_{k,n}^{\rm MHV}$ 
using the same prescription as in \cite{Cachazo:2004kj}. In the following we will illustrate the application of this technique by computing a few examples, but we comment that the approach can be used in general to obtain form factors with higher MHV degree and number of loops, as was done in \cite{Brandhuber:2004yw} for one-loop MHV amplitudes. 

\subsubsection{Four-particle bosonic NMHV form factor}

\begin{figure}[htb]
\centering
\includegraphics[width=0.8\textwidth]{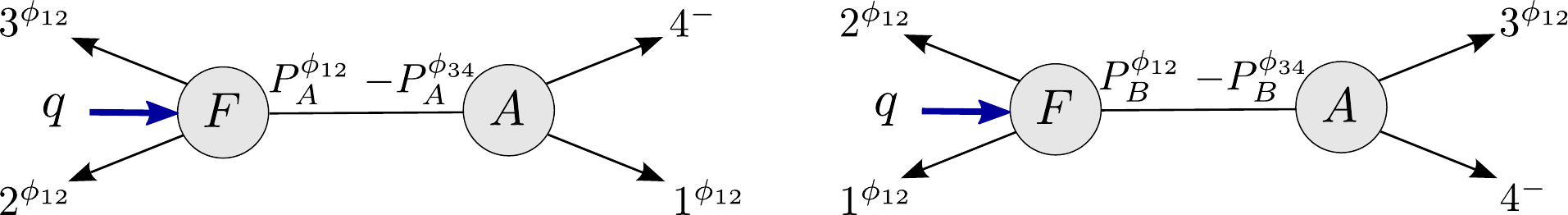}
\caption{\it Expansion of $ F^{\text{\rm NMHV}}(1^{\phi_{12}},2^{\phi_{12}},3^{\phi_{12}},4^-; q) $ using MHV rules.}
\label{fig:NMHV}
\end{figure}

As a first example, we  consider the bosonic form factor $ F^{\text{NMHV}}(1^{\phi_{12}},2^{\phi_{12}},3^{\phi_{12}},4^-; q) $ and compute it with MHV rules.  There are two diagrams that contribute to this, shown in Figure \ref{fig:NMHV}.
\\
These result in the respective expressions
\begin{align}
\label{eq:MHV41}
\big[ F_{3,4}^{\text{NMHV}}(1^{\phi_{12}},2^{\phi_{12}},3^{\phi_{12}},4^-;q)\big]^{(1)}  \ &= \ F^{\text{MHV}}_{3,3}(2^{\phi_{12}},3^{\phi_{12}},P_A^{\phi_{12}};q)\,A_3^{\text{MHV}}(-P_A^{\phi_{34}},4^-,1^{\phi_{12}}) \\
 \label{eq:MHV42}
 \big[ F^{\text{NMHV}}_{3,4}(1^{\phi_{12}},2^{\phi_{12}},3^{\phi_{12}},4^-;q)\big]^{(2)} \ &= \ 
 F^{\text{MHV}}_{3,3}(1^{\phi_{12}},2^{\phi_{12}},P_B^{\phi_{12}};q)\,A_3^{\text{MHV}}(-P_B^{\phi_{34}},3^{\phi_{12}},4^-)\ , 
\end{align}
with
\begin{align}
\begin{split}
P_A\ &= \ p_1+p_4\ , \qquad  |A\ra \,=\,(p_1+p_4)|\xi]\ , \\
P_B\ &=\ p_3+p_4 \ , \qquad  |B\ra \, =\,(p_3+p_4)|\xi]\ , 
\end{split}
\end{align}
where $|\xi]$ is the reference spinor used in the  off-shell continuation needed in order to define spinors associated to the internal momenta $P_{A,B}$, cf. \eqref{eq:off-shell-spinor} \cite{Cachazo:2004kj}. 
A crucial check of the correctness of the procedure is to confirm that the final answer for an amplitude or form factor evaluated with MHV diagrams is independent of the choice of the reference spinor $|\xi]$.

Using the fact that  that $F(a^{\phi_{12}},b^{\phi_{12}},c^{\phi_{12}};q) = 1 $ (omitting a delta function of momentum conservation), the first contribution \eqref{eq:MHV41} is simply given by 
\begin{align}
\frac{1}{\b{14}[14]}\times \frac{\b{A4}^2\b{14}^2}{\b{A1}\b{A4}\b{41}}=-\frac{\b{4A}}{[14]\b{A1}}= -\frac{\la 4|1|\xi]}{[14][\xi|4|1\ra}\ .
\end{align}
Analogously, the second contribution \eqref{eq:MHV42} is 
\begin{align}
\frac{1}{\b{34}[34]}\times \frac{\b{4B}^2\b{34}^2}{\b{34}\b{4B}\b{B3}}=\frac{\b{4B}}{[34]\b{B3}}= \frac{\la 4|3|\xi]}{[34][\xi|4|3\ra}\ .
\end{align}
Summing these, we get
\begin{align}
\frac{\la 4|3|\xi] [\xi|4|1\ra  [14]  - \la 4|1|\xi][34][\xi|4|3\ra }{[14][\xi|4|1\ra [34][\xi|4|3\ra }= \frac{[\xi| 
{p}_4 \, {p}_1 \, {p}_4 \, {p}_3 |\xi]-[\xi| {p}_4 \,  {p}_3 \, {p}_4 \, {p}_1 |\xi]}{[\xi 4] \b{43} [34] [41] \b{14}[4 \xi]}\ .
\end{align}
The numerator can be rewritten as
\beq
 [\xi 4]\b{41}\b{43}([14][3\xi]-[34][1\xi])= [\xi 4]^2\b{41}\b{43}[31]\ , 
\eeq
thus the final result is independent of the choice of $|\xi]$ and is given by
\begin{align}
F^{\text{NMHV}}_{3,4}(1^{\phi_{12}},2^{\phi_{12}},3^{\phi_{12}},4^-; q) = \frac{[31]}{[34][41]}\ , 
\end{align}
which is the $k$-increasing inverse soft factor, as expected.


\subsubsection{Four-particle super form factors}
\begin{figure}[htb]
\centering
\includegraphics[width=0.5\textwidth]{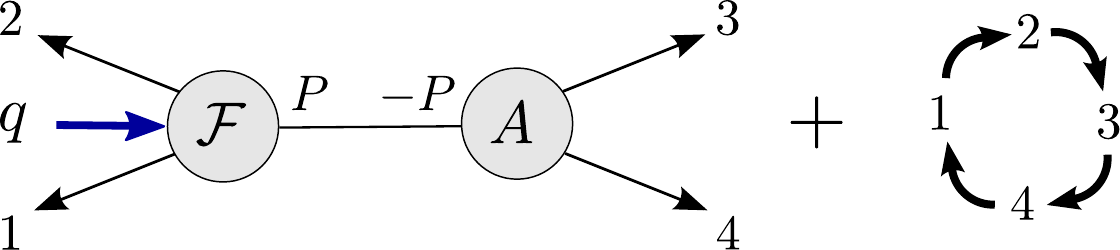}
\caption{\it Expansion of $\F^{\text{\rm NMHV}}_{3,4}$ using supersymmetric MHV rules.}
\label{fig:NMHVsusy}
\end{figure}
In this section we compute the supersymmetric form factor $\F^{\text{NMHV}}_{3,4}$ using MHV diagrams. The diagrams contributing are  shown in Figure \ref{fig:NMHVsusy}, which can be written as
\beq
\label{eq:4ptMHVsusy}
\F^{\text{NMHV}}_{3,4} \ = \ 
\F^{\text{MHV}}_{3,3}(1,2,P;q, \gamma^{+}) \dfrac{1}{\b{34}[34]}A_3^{\text{MHV}}(-P,3,4)\ + \ {\rm cyclic} (1,2,3,4)\,  ,
\eeq
where 
\beq
P=p_3+p_4\ , \qquad |P\ra = (p_3+p_4)|\xi]\ , 
\eeq
while the MHV superamplitude is
\beq
\label{MHVsupera}
A_n^{\rm MHV}(1, \ldots, n ) \ = \ \frac{\delta^{(8)}\big(\sum_{i=1}^{n}\lambda^i\eta^i \big)  }{\b{12}\cdots\b{n1}}\ .
\eeq
We consider first the term on the left of \fref{fig:NMHVsusy}. Writing the form factor as $\dfrac{(\eta^{-, P})^2}{\b{12}^2}$, the integration over $\eta^{-, P}$ becomes simply
\beqa
\begin{split}
& \int d^2\eta^{-, P} (\eta^{-, P})^2 \delta^{(4)} \big(  \lambda^1\eta^{-, 1} +\lambda^2\eta^{-, 2} + \lambda^P\eta^{-, P}\big) \delta^{(4)} 
\big(\lambda^3\eta^{-, 3} +\lambda^4\eta^{-, 4} - \lambda^P\eta^{-, P}\big)  \\  
 = & \b{12}^2\b{34}^2\prod_{i=1}^4(\eta^{-, i})^2\ .
\end{split}
\eeqa
Integrating over $\eta^{+, P} $ gives 
\begin{align}
\begin{split}
&\int d^2\eta^{+, P} \, \delta^{(4)} \big(\gamma^{+} - \lambda^1\eta^{+, 1} -\lambda^2\eta^{+, 2} - \lambda^P\eta^{+, P}\big) \delta^{(4)}\big(\lambda^3\eta^{+, 3} +\lambda^4\eta^{+, 4} - \lambda^P\eta^{+, P}\big)\\
= \ \ &\delta^{(4)}\big(\gamma^{+} - \sum\limits_{i=1}^4\lambda^i\eta^{+, i} ) \delta^{(2)} \big( \b{3P}\eta^{+, 3} + \b{4P} \eta^{+, 4} \big) \ . 
\end{split}
\end{align}
substituting this into \eqref{eq:4ptMHVsusy}, we get
\begin{align}
\label{eq:NMHVsusy}
\F^{\text{NMHV}}_{3,4}\  = \ \prod_{i=1}^4(\eta^{-, i})^2 \delta^{(4)}\big(\gamma^{+} - \sum\limits_{i=1}^4\lambda^i\eta^{+, i} \big)  \delta^{(2)}\big( \la 3|4|\xi]\eta^{+, 3} + \la 4 |3|\xi] \eta^{+, 4} \big) \frac{1}{[\xi|4|3\ra [34]\la 4|3 |\xi]} \ .
\end{align}
We note that \eqref{eq:NMHVsusy} does not scale with the reference spinor $|\xi]$. Also, we see that all the dependence on $|\xi]$ cancels out for all coefficients of  $\eta^{+, i}\cdot \eta^{+, j}$  as follows. For the cross terms $i\neq j$, the only contribution comes from the diagram with particles $i$ and $j$ on the amplitude side, for example the diagram in Figure \ref{fig:NMHVsusy} is the only one which carries  $\eta^{+, 3}\cdot \eta^{+, 4}$  with a coefficient
\begin{equation}
\frac{\la 3|4|\xi] \la 4 |3|\xi]}{[\xi|4|3\ra [34]\la 4|3 |\xi]} = \frac{1}{[34]}\ .
\end{equation}
For the terms with $(\eta^{+, i})^2$, the contribution comes from two diagrams with particle $i$ on the amplitude side. Taking as an example the $(\eta^{+, 4})^2$ coefficient, we must also take into account the  following particular diagram, 
\beqa
\begin{split}
& \F^{\text{MHV}}_{3,3}(2,3,P;q, \gamma^{+}) \dfrac{1}{\b{41}[41]}A_3^{\text{MHV}}(-P,4,1)
 \\
=\, &   \prod_{i=1}^4(\eta^{i,-})^2 \delta^{(4)}(\gamma^{+} - \sum\limits_{i=1}^4\lambda^i\eta^{+,i} ) \delta^{(2)}\left( \la 4|1|\xi]\eta^{+,4} + \la 1 |4|\xi] \eta^{+,1} \right) \frac{1}{[\xi|1|4\ra [41]\la 1|4 |\xi]}\ ,
\end{split}
\eeqa
where 
\beq
P=p_4+p_1\ , \qquad |P\ra = (p_4+p_1)|\xi]
\ . 
\eeq
Thus, summing the coefficients of $(\eta^{+,4})^2$  we get:
\begin{align}
\frac{[\xi|1|4\ra }{[41] \la 1|4 |\xi]}+ \frac{\la 4|3 |\xi]}{[\xi|4|3\ra [34]} = \frac{[13]}{[14][43]}\ .
\end{align}
This cancellation of the reference spinor clearly happens for all $i=1,\ldots,4$. Our final result for this form factor is
\begin{align}
\label{NMHV4pt1}
\F^{\text{NMHV}}_{3,4}&= \Delta^{4|4+} \prod_{i=1}^4(\eta^{-, i})^2
\times \sum_{i=1}^4 \left((\eta^{+, i})^2\frac{[i+1\,i-1]}{[i+1\,i][i\,i-1]}+\frac{\eta^{+, i} \cdot \eta^{+, i+1}}{[i\,i+1]}\right)\ ,
\end{align}
where we have defined
$
\Delta^{4|4+} \equiv    \delta^{(4)}\big (q - \sum\limits_{i=1}^4\lambda^i\tl^i \big)  \delta^{(4)}\big(\gamma^{+} - \sum\limits_{i=1}^4\lambda^i\eta^{+, i} \big)
$.
As mentioned earlier, this result agrees with what we have obtained from non-adjacent BCFW shifts.

\section{One loop}

\label{sec:1loopTk}

In this section we move on to the one-loop level. We begin by deriving the universal form of the IR-divergent part of generic form factors in $\N=4$ SYM.  This is determined by a single two-particle diagram where a four-point amplitude sits on one side of the cut.  We then compute the three-point form factor of $\cT_3$ at one loop, and then extend this result to $n$ points using supersymmetric quadruple cuts \cite{Drummond:2008bq}. Finally, we present the expression for the infinite sequence of $n$-point MHV form factors of $\cT_k$ for arbitrary $k$  and $n$. 

\noindent On general grounds, 
we can  expand $\F^{(1)}_{k,n}$ as%
\footnote{The precise definitions of the various  triangle  and box  integrals  can be found in Appendix \ref{app:scalar-integrals}.}
\begin{align}
\label{eq:all-1loop}
\F_{k,n}^{(1)}\ =\  -\F_{k,n}^{(0)} \sum_{i=1}^n s_{i\,i+1} I_{3;i}^{1m}(s_{i\,i+1}) \, + \, \text{finite boxes} \, + \, \text{three-mass triangles}\, , 
\end{align}
where $I_{3;i}^{1m}$ is a one-mass triangle, and $s_{i\, i+1} \equiv (p_i +p_{i+1})^2$.
We can motivate \eqref{eq:all-1loop}  by knowing that the answer should be expressed in terms of triangles and boxes (bubbles are absent since the theory is finite in the UV). Furthermore, the IR-divergent part of  any one-loop form factor must be proportional to its  tree-level counterpart in order to guarantee the correct exponentiation of these divergences, as we will explicitly show in the next section. This explains the first term in \eqref{eq:all-1loop}.  In practice, all the IR divergences contained in the box functions which do not contain two-particle invariants $s_{i\, i+1}$ have to cancel with corresponding divergences from one-mass triangles, leaving behind only finite boxes and a collection of one-mass triangles where the massless legs are $p_i$ and $p_{i+1}$. In Appendix \ref{app:1loopComponent} we explicitly compute the bosonic form factor $ F^{(1)}_{\Tr\phi_{12}^3}(1^{\phi_{12}},2^{\phi_{12}},3^{\phi_{12}},4^+;q)$ and show that the above structure holds, i.e. the IR divergent parts of the box functions cancel against two-mass triangles, leaving only the finite part of the boxes and one-mass triangles.

The above discussion leaves room for three-mass triangles, and does not put any constraints on what finite boxes will appear. However, the form factors with MHV helicity configuration which we will consider are special in two ways:

\begin{enumerate}
\item Three-mass triangles are in fact absent.  This can easily be understood by counting the fermionic degree of the cut diagram. Consider a triple cut contributing to this form factor, with two amplitudes and one form factor participating to the cut. The MHV form factor $\F_{k,n}$ has fermionic degree  $2( k-2) +8$, and hence one of the two superamplitudes must be a three-point $\overline{\rm MHV}$ superamplitude, so that the overall fermionic degree is $2(k-2)+8 + 8 + 4 - 4\times 3 = 2(k-2)+8$. Thus, at most two-mass triangles can be present.
\item Only two-mass easy boxes can appear (or one-mass for $n\leq 3$), similarly to the one-loop MHV superamplitudes. The reason is the same as for the MHV superamplitudes: in order to obtain the correct fermionic degree there must be two three-point $\overline{\rm MHV}$ superamplitudes participating in the cut (the overall fermionic degree being $2(k-2)+8 + 4 + 4 + 8 - 4\times 4 = 2(k-2)+8$), and these two three-point $\overline{\rm MHV}$ superamplitudes must not be adjacent in order not to constrain the external kinematics. 
Of course already at the NMHV case we expect to find  two-mass hard, three-mass and four-mass boxes as well as three-mass triangles, as indicated in \eqref{eq:all-1loop}. 
\end{enumerate}

The strategy we will follow will consist in computing the coefficient of the {\it finite} box functions using quadruple cuts. The complete result for the one-loop MHV super form factor will then be given by the sum of these finite box functions with the one-mass triangles accounting for the expected IR divergences. 

In the remaining part of this section we will first derive the IR-divergent part of general one-loop form factors. Next, we will consider the simplest case, that of $k=n$, which we call Sudakov in analogy with $\T_2$ (we also call these form factors \emph{minimal}), which we will compute using two-particle cuts. Finally, we will derive the expression of MHV form factors for general $n$ and $k$ using quadruple cuts.

\subsection{General  IR-divergent structure  of form factors}

As noted in \cite{Bena:2004xu},
the IR divergences of generic one-loop amplitudes in $\mathcal{N}=4$ SYM are  captured by a particular two-particle cut diagram where on one side of the cut there is a four-point amplitude.%
\footnote{See also \cite{Brandhuber:2009kh} for an application of the same ideas to dual conformal anomalies at one loop.} 
\begin{figure}[h]
\centering
\includegraphics[width=0.7\textwidth]{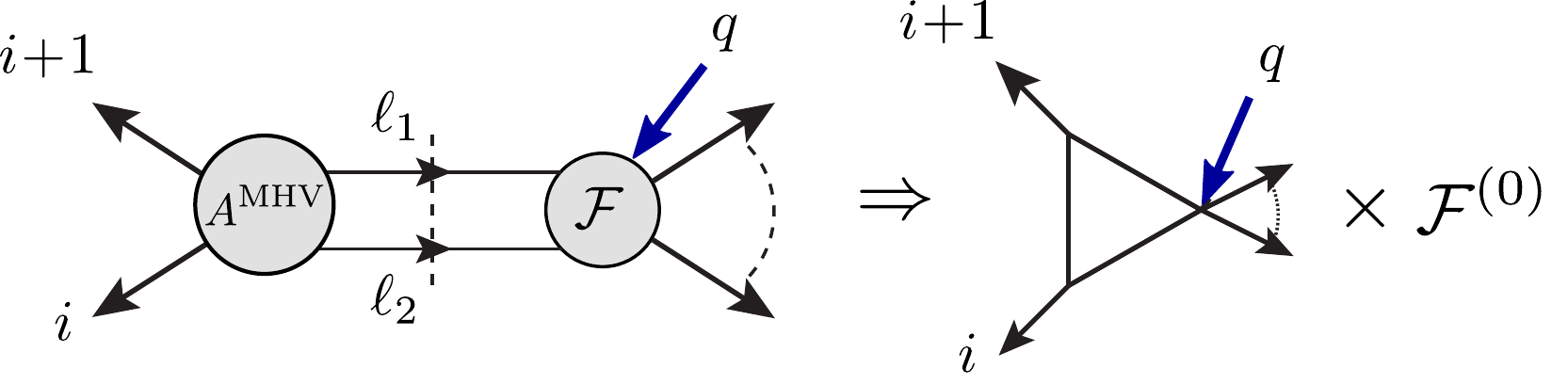}
\caption{\it The two-particle cut diagram which captures the IR divergences of general one-loop form factor. The integration region responsible for the IR divergences is the forward-scattering region, where   $\ell_1 \rightarrow -p_{i+1}$ and $\ell_2 \rightarrow -p_i$.}
\label{fig:IRcut}
\end{figure}
The same is true for form factors, and their IR divergences are fully captured by a two-particle cut diagram where the participating amplitude is a four-point amplitude. IR divergences arise from a particular region in the space of internal momenta  $\ell_1$ and $\ell_2$, namely the forward scattering region (see Figure \ref{fig:IRcut}). Indeed, when  $\ell_1 \rightarrow -p_{i+1}$, the four-point kinematics also forces  $\ell_2 \rightarrow -p_{i}$, and this creates a simple pole which is responsible for the IR divergence of the amplitudes.  Following the same proof as in \cite{Brandhuber:2009kh}, it is easy to show that in the limit $\ell_1 \rightarrow -p_{i+1}$ and $\ell_2 \rightarrow -p_{i}$, the two-particle cut in question can be uplifted to a one-mass triangle integral multiplied by the tree-level form factor. Summing over all the channels, we obtain the leading IR divergence of generic form factors%
\footnote{In writing the second equality we have dropped a factor of $e^{ \gamma_{\rm E} \epsilon}\, r_\Gamma = 1+ \cO(\epsilon^2)$, where $r_\Gamma$ is defined in~\eqref{eq:rgamma}.} 
\begin{align}
\label{eq:all-1loopIR}
\F_{k,n}^{\rm IR} \ = \ -\F_{k,n}^{(0)} \sum_{i=1}^n s_{i\,i+1} I_{3;i}^{1m}(s_{i\,i+1})  \ = \   \F_{k,n}^{(0)}\
\sum_{i=1}^n { (-s_{i\,i+1} )^{-\epsilon} \over \epsilon^2 }   \ .
\end{align}
\subsection{Three-point super form factor of $\T_3$}

\begin{figure}[htb]
\centering
\includegraphics[scale=0.7]{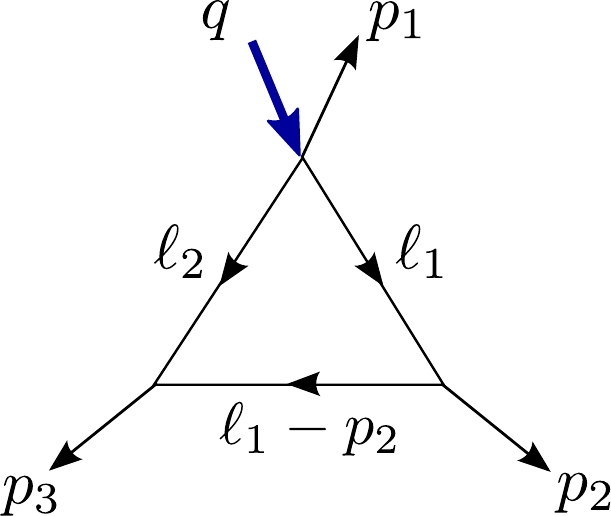}
\caption{\it The result for the $(q-p_1)^2$ cut of the one-loop Sudakov form factor of $\T_3$.}
\label{fig:1loop3}
\end{figure}

As a warm-up, we start by computing the simplest form factor at one loop, namely the Sudakov form factor.
\begin{figure}[h]
\centering
\includegraphics[width=0.4\textwidth]{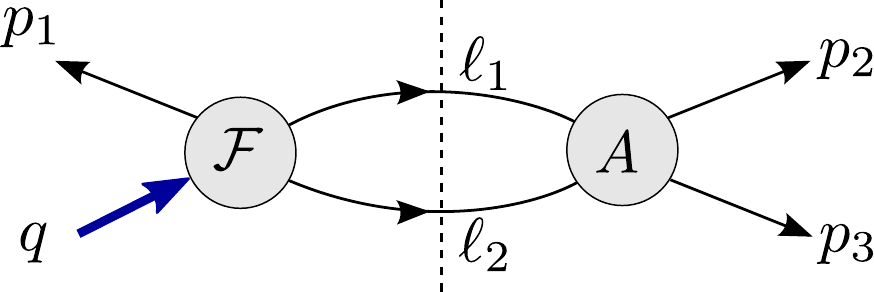}
\caption{\it $(q-p_1)^2$ two-particle cut for the Sudakov super form factor of $\T_3$.}
\label{fig:cut3}
\end{figure}\\
The cut of $\F^{(1)}_{3,3}$ across the $(q-p_1)^2$ channel, shown in Figure \ref{fig:cut3}, is given by
\beqa
\label{eq:1loop3}
\begin{split}
 &  \int\!\! d\text{LIPS}(\ell_1,\ell_2;P)\, \F_{3,3}(1,\ell_1,\ell_2;q,\gamma^{+})\ A^{\text{MHV}}(-\ell_1,2,3,-\ell_2) \\  
 =\, &  \int\!\! d\text{LIPS}(\ell_1,\ell_2;P) \frac{(\eta^{-,1})^2}{\b{\ell_1\,\ell_2}^2}\frac{\b{\ell_1\,\ell_2}^4}{\b{23}\b{3\,\ell_2}\b{\ell_2\,\ell_1}\b{\ell_1\,2}}  \\
 =\, &\frac{(\eta^{-,1})^2}{\b{23}^2}  \int\!\! d\text{LIPS}(\ell_1,\ell_2;P) \frac{\b{23} [3\,\ell_2] \b{\ell_2\,\ell_1} [\ell_1\,2]}{4(p_3\cdot \ell_2)(p_2\cdot \ell_1)} \ , 
\end{split}
\eeqa
where $P=q-p_1$, the MHV superamplitude is given in \eqref{MHVsupera}, and $d\text{LIPS}(\ell_1,\ell_2;P) $ stands for \emph{Lorentz Invariant Phase Space} measure, which is in general defined as
\begin{align}
\label{eq:LIPS}
d\text{LIPS} (\ell_1,\ell_2,\dots,\ell_n;P)\,\equiv\, \prod_{i=1}^n d^4\ell_i\delta^+(\ell_i^2)\times\delta^{(4)}\Big(\sum_{j=1}^n \ell_i - P\Big)\ .
\end{align}
Using $\ell_1 + \ell_2 = p_2+p_3$, the numerator of \eqref{eq:1loop3} can be written as  
$2 s_{23}(p_2\cdot \ell_1)  $, 
 thus the result is a one-mass triangle with massive corner $P$, as shown in Figure \ref{fig:1loop3}.
There is no ambiguity in lifting this cut to a full integral \cite{Kosower:1999xi}. Summing over the contribution of all cuts we arrive at  
the complete result for $\F^{(1)}_{3,3}$, 
\begin{align}
\label{eq:1loop3ff}
\F^{(1)}_{3,3}\ = \ 
\F^{(0)}_{3,3} \, \sum_{i=1}^3 { ( -  s_{i\,i+1} )^{-\epsilon} \over \epsilon^2} \ . 
\end{align}
We mention that this  one-loop Sudakov form factor of $\Tr\big[ (\phi_{12})^3\big] $ was computed earlier in \cite{Bork:2010wf} and our result agrees with theirs.

\subsection{$n$-point MHV super form factors of $\T_3$}

As stated earlier, we only need to compute the quadruple cut diagrams of the one-loop MHV super form factor of $\T_3$. The final result will then be expressed as a sum of the IR-divergent expression \eqref{eq:all-1loopIR} plus finite two-mass easy boxes, whose coefficients we are going to determine now using supersymmetric quadruple cuts  \cite{Drummond:2008bq}. 

\begin{figure}[htb]
\centering
\includegraphics[width=0.28\textwidth]{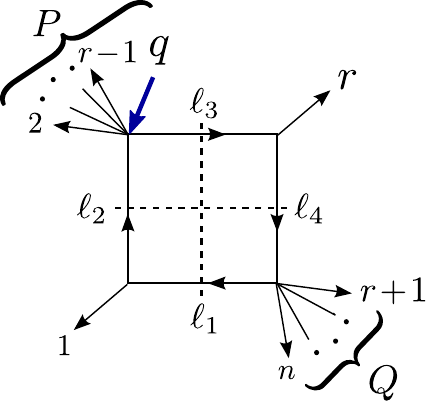}
\caption{\it Quadruple cut of the super form factor $\F_{3,n}^{\text{\rm MHV}(1)}$.}
\label{fig:quadcut}
\end{figure}

The two-mass easy quadruple cuts we consider are shown in Figure \ref{fig:quadcut}, where for convenience we label the massless legs $1$ and $r$.
\\
The coefficient of the corresponding box is given by 
\begin{align}
\begin{split}
\label{eq:qc-coeff}
\cC (1, P, r, Q) \ = \  {1\over 2} \sum_{\mathcal{S}_{\pm}} \int\!\prod_{i=1}^4 d^4 \eta^i \ 
 \F^{\text{MHV}}_{3,r}(2,\dots,r-1,\ell_3,-\ell_2;q,\gamma^{+})\times A^{\MHVb}(-\ell_3,r,\ell_4)& \\
\times \, A^{\text{MHV}}(-\ell_4,r+1,\dots,n,\ell_1)\times A^{\MHVb}(-\ell_1,1,\ell_2)& \ , 
\end{split}
\end{align}
where the sum is over the solutions to the cut equations. Since only one solution to the cut equations $\ell_1^2=\ell_2^2=\ell_3^2=\ell_4^2=0$ contributes to \eqref{eq:qc-coeff}, one can drop the sum over $\mathcal{S}_{\pm}$, leaving an overall factor of $1/2$.
The form factor $ \F^{\text{MHV}}_{3,r}$  is given in \eqref{eq:FF3MHVsusy}, and the $\text{MHV}$ and $\overline{\rm MHV}$ superamplitudes entering this expression are 
given in \eqref{MHVsupera} and  \eqref{MHVbsupera}, respectively. 
Because of the presence of $\MHVb$ three-particle amplitudes on the massless corners, we have
\begin{equation}
\label{eq:proplambdas}
\lambda^{\ell_3}\propto\lambda^{\ell_4}\propto\lambda^r\ ,\qquad\qquad \lambda^{\ell_1}\propto\lambda^{\ell_2}\propto\lambda^1\ .
\end{equation}
Using  the delta-functions contained in the $\MHVb$  and MHV amplitudes, together with the conditions \eqref{eq:proplambdas} one can quickly determine the fermionic variables associated to the internal supermomenta, 
\begin{equation}
\eta^{\ell_4}=  \sum_{i=r+1}^n \frac{\b{1\,i}}{\b{1\,\ell_4}}\,\eta^i \ , \quad
\eta^{\ell_1}= - \sum_{i=r+1}^n \frac{\b{i\,r}}{\b{\ell_1\,r}}\,\eta^i\ , 
\end{equation}
and 
\begin{align}
\label{eq:etal3b}
\eta^{\ell_3} = \frac{[\ell_4\,\ell_3]}{[\ell_4\,r]}\,\eta^r +\sum_{i=r+1}^n \frac{\b{1\,i}[\ell_3\,r]}{\b{1\,\ell_4}[\ell_4\,r]}\,\eta^i \ , \\
\label{eq:etal2b}
\eta^{\ell_2}= \frac{[\ell_2\,\ell_1]}{[1\,\ell_1]}\,\eta^r +\sum_{i=r+1}^n \frac{\b{1\,r}[1\,\ell_2]}{\b{\ell_1\,r}[1\,\ell_1]}\,\eta^i\ . 
\end{align}
Integrating out the internal $\eta$ variables produced  the two expected supermomentum conservation delta-functions 
$\delta^{(4)} ( \gamma^+ - \sum_{i=1}^n \lambda^i \eta^{+,i}) \, \delta^{(4)}  (  \sum_{i=1}^n \lambda^i \eta^{-,i}) $ as well as a Jacobian 
\begin{equation} 
\label{eq:jac1}
J\ =\ (\b{\ell_1\,\ell_4}[r\,\ell_4][\ell_1\,1])^4\, =\,   [1|\ell_1\,\ell_4|r]^4\ .
\end{equation}
Let us now manipulate the Parke-Taylor prefactors coming from \eqref{eq:qc-coeff} together with \eqref{eq:jac1}:
\begin{align}
\begin{split}
&\frac{1}{\b{\ell_2\,2}\ldots\b{r-1\,\ell_3}\b{\ell_3\,\ell_2}} \times \frac{1}{[\ell_3\,r][r\,\ell_4][\ell_4\,\ell_3]} \\ 
\times \ & \frac{1}{\b{\ell_4\,r+1}\dots\b{n\,\ell_1}\b{\ell_1\,\ell_4}} \times \frac{1}{[\ell_1\,1][1\,\ell_2][\ell_2\,\ell_1]} \times (\b{\ell_1\,\ell_4}[r\,\ell_4][\ell_1\,1])^4
\\
\label{eq:ptman1}
= \ & \text{PT}_n [1|\ell_1\,\ell_4|r]^3 \frac{\b{n1}\b{12}\b{r-1\,r}\b{r\,r+1}}{\la r-1|\ell_3\,\ell_4|r+1\ra \la 2|\ell_2\,\ell_1|n\ra [ 1|\ell_2\,\ell_3|r]}\ , 
\end{split} 
\end{align}
where $\text{PT}_n\equiv   1 / ( {\b{12}\b{23}\cdots\b{n1}})$.
This expression can be considerably simplified by using momentum conservation and  the replacements  \eqref{eq:proplambdas} inside expressions which are homogeneous functions of degree zero of the spinors associated to the cut loop momenta. In this way one can rewrite this product of amplitudes as 
\begin{align}
-\text{PT}_n \frac{[1\,\ell_2]\b{\ell_2\,r} [\ell_3\,r]\b{\ell_3\,1} \b{r\,r+1}}{\b{r\,r+1}}=\text{PT}_n\,  \Tr_+(\ell_2  \, p_r \,   \ell_3 \,  p_1) \ .
\end{align}
Using again momentum conservation and $(p_1\cdot\ell_2)=0$ we can rewrite the trace as 
\begin{align}
\begin{split}
&\Tr_+(\ell_2  \, p_r \,   \ell_3 \,  p_1) \ = \ \Tr_+(Q\, p_r \, P \,  p_1) \ = \ 2 (p_1\cdot P) (p_r\cdot Q) + 2 (p_r\cdot P) (p_1\cdot Q) -s_{1r} (Q\cdot P) 
\ . 
\end{split} \nonumber \\[5pt]
\end{align}
Introducing the kinematic variables 
\begin{align}
s\ \equiv\ (p_r+Q)^2\ , \qquad t\ \equiv\ (p_r+P)^2\ , 
\end{align}
we can write  $s_{1r}\ =\  - (s+t-P^2-Q^2)$. With that we can finally rewrite the trace as
\begin{align}
\Tr_+(\ell_2  \, p_r \,   \ell_3 \,  p_1)  \ = \ P^2 Q^2-st\ .
\end{align}
Substituting this back into \eqref{eq:qc-coeff},  we arrive at the result for the supercoefficient, 
\beq
\label{eq:result-qc}
\cC(1,P, r, Q) \ = \ 
\,
\F_{2,n}^{\text{MHV}(0)} 
 \left( P^2 Q^2-st \right)  
\  \delta\Big(\dfrac{1}{2}\, \sum\limits_{i<j=2}^{r-1}(2-\delta_{ij})\dfrac{\b{1\,i}\b{j\,r}}{\b{r\,1}}\eta^{-,i} \cdot \eta^{-,j}\Big)\ .
\eeq
We note that the delta-function appearing above corresponds precisely to that of the form factor entering the quadruple cut, where we conveniently singled out the two internal loop legs $\ell_2$ and $\ell_3$ (the corresponding 
spinor variables being in turn proportional to the two external momenta entering the adjacent massless corners, $\lambda^{\ell_2}\propto \lambda^1,\, \lambda^{\ell_3}\propto \lambda^r$ , cf. \eqref{eq:proplambdas}). We can therefore rewrite \eqref{eq:result-qc}~as
\begin{align}
\label{ccpp}
\cC(1,P, r, Q) = \F_{2,n}^{\text{MHV}(0)}   \left( P^2 Q^2-st \right) f_{3,r}(2,\dots , r-1,r,1)\, , 
\end{align}
where $f_{3,r}$ is defined in \eqref{eq:little-f}%
\footnote{We stress that,  in \eqref{ccpp},  we should use the form of the quantity  $f_{3,r}$ (defined in  \eqref{eq:little-f}) given in \eqref{eq:FF3MHVsusy} and not \eqref{eq:FF3MHVsusy-simple}. The reason is that these two expressions are  only equivalent on the support of the delta-function $\delta\big(\sum_{i=1}^rp_i-q\big)$, which is not true in this case.}.

\begin{figure}[htb]
\centering
\includegraphics[width=0.65\textwidth]{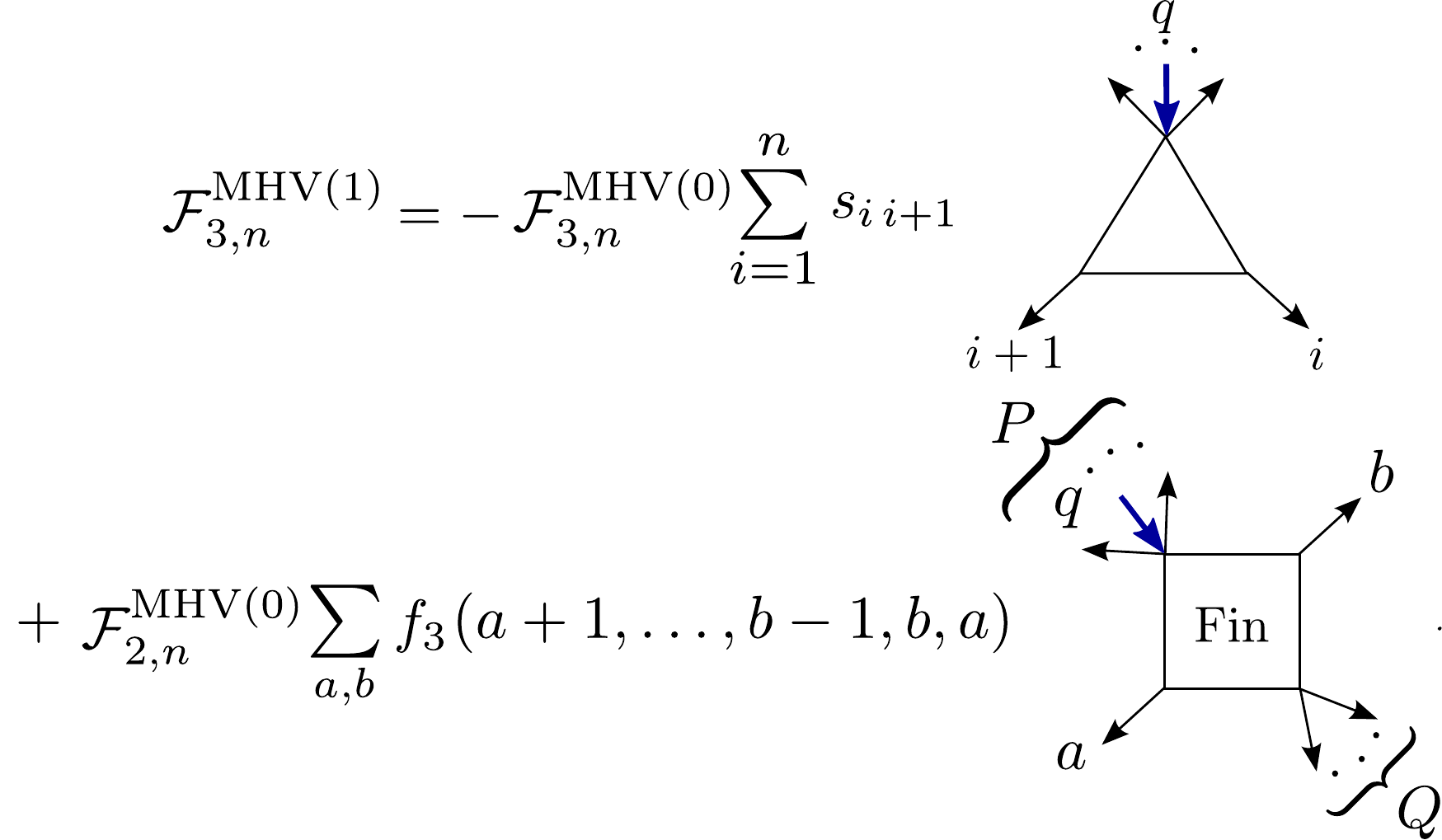}
\caption{\it One-loop result for $\F_{3,n}^{\text{\rm MHV} (1)}$. Here $P$ and $Q$ stand for the momenta of the massive corners and, as usual, $s\equiv(P+p_a)^2,\;t\equiv(Q+p_a)^2$ .
}
\label{fig:1loopfullanswer}
\end{figure}

We are now ready to write down the full result for the one-loop MHV super form factor $\F_{3,n}^{\text{MHV} (1)}$ for general $n$. It is given by 
\begin{align}
\begin{split}
\label{eq:1-loop-MHV-BIS}
\F_{3,n}^{\text{MHV} (1)}\ &=  \  \F_{3,n}^{\text{MHV}(0)}\,    \sum_{i=1}^n {  ( - s_{i\,i+1})^{- \epsilon}   \over \epsilon^2}  \\
   &+ \F_{2,n}^{\text{MHV}(0)} \sum_{a, b} f_{3}(a+1,\dots,b-1,b,a)  \text{Fin}^{\rm 2me}(p_a, p_b, P, Q)\, .   
\end{split}
\end{align}
%
For clarity, we illustrate \eqref{eq:1-loop-MHV-BIS} graphically in Figure \ref{fig:1loopfullanswer}.

\subsection{$n$-point MHV super form factors of $\T_k$}

The one-loop result for general $k$ is not qualitatively different from that for $k=3$ computed in the previous section; the only undetermined coefficients are those of finite two-mass easy box-functions, which we find using quadruple cuts.
Indeed,  once we know the result for $\F_{3,n}^{\text{MHV}(1)}$, the generalisation for $\F_{k,n}^{\text{MHV}(1)}$ is almost immediate. This is because  the tree-level result  \eqref{conj}  for $\F_{k,n}^{\text{MHV}(0)}$ has  the same trivial dependence on legs $n-1$ and $n$ as $\F_{3,n}^{\text{MHV}(0)}$. The answer is then an immediate generalisation of \eqref{eq:1-loop-MHV-BIS}:
\beqa
\label{eq:1-loop-MHV-all-k}
\begin{split} 
\F_{k,n}^{\text{MHV} (1)} &=   \    
\F_{k,n}^{\text{MHV}(0)}\, 
\sum_{i=1}^n 
{  ( - s_{i\,i+1})^{- \epsilon}   \over \epsilon^2}  \\
   &+  \, \F_{2,n}^{\text{MHV}(0)} \sum_{a, b} f_{k}(a+1,\dots,b-1,b,a)  \ \text{Fin}^{\rm 2me}(p_a, p_b, P, Q)\ .   
\end{split} 
\eeqa
This is our final, compact  expression for the  $n$-point  form factor of $\cT_k$ at one loop with arbitrary $k$ and $n$.

\label{sec:ff-one-loop}

\section{Two loops}
\label{sec:ff-two-loops}

In this section we proceed to computing the minimal form factors of $\T_k$ at two loops. The first step consists in using generalised unitarity to construct the  two-loop form factors in terms of a basis of integral functions. Here we are in the fortunate situation where  all  the required integral functions are known analytically from the work of \cite{Gehrmann:1999as,Gehrmann:2000zt} in terms of classical and Goncharov polylogarithms. Such expressions are typically rather long,  but past experience \cite{DelDuca:2010zg, Goncharov:2010jf, Heslop:2011hv, Prygarin:2011gd, Brandhuber:2012vm, Dixon:2011pw,Dixon:2014voa}
suggests that for appropriate finite  quantities,  the final result can be condensed to a much   simpler and compact form. 

Following this line of thought, and also inspired by the well-known exponentiation of IR divergences, we will introduce finite remainder functions \cite{Anastasiou:2003kj,Bern:2005iz, Bern:2008ap, Drummond:2008aq}. These remainders are defined in terms of two important universal constants, and our calculation confirms that they coincide with the cusp anomalous dimension and collinear anomalous dimension which appear in the definition of remainders of amplitudes \cite{Bern:2008ap, Drummond:2008aq} and form factors of $\cT_2$ \cite{vanNeerven:1985ja, Gehrmann:2011xn, Brandhuber:2012vm}.%
\footnote{This result disagrees with the findings of \cite{Bork:2010wf}, where a different result for the collinear anomalous dimension was obtained, see\sref{remainder-constants} for more details. } 

Finally, we use the symbol of transcendental functions and the related, refined notion of the coproduct  \cite{Golden:2014xqa}
to construct the remainders in an extremely compact form.
For the remainder of $\cF_{3,3}$ we find a three-line expression containing only classical
polylogarithms, while the answer for $\cF_{k,k}$ is a combination of universal, compact building blocks which contain classical polylogarithms supplemented by just two Goncharov polylogarithms.

We now present a brief outline of the rest of the chapter. 
In \sref{sec:remainder} we define finite remainder functions of the minimal two-loop form factors, and use the concept of the symbol of transcendental functions, revised in \sref{sec:Trans-func-symbols}, to rewrite the result in terms of classical polylogarithms only. 
In \sref{sec:allk} we work out the analytic results for 
form factors of $\cT_k$ with $k>3$ and are able to express them in terms of a single universal building block that depends on three scale-invariant ratios of Mandelstam variables. Again, using the symbol and coproduct of transcendental functions we find a compact answer which, in addition to classical polylogarithms,  contains also two Goncharov polylogarithms. 
Finally, in \sref{sec:discussion} we analyse in some detail the behaviour of form factors in collinear and soft limits, and note that minimal form factors have unconventional factorisation properties compared to amplitudes and non-minimal form factors.

\subsubsection{Colour decomposition and planarity}
\label{sec:colour-planar}

In this section we briefly consider the colour decomposition of form factors and its implications for the calculation of  two-loop  form factors of $\T_k$. 

Following the same procedure  as for scattering amplitudes, a planar $n$-point  form factor of a certain  single-trace operator $\O$ can be expressed as
\begin{equation} \label{eq:ff-colordecomp}
\F^{\, a_1\cdots a_n}_{\O,n}=\sum_{\sigma\in S_n/\mathbb{Z}_n}\mathrm{Tr}(t^{a_{\sigma(1)}}t^{a_{\sigma(2)}} \cdots t^{a_{\sigma(n)}})\F_{\O,n}(\sigma(1),\sigma(2),\dots, \sigma(n);q,\gamma^+)\ ,
\end{equation}
where $t^a$ are  fundamental generators of  $SU(N)$, and the  $\F_{\O,n}$  are colour-ordered form factors. 

An important remark is in order here. For the case of $\T_2$, the minimal ({\it i.e.}~two-point) form factor has the colour factor $\Tr(t^a t^b)$, which is simply $\delta^{ab}$. As noticed in \cite{Brandhuber:2012vm}, this simple fact has striking consequences for the two-loop calculations. Consider for instance a two-particle cut of the form 
\beq
\int\!d{\rm LIPS} (\ell_1, \ell_2; p_1 + p_2 ) \ \F_{2,2}^{(0)}(\ell_1,\ell_2;q,\gamma^+) \times \A^{(1)}_4(-\ell_1,-\ell_2, 1,  2)\ .
\eeq
The colour factor $\delta^{a_{\ell_1} a_{\ell_2}}$ arising from  the form factor can  contract with a double-trace term from the complete  one-loop amplitude $\A^{(1)}_4$,  generating extra powers of $N$. Hence, these double-trace terms, which are normally subleading in colour, are lifted to leading order in $N$. As a consequence, one has to keep double-trace contributions from  $\A^{(1)}_4$. 
This is the reason why planar two-loop form factors of $\T_2$ receive contributions from non-planar integral topologies \cite{vanNeerven:1985ja}. In \cite{Brandhuber:2012vm} it was shown that this also applies to non-minimal form factors of $\T_2$.

Fortunately this is not the case for $k>2$ at two loops. This  is because now one can only have three- or higher-point form factors entering the cuts, which are never dressed with $\delta^{ab}$ colour factors. This  situation is very similar to the case of planar scattering amplitudes, where only planar integrals contribute.
We will make use of this fact in the two-loop calculation of $\F_{k,k}$ in the following sections.

Note that form factors are still intrinsically non-planar quantities since the operators are colour singlets. In particular, for form factors $\F_{k,n}$ non-planar integral topologies arise starting at $k^{\rm th}$ loop order. Moreover, even for one- and two-loop form factors of $\T_2$, where only planar integrals contribute, one cannot define a consistent set of region momenta for all integrals contributing to a certain form factor.

\subsection{Minimal form factor of $\T_3$ at two loops}
\label{sec:T3}
\subsubsection{Unitarity cuts}
In this section we calculate the two-loop form factor $F_{3,3}^{(2)}$ using
generalised unitarity. 
In particular we show that two-particle cuts combined with two different types of three-particle cuts are sufficient to fix the result uniquely and express it as a linear combination of planar two-loop master integrals.

\noindent We start by considering the two-particle cuts. 
\begin{figure}[htb]
\centering
\includegraphics[width=0.8\textwidth]{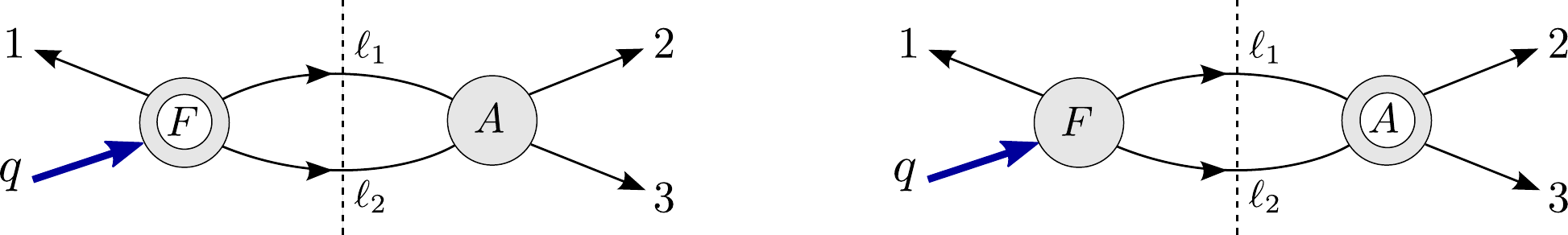}
\caption{\it Two-particle cuts of $F^{(2)}_{3,3}$ in the kinematic channel $s_{23}=(p_2+p_3)^2$. There are two possible factorisations: $F^{(1)} \times A^{(0)}$ (left) or $F^{(0)} \times A^{(1)}$ (right). Cyclic permutations of the legs 123 generate the remaining two-particle cuts.}
\label{fig:twoloopcut1}
\end{figure}
\noindent

At two-loop level there are two such cuts as shown in Figure \ref{fig:twoloopcut1}.
First, we consider the cut on the left-hand side of  Figure \ref{fig:twoloopcut1}, where a one-loop form factor is merged with a tree-level four-point amplitude. The cut integrand is given by
\beq
\mathcal{C}_{s_{23}}^{(1)} \, = \, \int  \! d\text{LIPS}(\ell_1,\ell_2;P)\, F^{(1)}_{3,3}(1, \ell_1, \ell_2; q) \, A^{(0)}_4(-\ell_2, -\ell_1, 2, 3)
\ , 
\eeq
where $P= p_2+p_3$ and, making the helicities explicit,
\beqa \label{cut1}
A^{(0)}_4(-\ell_2^{\phi_{34}}, -\ell_1^{\phi_{34}}, 2^{\phi_{12}}, 3^{\phi_{12}}) &=&
{\b{ \ell_2 \, \ell_1 } \b{ 23 }  \over \b{3 \, \ell_2 } \b{ \ell_1 \, 2 } }
= {s_{23} \over 2 (\ell_1 \cdot p_2) } \, ,\\[5pt] \label{cut12}
F^{(1)}_{3,3}(1^{\phi_{12}}, \ell_1^{\phi_{12}}, \ell_2^{\phi_{12}}; q) &=&
s_{23} \, I_{3}^{1 \rm m}(\ell_1, \ell_2; p_2+p_3) + (q- \ell_2)^2\, I_{3}^{1\rm m}(p_1, \ell_1; q- \ell_2) \nonumber 
\\ [5pt]
&+& (q- \ell_1)^2\, I_{3}^{1 \rm m}(\ell_2, p_1; q-\ell_1) \, .
\eeqa
Here $I_{3}^{1 \rm m}(a, b; c)$ is a  one-mass triangle integral, 
\begin{equation*}
\includegraphics[width=0.35\linewidth]{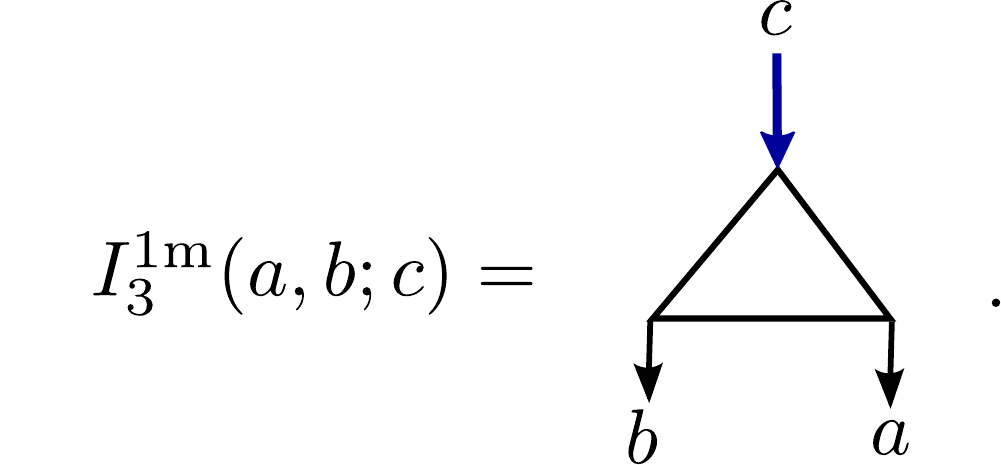}
\end{equation*}
From \eqref{cut1}, it is clear that the effect of $A^{(0)}_4(-\ell_2^{\phi_{34}}, -\ell_1^{\phi_{34}}, 2^{\phi_{12}}, 3^{\phi_{12}})$ is simply to attach the following three-propagator object with numerator $s_{23}$ to the one-loop form factor: 
\beq
\label{eq:attach}
\vcenter{\hbox{\includegraphics[width=0.16\linewidth]{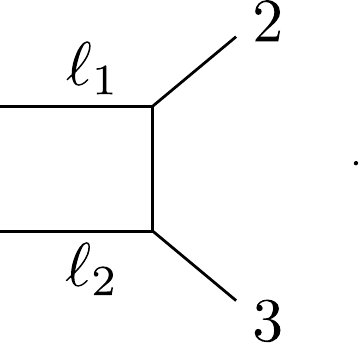}}} 
\eeq
By attaching the structure in \eqref{eq:attach}  to all the triangles appearing in the one-loop form factor \eqref{cut12}, we find that the cut integrand $\cC_{s_{23}}^{(1)}$ is given by the following sum, 
\beq
\label{eq:integrals-cut-1}
\vcenter{\hbox{\includegraphics[width=0.95\linewidth]{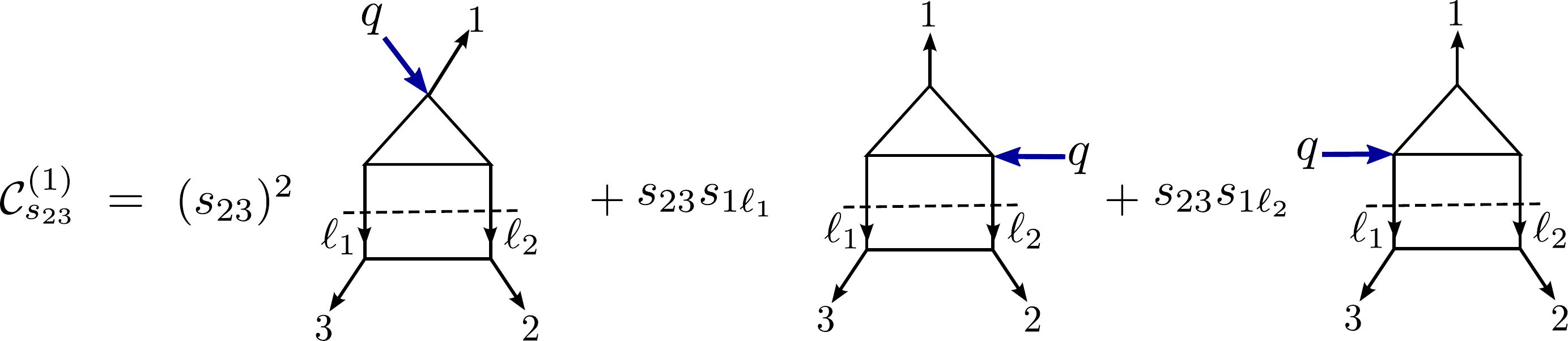}}}\, .
\eeq\\[3pt]
The straight dashed lines in the integrals above indicate that the momenta $\ell_1$ and $\ell_2$ are cut. We now introduce a more concise notation for numerators which will be used in the following. To indicate a factor of $s_{i_1i_2\cdots i_n} $ in the numerator of an integral, we draw a curved dashed line passing through $n$ propagators whose momenta sum to $\sum_{j=1}^n p_{i_j}$. In this notation, \eqref{eq:integrals-cut-1} can be represented as 
\begin{figure}[h]
\centering
\includegraphics[width=0.6\linewidth]{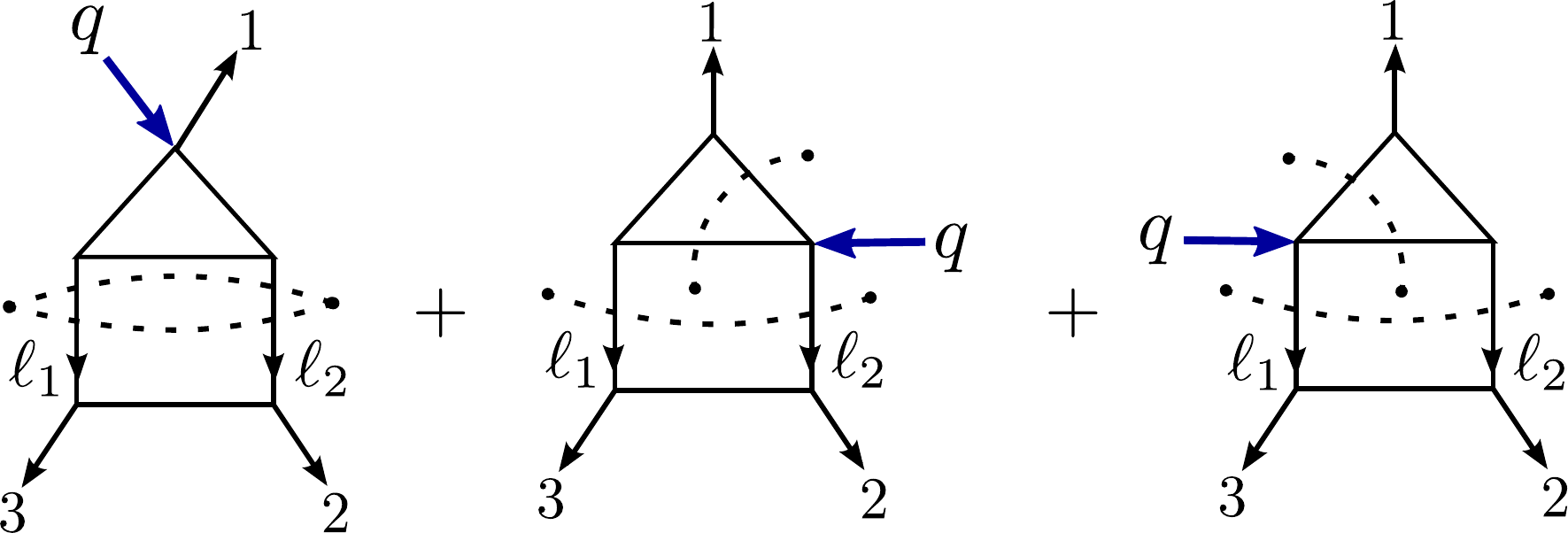}
\caption{\it Integrals detected by two-particle cuts in the two-loop form factor $F_{3,3}^{(0)}$.}
\label{fig:integrals-cut-2}
\end{figure}

Note that at this stage we have also uplifted the cut integrals to full Feynman integrals by replacing the cut legs by propagators.
We stress that this procedure induces ambiguities in the numerators, since on the cut $\ell_1^2=\ell_2^2=0$, and hence we cannot distinguish $s_{1\ell_1}$ from  $2(p_1 \cdot \ell_1)$ or $s_{1\ell_2}$ from $2(p_1 \cdot \ell_2)$. Such ambiguities will be eliminated later using
three-particle cuts.

\noindent The second two-particle cut, depicted on the right-hand side of Figure \ref{fig:twoloopcut1}, is given by
\beq
\mathcal{C}_{s_{23}}^{(2)} \, = \, \int \! d\text{LIPS}(\ell_1,\ell_2;P)\, F^{(0)}_{3,3}(1, \ell_1, \ell_2; q) \, A^{(1)}_4(-\ell_2, -\ell_1, 2, 3)
\ , 
\eeq
 where
\begin{align}
\begin{split}
F^{(0)}_{3,3}(1^{\phi_{12}}, \ell_1^{\phi_{12}}, \ell_2^{\phi_{12}}; q) =&
1\, , \\[5pt]
A^{(1)}_4(-\ell_2^{\phi_{34}}, -\ell_1^{\phi_{34}}, 2^{\phi_{12}}, 3^{\phi_{12}}) =&
{s_{23} \over 2\,(\ell_1 \cdot p_2) } \big[ s_{23}\, (p_2-\ell_1)^2\, I_{4}^{0\rm m} (-\ell_2, -\ell_1, 2, 3) \big] \, ,
\end{split}
\end{align}
where $I_{4}^{0\rm m}$ stands for the zero-mass scalar box integral, 
\begin{equation}
\includegraphics[width=0.35\textwidth]{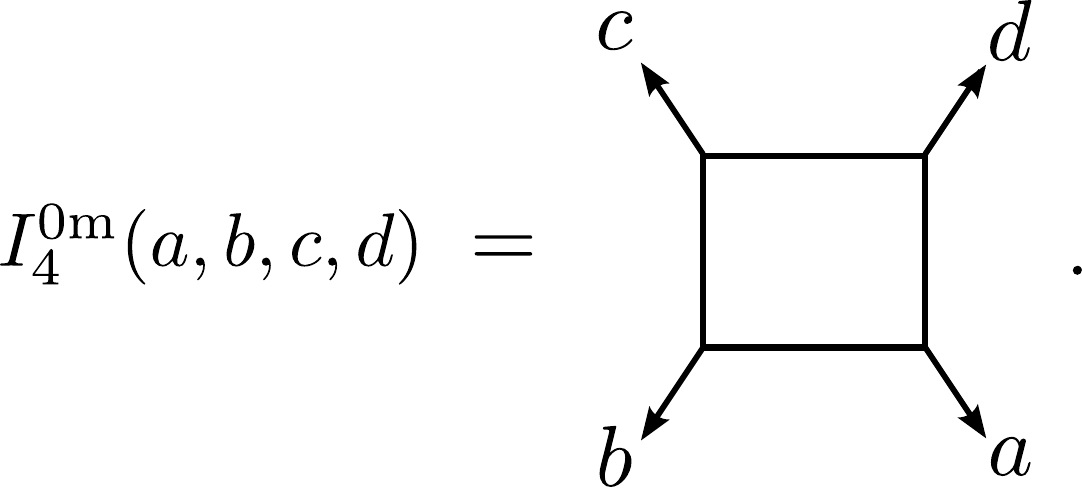}
\end{equation}
Uplifting $\mathcal{C}_{s_{23}}^{(2)}$ to a full Feynman integral we obtain the contribution depicted in \eqref{eq:integrals-cut-2-2}, 
\beq
\label{eq:integrals-cut-2-2}
\vcenter{\hbox{\includegraphics[width=0.14\linewidth]{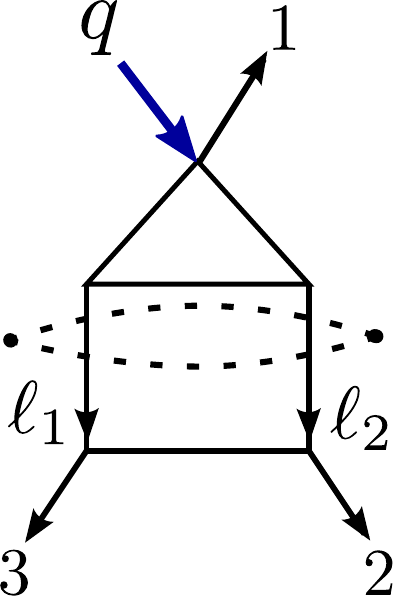}}}\ ,
\eeq
which was already detected in the first two-particle cut.
Therefore the integrals of \fref{fig:integrals-cut-2} alone comprises the full result for this cut.

\begin{figure}[h]
\centering
\includegraphics[width=0.38\linewidth]{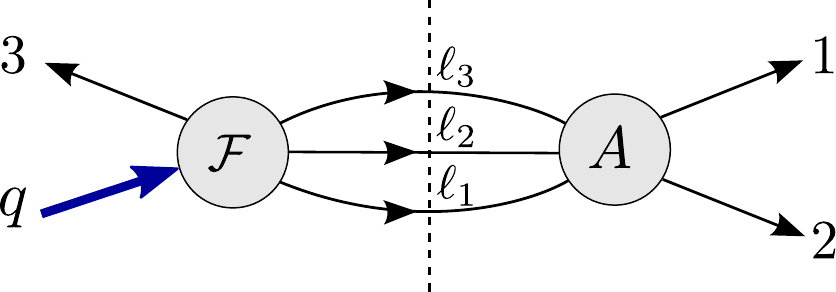}
\caption{\it A possible three-particle cut of $\F_{3,3}^{(2)}$.}
\label{fig:tricut1}
\end{figure}
We now move on to investigate three-particle cuts. The first case we want to consider is shown in Figure \ref{fig:tricut1}.
This three-particle cut is given by
\begin{equation}
\label{eq:triplecut1}
\cC^{(3)}_{s_{12}}\, =\, \int \! d\text{LIPS}(\ell_1,\ell_2,\ell_3 ;P) \int d^{12}\eta \ \F^{(0)}_{3,4}(3,\ell_3,\ell_2,\ell_1;q,\gamma^+)\, A^{(0)}_{5} (1,2,-\ell_1,-\ell_2,-\ell_3)\ ,
\end{equation}
where $P=p_1+p_2$ and $d^{12}\eta = d^4\eta^{\ell_1} d^4\eta^{\ell_2} d^4\eta^{\ell_3}$. Importantly,
in order to perform the sum over internal helicities efficiently
we use the supersymmetric formalism for form factors developed in \cite{Brandhuber:2011tv}, and adapted in \cite{Penante:2014sza} to the case of the operators $\cT_k$, see \sref{sec:ff-tree}. At the end of the calculation we will select all external particles to be $\phi_{12}$. 

There are two distinct choices of $R$-charge sectors for the form factor and  amplitude participating in the cut, namely 
\begin{equation}
\F^{\rm NMHV}_{3,4}\times A^{\rm MHV}_5\quad \text{and} \quad \F^{\rm MHV}_{3,4} \times A^{\overline{\rm MHV}}_5 \ .
\end{equation}
We consider first the case $\F^{\rm NMHV} \times A^{\rm MHV}$. The tree-level expressions entering \eqref{eq:triplecut1} are given by (omitting  a trivial delta-function of momentum conservation)\\
\begin{align}
\label{eq:FFNMHV}
\begin{split}
\F^{\text{NMHV}}_{3,4}
&=\delta^{(4)}\big(\gamma^+ - \sum_i \lambda^i \eta^{+, i} \big) \delta^{(4)}
\big(\sum_i \lambda^i \eta^{-, i} \big)
\Bigg[
{\delta^{(4)}\big( [\ell_3\,\ell_2]\eta^3 + [\ell_2\,3] \eta^{\ell_3} + [3\,\ell_3] \eta^{\ell_2} \big)
(\eta^{-,1})^2 \over [\ell_3\,\ell_2][\ell_2\,3][3\,\ell_3] (s_{3\ell_3\ell_2})^2 } 
 \\ 
&- (\ell_1 \leftrightarrow \ell_3 )
\Bigg] \, ,
\end{split} \\
A^{\rm MHV}_5 & =\frac{ \delta^{(8)}\big( \lambda^1 \eta^1 + \lambda^2 \eta^2 -\lambda^{\ell_1} \eta^{\ell_1} -\lambda^{\ell_2} \eta^{\ell_2}-\lambda^{\ell_3} \eta^{\ell_3}\big)}{\b{12}\b{2\,\ell_1}\b{\ell_1\,\ell_2}\b{\ell_2\,\ell_3}\b{\ell_3\,1} }\ ,
\end{align}\\
where the NMHV form factor of $\cT_3$ is given in \eqref{sss}. After performing the integrations over the internal $\eta^{\ell_i}$'s, we arrive at the result 
\begin{align}
\label{eq:res-triplecut-1}
\left. 
\cC_{s_{12}}^{(3)}\right|_{\rm A}\,=\,\F_{3,3}^{(0)} \frac{\b{12}}{\b{2\,\ell_1}\b{\ell_1\,\ell_2}\b{\ell_2\,\ell_3}\b{\ell_3\,1}} \left(\frac{[3|q|\ell_1\ra^2}{[\ell_3\,\ell_2][\ell_2\,3][3\,\ell_3]}-  (\ell_1 \leftrightarrow \ell_3 )\right)\, .
\end{align}
The second case is $\F^{\rm MHV} \times A^{\overline{\rm MHV}}$. The expressions entering \eqref{eq:triplecut1} can be  written as\\
\begin{align}
\F^{\rm MHV}_{3,4} &= \frac{1}{\b{3\,\ell_3} \b{\ell_3\,\ell_2}\b{\ell_2\,\ell_1}\b{\ell_1\, 3} } \left[(\eta^{3,-})^2\frac{\b{3\,\ell_3}\b{\ell_1\,3}}{\b{\ell_3\,\ell_1}}- (\eta^{\ell_2,-})^2\frac{\b{\ell_2\,\ell_3}\b{\ell_1\,\ell_2}}{\b{\ell_3\,\ell_1}}\right]\ ,\\[5pt]
A^{\overline{\rm MHV}}_5 &= \frac{1}{[2\,\ell_1][\ell_1\,\ell_2][\ell_2\,\ell_3][\ell_3\,1][12]^9} \prod_{i=1}^3 \, \delta^{(4)}\big([12]\,\eta^{\ell_i}+  [2\,\ell_i]\, \eta^1 +  [\ell_i\,1] \,\eta^2 \big)\ .
\end{align}\\
After summing over internal helicities, we get
\begin{align}
\label{eq:res-triplecut-2}
\begin{split}
\left. 
\cC_{s_{12}}^{(3)}\right|_{\rm B}  \,=\,\F_{3,3}^{(0)}\, &  \frac{(s_{12})^2 [12]}{\b{3\,\ell_3} \b{\ell_3\,\ell_2} \b{\ell_2\,\ell_1}\b{\ell_1\, 3}[2\,\ell_1][\ell_1\,\ell_2][\ell_2\,\ell_3][\ell_3\,1] } \\[8pt]
\times\, & \left(\frac{\b{3\,\ell_3}\b{\ell_1\,3}}{\b{\ell_3\,\ell_1}}-\frac{[\ell_2|q|3\ra ^2 }{(s_{12})^2}\frac{\b{\ell_2\,\ell_3}\b{\ell_1\,\ell_2}}{\b{\ell_3\,\ell_1}} \right)\ .
\end{split}
\end{align}\\
Summarising, the total result for the cut \eqref{eq:triplecut1} is the sum of \eqref{eq:res-triplecut-1} and \eqref{eq:res-triplecut-2},
\begin{align}
\label{3pcab}
\cC_{s_{12}}^{(3)}\,=\,\left. \cC_{s_{12}}^{(3)}\right|_{\rm A} + \left. \cC_{s_{12}}^{(3)}\right|_{\rm B}\ .
\end{align}
Taking the purely scalar component of this cut amounts simply to performing the replacement $\F_{3,3}^{(0)} \to 1$. 
\begin{figure}[h]
\centering
\includegraphics[width=0.42\linewidth]{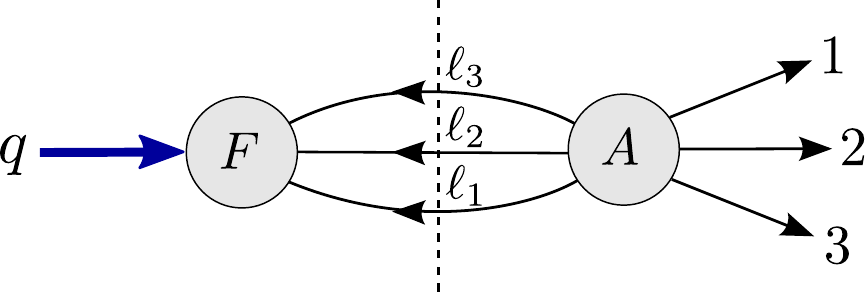}
\caption{\it A second possible three-particle cut of $\F_{3,3}^{(2)}$.}
\label{fig:tricut2}
\end{figure}

\noindent The next cut we wish to consider is shown in Figure \ref{fig:tricut2},  and is given by
\begin{equation}
\label{eq:triplecut2}
\cC_{s_{123}}^{(4)} \, =\, \int \! d\text{LIPS}(\ell_1,\ell_2,\ell_3 ;-q)\, \int\!d^{12} \eta \, \cF^{(0)}_{3,3}(-\ell_1,-\ell_2,-\ell_3;q)\, A^{{\rm NMHV }(0)}_{6} (1,2,3,\ell_1,\ell_2,\ell_3)\ .
\end{equation}
Selecting the external particles to be all scalars $\phi_{12}$, we see that the only non-vanishing form factor contributing to the cut is  $F^{(0)}_{3,3} (-\ell_1^{\phi_{12}},-\ell_2^{\phi_{12}},-\ell_3^{\phi_{12}};q) = 1$ (again omitting a momentum conservation delta-function). 
This is the only internal helicity assignment we need to consider, thus the single amplitude appearing on the right-hand side of the cut is the following six-scalar NMHV amplitude,  
\begin{align}
\begin{split}
A^{\rm NMHV}_6(1^{\phi_{12}},2^{\phi_{12}},3^{\phi_{12}},\ell_1^{\phi_{34}},\ell_2^{\phi_{34}},\ell_3^{\phi_{34}})\, = \, &\frac{\b{\ell_2\,\ell_3}[23]\la 1 | \ell_2+\ell_3|\ell_1]}{\b{\ell_3\,1}[3\,\ell_1]\la \ell_2|\ell_3+1|2]s_{1\ell_2\ell_3}}\\[5pt]
+\, & \frac{\b{\ell_1\,\ell_2}[12]\la 3 | \ell_1+\ell_2|\ell_3]}{\b{3\,\ell_1}[\ell_3\,1] \la \ell_2|\ell_1+3|2]s_{3\ell_1\ell_2}}\ . 
\end{split}
\end{align}
Hence the result of this triple cut is given by   
\beq
\label{tcc}
\cC_{s_{123}}^{(4)} \, =\, \int \! d\text{LIPS}(\ell_1,\ell_2,\ell_3 ;-q)\,A^{\rm NMHV}_6(1^{\phi_{12}},2^{\phi_{12}},3^{\phi_{12}},\ell_1^{\phi_{34}},\ell_2^{\phi_{34}},\ell_3^{\phi_{34}})
\ . 
\eeq

\subsubsection{Two-loop result}
\label{sec:result}
The two-particle cuts employed earlier show that the full two-loop result contains the combination of integrals shown if \fref{fig:integrals-cut-2}. As discussed earlier, this set of cuts does not uniquely determine the numerators of these integrals, and furthermore does not probe the presence of any integral function which  only has  three-particle cuts. 

Using the result of the three-particle cuts \eqref{3pcab} and \eqref{tcc}, we can fix all such ambiguities. In particular, we have identified two additional integral topologies without two-particle cuts contributing to the final result. 
The unique function with the correct two- and three-particle cuts turns out to be
\beq \label{eq:basisT3}
\cF^{(2)}_{3,3} = \sum^3_{i=1} \Big[  I_1(i) + I_2(i) + I_3(i) +I_4(i) - I_5(i)  \Big]  \, ,
\eeq
where the integrals $I_k$ are  given by
\begin{figure}[h]
\centering
\includegraphics[scale=0.50]{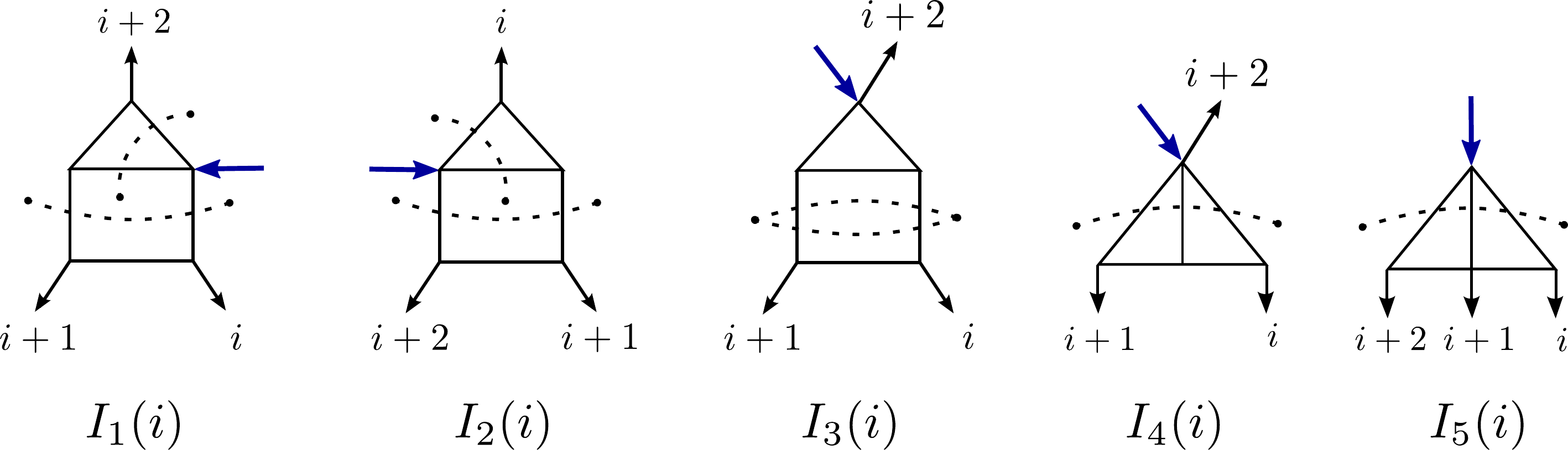}
\caption{\it Integral basis for the complete two-loop form factor $F_{3,3}^{(2)}$.}
\label{fig:basis}
\end{figure}\\
Explicit expressions for most of the integrals that appear in \fref{fig:basis} can be found in  \cite{Gehrmann:2000zt}. The ones that cannot be found there are $I_1$ and $I_2$, which have the same topology. As an example
we focus on $I_2$, i.e. the second integral in \fref{fig:basis}, 
and employ the \texttt{FIRE} algorithm \cite{Smirnov:2008iw} in order to  decompose it in terms of scalar two-loop master integrals, with the  result
\begin{equation} 
\label{uuu} 
\includegraphics[width=\linewidth]{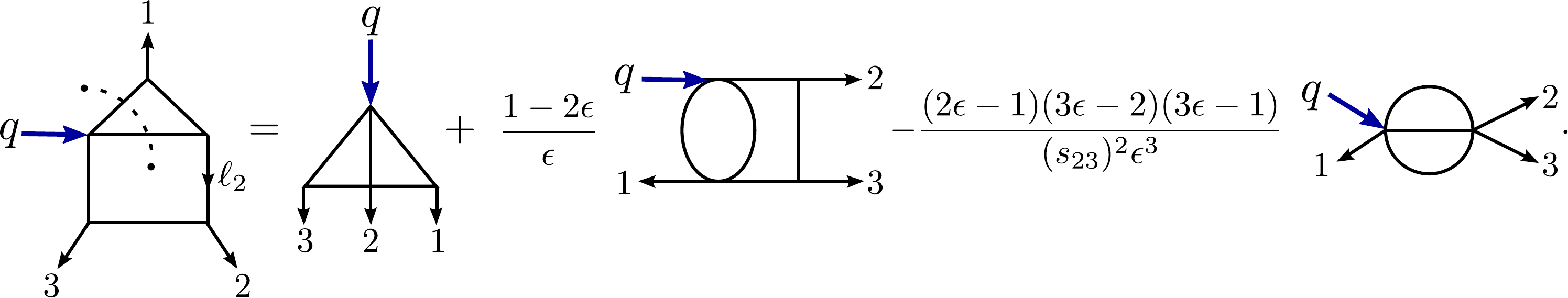}
\end{equation}
The dashed line in the integral on the  left-hand side of \eqref{uuu}  represents the numerator $s_{1 \ell_2}$ (for simplicity, we divided the whole expression by $s_{23}$ when compared to $I_2(1)$).

A few comments are in order here. 

\begin{enumerate}
\item The first integral on the right-hand side of  \eqref{uuu} can naturally be combined with $I_5(1)$ in \fref{fig:basis}. This is important as it ensures that the contribution to the final answer from this topology is a linear combination of multiple polylogarithms with purely numerical, i.e. ~momentum-independent coefficients. The explicit expressions of the first and second integrals in terms of two-dimensional Goncharov polylogarithms can be found in \cite{Gehrmann:2000zt}, 
Eqns.~(4.32)--(4.37) and Eqns.~(4.26)--(4.31), respectively. Also note that the $\epsilon$-dependent prefactor of the second integral ensures that the expanded result has homogenous degree of transcendentality. Finally, the third integral in \eqref{uuu} multiplied with its $\epsilon$-dependent coefficient turns out be $-(1/2)\,  I_4(2)$ which follows from Eqn.~(5.15) of 
\cite{Gehrmann:1999as} which also has homogenous degree of transcendentality.

\item Once the reduction \eqref{uuu} is substituted into \eqref{eq:basisT3}
the final result is expressed as a linear combination of transcendental functions with numerical coefficients.
We refrain from writing explicitly the result at this stage because of its considerable length. Instead in the next section we will identify its universal IR divergences and construct the  finite remainder function. This remainder is a transcendental function of degree four and, as we will show, can be brought to an extremely compact form that involves only classical polylogarithms.

\item As noted in \cite{Bork:2010wf}, the elements of the integral basis of \fref{fig:basis} can be obtained from dual conformal integrals upon taking certain external region momenta to infinity. Consider for instance the simpler one-loop form factor, which may be obtained by taking one of the region momenta $x_i$ of a box integral to infinity,  as shown in \fref{fig:dualconf1},
\begin{figure}[h]
\centering
\includegraphics[width=0.5\linewidth]{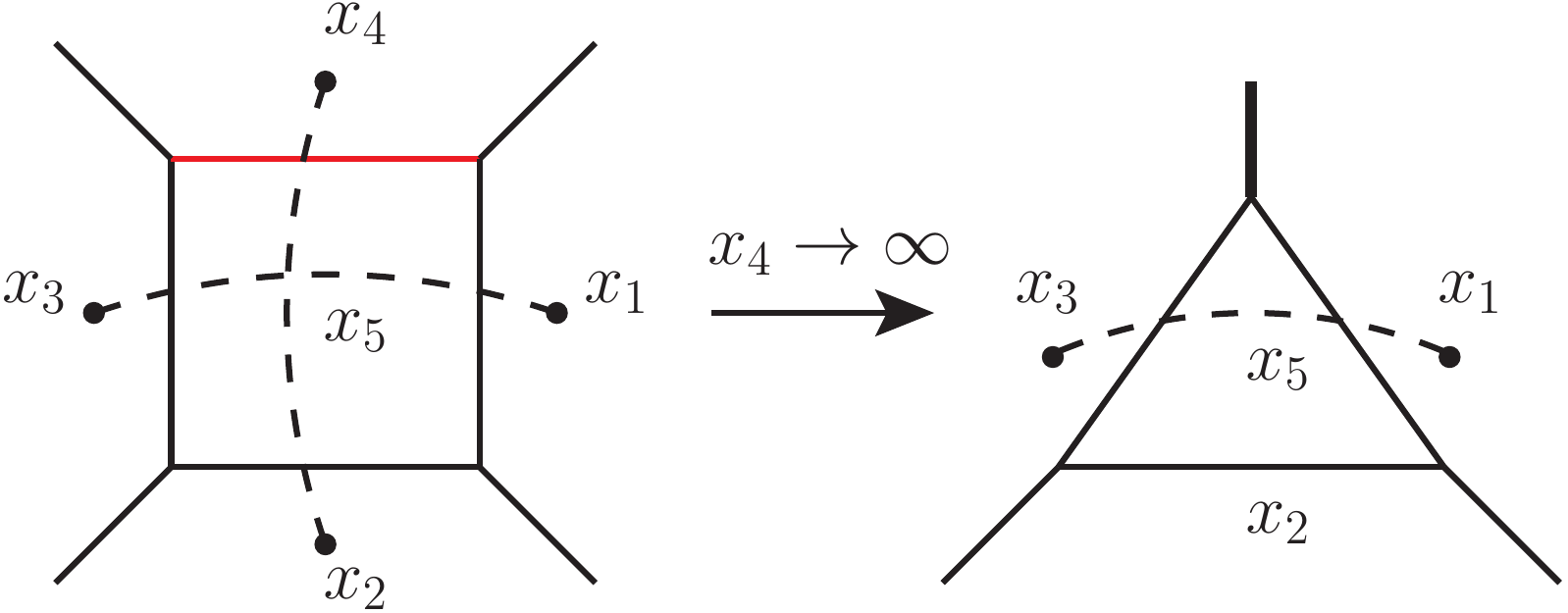}
\caption{\it The one-mass triangle integral obtained from a zero-mass box under the limit where one dual momentum variable is taken to infinity.}
\label{fig:dualconf1}
\end{figure}\\
In this example, as $x_4  \rightarrow \infty$, the propagator marked in red is cancelled by the numerator of the integral, and the box reduces to a triangle. If the external legs were all massive, the above two integrals would be identical due to  dual conformal symmetry \cite{Broadhurst:1993ib}. However, when there are  massless legs as in the present case, both integrals are IR  divergent, and the symmetry is broken. Even though the symmetry is generally broken, interestingly, one could still use this ``pseudo" dual conformal symmetry to fix unambiguously the numerators of each elements of the integral basis of \fref{fig:basis}. In what follows we show how this basis emerges from ``pseudo" dual conformal double-box and penta-box integrals,
\begin{figure}[h]
\centering
\includegraphics[scale=0.75]{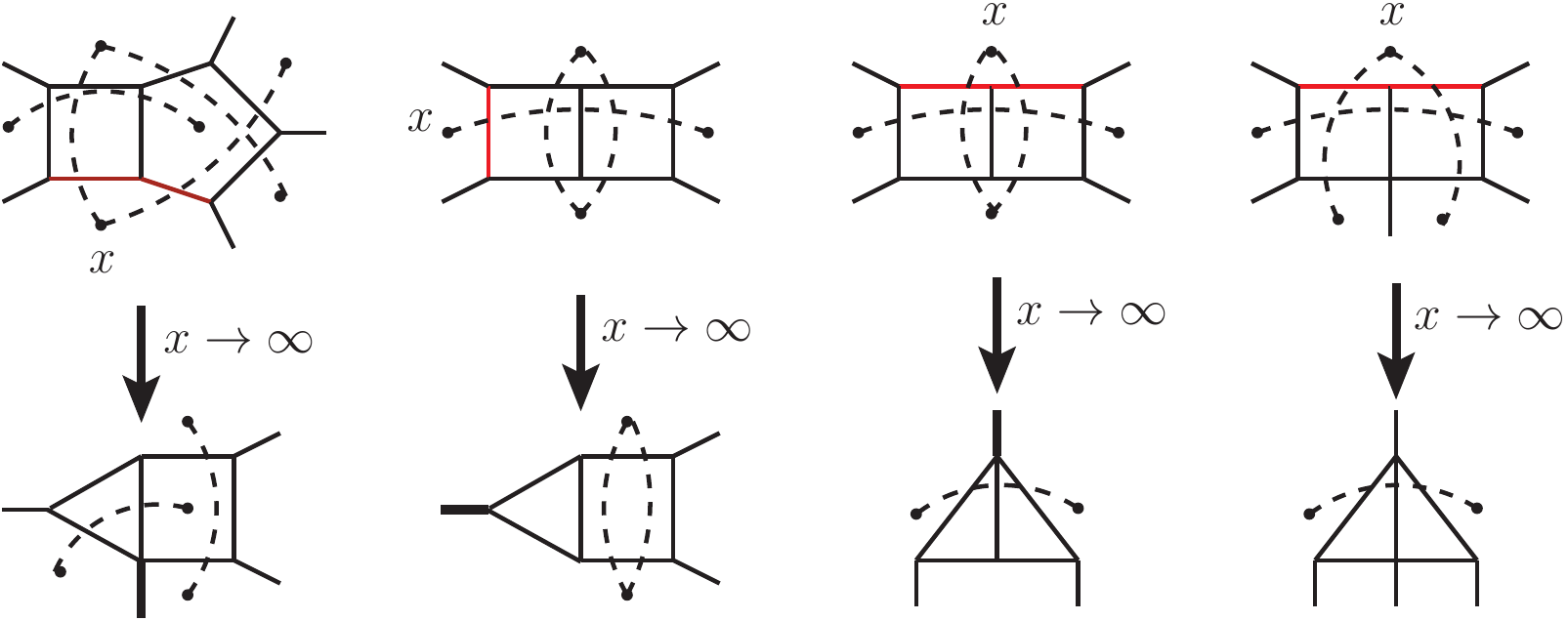}
\caption{\it All integrals in the basis of \fref{fig:basis} with appropriate numerators can be obtained from dual-conformal integrals under a limit where one dual momentum variable $x$ is taken to infinity, similar to \fref{fig:dualconf1}.}
\label{fig:dualconf2}
\end{figure}\\
However, beyond two loops we find that not  all integrals can be obtained by using this procedure. Furthermore,  non-planar   integrals (which do not transform covariantly under the symmetry even at the integrand level) start to appear at three loops.

\item We observe a disagreement between our result \eqref{eq:basisT3} and the result for the same quantity as computed in Eqn.~(4.44) of \cite{Bork:2010wf}. Specifically, in our derivation the integral $G_3$ of \cite{Bork:2010wf} is missing. Our cut analysis did not detect such an integral topology and we also argue that it is  in fact not allowed  for  form factors in $\cN=4$ SYM, as it contains a triangle sub-integral which is not connected to the off-shell leg $q$, thus violating the no-triangle property of $\cN=4$ SYM, see also \cite{Boels:2012ew}. 
\end{enumerate}

\subsection{The three-point remainder function}
\label{sec:remainder}

In this section we construct a finite remainder function associated to the two-loop form factor of the operators ${\cal T}_3$, similarly to what was done in \cite{Brandhuber:2012vm} for the form factor of the stress-tensor multiplet operator ${\cal T}_2$. The result expressed in terms of the explicit form of the integral functions is very complicated, and in order to simplify it we determine its symbol.
From this we will finally derive a very compact form of the three-point remainder containing only classical polylogarithms. 

\subsubsection{Defining a form factor remainder function}
\label{remainder-constants}

We begin by defining the remainder function.  Its expression is given in complete analogy with the amplitude remainder \eqref{eq:2-loop-remainder}\footnote{In our conventions the 't Hooft coupling is defined as $a\equiv\lambda/(e^{\epsilon \gamma} (4 \pi )^{2-\epsilon })$ cf. \eqref{eq:a}. }
\beq
\label{eq:remainder}
\cR_{k,k}^{(2)} \ \equiv \    \cG^{(2)}_{k,k}(\epsilon )\, - \big[\tfrac{1}{2} \big( \cG^{(1)}_{k,k} (\epsilon) \big)^2 -  f^{(2)} (\epsilon)\  \cG^{(1)}_{k,k}  ( 2 \epsilon ) - C^{(2)}\big]
\, + \cO (\epsilon ) \
\ , 
\eeq
where $\cG^{(L)}_{k,k}$ is the helicity-independent form factor $L$-loop ratio function, defined in the same fashion as for amplitudes \eqref{eq:ratio-function},
\beq
\cG^{(L)}_{k,k}\, \equiv\, {\F^{{\rm MHV}(L)}_{k, k}\over \F^{{\rm MHV}(0)}_{k, k}}\, ,
\eeq
and
$f^{(2)} (\eps)\equiv f_0^{(2)} + f_1^{(2)}  \eps+ f_2^{(2)} \eps^2$.

Using \eqref{eq:basisT3} for $\F^{{\rm MHV}(2)}_{3, 3}$, we find that the $1/\epsilon^4$ and $1/\epsilon^3$ poles cancel between the first two terms of \eqref{eq:remainder}. Next
we require that the remainder is finite, and  hence that the remaining $1/\epsilon^2$ and $1/\epsilon$ poles vanish. This fixes two coefficients in the $\eps$-expansion of  $f^{(2)}$,  
\beq
f_0^{(2)} = - 2 \zeta_2 \, , \qquad f_1^{(2)} = - 2 \zeta_3 \, . 
\eeq
We note that these results for $f_0^{(2)} $ and $f_1^{(2)} $ agree with the corresponding quantities found in the case of the remainder function of the stress-tensor multiplet operator computed in 
\cite{Brandhuber:2012vm}.%
\footnote{We observe a disagreement between our result $f_1^{(2)} = - 2 \zeta_3$ and the computation of  \cite{Bork:2010wf}, where the result $\tilde{f}_1^{(2)} = - 14 \,  \zeta_3$ was found. } 
At this stage, however, we cannot make any prediction for $f^{(2)}_2$ and $C^{(2)}$. In the following we will set $f^{(2)}_2 = -2 \zeta_4$ and $C^{(2)}=0$ so that
\beq
\label{eq:f2}
f^{(2)}= -2 \zeta_2 -2\zeta_3\epsilon -2 \zeta_4\epsilon^2 \ .
\eeq
In this way $f^{(2)}$ matches a closely related quantity appearing in the definition of finite remainders of MHV amplitudes in $\cN=4$ SYM \cite{Bern:2008ap, Drummond:2008aq} (see \eqref{eq:f-amp-2-loops}) and form factors with $k=2$ \cite{Brandhuber:2012vm}. In order to fix $f^{(2)}_2$ and $C^{(2)}$ individually we would have to calculate also $\F^{{\rm MHV}(2)}_{3, 4}$ and impose that in a collinear limit, where two adjacent momenta $p_i$, $p_{i+1}$ become parallel, the four-point remainder morphs smoothly into the three-point remainder,  
$\cR_{3,4}^{(2)} \to \cR_{3,3}^{(2)}$, without any additional constant. We briefly note here that collinear (and soft) limits
of minimal form factors exhibit novel subtleties compared to amplitudes, and we defer a detailed discussion to \sref{sec:discussion}.

Finally, we notice that the $n$-point remainder function  depends on $3n-7$ simple ratios of Mandelstam variables. For $n=3$, we will choose the following variables:
\begin{align}
\label{eq:uvw}
u\ =\ \frac{s_{12}}{q^2},\quad v\ =\ \frac{s_{23}}{q^2},\quad w\ =\ \frac{s_{31}}{q^2},\quad\qquad u+v+w=1\ .
\end{align}

\subsubsection{The three-point remainder: from symbols to simple functions}
\label{sec:symbol3}

In the previous two sections we derived the three-point, two-loop form factor and defined its corresponding remainder function. Using the results for the integral functions given in 
\cite{Gehrmann:1999as,Gehrmann:2000zt} we find that the remainder is   a complicated sum of functions of homogeneous degree of transcendentality equal to four which include Goncharov polylogarithms. The expression is rather lengthy and we refrain from presenting it here. However, past experience \cite{Goncharov:2010jf} suggests that one can do much better by studying the symbol of the function. Indeed, this is the strategy we will follow, and at the end we will be able to present a greatly simplified result. 

We find that the symbol of the remainder function is given by the following, strikingly simple expression:
\beq
\label{eq:symbol}
\mathcal{S}_{3,3}^{(2)}(u,v,w)\ =\ 
u \otimes v \otimes  \left[ \frac{u}{w}\otimes_S \frac{v}{w} \right] + 
{1 \over 2}  u \otimes {u \over   (1-u)^3 }  \otimes \frac{v}{w}\otimes
   \frac{v}{w} \, \,   + \,\, \text{perms}\, (u,v,w)\ ,
  \eeq
where $\otimes_S$ in the expression above stands for the symmetrised tensor product
\begin{equation}
x \otimes_S y\,\equiv\, x \otimes y + y\otimes x \ .
\end{equation}
Before reconstructing the remainder from its symbol \eqref{eq:symbol},  we wish to describe a few general properties
of this remainder and compare them with the properties of symbols of other known remainders of amplitudes and form factors.

\begin{enumerate}
\item All entries are taken from the list $\{u,v,w,1-u,1-v,1-w\}$. This is the same list found
for the three-point, two-loop form factor remainder of $\Tr(\phi^2)$ \cite{Brandhuber:2012vm} but does not include the square-root arguments $y_u,y_v,y_w$ present in the case of the two- and three-loop six-point amplitudes in $\N=4$ SYM \cite{Goncharov:2010jf, Dixon:2011pw, Dixon:2011nj}.
\item The first entries of the symbol describe the locations of discontinuities of the remainder and from unitarity we know that cuts should originate at $P_J^2=0$ or $P_J^2=\infty$, where $P_J^2$ are appropriate kinematic invariants~---~in our case $s_{12},s_{23},s_{31}$ and $q^2$.
Hence, the first entry condition \cite{Gaiotto:2011dt} implies in our case that the first entries must be taken from the list $\{u,v,w\}$, which is obviously the case for \eqref{eq:symbol}.
\item In the literature on amplitudes various other conditions on e.g.  the second and final entries were put forward. However the symbol \eqref{eq:symbol} does  not follow the
pattern observed for two-loop amplitudes or two-loop form factors of $\Tr(\phi^2)$.
For the second entries we observe that if the first entry is $u$ then the second entry is taken from the list $\{u,v,w,1-u \}$,  while the last entry is always an element of the list
$\{u/v, v/w, w/u\}$. We note that the same entry conditions are true for the building blocks of the $k$-point, two-loop form factors of $\Tr( \phi_{12}^k)$, which we will discuss in the next sections.
A possible reason why these entry conditions deviate from those of amplitudes and form factors of $\Tr(\phi_{12}^2)$ is related to the fact that the form factors we study here have unconventional factorisation properties in collinear and soft limits,  as discussed in \sref{sec:discussion}.
\end{enumerate}

We now move on to reconstructing  the remainder from its symbol \eqref{eq:symbol}.
In our case the original expression of the remainder contains many Goncharov polylogarithms, as well as classical (poly)logarithms. However, there is a sharp criterion proposed by Goncharov \cite{gonch, Goncharov:2010jf} that allows us to test if a function of transcendentality four can be rewritten in terms of classical polylogarithms $\text{Li}_k$ with $k\leq 4$ only. This criterion is  expressed at the level of the symbol as \eqref{eq:gonch-cond} and can be rephrased in terms of the symbol coproduct as \cite{Golden:2014xqa}
\beq 
\label{eq-gonch-cond2}
\delta (\cS )\Big|_{\Lambda^2 B_2} = 0 \, ,
\eeq
where the $\Lambda^2 B_2$ component of a symbol (coproduct) is defined as \cite{Golden:2014xqa}%
\footnote{To be more precise we should note that
$\Lambda^2 B_2$ is defined in \cite{Golden:2014xqa} as a particular component of the coproduct 
$\delta$ of a function, but here we will work always at the level of the symbol of the function. The same comment applies to the ${B_3 \otimes \mathbb{C}}$ component of the coproduct introduced later.}
\beq  \label{eq:delta-B2}
\delta(a \otimes b \otimes c \otimes d)\Big|_{\Lambda^2 B_2} \equiv \left( a \wedge b \right) \wedge \left( c \wedge d \right) 
\ , 
\eeq
and $\wedge$ stands for the anti-symmetrised tensor product \eqref{eq:wedge}.
Interestingly, our symbol $\mathcal{S}_{3,3}^{(2)}(u,v,w)$ \eqref{eq:symbol} satisfies Goncharov's constraint \eqref{eq:gonch-cond}, or equivalently \eqref{eq-gonch-cond2}.

A strategy to accomplish this goal was outlined in \cite{Goncharov:2010jf} and starts
by investigating the symmetry properties of the symbol under pairwise (anti)symmetrisation of the entries. In this fashion  one can decompose the symbol into four terms,
\beq
\cS_{3,3}^{(2)}(u,v,w)=\text{A}\otimes\text{A}+\text{S}\otimes\text{A}+\text{A}\otimes\text{S} +\text{S}\otimes\text{S}\ ,
\eeq
where {\it e.g.}~$\text{S}\otimes\text{A}$ means symmetrisation of the first two entries and antisymmetrisation of the last two entries.
Next one scans the symmetry properties of the functions that may appear in the answer, as shown in Table \ref{tab:symmetries} (taken from \cite{Goncharov:2010jf}).
\begin{table}[h]
\centering
\begin{tabular}{|c|c|c|c|c|}
\hline
Function & $\text{A}\otimes\text{A}$ & $\text{S}\otimes\text{A}$ & $\text{A}\otimes\text{S}$ & $\text{S}\otimes\text{S}$ \\
\hline
$\text{Li}_4(z_1)$ & $\times$ & $\times$ & \checkmark & \checkmark \\
\hline
$\text{Li}_3(z_1)\, \log(z_2)$ & $\times$ & $\times$ & \checkmark & \checkmark \\
\hline 
$\text{Li}_2(z_1)\, \text{Li}_2(z_2)$ & \checkmark & \checkmark & \checkmark & \checkmark \\
\hline
$\text{Li}_2(z_1) \, \log(z_2)\, \log(z_3) $ & $\times$ & \checkmark & \checkmark & \checkmark \\
\hline
$\log(z_1)\, \log(z_2)\, \log(z_3)\, \log(z_4) $ & $\times$ & $\times$ & $\times$ & \checkmark \\
\hline
\end{tabular}
\caption{\it Symmetry properties of the symbol of transcendentality four functions.}
\label{tab:symmetries}
\end{table}

Remarkably, we find that our symbol \eqref{eq:symbol}  satisfies even more stringent constraints than \eqref{eq:gonch-cond}, namely
its $\text{A}\otimes\text{A}$ and $\text{S}\otimes\text{A}$ components both vanish. Inspecting Table \ref{tab:symmetries}, we see that $\text{Li}_2$ functions are absent and
only the following functions
\beq
\label{eq:listoffunctions}
\big\{\,\text{Li}_4(z_1),\;\text{Li}_3(z_1) \log(z_2),\;\log(z_1)\log(z_2)\log(z_3)\log(z_4) \, \big\}
\eeq
can appear in the answer.
Goncharov's theorem does not predict what the possible arguments of these functions should be.  
We find that with the following list of arguments 
\beq
\label{eq:listofarguments}
\left\{u,v,w,1-u,1-v,1-w,-\frac{u}{v},-\frac{u}{w},-\frac{v}{u},-\frac{v}{w},-\frac{w
   }{u},-\frac{w}{v},-\frac{u v}{w},-\frac{u w}{v},-\frac{v w}{u}\right\}\ ,
\eeq 
we can construct an ansatz for the result which reproduces the symbol of the remainder
\eqref{eq:symbol}.

Following this procedure we find that the result for the integrated symbol is a remarkably compact two-line function: 
\begin{align}
\label{eq:integratedsymbol}
\begin{split}
\cS_{3,3}^{(2)\,\text{Int}}\ &=\
  \frac{3}{4} \, \text{Li}_4\left(-\frac{u v}{w}\right)-\frac{3}{2} \, \text{Li}_4(u)
-\frac{3}{2} \, \log(w) \text{Li}_3 \left(-\frac{u}{v} \right)\\
&
+\,\frac{ {\log}^2(u)}{32} \Big[ {\log}^2(u) +2 \log^2(v)-4  \log(v) \log(w) \Big]
+ {\rm perms}\, (u,v,w) \ .
   \end{split}
\end{align}
The appearance of the  combination of ${\rm Li}_4$ functions  in \eqref{eq:integratedsymbol} with their particular arguments can in fact be inferred by analysing  the $B_3 \otimes \mathbb{C}$ component of the coproduct $\delta$ \cite{Golden:2014xqa}. At the level of the symbol, this component projects out terms which can be written as symbols of products of functions of lower transcendentality. It is defined as 
\begin{align} 
\label{eq:B3C}
\begin{split}
\delta(a \otimes b \otimes c \otimes d)\Big|_{ B_3 \otimes \mathbb{C} } \equiv 
\left( (a \wedge b) \otimes c  -  (b \wedge c) \otimes a  \right) \otimes d
-
\left( (b \wedge c) \otimes d  -  (c \wedge d) \otimes b  \right) \otimes a  \, , \\[6pt]
\end{split}
\end{align}
which is identical to the definition of the projection operator $\rho$ introduced in 
\cite{Duhr:2011zq} (see also \eqref{eq:projector-rho}). For our three-point remainder, which consists only of classical polylogarithms, this implies that we  project onto the ${\rm Li}_4$ part of the 
remainder. We find that 
\beq  \label{eq: B3R3}
\delta ( \mathcal{S}_{3,3}^{(2)}(u,v,w) )\Big|_{B_3 \otimes \mathbb{C}} 
= 
-{3 \over 2} \{ u \}_3 \otimes u + { 3 \over 4 } \left\{ - {u v \over w} \right\}_3 \otimes { u v \over w} \, ,
\eeq
where we introduced the  shorthand notation: 
\beq
\{ x \}_k \equiv  \{ x \}_2 \otimes^{k-2} x \, , \quad { \rm with} \quad \{ x \}_2 \equiv - (1-x) \wedge x \, . 
\eeq
Noting that 
\beq
\delta ( {\rm Li}_4 (x)  )\Big|_{B_3 \otimes \mathbb{C}} \ = \ \{ x\}_4 \, = \, \{ x\}_3 \otimes x
\ , 
\eeq
we immediately infer from \eqref{eq: B3R3} that the remainder   $\mathcal{R}^{(2)}_{3,3}$
is given by  
$-{3 \over 2} {\rm Li}_4 ( u  ) + \frac{3}{4} {\rm Li}_4 \left( - \dfrac{uv}{w}  \right) $, modulo products of lower transcendentality functions, in accordance with \eqref{eq:integratedsymbol}.

\begin{figure}[htb]
\centering
\includegraphics[width=1\textwidth]{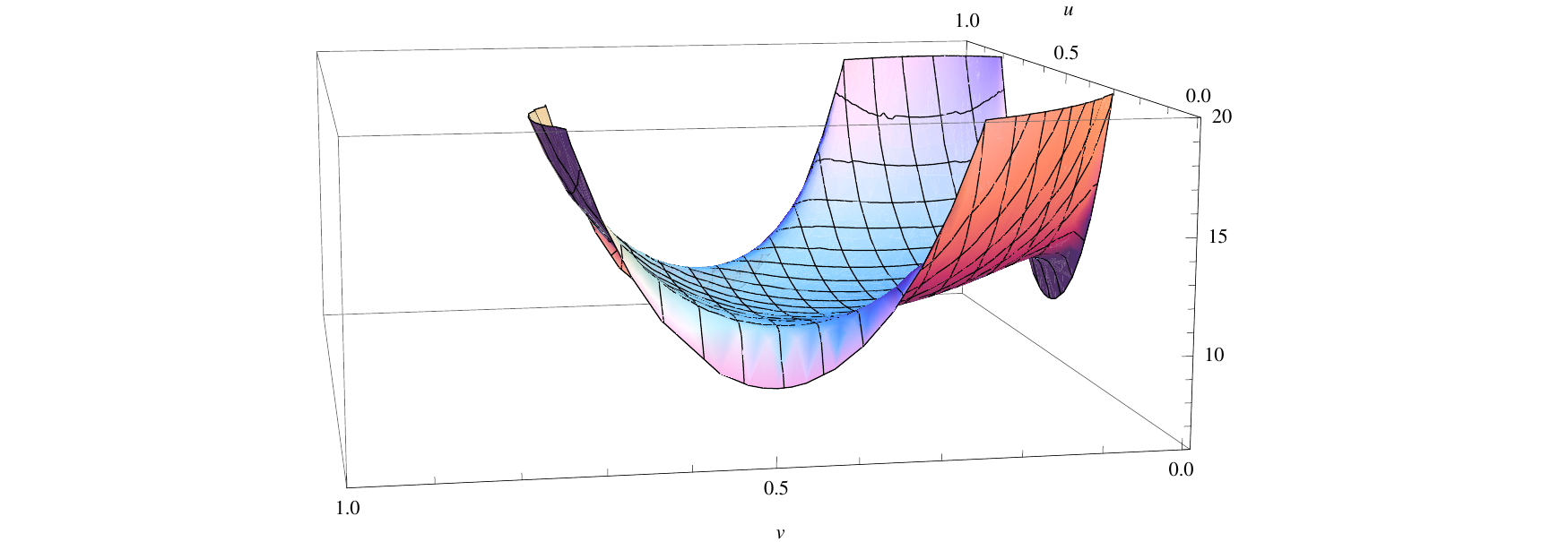}
\caption{\it Plot of the remainder function $\cR_{3,3}^{(2)} (u,v,1-u-v)$, where $u$ and $v$ live in a triangular region bounded by $u=0.01$, $v=0.01$ and $u+v=0.99$. As we approach the edges, for instance $u=0$, the remainder diverges as $\log^2 u$, as explained in the text.}
\label{fig:plotR3}
\end{figure}

The function \eqref{eq:integratedsymbol} is not yet the full remainder because the symbol is blind to transcendentality four functions containing powers of $\pi$ (or $\zeta_i$). In order to fix these ambiguities, we subtract \eqref{eq:integratedsymbol} from the full remainder function and inspect what is left over. These so-called ``beyond the symbol'' terms 
 are a linear combination of terms of the form $\pi^2 \log x \log y$, $\pi^2 \, \text{Li}_2(x)$, $\zeta_3 \log x$ and $\zeta_4$ and their coefficients can be determined numerically.
In order to perform the numerical comparison with the original remainder we have used the \texttt{GiNaC} software \cite{DBLP:journals/corr/cs-SC-0004015}. We find the following  result  for the beyond the symbol terms: 
\begin{align}
\label{eq:beyondthesymbol}
\cR_{3,3}^{(2)  \text{bts}} =\   {\zeta_2 \over 8}  \log(u)\, \Big[ 5 \log(u)
    -2 \log (v) \, \Big] + \frac{\zeta_3}{2}  \log (u) +\frac{7}{16}\zeta_4  + {\rm perms}\, (u,v,w)  \ .
\end{align}
Summarising, the final result for the remainder function  $\cR_{3,3}^{(2)}$ is the sum of \eqref{eq:integratedsymbol} and \eqref{eq:beyondthesymbol}, 
\begin{align}
\label{eq:remainderT3}
\begin{split}
\cR_{3,3}^{(2)} \ \equiv \ & 
 -\frac{3}{2}\, \text{Li}_4(u)+\frac{3}{4}\,\text{Li}_4\left(-\frac{u v}{w}\right)  
-\frac{3}{2}\log(w) \, \text{Li}_3 \left(-\frac{u}{v} \right)   
+ \frac{ 1}{16}  {\log}^2(u)\log^2(v)
\\
&
+ {\log^2 (u) \over 32} \Big[ \log^2 (u) - 4 \log(v) \log(w) \Big]  
+  {\zeta_2 \over 8 }\log(u) [  5\log(u)- 2\log (v) ]
 \\
&
+  {\zeta_3 \over 2} \log(u) + \frac{7}{16}\, \zeta_4  + {\rm perms}\, (u,v,w) \ .
\end{split}
\end{align}
We plot the remainder function $\cR_{3,3}^{(2)} (u,v,1-u-v)$ in Figure \ref{fig:plotR3}.

One important feature which stands out is that the remainder  blows up at the boundaries of the Euclidean kinematic region $u=0$, $v=0$ and $u+v=1$. We need to distinguish here two types of limits: 
\begin{enumerate}
\item The situation where we approach a generic point on one of the three edges corresponds to a collinear limit. For instance,  taking $u\to 0$ (and $v+w\to 1$) is equivalent to the collinear limit $p_1 \, ||\, p_2$. In this situation
the remainder diverges as $\log^2 (u)$. The derivation of this result can be found in  \sref{sec:discussion}. 
\item The case where we approach one of the corners, for instance $u=w=0$, corresponds to the soft limit $p_1 \to 0$. As will be discussed in \sref{sec:discussion}, this soft limit can be parametrised as 
$ u=x \, \delta$, $v = 1 - \delta$, $w=y\,\delta$ with $x+y=1$ and $\delta \to 0$ and the remainder diverges as $(1/4) \log^4 (\delta)$, explaining the spikes in  \fref{fig:plotR3} in the positive vertical direction.
\end{enumerate}

This behaviour might appear  unexpected for  remainder functions, which usually have smooth  collinear and soft limits. However one has to appreciate that here we are considering a special form factor, with the minimal number of external legs. 
Hence we cannot extrapolate the usual intuition about factorisation since there is no form factor with fewer legs  this minimal form factor could factorise on, as we discuss in more
detail in \sref{sec:discussion}.

\subsection{The two-loop remainder function for all $k>3$}
\label{sec:allk}

Having obtained and described in detail the three-point remainder  $\cR_{3,3}^{(2)}$ of the form factor of the operator ${\rm Tr }[ (\phi_{++})^3]$ at two loops,  we now move on to study the $k$-point form factors  $F_{k,k}^{(2)}$ of  ${\rm Tr }[ ( \phi_{++})^k]$ for arbitrary $k>3$.

\subsubsection{The $k$-point minimal form factors  from cuts}

The study of the cuts of these form factors proceeds in an almost identical way compared to the $k=3$ case, with one important exception, namely the appearance of a new integral function which is the product of two one-loop triangle functions. Specifically, our result for the  
minimal form factor of ${\rm Tr }[ ( \phi_{++})^k]$ for $k>3$  at two loops is given by the following simple  extension of that of $k=3$,
\beq 
\label{eq:basisTk}
F^{(2)}_{k,k}\,=\, \sum_{i=1}^n\Big[ I_1(i) + I_2(i) + I_3(i) +I_4(i) - I_5(i) +  {1\over 2} \sum_{j=i+2}^{i-2} I_6(i, j) \Big] \, ,
\eeq
where the integral basis is the same as that of \fref{fig:basis} augmented by one new integral, namely $I_6$:
\begin{figure}[h]
\centering
\includegraphics[width=.92\linewidth]{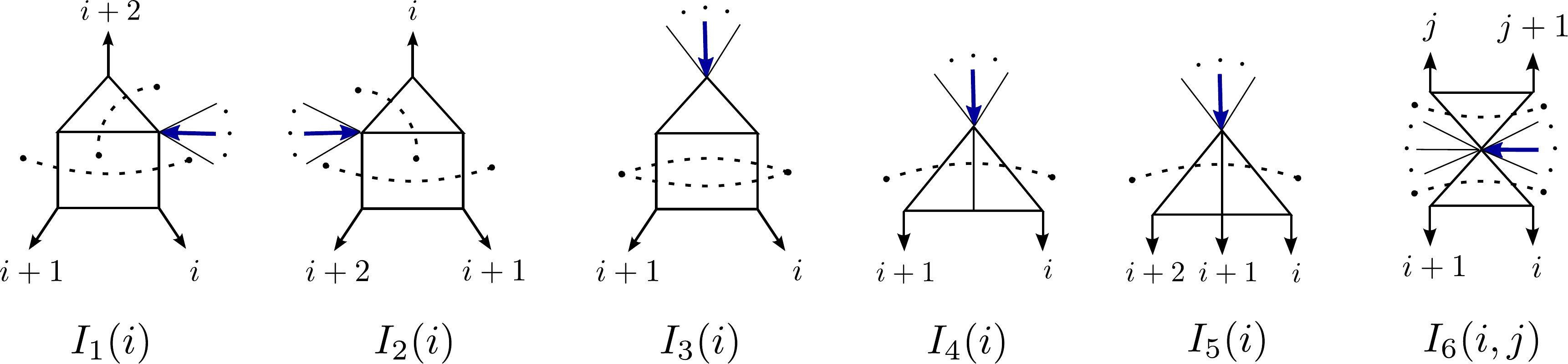}
\caption{\it Integral basis for $F_{k,k}^{(2)}$.}
\label{fig:basis2} 
\end{figure}
The factor of $1/2$ in front of $I_6$ is present in order to remove double counting.%
\footnote{A similar but not identical result for the same quantities was presented in \cite{Bork:2010wf}. As in the three-point case discussed earlier, our result differs from theirs by the absence of the function $G_3$ appearing in Eqn.~(4.44) of \cite{Bork:2010wf}. }

Note that the appearance of the extra integral function $I_6$ can be inferred easily from two-particle cuts, specifically   by attaching the three-level amplitude \eqref{eq:attach} (with $2$ and $3$ replaced by $i$ and $i+1$) to the following one-loop triangle integral, 
\begin{figure}[h]
\centering
\includegraphics[scale=0.42]{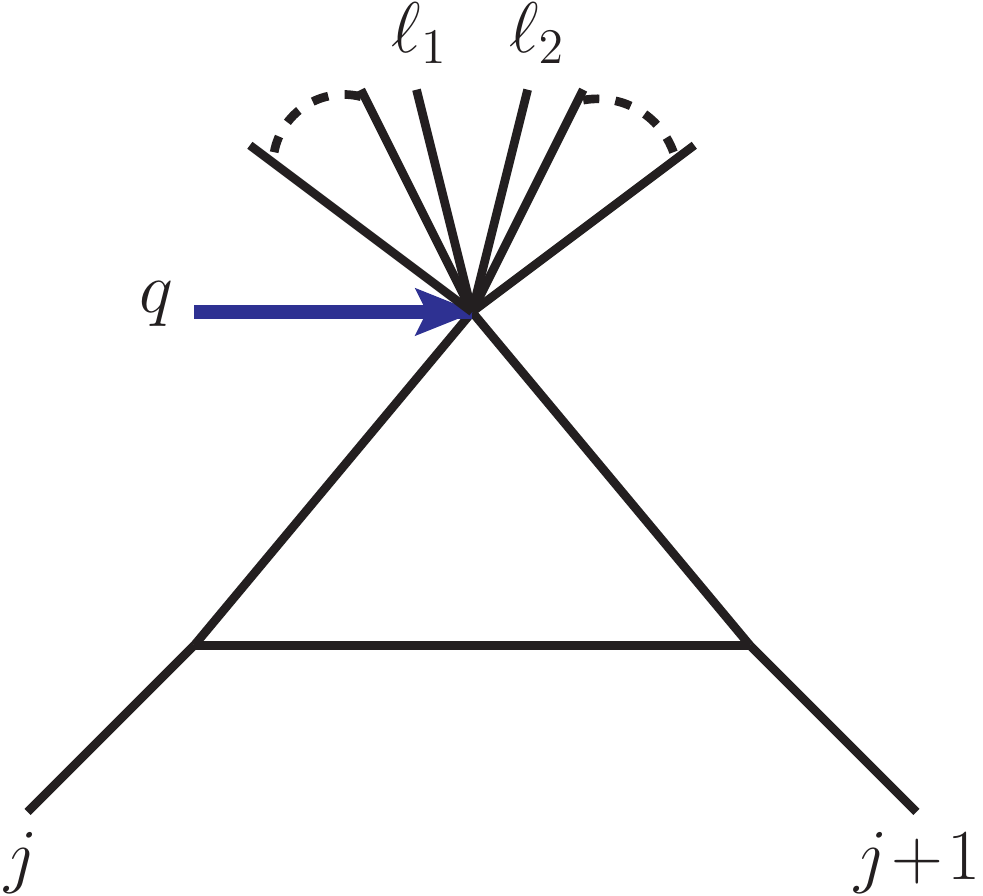}
\caption{\it Integral present in the one-loop form factor $F_{k,k}^{(1)}$ which produces the topology $I_6$ of \fref{fig:basis2} under a two-particle cut.}
\label{fig:double-triangle}
\end{figure}

Clearly, the integral of \fref{fig:double-triangle} is  present only when   $k>3$. With this additional term, the basis shown in \fref{fig:basis2} has  all the correct two-particle cuts.

Let us now discuss how the  following triple cuts might get altered when compared to the $k=3$ case studied earlier.
\begin{figure}[h]
\centering
\includegraphics[width=0.85\linewidth]{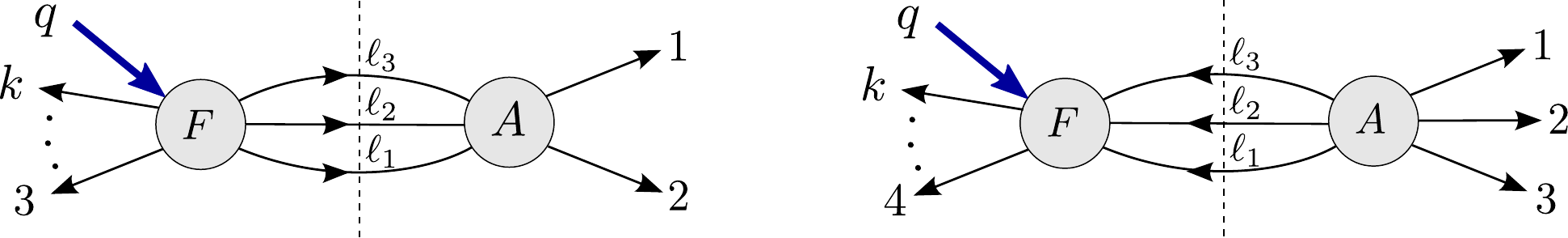}
\caption{\it Triple cuts employed in the derivation of  $
F_{k,k}^{(2)}$.}
\label{fig:tri-cut-k}
\end{figure}
To begin with, we note that $I_6$ does not contribute to any triple cut. Since the remaining integrals in  \fref{fig:basis2} are identical to those contributing to  $F_{3,3}^{(2)}$, we  only need to confirm that the results of the above triple cuts are the  same as those of $F_{3,3}^{(2)}$. 

This agreement is immediate for the diagram on the right-hand side of \fref{fig:tri-cut-k} since the form factor appearing there is minimal, thus simply $1$. For the diagram on the left-hand side, a simple way to show this is to note that the tree-level form factors which  enter  the cut are actually identical for $k=3$ and for $k>3$. They are $(k+1)$-point NMHV form factors with one negative-helicity  gluon, $g^-$, or two fermions, $\bar{\psi}$, which indeed take the same form for any $k$, namely
\begin{align} 
\label{fff}
 & F_{k,k+1}^{\rm NMHV}(\phi_1, \ldots, \phi_{i-1}, g^-_i, \phi_{i+1}, \ldots, \phi_{k+1} )
= {[i\!-\! 1 \, i\!+\! 1] \over [i\!-\! 1 \, i] [i \, i\!+\! 1 ]} \, , \nonumber \\
& F_{k,k+1}^{\rm NMHV}(\phi_1, \ldots, \phi_{i-1}, \bar{\psi}_{i}, \bar{\psi}_{i+1}, \phi_{i+2}, \ldots, \phi_{k+1} )
= {1 \over  [i \, i\!+\! 1 ]} \, ,  \\ \nonumber
& F_{k,k+1}^{\rm NMHV}(\phi_1, \ldots, \bar{\psi}_{i-1}, \phi_{i}, \bar{\psi}_{i+1}, {\phi}_{i+2}, \ldots, \phi_{k+1} )
= 0 \, .
\end{align}
The above results \eqref{fff} can be obtained simply by taking the conjugate of the $(k+1)$-point MHV form factors of ${\rm Tr}(\phi_{12}^k)$.

In conclusion, compared to the case  $k=3$, the only difference in the result is that now we need to include the double-triangle integrals $I_6(i,j)$.

\subsubsection{The symbol of the $k$-point remainder}

In this section we construct the two-loop remainder function and its symbol for the case of general $k$. The remainder is defined in \eqref{eq:remainder}, where now $k=n>3$. The ingredients of this formula are the one-loop minimal form factor defined in \eqref{eq:1-loop-MHV-all-k}, and the two-loop form factor derived earlier in this section. A few  comments are in order. 

\begin{enumerate}
\item We find that the cancellation of the IR poles in $\epsilon$ proceeds exactly as in the three-point case, and as a result the remainder function is defined with the same universal function $f^{(2)} (\eps)$ defined in \eqref{eq:f2}.
\item  As noticed earlier, the two-loop form factor contains an extra integral topology $I_6$ if $n\!=\!k\!>\!3$. This topology is exactly cancelled by the cross terms coming from the square of the one-loop form factor appearing in the definition of the remainder. There is an important consequence of this cancellation, namely all the remaining integral topologies  contributing to the remainder depend only on either triplets of adjacent momenta $p_i$, $p_{i+1}$ and $p_{i+2}$  ($I_1$, $I_2$,  and $I_5$) or pairs of adjacent momenta ($I_3$ and $I_4$).  As a result the remainder function can be written as a cyclic sum over universal sub-remainders which depend on three momenta, 
\begin{align}
\label{eq:3part-bb}
\cR_{k,k}^{(2)}\,=\, \sum_{i=1}^k r^{(2)}(u_i,v_i,w_i) \ ,
\end{align}
as we will show in detail below. Here the parameters $u_i, v_i, w_i$ are generalisations of the $u, v, w$ ratios of the $k=3$ case, and are defined  as
\begin{align}
\label{eq:uvw2}
u_i\,=\,\frac{u_{i\,i+1}}{u_{i\,i+1 \, i+2}}\, ,\quad
& v_i\,=\,\frac{u_{i+1\,i+2}}{u_{i\,i+1 \, i+2}}\, ,\quad
w_i\,=\,\frac{u_{i+2\,i}}{u_{i\,i+1 \, i+2}} \ ,
\end{align}
with 
\beq
u_{i\,i+1 \, i+2} \equiv u_{i\,i+1}+u_{i+1\,i+2}+u_{i+2\,i} \, .
\eeq 
Note that we have  defined $u_{ij}\equiv s_{ij}/q^2$, and  $u_i+v_i+w_i = 1$.
For notational simplicity we will in the following replace $r^{(2)}(u_i,v_i,w_i)$ by
$r^{(2)}_{i}$. We should stress at this point that these are the basic building blocks of
$\cR_{k,k}^{(2)}$ and do not depend on the value of $k$.
\end{enumerate}

Using the explicit expressions of the integral functions $I_1, \ldots, I_6$ we have computed the remainder function  in terms of multiple polylogarithms. As in the three-point case, this expression is quite lengthy and we will only present it after simplifying it using its symbol. 

Again it turns out that the symbol is extremely simple. As anticipated above, it is written as a sum of  building blocks which depend on $u_i$, $v_i$ and $w_i$: 
\begin{align}
\cS_{k,k}^{(2)}\,=\, \sum_{i=1}^k s^{(2)}(u_i,v_i,w_i) \equiv \sum_{i=1}^k s^{(2)}_i \ ,
\end{align}
where
\begin{align}  \nonumber
s^{(2)}_i
\ &= \
 u_i\otimes (1-u_i)\otimes \left[ \frac{u_i-1}{u_i} \otimes \frac{v_i}{w_i}+ \frac{v_i}{w_i}\otimes \frac{w_i^2}{u_i v_i} \right] \\ 
& +u_i\otimes u_i\otimes {1-u_i \over v_i} \otimes \frac{w_i}{v_i}  
 +u_i\otimes v_i\otimes  \left[ \frac{v_i}{w_i} \otimes_S \frac{u_i}{w_i} \right]
 + (u_i \leftrightarrow v_i ) \, .
\end{align}
As was done earlier for the $k=3$ case, it is useful to study the coproduct of the remainder function. A key difference is that, unlike  the case of the form factor of $\T_3$, the symbol $s^{(2)}_i$ does not obey Goncharov's condition \eqref{eq:gonch-cond}. Instead  we find that the corresponding component of the coproduct is
\beq 
 \label{eq:BkB2B2}
\delta ( s^{(2)}_i )\Big|_{\Lambda^2 B_2} \ = \ 
\left\{- { w_i \over v_i } \right\}_2 \wedge \left\{ u_i \right\}_2 \ + \ ( u_i \leftrightarrow v_i )
\ . 
\eeq 
We also quote its  $B_3 \otimes \mathbb{C} $ component, given by 
\begin{align}  \label{eq:BkB3C}
\delta ( s^{(2)}_i )\Big|_{B_3 \otimes \mathbb{C}} 
&= 
\{ 1- u_i  \}_3 \otimes {  w_i \over v_i } \ + \ \{ u_i \}_3 \otimes { u_i \over w_i}
\ + \ \left\{ - {w_i  \over u_i v_i} \right\}_3 \otimes { w_i \over u_i} 
\cr
& + \ \left\{ -{ w_i \over v_i} \right\}_3  \otimes { v_i \over u_i w^2_i} 
\ - \  \left\{{ v_i \over 1-u_i } \right\}_3 \otimes u_i
\ + \  \left( u_i \leftrightarrow v_i \right)
\ .
\end{align}
Because of the non-vanishing of the  ${\Lambda^2 B_2}$ component \eqref{eq:BkB2B2}, $s^{(2)}_i$ cannot be integrated to purely classical polylogarithms. However, it is not difficult to recognise what multiple polylogarithms can give rise  to \eqref{eq:BkB2B2}. For instance $ {\rm Li}_{1,3}\Big(u_i, -\dfrac{w_i}{u_i v_i}\Big) \ + \ ( u_i \leftrightarrow v_i )$, or the cluster algebra inspired  function $L_{2,2}\Big(u_i, -\dfrac{w_i}{v_i}\Big) \ + \ ( u_i \leftrightarrow v_i ) $ defined in \cite{Golden:2014xqa} can do the job.

In the present case,  it turns out to be more convenient to consider the following combination of Goncharov polylogarithms, as we will explain shortly,%
\footnote{These Goncharov polylogarithms already appear in the explicit expressions of the integrals $I_1(i)$ and $I_2(i)$ belonging to the basis of \fref{fig:basis2}.}
\begin{align}
\label{eq:extragonch}
\begin{split}
r_{{\rm nc},i}^{(2)}  \, &\equiv\,-  G\left(\left\{1-u_i,1-u_i,1,0\right\},v_i\right)\ - \ ( u_i \leftrightarrow v_i ) \, , 
\end{split}
\end{align}
where the symbol of $G_{v}\,\equiv\, G\left(\left\{1-u,1-u,1,0\right\},v\right)$ is given by
\begin{align}
\begin{split}
\cS[G_v]\ = \ &   
v \otimes w \otimes \big[ w \otimes_S u -  u \otimes u \big]
+ v \otimes(1-v) \otimes {u \over w} \otimes {  u \over w } -v \otimes (1-u) \otimes (1-u) \otimes u   
\\
&
 + {w \over v(1-u)} \otimes  (1-u) \otimes u \otimes {w \over u}
 + {  v(1-u) \over w } \otimes \left[  (1-u) \otimes_S {  1-u \over w }  \right] \otimes u
\ .
\end{split}
\end{align}
As for the functions ${\rm Li}_{1,3}$ and $L_{2,2}$ mentioned previously, the $\Lambda^2 B_2$ component of the coproduct of  
$\cS[r_{{\rm nc},i}^{(2)}]$ is  equal to \eqref{eq:BkB2B2}. Hence we can decompose the symbol of the remainder  $s^{(2)}_i$  into a non-classical and a classical contribution:  
\beqa
s^{(2)}_i\, =\, s_{{\rm nc},i}^{(2)}  +  s_{{\rm cl},i}^{(2)}\ ,
\eeqa
where $s_{{\rm nc},i}^{(2)}$ is the symbol of $r_{{\rm nc},i}^{(2)}$. Hence  $s_{{\rm cl},i}^{(2)}$  has now a vanishing $\Lambda^2 B_2$ component, or equivalently satisfies Goncharov's condition  \eqref{eq-gonch-cond2}, and thus can be rewritten in terms of classical polylogarithms only.  

We now move on to determining the  classical part of the remainder.  In order to do so, it is convenient to  first examine
the  $B_3 \otimes \mathbb{C}$ component of the non-classical remainder $r_{{\rm nc},i}^{(2)} $. It is given by 
\begin{align}
\begin{split}
\delta( s_{{\rm nc},i}^{(2)} )\Big|_{B_3 \otimes \mathbb{C}} 
&=  \{ 1- u_i  \}_3 \otimes { u_i w_i  \over (1- u_i)^2  v_i } \ + \ \{ u_i \}_3 \otimes { u_i \over w_i (1-u_i) }
\ + \ \left\{ - {w_i  \over u_i v_i} \right\}_3 \otimes {w_i \over u_i } 
\cr
& + \ \left\{ -{ w_i \over v_i} \right\}_3  \otimes {v_i \over u_i w^2_i} 
\ - \  \left\{{ v_i \over 1-u_i  } \right\}_3 \otimes u_i
\ + \ 
\left( u_i \leftrightarrow v_i  \right) 
\ .
\end{split}
\end{align}
This is a somewhat complicated expression, however the particular  choice of $r_{{\rm nc},i}^{(2)}$ we made  in \eqref{eq:extragonch} is such that 
the  $B_3 \otimes \mathbb{C}$ component of the coproduct of $r_{{\rm cl},i}^{(2)}$ turns out to be very simple~---~in fact this was the motivation behind choosing our particular form of $r_{{\rm nc},i}^{(2)}$. Furthermore, $r_{{\rm nc},i}^{(2)}$ does not develop any singularity  in the soft or collinear limits (this is shown explicitly in \sref{mff},  see  \eqref{cp}). 
For the $B_3 \otimes \mathbb{C}$ component of the classical remainder $r_{{\rm cl},i}^{(2)} $
we find on the other hand
\beq 
\delta\big(s_{{\rm cl},i}^{(2)}\big)\Big|_{ B_3 \otimes \mathbb{C}} 
\ =\
\{1- u_i \}_3 \otimes { (1- u_i )^2 \over u_i } + \{ u_i \}_3 \otimes (1- u_i) 
+  
\left( u_i \leftrightarrow v_i  \right) \, .
\eeq
By applying the  identity $\{ 1- 1/u \}_3 =  - \{ 1- u \}_3 - \{ u \}_3 $, this expression can be recast as 
\beq
\delta\big(s_{{\rm cl},i}^{(2)}\big)\Big|_{ B_3 \otimes \mathbb{C}} 
\ =\
\{1- u_i \}_3 \otimes (1- u_i)  + 
\{ u_i \}_3 \otimes u_i
- 
\{1- {1 \over u_i} \}_3 \otimes \Big(1- {1 \over u_i} \Big) +  
\left( u_i \leftrightarrow v_i  \right)\, . 
\eeq
From the above result, we see immediately that the classical part of the remainder  $r_{{\rm cl},i}^{(2)}$  is  given by  ${\rm Li}_4(1-u_i) + {\rm Li}_4(u_i) - {\rm Li}_4(1-1/u_i)+  
\left( u_i \leftrightarrow v_i  \right)$ modulo  products of functions of lower degree of transcendentality, which  can be fixed by following the same strategy as in the  $k=3$ case. 
Doing so, we find that the classical part of the remainder is: 
\begin{align}
\label{eq:intmodsymbolk}
\begin{split}
r_{{\rm cl},i}^{(2)} =\, &    \text{Li}_4(1-u_i)+\text{Li}_4(u_i)-\text{Li}_4\left(\frac{u_i - 1}{u_i}\right)
+  \log \left( {1-u_i \over w_i }\right) 
\left[  \text{Li}_3\left(\frac{u_i - 1}{u_i}\right) - \text{Li}_3\left(1-u_i\right) \right] \\
 +\, & \log \left(u_i\right) \left[\text{Li}_3\left(\frac{v_i}{1-u_i}\right)+\text{Li}_3\left(-\frac{w_i}{v_i}\right) + \text{Li}_3\left(\frac{v_i-1}{v_i}\right)
 -\frac{1}{3}  \log ^3\left(v_i\right) -\frac{1}{3} \log ^3\left(1-u_i\right)  \right]
 \\   
+\, & \text{Li}_2\left(\frac{u_i-1}{u_i}\right) \text{Li}_2\left(\frac{v_i}{1-u_i}\right)-  \text{Li}_2\left(u_i\right) \left[
   \log \left({1-u_i \over w_i }\right) \log \left(v_i \right) +\frac{1}{2} \log ^2\left( { 1-u_i \over w_i }\right) \right] 
\\
-\, & \frac{1}{24} \log ^4\left(u_i\right)
+\frac{1}{8} \log ^2\left(u_i\right) \log ^2\left(v_i\right)  + \frac{1}{2} \log ^2\left(1-u_i\right) \log \left(u_i\right) \log \left( { w_i \over v_i}\right)\\
+\, & \frac{1}{2} \log \left(1-u_i\right) \log ^2\left(u_i\right) \log \left(v_i\right)
+ \frac{1}{6} \log ^3\left(u_i\right) \log \left(w_i\right) \ +\  (u_i\, \leftrightarrow\, v_i) \ .
\end{split}
\end{align}
The beyond the symbol terms (obtained in the same way as for $\cR^{(2)}_{3,3}$) are
\begin{align}
\label{eq:btsk}
\begin{split}
r_{\text{bts},i} ^{(2)} = \ & \zeta_2 \, \Big[ \log \left(u_i\right) \log \left(1-v_i \over v_i \right)
+ \frac{1}{2}\log ^2\left( { 1-u_i \over w_i }\right) - \frac{1}{2}\log ^2\left(u_i\right)   \Big]
\\
-\, &   \zeta_3 \log (u_i)  - {\zeta_4\over 2} 
+\  (u_i\, \leftrightarrow\, v_i)
 \ .
\end{split}
\end{align}
\renewcommand{\arraystretch}{1.5}
\begin{table}[t]
\centering
\begin{tabular}{|c||c|}
\hline
$  k $ &  Estimated error\hspace{10pt} \\
\hline  $4$ & $\cO(10^{-17})$
\\
\hline  $5$ & $\cO(10^{-14})$ \\
\hline  $6$ & $\cO(10^{-15})$ \\
\hline
\end{tabular}
\caption{\it Numerical checks of the remainders $\cR^{(2)}_{k,k} $ for $k\,=\,4,\,5,\,6$.}
\label{tab:num-checks}
\end{table}
Finally, the two-loop remainder function for general $k$ is given by
\begin{equation}
\label{eq:remainderk} 
\cR^{(2)}_{k,k} \ = \ \sum_{i=1}^k \Big[
r_{{\rm nc},i}^{(2)}  +  r_{{\rm cl},i}^{(2)}+r_{\text{bts},i} ^{(2)} \Big]\ ,
\end{equation}
where $r_{{\rm nc},i}^{(2)}$, $ r_{{\rm cl},i}^{(2)}$ and $r_{\text{bts},i}^{(2)}$, and  are defined  in  \eqref{eq:extragonch},  \eqref{eq:intmodsymbolk} and  \eqref{eq:btsk},  respectively.

We have also checked  our result \eqref{eq:remainderk}  against numerical evaluations of the remainder for several  values of $k$ and sets of kinematical data,  finding excellent agreement (see  Table~\ref{tab:num-checks}).

\subsection{Collinear and soft limits}
\label{sec:discussion}

In this section we discuss some general properties of the form factors
under soft and collinear limits. This discussion is somewhat beyond the main
line of the work presented so far, but will be relevant for future studies of non-minimal form factors.

When discussing collinear or soft limits it is crucial to distinguish the cases of minimal
and non-minimal form factors. In the latter case, the number of external on-shell particles
is larger than the number of fields in the operator, and the factorisation properties 
are identical to those of amplitudes. 
This follows from a slight generalisation of arguments presented in \cite{Brandhuber:2012vm}
for form factors of $\Tr (\phi_{12}^2)$ with three or more external particles, which in turn are inspired by the original proof for amplitudes given in \cite{Kosower:1999xi}. For minimal form factors, which are the main focus of this work,  the story is more interesting since they cannot factorise into form factors with fewer legs\footnote{Technically there could exist \emph{sub-minimal} form factors starting at two loops, as can be seen from the three-particle cut $F_{k,k}(1,\dots,k-3,\ell_1,\ell_2,\ell_3;q)\times A_5(-\ell_1,-\ell_2,-\ell_3,k-2,k-1)$. However, one can convince oneself that this is not the case for $\O_k$ since there is no consistent helicity assignment that leads to a non-vanishing result.}. Hence, the argument of \cite{Kosower:1999xi} does not apply and the factorisation
properties deviate dramatically from those of amplitudes.

\subsubsection{Minimal form factors}
\label{mff}

We begin by looking at minimal form factors, and specifically we wish to study the collinear and soft behaviour  of their remainder functions derived in the previous sections.  

For non-minimal form factors, one can define a properly normalised  $n$-point remainder function%
\footnote{At two-loop level the appropriate normalisation is obtained by introducing the   $n$-independent, transcendentality-four constant  $C^{(2)}$ in the definition \eqref{eq:remainder}.}
such that, under a collinear limit one has
\beq
\label{usual}
\cR_n \to \cR_{n-1}
\ . 
\eeq
Note that \eqref{usual} is the usual behaviour of remainders of loop amplitudes in $\cN=4$ SYM as discussed in \cite{Bern:2008ap, Anastasiou:2009kna}, and confirmed for the case of form factors of $\Tr \, (\phi_{12}^2)$ in \cite{Brandhuber:2012vm}.

As already mentioned in \sref{sec:symbol3} (see Figure \ref{fig:plotR3}), this is not possible for the case of a minimal remainder function. 
This is caused by the simple fact that tree-level minimal form factors are $1$,  and remain $1$ under collinear/soft limits.  In what follows we will  quantify the failure to obey conventional factorisation. It is worth stressing that this failure only affects finite terms,  while the universality of IR divergences also extends to the minimal form factors. This
is related to the fact that we were able to define a finite remainder function for minimal form factors \eqref{eq:remainder} in complete analogy with scattering amplitudes in $\cN=4$ SYM and non-minimal form factors of $\Tr (\phi^2)$ in \cite{Brandhuber:2012vm}.

We begin our study with the simplest remainder function,  namely $\cR^{(2)}_{3,3}$ given in  \eqref{eq:remainderT3}. We consider  the collinear limit $p_1\,||\,p_2$, which we parameterise as 
\begin{equation} 
\label{collinear12}
p_1 \rightarrow z P \,, \quad  p_2  \rightarrow (1 - z) P \,, \qquad P^2=0\ .
\end{equation}
In terms of the $u,v,w$ variables, this is equivalent to
\begin{equation}
u \rightarrow 0 \, , \quad  v \rightarrow (1-z) \, ,  \quad w \rightarrow z \, . 
\end{equation} 
In the limit \eqref{collinear12}, an explicit calculation shows that  
\begin{equation}
\cR^{(2)}_{3,3}(u, v, w) \,{\buildrel 1 \parallel 2\over
{\relbar\mskip-1mu\joinrel\longrightarrow}}
\, 
 \sum^{2}_{m=1} \log^m(u) \ C_{3;m}(z)\, , 
\end{equation}
where the coefficients $C_{3;m}(z)$ are given by
\begin{align}
\begin{split}
C_{3;2}(z) &= {1 \over 4} \left[ \log^2\left( {z \over 1-z} \right) - 2 \zeta_2 \right] \, ,\\[5pt]
C_{3;1}(z) &= -C_{3;2}(z) \log \left[ z (1 - z) \right] + 
 {3 \over 2} \left[ {\rm Li}_3\left({ z \over z-1}\right) + {\rm Li}_3 \left({z-1 \over z}\right) \right] - \zeta_3 \, .
\end{split}
\end{align}
Next, we consider  the soft limit $p_1 \rightarrow 0$, where we have to take $z \rightarrow 0$
in addition to $u \rightarrow 0$. 
Equivalently, one can parametrise the soft limit as
\begin{align}
u = x \, \delta\, ,\qquad v = 1-  \delta\, ,\qquad w = (1 -x )\, \delta\, ,
\end{align}
with $\delta \rightarrow 0$. In this  limit we find
\begin{equation}
\cR^{(2)}_{3,3}(u, v, w) 
\ {\buildrel p_1 \to \, 0\over
{\relbar\mskip-1mu\joinrel\longrightarrow}}
\  
 \sum^{4}_{m=1} \log^m( \delta ) \ S_{3;m}(x)\ ,
\end{equation}
where the coefficients $S_{3;m}(x)$ at each order are given by
\begin{align}
\begin{split}
S_{3;4}(x) &\,=\, {1 \over 4} \, ,\\
S_{3;3}(x) &\,=\,  {1 \over 2 } \log \big[ x(1 - x)   \big] \, , \\
S_{3;2}(x) &\,=\,  [S_{3;3}(x)]^2 +  {1 \over 2} \log(1 - x) \log(x) + \zeta_2 \, ,  \\ 
S_{3;1}(x) &\,=\, 2 \Big( S_{3;3}(x) S_{3;2}(x)- [S_{3;3}(x)]^3 - \zeta_3   \Big) \, .
\end{split}
\end{align}
Now we turn our attention to the study of $\cR^{(2)}_{k,k}$ with $k>3$, in particular we will analyse the behaviour of the three-particle building blocks $r^{(2)}_i$ defined in \eqref{eq:3part-bb}. For the collinear limit $p_1 \,||\, p_2$  introduced in \eqref{collinear12}, both $r^{(2)}_i$ and $r^{(2)}_k$ contribute. Here we focus on $r^{(2)}_1$ only, since $r^{(2)}_k$ behaves in a similar  way.

\noindent We begin by observing that  $r_{{\rm nc},i}^{(2)}$ is regular as $u_i \rightarrow 0$, specifically
\begin{align}
\label{cp}
\begin{split}
 \lim_{u_i \rightarrow  0 }  r_{{\rm nc},i}^{(2)} =\, & 0- G\left(\left\{1,1,1,0\right\},v_i\right)\\
  = & - {1 \over 6}\log^2(1-v_i)\big[ \log(v_i) \log(1-v_i) 
 + 
3{\rm Li}_2(v_i) \big]
\\
& -\, 
\log(1-v_i)\, {\rm S}_{1,2}(v_i)
+{\rm S}_{1,3}(v_i) \, , 
\end{split}
\end{align}
where $\mathrm{S}_{n,p}(z)$ denotes a Nielsen polylogarithm,
\begin{align}
\label{eq:Nielsen}
\mathrm{S}_{n,p}(z)\,=\,\frac{(-1)^{n+p-1}}{(n-1)!p!}\int_0^1\!\frac{dt}{t}(\log t)^{n-1}\big[\log(1-zt)\big]^p\ .
\end{align}
On the other hand, if we now consider  the limit $w_i \rightarrow 0$ we observe that   this function develops a $\log^2(w_i)$ singularity. 
This singularity is required in order to cancel an identical and opposite  singularity arising  from $r^{(2)}_{{\rm cl}, i} + r_{\text{bts},i} ^{(2)}$. 
This is expected since $w_i \rightarrow 0 $ corresponds to two non-adjacent legs becoming collinear, which is not a physical singularity.

\noindent Setting in the collinear limit
\begin{equation}
u_1 \rightarrow 0 \, , \quad  v_1 \rightarrow (1-z) \, ,  \quad w_1 \rightarrow z \, ,
\end{equation}  
we obtain
\begin{equation}
r^{(2)}_1 
\,{\buildrel 1 \parallel 2\over
{\relbar\mskip-1mu\joinrel\longrightarrow}}
\, 
\sum^{2}_{m=1} \log^m(u_1)\, C^{1 \parallel 2}_{k;m}(z)\ ,
\end{equation}
where
\begin{align}
\begin{split}
C^{1 \parallel 2}_{k;2}(z) &= {1 \over 2} \left( {1\over 2} \, {\log^2(1-z)} +  {\rm Li}_2(z) - \zeta_2 \right)\ ,\\
C^{1 \parallel 2}_{k;1}(z) &= {1 \over 2} \log^2(1-z) \log(z)  -  {1 \over 3} \log^3(1-z) + 2 \, {\rm Li}_3 \left( {z \over z-1} \right) + 
 {\rm Li}_3(1-z) - \zeta_3 \ .
\end{split}
\end{align}
Finally, we consider the soft limit  for $r^{(2)}_1$. Because of the lack of permutation symmetry, $r^{(2)}_1$ behaves differently under the limits $p_1 \rightarrow 0$ and $p_2 \rightarrow 0$. In the limit $p_2 \rightarrow 0$, or equivalently
\beq
u_1 \,=\, x \,  \delta\, ,\quad v_1 \,=\, (1-x)\, \delta\, ,\quad  w_1\,=\, 1- \delta\, , 
\eeq
with $\delta \rightarrow 0$, we have
\begin{equation} \label{soft123}
r^{(2)}_1
\,{\buildrel p_2 \to \, 0 \over
{\relbar\mskip-1mu\joinrel\longrightarrow}}
\, 
 \sum^{4}_{m=1}  \log^m( \delta ) \, S^{p_2}_{k;m}(x)\ ,
\end{equation}
with
\begin{align}
\begin{split}
S^{p_2}_{k;4}(x) &\,=\, {1 \over 4} \, ,\\
S^{p_2}_{k;3}(x) &\,=\, {1 \over 2} \log \big[ x (1-x) \big] \, ,
\cr
S^{p_2}_{k;2}(x) &\,=\,  2 \zeta_2 + [S^{p_2}_{k;3}(x)]^2 + {1 \over 2} \log(x) \log(1-x)  \, , \cr 
S^{p_2}_{k;1}(x) &\,=\, S^{p_2}_{k;3}(x) \Big[ 4 \zeta_2 +\log(x) \log(1-x)  \Big] - 2\zeta_3 \, 
.
\end{split}
\end{align}
For $p_1 \rightarrow 0$, or equivalently
\beq
u_1\, =\, x\, \delta\, ,\quad  v_1\, =\,1 - \delta\, ,\quad w_1\,=\, (1-x)\, \delta\ ,
\eeq 
with $\delta \rightarrow 0$, we have
\begin{equation}
r^{(2)}_1
\,{\buildrel p_1 \to 0\over
{\relbar\mskip-1mu\joinrel\longrightarrow}}
\, 
 \sum^{4}_{m=1} \log^m( \delta ) S^{p_1}_{k;m}(x)\ ,
\end{equation}
where
\begin{align}
S^{p_1}_{k;4}(x) &\,=\,S^{p_1}_{k;3}(x) \,=\, S^{p_1}_{k;2}(x)\,=\,0  \, , \cr 
S^{p_1}_{k;1}(x) &\,=\, \zeta_2 \log \left(  {1 - x \over x }  \right) - \zeta_3 \, .
\end{align}
Note that  $r^{(2)}_1$ is less singular as $p_1 \rightarrow 0$ compared to the previous case where   $p_2 \rightarrow 0$. However, the full remainder is completely symmetric and should behave in the same way for arbitrary $p_i \rightarrow 0$. Indeed it is the building block with external legs $p_{k}, p_1, p_2$, namely $r^{(2)}_k$, that carries the leading divergence when $p_1 \rightarrow 0$, and the behaviour is precisely the same as \eqref{soft123}. 

\subsubsection{Non-minimal form factors}

In this section, we verify in an explicit example that $n$-point   form factors of the operator $\cT_k$  (with $k<n$)   obey the same universal factorisation properties that hold for scattering amplitudes in general gauge theories 
\cite{Bern:1994zx,Bern:1993qk}, as well as for form factors of the stress tensor operator as shown in \cite{Brandhuber:2012vm}. 
This relation states that under the limit where two adjacent particles $a$ and $b$ with helicities $\sigma_a,\,\sigma_b$ become collinear, the $L$-loop $n$-point colour-ordered form factor (or amplitude) factorises into a sum of $(n-1)$-point form factors of equal or lower loop order, and the collinear divergences are encoded into the coefficients of each term~---~the \emph{splitting amplitudes}. For a general form factor we have
\begin{align}
\label{eq:coll-general}
\begin{split}
\F_{\O,n}^{(L)}(1^{\sigma_1},\dots, a^{\sigma_a},b^{\sigma_b}, \dots, n^{\sigma_n}) 
\,{\buildrel a \parallel b\over
{\relbar\mskip-1mu\joinrel\longrightarrow}}
\, \sum_{\ell=0}^L\sum_{\sigma} \Big[  & \F^{(\ell)}_{\O,n-1} (1^{\sigma_1},\dots, (a+b)^{\sigma}, \dots, n^{\sigma_n})\\
& \times  \text{Split}_{-\sigma}^{(L-\ell)}(a^{\sigma_a},b^{\sigma_b}) \Big]\ , 
\end{split}
\end{align}
where $\sigma_i$ denote physical polarisations, and the sum is over all possible internal helicities~$\sigma$. 

\begin{figure}[h]
\centering
\includegraphics[width=0.4\textwidth]{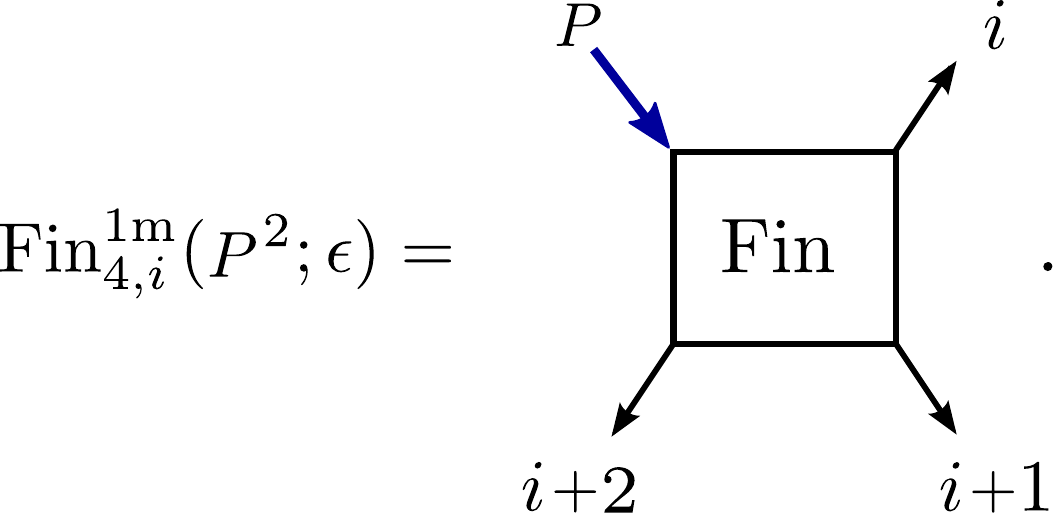}
\caption{\it One-mass finite box function with massless corner with momentum $P$.}
\label{fig:one-mass-fin-box}
\end{figure}

To confirm these factorisation properties, 
we will take as a representative example the particular component  form factor $F_{3,4}^{(1)}(1^+,2^{\phi_{12}},3^{\phi_{12}},4^{\phi_{12}};q)$, and will consider  the collinear limit $p_1\,||\,p_2$ defined in \eqref{collinear12}.  
For this case,  \eqref{eq:coll-general} predicts that 
\begin{align}
\label{eq:coll-one-loop}
\begin{split}
F_{3,4}^{(1)}(1^+,2^{\phi_{12}},3^{\phi_{12}},4^{\phi_{12}};q)\Big|_{1||2} \,&=\, F_{3,3}^{(0)}(P^{\phi_{12}},3^{\phi_{12}},4^{\phi_{12}};q)\,\text{Split}_{-\phi_{12}}^{(1)}(1^+,2^{\phi_{12}})\\
& +\,F_{3,3}^{(1)}(P^{\phi_{12}},3^{\phi_{12}},4^{\phi_{12}};q)\,\text{Split}_{-\phi_{12}}^{(0)}(1^+,2^{\phi_{12}})\ ,
\end{split}
\end{align}
where the tree-level and one-loop splitting functions with the helicities specified above are given by
\begin{align}
\label{eq:split-tree}
\text{Split}_{-\phi_{12}}^{(0)}(1^+,2^{\phi_{12}}) \, & = \,  \frac{1}{\b{12}} \sqrt{\frac{1-z}{z}}\ ,\\
\label{eq:split-1loop}
\text{Split}_{-\phi_{12}}^{(1)}(1^+,2^{\phi_{12}}) \, & = \, \frac{c_{\Gamma}}{\epsilon^2} (-s_{12})^{-\epsilon} \Big[1 - F\Big(\frac{z-1}{z}\Big) - F\Big(\frac{z}{z-1}\Big) \Big]\frac{1}{\b{12}} \sqrt{\frac{1-z}{z}}\ ,
\end{align}
and where we have introduced the shorthand notation $F(x)\,\equiv\,{_{2\!}F_1}(1,-\epsilon,1-\epsilon;x)$. The form factors appearing on the right-hand side of \eqref{eq:coll-one-loop} are given by
\begin{align}
\label{eq:RHS-FFs}
\begin{split}
F_{3,3}^{(0)}(P^{\phi_{12}},3^{\phi_{12}},4^{\phi_{12}};q) \,=\,& 1\ , \\[6pt]
F_{3,3}^{(0)}(P^{\phi_{12}},3^{\phi_{12}},4^{\phi_{12}};q) \,=\,&  -\frac{c_{\Gamma}}{\epsilon^2} \big[(-s_{P3})^{-\epsilon}+(-s_{34})^{-\epsilon}+(-s_{4P})^{-\epsilon}\big] \ .
\end{split}
\end{align}
In order to check \eqref{eq:coll-one-loop}, we use the general expression for the super form factors of $\T_3$ given in \eqref{eq:1-loop-MHV-BIS}. For the case of $F_{3,3}^{(1)}(1^+,2^{\phi_{12}},3^{\phi_{12}},4^{\phi_{12}};q)$, \eqref{eq:1-loop-MHV-BIS} reduces to
\begin{align}
\label{eq:F41loop}
\begin{split}
F^{(1)}_{3,4}(1^+,2^{\phi_{12}},3^{\phi_{12}},4^{\phi_{12}};q) =\  -\frac{c_{\Gamma}}{\epsilon^2} \frac{\b{24}}{\b{12}\b{14}} \big[(-s_{12})^{-\epsilon}+(-s_{23})^{-\epsilon}+(-s_{34})^{-\epsilon}+(-s_{41})^{-\epsilon}\big]& \\
 + \frac{\b{34}}{\b{31}\b{41}}\text{Fin}^{1{\rm m}}_{4,3}(s_{341};\epsilon)+ \frac{\b{24}}{\b{12}\b{14}}\text{Fin}^{1{\rm m}}_{4,4}(s_{412};\epsilon)+ \frac{\b{23}}{\b{13}\b{12}}\text{Fin}^{1{\rm m}}_{4,2}(s_{234};\epsilon)&\ ,
\end{split}
\end{align}
where $\text{Fin}_{4,i}^{1{\rm m}}(P^2;\epsilon)$ stands for the one-mass finite box function shown in Figure \ref{fig:one-mass-fin-box}, and is given by
\begin{align}
\begin{split}
&\text{Fin}_{4,i}^{1{\rm m}}(P^2;\epsilon)\,=\,-\frac{c_\Gamma}{\epsilon^2}\big[ (-s)^{-\epsilon}h(a\,s) + (-t)^{-\epsilon}h(a\,t)-(-P^2)^{-\epsilon} h(a\,P^2)\big]\ ,\\[5pt]
&s=s_{i\,i+1},\quad t=s_{i+1\,i+2},\quad u=s_{i\,i+2},\quad P^2=s_{i\,i+1\,i+2},\quad a\,\equiv\,-\frac{u}{st}\ ,
\end{split}
\end{align}
where we have defined  $ h(x)\, \equiv\, {_{2\!}F_1} \left(1,-\epsilon,1-\epsilon,x\right)-1 $ . 
\\
Under the collinear limit $p_1\,||\,p_2$ given in \eqref{collinear12}, we first notice that the term \newline $\dfrac{\b{34}}{\b{31}\b{41}}\text{Fin}^{1{\rm m}}_{4,3}(s_{341};\epsilon) $ in \eqref{eq:F41loop} is subleading,  and the remaining terms give
\begin{align}
\label{eq:F4coll}
\begin{split}
& F_{3,4}^{(1)}(1^+,2^{\phi_{12}},3^{\phi_{12}},4^{\phi_{12}};q)\Big|_{1||2}\, =\   -\frac{c_{\Gamma}}{\epsilon^2} \sqrt{\frac{1-z}{z}}\frac{1}{\b{12}}  \Big[ (-s_{12})^{-\epsilon}+(-s_{3P}(1-z))^{-\epsilon}+(-s_{34})^{-\epsilon}\\
& +(-s_{4P}z)^{-\epsilon} +(-s_{4P}z)^{-\epsilon}h\Big((z-1)\frac{s_{4P}}{s_{12}}\Big)+ (-s_{12})^{-\epsilon}h\Big(\frac{z-1}{z}\Big)-(-s_{4P})^{-\epsilon}h\Big(\frac{z-1}{z}\frac{s_{4P}}{s_{12}}\Big)\\
&+ (-s_{12})^{-\epsilon}h\Big(\frac{z}{z-1}\Big)+ (-s_{3P}(1-z))^{-\epsilon}h\Big(-z\frac{s_{3P}}{s_{12}}\Big)-(-s_{3P})^{-\epsilon}h\Big(\frac{z}{z-1}\frac{s_{3P}}{s_{12}}\Big)\Big]\ .
\end{split}
\end{align}
From \eqref{eq:F4coll} we already see the tree-level splitting amplitude \eqref{eq:split-tree} appearing as an overall prefactor. The terms with $(-s_{12})^{-\epsilon}$ combine to give the one-loop splitting amplitude \eqref{eq:split-1loop} as expected, 
\begin{align}
\begin{split}
F_{3,4}^{(1)}(1^+,2^{\phi_{12}},3^{\phi_{12}},4^{\phi_{12}};q)\Big|_{1||2}^{(-s_{12})^{-\epsilon}}\, =\  &\,\frac{c_{\Gamma}}{\epsilon^2} (-s_{12})^{-\epsilon} \sqrt{\frac{1-z}{z}}\frac{1}{\b{12}} \Big[1- F\Big(\frac{z-1}{z}\Big) - F\Big(\frac{z}{z-1}\Big) \Big]\\[6pt]
=&\ \text{Split}^{(1)}_{-{\phi_{12}}}(1^{\phi_{12}},2^{\phi_{12}})\,F^{(0)}_{3,3}(P^{\phi_{12}},3^{\phi_{12}},4^{\phi_{12}};q)\ ,\\[8pt]
\end{split}
\end{align}
whereas performing an expansion in $\epsilon$  of the remaining terms shows that it matches precisely $\,F_{3,3}^{(1)}(P^{\phi_{12}},3^{\phi_{12}},4^{\phi_{12}};q)\,\text{Split}_{-\phi_{12}}^{(0)}(1^+,2^{\phi_{12}})$. Thus we conclude that the universal collinear factorisation structure \eqref{eq:coll-general} is obeyed for the particular  one-loop form factor  we considered. Confirming the collinear factorisation at two-loop order would require the
calculation of the non-minimal form factor $F^{(2)}_{3,4}$,  which we leave for future investigation.

\chapter{The dilatation operator and on-shell methods}
\label{ch:dilatation}

\section{Introduction and motivation}
\label{sec:dilatation-introduction}

So far we have studied the application of on-shell methods to form factors, which are partially off-shell quantities. In this chapter we move on to the study of a completely off-shell quantity: the one-loop dilatation operator in the $SO(6)$ and $SU(2|3)$ sectors in planar $\N=4$ SYM.

The complete one-loop dilatation operator is known from the work of Beisert and Staudacher \cite{Beisert:2003jj,Beisert:2003yb} and it has led to the discovery of integrability in the planar sector, which allows for the computation of anomalous dimensions for finite values of the 't Hooft coupling. Such computations are of great importance as they produce results at strong coupling which can be compared to string theory predictions and may shed some light into the strong coupled regimes of field theories. In the scattering amplitudes context, it is known that they are invariant under the Yangian of $PSU(2,2|4)$ \cite{Drummond:2009fd} which arises from the combination of superconformal and dual superconformal symmetries \cite{Drummond:2008vq}. Therefore an important question is what is the connection between the realisation of Yangian symmetry on amplitudes and the integrability of the dilatation operator. 
In this chapter we perform a few steps in connecting the two approaches and apply two on-shell methods to the dilatation operator: MHV rules and generalised unitarity.

An additional  motivation for our work is provided by the interesting papers \cite{Koster:2014fva,Wilhelm:2014qua}. In particular, \cite{Koster:2014fva} successfully computes the one-loop dilatation operator $\Gamma$ in the $SO(6)$ sector using $\cN=4$ supersymmetric twistor actions  \cite{Boels:2006ir,Boels:2007qn,Adamo:2011cb}. It is known that such  actions, in conjunction with a particular axial gauge choice,  generate the MHV rules in twistor space \cite{Boels:2007qn}, and the question naturally arises as to whether one could derive the dilatation operator directly using MHV diagrams in momentum space, without passing through twistor space. The answer to this question is positive and this is the subject of \sref{sec:dilatation-MHV-diagrams}. Furthermore, we find that the calculation is very simple~---~it amounts to the evaluation of a single MHV diagram in dimensional regularisation, leading to a single UV-divergent integral, identical to that appearing in  \cite{Minahan:2002ve} and reviewed in \sref{sec:dilatation}.

The MHV diagram expansion can be obtained from the $\N=4$ SYM action using a particular axial gauge choice, followed by a field redefinition \cite{Mansfield:2005yd,Gorsky:2005sf}, thus its validity not only applies to on-shell amplitudes, but also to off-shell quantities such as correlation functions.

There are several reasons to pursue an approach based on MHV diagrams. Firstly, it is interesting to consider the application of this method to the computation of fully off-shell quantities such as correlation functions. Secondly, in the MHV diagram method  there is a natural way to regulate the divergences arising from loop integrations, namely dimensional regularisation, used  in conjunction with the  four-dimensional expressions for the vertices. In this respect, we recall that  one-loop amplitudes were calculated with MHV diagrams in  \cite{Brandhuber:2004yw}, where the infinite sequence of MHV amplitudes in $\cN=4$ SYM was rederived.  One-loop amplitudes in $\cN=1$ SYM were subsequently computed  in \cite{Bedford:2004py, Quigley:2004pw}, while in \cite{Bedford:2004nh} the cut-constructible part of the infinite sequence of MHV amplitudes in pure Yang-Mills  at one loop was presented. The $\cN=1$ and $\cN=0$ amplitudes have ultraviolet (UV) divergences (in addition to infrared ones), which are also regulated in dimensional regularisation. The two-point correlation function relevant for the dilatation operator also exhibits  UV divergences, which we regulate in exactly the same way as in the case of amplitudes.%
\footnote{The reader may consult \cite{Brandhuber:2005kd,Brandhuber:2011ke} for further applications of the MHV diagram method to the calculation of loop amplitudes.}

In the second part of this chapter, \sref{sec:dilatation-unitarity}, we move on and apply generalised unitarity \cite{Bern:1997sc,Britto:2004nc} which, as we shall see, allows for an even more efficient calculation of the dilatation operator. The use of generalised unitarity will further simplify the already remarkably simple calculation of the dilatation operator performed with MHV rules and will allow us to easily do the computation also in the $SU(2|3)$ sector. The application of unitarity to the derivation of the dilatation operator is welcome also from a conceptual point of view, since the only ingredients of  the calculation are on-shell amplitudes~---~with no off-shell information being introduced. This supports the hope that using this approach one may be able to connect directly the amplitudes and their  hidden structures and symmetries to the integrability of the dilatation operator in $\N=4$ SYM. It is important to mention that other applications of unitarity to the calculation of $n$-point correlators and correlation functions of Wilson lines have appeared in
\cite{Engelund:2012re, Laenen:2014jga, Engelund:2015cfa}, apart from the already mentioned \cite{Wilhelm:2014qua,Nandan:2014oga}  in the specific context of the dilatation operator.

\subsection*{General Strategy}

Recall from \sref{sec:dilatation} that the one-loop dilatation operator can be obtained by computing the UV divergent part of the two-point function 
\begin{equation}
\label{eq:two-point-function}
\left\langle \O(x_1) 
\bar{\O}(x_2)\right\rangle\Big|_{\rm UV}^{\text{one-loop}}\ 
\end{equation}
of the appropriate operators belonging to the sector under study. For both MHV diagram and generalised unitarity methods, the extraction of the UV divergent part of two-point correlation functions lands on only one integral which is the same as \eqref{eq:X-position} in the MZ calculation, shown in Figure \ref{fig:MZintegral}.
%

\begin{figure}[h]
\centering
\includegraphics[width=0.3\linewidth]{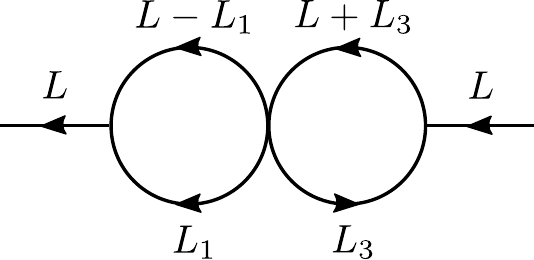}
\caption{\it The double-bubble integral relevant for the computation of $I(x_{12})$.}
\label{fig:double-bubble}
\end{figure}

Since we will be using methods inspired on scattering amplitude, it is useful to present this integral in momentum space, where it is a simple, single-scale integral~---~the double bubble shown in Figure \ref{fig:double-bubble}. It is given by
\begin{align}
\label{eq:double-bubble}
\begin{split}
I(x_{12})\,=&\,\int\!\prod_{i=1}^4 \frac{d^d
  L_i}{(2\pi)^d}\frac{e^{i(L_1+L_2)\cdot x_{12}}}{L_1^2\,L_2^2\,L_3^2\,L_4^2}\, (2 \pi)^d \, \delta^{(d)}\Big(\sum_{i=1}^4L_i\Big)\ \\[5pt]
=&\,\int\!\frac{d^d L}{(2\pi)^d}\; e^{iL\cdot x_{12}}\int\!\frac{d^d L_1}{(2\pi)^d}\frac{d^d L_3}{(2\pi)^d}\frac{1}{L_1^2\,(L-L_1)^2\,L_3^2\,(L+L_3)^2}\ , 
\end{split}
\end{align}
where $L \equiv L_1 + L_2$.
\\
The integral over $L_1$ and $L_3$  is the product of two bubble integrals with momenta as in Figure \ref{fig:double-bubble}, which are separately UV divergent. In momentum space, the UV divergence arise from the regions $L_1 \rightarrow \infty $ and $L_3 \rightarrow \infty $. The leading UV divergence of \eqref{eq:double-bubble} computed in dimensional regularisation ($d=4-2\eps$) is equal to 
\begin{align}
\label{eq:double-bubble-divergence}
\left. I(x_{12}) \right|_{\rm UV} \, = \, \frac{1}{\epsilon}\cdot  \frac{1}{8 \pi^2} \cdot \frac{1}{(4 \pi^2 x_{12}^2)^2} \ , 
\end{align}
where we have performed an inverse Fourier transform to position space using
 \beq
 \int\!{d^d p\over (2\pi)^d} \, {e^{i p \cdot x} \over (p^2)^s} \ = \   { \Gamma( {D\over 2} - s) \over 4^s \, \pi^{D\over 2}\, \Gamma (s) } \, {1\over (x^2)^{ {D\over 2}  - s}}
 \ . 
 \eeq
In all our computations only single-scale integrals appear. In the cases where they have tensor numerators, we employ the Passarino-Veltman (PV) reduction method to write them in terms of scalar integrals \cite{Passarino:1978jh}. The reductions will be shown explicitly in Appendix \ref{app:PV}.

\begin{figure}[h]
\centering
\includegraphics[width=0.3\linewidth]{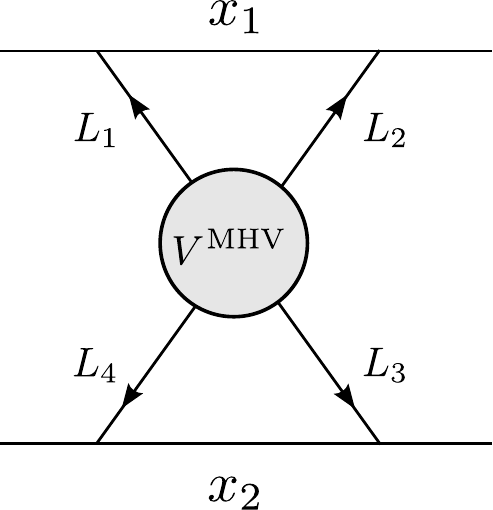}
\caption{\it
The single MHV diagram contributing to the dilatation operator at one loop. 
}
\label{MHVdiagram}
\end{figure}

For the MHV diagram computation, there is a single MHV diagram to compute, represented in Figure  \ref{MHVdiagram}. 
It consists of one  supersymmetric four-point MHV vertex, 
\begin{equation}
V^{\rm MHV} (1,2,3,4) \, = \,
\frac{ \delta^{(4)} \left(\sum\limits_{i=1}^4 L_i \right) \delta^{(8)} \left(\sum\limits_{i=1}^4 \ell_i \eta_i   \right) }{ \lan 12\ran \lan 23\ran \lan 34\ran \lan 41\ran} \,  , 
\end{equation}
and four scalar propagators  $1 / (L_1^2 \cdots L_4^2)$ connecting it to the four scalars in the operators. 
Here 
$L_i$ are the (off-shell) momenta of the four particles in the vertex. The off-shell continuations of the spinors associated to the internal  legs are defined using the prescription of \cite{Cachazo:2004kj} (also shown in \eqref{eq:off-shell-spinor}), namely 
\begin{equation}
\ell_{i \alpha}\, \equiv\,   L_{i \alpha \dot{\alpha}} \xi^{\dot \alpha} 
\ . 
\end{equation}
Here $\xi^{\dot \alpha}$ is a constant reference spinor%
\footnote{As we mentioned earlier, MHV diagrams were derived in \cite{Mansfield:2005yd,Gorsky:2005sf} from a change of variables in the Yang-Mills action quantised in the lightcone gauge. The  spinor  $\xi^{\dot \alpha}$ is precisely related to this gauge choice.}. The final result must be independent of the choice of $\xi^{\dot \alpha}$.

\begin{figure}[h]
\centering
\includegraphics[width=0.3\linewidth]{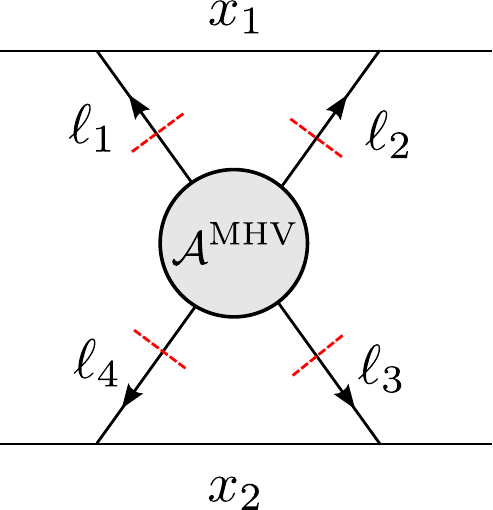}
\caption{\it
The single cut diagram contributing to the dilatation operator at one loop. Notice that the amplitude is colour dressed.
}
\label{fig:cut-diagram}
\end{figure}

For the unitarity computation, we go one step further and consider the quadruple cut of the two-point function \eqref{eq:two-point-function}, thus the four momenta $L_1,\,\dots,L_4$ are taken to be on-shell, denoted as $\ell_1,\,\dots,\,\ell_4$ and represented in Figure \ref{fig:cut-diagram}.

At one loop, the no-triangle property \cite{Bern:1994zx} of the
one-loop $S$-matrix of $\N=4$ SYM implies that maximal cuts employed in \cite{Britto:2004nc} are enough to completely determine all amplitudes of the theory. Similarly, we identify certain quadruple cuts which are sufficient to determine the dilatation operator at one loop.  




\subsection{The one-loop dilatation operator in the $SO(6)$ sector}

Recall from \sref{sec:dilatation} that operators in the $SO(6)$ sector are of the form 
\beq
\cO_{A_1 B_1, A_2 B_2, \ldots, A_L B_L} (x) \ \equiv \ \Tr \big( \phi_{A_1 B_1} (x) \cdots  \phi_{A_L B_L} (x) \big) \ ,
\eeq
and at one loop and in the planar limit, only nearest neighbour scalar fields can be connected by vertices. This simplifies the calculation to that of 
$\left\langle (\phi_{AB} \phi_{CD})(x_1)(\phi_{A^\prime B^\prime} \phi_{C^\prime D^\prime})(x_2) \right\rangle $, where colour indices are suppressed. The expected flavour structure of this correlation function is 
\begin{align}
\label{eq:ABC}
\begin{split}
\big\langle (\phi_{AB} \phi_{CD})(x_1)&(\phi_{A^\prime B^\prime} \phi_{C^\prime D^\prime})(x_2) \big\rangle  \\ &\!=\,\cA\,\epsilon_{ABCD} \epsilon_{A^\prime B^\prime C^\prime D^\prime} + \cB\,  \epsilon_{AB A^\prime B^\prime} \epsilon_{CD  C^\prime D^\prime} + \cC\, \epsilon_{AB C^\prime D^\prime} \epsilon_{A^\prime B^\prime CD}\ .
\end{split}
\end{align}
These three terms are usually referred to  as  trace, permutation and identity as shown in Table \ref{tab:R-contractions}. We are only interested in the leading UV-divergent contributions to the coefficients $\cA$, $\cB$ and $\cC$, which we denote by $\cA_{\rm UV},\,\cB_{\rm UV}$ and $\cC_{\rm UV}$. According to \cite{Minahan:2002ve} they are expected to be
\beq
\cA_{\rm UV} \ = \ {1\over 2}\, , \qquad  \cB_{\rm UV} \ = \ -1 \, , \qquad \cC_{\rm UV} \ = \ 1 \ . 
\eeq
In the definitions of  $\cA_{\rm UV}$, $\cB_{\rm UV}$, and $\cC_{\rm UV}$ we omit a factor of $\lambda/(8 \pi^2) \times \big(1 / ( 4 \pi^2 x_{12}^2)\big)^2 \times (1/ \epsilon)$ arising from the UV-divergence \eqref{eq:double-bubble-divergence} and the colour contractions. These factors will be reinstated at the end. This leads to the famous result  of \cite{Minahan:2002ve} for the one-loop dilatation operator $\Gamma$  in the $SO(6)$ sector, 
\begin{equation}
\label{eq:pss}
\Gamma^{SO(6)}  \ = \  {\lambda \over 8 \pi^2} \sum_{n=1}^{L} \left( \uno \, - \, \mathbb{P}_{n, n+1} \, + \, {1\over 2} \, {\rm Tr}_{n, n+1}  \right) 
\ , 
\end{equation}
where $\mathbb{P}$ and ${\rm Tr}$ are the permutation and trace operators, respectively, $L$ is the number of scalar fields in the operator, and $\lambda$ the 't Hooft coupling. 

The strategy for both MHV diagram and unitarity computations amount to choosing the $SU(4)$ $R$-symmetry assignments such that only one term in $\Gamma^{SO(6)}$ survives ($\uno$, $\mathbb{P}$ or $\Tr$). These representative assignments can be seen in Table \ref{tab:R-assignments}.
\begin{table}[h]
\centering
\begin{tabular}{c|cc}
& $ABCD$ & $A'B'C'D'$ \\
 \hline
 $\text{Tr}$ & $1234$ & $2413$ \\
 $\mathbb{P}$ & $1213$ & $3424$\\
 $\uno$ & $1213$ & $2434$ 
\end{tabular}
\caption{\it $R$-symmetry assignments for each representative term in the $SO(6)$ one-loop dilatation operator.}
\label{tab:R-assignments}
\end{table}
The computation of $\Gamma^{SO(6)}$ using MHV diagrams is shown in \sref{sec:dilatation-MHV-diagrams} and using generalised unitarity in \sref{sec:unitarity-SO(6)}.

\subsection{The one-loop dilatation operator in the $SU(2|3)$ sector} 

The $SU(2|3)$ sector is particularly interesting, as it involves also fermions. Indeed, operators in this sector are formed by letters taken from the set  
 $\big\{\psi_{1\,\alpha}, \phi_{1A}\big\}$, with $\alpha=1,2$ transforming under one $SU(2)_{L}$ Lorentz group and $A=2,3,4$ transforming under $SU(3)\subset SU(4)_R$ $R$-symmetry group. We thus have one  fermion and three scalar fields. The dilatation operator in this sector was derived in \cite{Beisert:2003ys}. 
Its expression is given by
\begin{align}
\label{eq:dilsu23}
\begin{split}
\Gamma^{SU(2|3)}\,=\,\dfrac{\lambda}{8\pi^2}\bigg[
  &\left\{\begin{smallmatrix}A\,B\\\cr A\,B\end{smallmatrix}\right\} \,
  -\, \left\{\begin{smallmatrix}A\,B\\\cr
  B\,A\end{smallmatrix}\right\} \, + \,
  \left\{\begin{smallmatrix}A\,\beta\\\cr
  A\,\beta\end{smallmatrix}\right\}\,+\,
  \left\{\begin{smallmatrix}\alpha\,B\\\cr
  \alpha\,B \end{smallmatrix}\right\} \, \\[5pt]
  - \, \Big(
  &\left\{\begin{smallmatrix}A\,\beta\\ \cr
  \beta\,A\end{smallmatrix}\right\}\,+\,
  \left\{\begin{smallmatrix}\alpha\,B\\ \cr
  B\,\alpha \end{smallmatrix}\right\}\Big) \,+\,
  \left\{\begin{smallmatrix}\alpha\,\beta\\ \cr
  \alpha\,\beta\end{smallmatrix}\right\}
  \,+\,\left\{\begin{smallmatrix}\alpha\,\beta\\ \cr
  \beta\,\alpha\end{smallmatrix}\right\} \bigg] .
  \end{split}
  \end{align}
Here $A,B=2,3,4$ denote scalars and the notation $ \left\{\begin{smallmatrix} I \, J\\ \cr
  K \, L \end{smallmatrix}\right\} $ stands for the action of the dilatation operator $\Gamma \b{\Phi_I(x_1) \Phi_J(x_2)} \propto \b{\Phi_K(x_1) \Phi_L(x_2)} $ with each $\Phi $ being a field belonging to the $SU(2|3)$ sector.
In \sref{sec:unitarity-SU(2|3)} we will rederive \eqref{eq:dilsu23} using generalised unitarity.

\section{$\Gamma^{SO(6)}$ from MHV Rules}
\label{sec:dilatation-MHV-diagrams}

In this section we apply the MHV diagram method to the computation of the one-loop dilatation operator $\Gamma^{SO(6)}$. To do so, we  compute the UV-divergent part of the coefficients $\cA$, $\cB$, $\cC$ defined in \eqref{eq:ABC}, representing the trace, permutation and identity flavour structures, respectively.   
Next we extract the relevant component vertices for the three flavour assignments in Table \ref{tab:R-assignments}. These turn out to be: 
\begin{table}[h]
\centering
\begin{tabular}{c l}
${\rm Tr}: $ & $ 
\quad A(1^{\phi_{12}},4^{\phi_{13}},3^{\phi_{24}},2^{\phi_{34}})\, =\,   \dfrac{ \lan 13\ran \lan 24\ran}{\lan 12 \ran \lan 34 \ran} \quad
$ \\
& \\
$\mathbb{P}:$ & $ \quad A(1^{\phi_{12}},4^{\phi_{24}},3^{\phi_{34}},2^{\phi_{13}}) \, =\,  -1 \quad $
 \\
 & \\
$\uno:$ & $ \quad
A(1^{\phi_{12}},4^{\phi_{34}},3^{\phi_{24}},2^{\phi_{13}}) \, =\, \dfrac{ \lan 13\ran \lan 24\ran}{\lan 23 \ran \lan 14 \ran}\quad$\\
\end{tabular}
\caption{\it On-shell amplitudes corresponding to the $R$-symmetry assignments outlined in Table \ref{tab:R-assignments}.}
\label{tab:sss}
\end{table}

Hence in the case of $\mathbb{P}$ the resulting loop integral is precisely the double-bubble  integral $I(x_{12})$ of  \eqref{eq:double-bubble} (up to a sign), while in the other two cases the double-bubble integrand is dressed with the vertex factors in Table \ref{tab:sss}. In the following we discuss the additional contributions from the vertex  for the three 
configurations ${\rm Tr}$, $\mathbb{P}$ and $\uno$. 

\subsection*{The Tr integrand} 

We begin our analysis with the vertex factor  for  the trace configuration shown in Table \ref{tab:sss}.
Using the off-shell prescription for MHV diagrams we can rewrite  it as 
\beq
T\equiv { [\xi | L_1 L_3 |\xi]  \,  [\xi | L_2 L_4 |\xi] \over [\xi | L_1 L_2 |\xi] \,  [\xi | L_3 L_4 |\xi] }
\ . 
\eeq
Using momentum conservation to eliminate $L_2$ and $L_4$,  this can be recast as a sum of three terms, 
\beq 
\label{T2}
T \ = \  - { [\xi | L_1 L_3 |\xi]  \over   [\xi | L_3 L |\xi] } - { [\xi | L_1 L_3 |\xi]  \over   [\xi | L_1 L |\xi] }  - { [\xi | L_1 L_3 |\xi]^2   \over  [\xi | L_1 L |\xi] \,  [\xi | L_3 L |\xi] } 
\ , 
\eeq
where $L\equiv L_1 + L_2$. The first two terms correspond to linear bubble integrals in $L_1$ and $L_3$, respectively. We will study separately the contribution arising from the last term. 
The linear  bubble integral can be written in terms of a scalar bubble as (see Appendix \ref{app:PV} for a derivation)
\beq
\label{eq:Linear-Bubble}
\int\!{d^d K \over (2 \pi)^d} {K^\mu \over K^2 (K \pm L)^2} \ = \ \mp {L^\mu\over 2} {\rm Bub}(L^2)  \, ,
\eeq
where 
\beq
\label{eq:scalar-single-bubble}
{\rm Bub} (L^2) \equiv \int\!{d^d K \over (2 \pi)^d} {1 \over K^2 (K + L)^2}
\ . 
\eeq
This is one of the two scalar bubbles comprising the MZ integral \eqref{eq:double-bubble}. In the following we will then only quote the coefficient dressing the MZ integral. Doing so, the first term in \eqref{T2} becomes, after the reduction,  
\beq 
- {[ \xi | L L_3 | \xi] \over [ \xi | L_3 L | \xi ] }\cdot {1\over 2}  \ = \ {1\over 2} \, .
\eeq
Similarly, the second term in \eqref{T2} gives a result of $+ 1/2$. Next we move to the third term. To simplify its expression, we first notice that the bubble integral in $L_1$ is symmetric under the transformation $L_1 \to L - L_1$. The idea is then to simplify the integrand by using this symmetry. Thus, we rewrite the quantity  $[ \xi | L_1 L_3 | \xi ] $ in the numerator as 
$[ \xi | L_1 L_3 | \xi ] \ = \ [ \xi  | (L_1 - {1\over 2} L) L_3 | \xi ] + {1\over 2} [\xi | L L_3 | \xi] $. Doing so, we get 
\beq
 - { [\xi | L_1 L_3 |\xi]^2   \over  [\xi | L_1 L |\xi] \,  [\xi | L_3 L |\xi] } =
 - { [\xi |( L_1  - {L\over 2}) L_3 |\xi]^2   \over  [\xi | L_1 L |\xi] \,  [\xi | L_3 L |\xi] } + {1\over 4}   { [\xi | L L_3 |\xi]   \over  [\xi | L_1 L |\xi]  } +   { [\xi |(L_1 - {1\over 2}  L)  L_3 |\xi]   \over  [\xi | L_1 L |\xi]  }\ . 
 \eeq
We then notice that the first  and the second term are  antisymmetric under the transformation $L_1 \to L - L_1$ and hence vanish upon integration. 
The third term is a sum of two  linear bubbles in $L_3$, and the corresponding contributions  are quickly seen to be equal to $- 1/2$ and zero, respectively. 

Summarising, the trace integral gives a contribution of $1/2$ times the dimensionally regularised   MZ integral. Thus $\cA_{\rm UV} =  {1/ 2}$.

\subsection*{The $\mathbb{P}$ integrand}
In this case the vertex is simply  $-1$ and the corresponding result is  $-1$ times the MZ integral, or $\cB_{\rm UV} = -1$. 

\subsection*{The $\uno$ integrand}

The relevant vertex factor is written in Table \ref{tab:sss}. In this case we observe that 
\beq 
{\lan 13 \ran \lan 24 \ran \over \lan 23 \ran \lan 14 \ran } \ = \  1 + {\lan 12 \ran \lan 34 \ran \over \lan 23 \ran \lan 14 \ran } 
 \ . 
 \eeq
 The first term gives a contribution equal to the MZ integral, and we will now argue that the second term is UV finite, and hence does not contribute to the dilatation operator. Indeed, we can write 
 \beq
 \label{sopra}
  {\lan 12 \ran \lan 34 \ran \over \lan 23 \ran \lan 14 \ran }  \ =\  {[ \xi | L_1 L | \xi ] [ \xi | L_3 L | \xi ] \over [ \xi |(L -  L_1) L_3 | \xi ] [ \xi | L_1 (L + L_3) | \xi ] }
  \ . 
  \eeq
  The UV divergences we are after arise when  $L_1$ and $L_3$ are large. The integrand \eqref{sopra} provides one extra power of momentum per integration, which makes each of the two bubbles in the MZ integral finite.%
 \footnote{One may also notice that for large $L_1$ and $L_3$ the integrand becomes an odd function of these two variables, and thus the integral should be suppressed even further than  expected from power counting.}   
 Thus  $\cC_{\rm UV}=1$.  

We end this section with a comment regarding the independence of the integrals above on the reference spinor $\xi$. Since MHV diagrams are obtained from a particular axial gauge choice, combined with a field redefinition \cite{Mansfield:2005yd,Gorsky:2005sf},  it is  guaranteed that $\xi$-dependence drops out at the end of the calculation. In the present case one can see this directly as follows. Lorentz invariance ensures that the result of the $L_1$- and $L_3$-integrations can only  depend on  $L^2$, as the other Lorentz-invariant quantity $[\xi | L^2 | \xi]$ vanishes (note that $L\cdot \xi$ cannot appear as our integrands only depend on the anti-holomorphic spinor $\xi_{\dot\alpha}$).
 

\section{Unitarity}
\label{sec:dilatation-unitarity}

We now proceed to the computation of the one-loop dilatation operator in the $SO(6)$ and $SU(2|3)$ sectors using generalised unitarity. 

The computation here simplifies that of MHV diagrams considerably. There is only one quadruple cut to consider~---~that where the four propagators with momenta $L_1$, $L_2 \equiv L-L_1$, $L_3$ and $L_4 \equiv -(L+L_3)$ are set on shell. By computing these cuts we will be able to identify the coefficient of the double bubble \eqref{eq:double-bubble} in all relevant cases, once again without having to perform any integral. The cut double bubble can then be lifted to a full integral, and by picking its UV divergence \eqref{eq:double-bubble-divergence} we can immediately write down the dilatation operator.

\subsection{$\Gamma^{SO(6)}$ from unitarity}
\label{sec:unitarity-SO(6)}

In this section we will compute $\Gamma^{SO(6)}$ using generalised unitarity. This calculation is essentially the same of \sref{sec:dilatation-MHV-diagrams} with the important change that we no longer need to use off-shell continuations of amplitudes. The set up is shown in Figure \ref{fig:cut-diagram}.

For each case there is a single cut diagram to consider. The integrand
is constructed with four cut scalar propagators with momenta $L_i$,
$i=1, \ldots , 4$,  and one on-shell amplitude, as shown in Figure \ref{fig:cut-diagram}. The operators are
connected to the amplitude via appropriate form factors, which in the
scalar case are simply
\begin{align}
\label{eq:form-factor-scalars}
\begin{split}
F_{\phi^a \tilde\phi^b}(\ell_1^{\phi^{a'}},\ell_2^{\tilde\phi^{b'}};L)\,&\equiv\,
\int\!\!d^4x \  e^{iL\cdot x}\ 
 \big\langle \, 0\, |(\phi^a \tilde\phi^b)(x)|\, \phi^{a'}(\ell_1),\tilde\phi^{b'}(\ell_2)\, \big\rangle\\
\,&=\, (2 \pi)^4 \delta^{(4)}\big(L-\ell_1-\ell_2\big)\,  \delta^{aa'}\delta^{bb'}\  ,
\end{split}
\end{align}
where we have used $\phi$ and $\tilde\phi$ to denote two scalar fields
having distinct $R$-symmetry indices as is sufficient for our
purposes cf. Table \ref{tab:R-assignments}. Table \ref{tab:sss} shows the relevant amplitudes for the three flavour assignments considered in Table \ref{tab:R-assignments}. Note that the $\ell_i$ represent the on-shell (cut) versions of the loop momenta $L_i$.
%
%

Three observations are in order here. First, we note that the
same integrands as in the approach of \sref{sec:dilatation-MHV-diagrams} appear, with the important difference that there the
spinors associated with the on-shell momenta are given by the
appropriate off-shell continuation for MHV diagrams. Here the spinors for the cut loop momenta do not need any off-shell continuation. Furthermore, for the case of the $\mathbb{P}$ integrand there is obviously no difference between the two approaches, and the resulting integral is given by a double bubble where all the four propagators are cut. In the other two cases, this integral is dressed by the appropriate amplitude.  Finally, we note that the colour factor associated with all diagrams is obtained from the contraction 
\begin{equation}
\label{eq:NN}
\cdots (t^{b}t^a)^{i}_{\, j} \cdots  \Tr (t^a t^b t^c t^d) \cdots (t^d t^c)^l_{\, m} \cdots = \cdots N^2 \delta^i_{m} \delta ^l_j \cdots\ , 
\end{equation}
where the trace arises from the amplitude and the factors $\cdots (t^{b}t^a)^{i}_{\, j} \cdots $ and $\cdots (t^d t^c)^l_{\, m} \cdots$ from the operators (and we indicate only generators corresponding to the fields being contracted). 
We now proceed to construct the relevant integrands. 

\subsection*{The trace integrand} 

In this case the relevant amplitude (which multiplies four cut
propagators) can be rewritten as
\begin{align}
\frac{\b{13}\b{24}}{\b{12}\b{34}}\,=\,\frac{\Tr_+(\ell_1\,\ell_3\,\ell_4\,\ell_2)}{(\ell_1+\ell_2)^2(\ell_3+\ell_4)^2}
\,=\,-\frac{2(\ell_1\cdot\ell_3)}{L^2}
\ , 
\end{align}
where we have used  $\ell_1+\ell_2=-(\ell_3+\ell_4)\equiv L$ and the trace expansion \eqref{eq:trace-expand}. Having rewritten the amplitude in terms of products of momenta, we next lift the four cut momenta off shell. The resulting integral has  the  structure of a product of two linear bubbles,  
\begin{align}
\label{eq:trace}
-\frac{2}{L^2}\int\!\frac{d^d L_1}{(2\pi)^d}\frac{L_1^\mu}{L_1^2\,(L-L_1)^2}\int\! \frac{d^d L_3}{(2\pi)^d}\frac{L_{3\,\mu}}{L_3^2\,(L+L_3)^2}\ .
\end{align}
Using \eqref{eq:Linear-Bubble} we find that  \eqref{eq:trace} is equal to  $1/2$ times a double bubble. 
Using \eqref{eq:double-bubble-divergence} we finally get  $\A_{\rm UV}=1/2$.

\subsection*{The $\mathbb{P}$ integrand}
No calculation is needed in this case, and the result is simply given by minus a cut double-bubble integral. Lifting the cut integral to a full loop integral  we get  $\mathcal{B}_{\rm UV}=-1$.

\subsection*{The $\uno$ integrand}

\begin{figure}[h]
\centering
\includegraphics[width=0.31\linewidth]{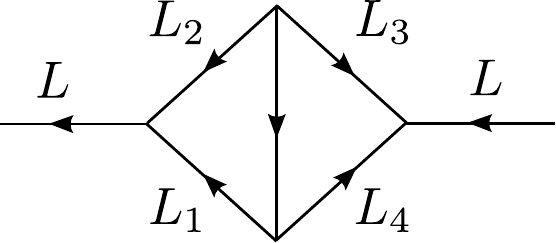}
\caption{\it The finite integral corresponding to the second term in \eqref{sem}. This integral is UV-finite and thus irrelevant for the calculation of the dilatation operator.}
\label{finiteintegral}
\end{figure}

The relevant amplitude in this case is
\begin{align}
\label{eq:amp-uno}
\frac{\b{13}\b{24}}{\b{23}\b{14}}\,=\,1+\frac{\b{12}\b{34}}{\b{23}\b{14}}\ .
\end{align}
Thus the first term in \eqref{eq:amp-uno} gives the cut double-bubble integral, whereas we can use  on-shell identities to rewrite the second term as
\begin{align}
\label{sem}
\frac{\b{12}\b{34}}{\b{23}\b{14}}\,=\,\frac{\b{12}\b{34}[34]}{\b{23}\b{14}[34]}=-\frac{L^2}{2(\ell_1\cdot\ell_4)}\ . 
\end{align}
Lifting the cut propagators of the second integral to full propagators,  it is immediate to see  that this term produces the integral represented in Figure \ref{finiteintegral}. This integral is finite in four dimensions and  
thus does not contribute to $\mathcal{C}_{\rm UV}$. We then conclude that $\mathcal{C}_{\rm UV}=1$. 
\\
A comment is in order here. In principle an ambiguity is still present corresponding to an integral such as that of Figure \ref{finiteintegral} but with one of the four propagators $L_1, \ldots , L_4$ collapsed (say $L_4$), which is UV divergent. This integral can be excluded by looking at a triple cut corresponding to  cutting the  propagators $L_1$, $L_3$ as well as the middle propagator  in  Figure~\ref{finiteintegral}. 


\subsection{$\Gamma^{SU(2|3)}$ from unitarity}
\label{sec:unitarity-SU(2|3)}

In this section we derive $\Gamma^{SU(2|3)}$ shown in \eqref{eq:dilsu23} using generalised unitarity. As for the $SO(6)$ case, in the planar limit only contractions between nearest-neighbour fields in $\O(x_1)$ and $\bar{\O}(x_2)$ have to be considered. The first two terms on the right-hand side of \eqref{eq:dilsu23} denote the scalar identity $\uno$ and permutation $\mathbb{P}$ structures already familiar from the $SO(6)$ case (the trace structure is absent given the restricted choice of scalar letters).  The novelty is that now we have to consider two additional types of contractions: scalar-fermion $\to$ scalar-fermion, and two-fermion $\to$ two-fermion, as indicated in the remaining terms in \eqref{eq:dilsu23}. We compute each of these processes separately in the following sections.

\subsection*{Scalar-fermion $\to$ scalar-fermion}

In this case we are interested in a fermion field $\psi_{1\,\alpha}$ and one of the scalars $\phi_{12},\,\phi_{13},$ or $\phi_{14}$. Without loss of generality we will consider $\phi_{12}$. There are two cases to consider, 
\beq   
\mathbf{U}:
\qquad \big\la( 
\phi^a_{12} \psi^b_{1\,\alpha})(x_1)(\psi^c_{234\,\dot{\alpha}}\phi^d_{34})(x_2)\big\ra\, , 
\eeq
and 
\beq
\mathbf{S}:
\qquad
\big\la( \phi^a_{12} \psi^b_{1\,\alpha})(x_1)(\phi^c_{34}\psi^d_{234\,\dot{\alpha}})(x_2)\big\ra\ , 
\eeq
where the letters {\bf U} and {\bf S} indicate whether the contractions between the two fields  are unswapped or swapped.
The relevant form factor  is 
\beq
\label{eq:form-factor-fermion-scalar}
\begin{split}
F_{ \phi^a_{12}  \psi_{1\,\alpha}^b}(\ell_1^{\phi_{12}^{a'}}, \ell_2^{\psi_{1\,\alpha}^{b'}};L)\,
&\equiv\,
\int\!\!d^4x \ e^{iL\cdot x} \, \la 0|( \phi^a_{12}  \psi_{1\,\alpha}^b)(x)|\phi_{12}^{a'}(\ell_1),\psi_{1}^{b'}(\ell_2)\ra\\
\,&=\, (2 \pi)^4 \delta^{(4)}\big(L-\ell_1-\ell_2\big)\lambda^2_\alpha\,\delta^{aa'}\delta^{bb'}\, ,
\end{split}
\eeq
and similarly for $\bar{\O}(x_2)$. 

We begin by considering the {\bf U} case. By 
contracting the two form factors with the four planar permutations of the full amplitude, we obtain\footnote{Two out of the six possible contractions, namely those where particles 1 and 2 are not adjacent, do not contribute at large $N$.} 
\begin{align}
\label{eq:contraction-fermion-scalar-1}
\begin{split}
&\lambda^2_\alpha\tl^3_{\dot{\alpha}}\,\delta^{aa'}\delta^{bb'}\delta^{cc'}\delta^{dd'}\\
\times \Big[&A( 1^{\phi_{12}}, 2^{\psi_1},3^{\psi_{234}},4^{\phi_{34}})\,\Tr(t^{a'}t^{b'}t^{c'}t^{d'}) +
 A(1^{\phi_{12}}, 2^{\psi_1}, 4^{\phi_{34}},3^{\psi_{234}})\,\Tr(t^{a'}t^{b'}t^{d'}t^{c'})\\
 - &A(1^{\phi_{12}}, 3^{\psi_{234}}, 4^{\phi_{34}},2^{\psi_1})\,\Tr(t^{a'}t^{c'}t^{d'}t^{b'}) - 
 A(1^{\phi_{12}}, 4^{\phi_{34}},3^{\psi_{234}},2^{\psi_1})\,\Tr(t^{a'}t^{d'}t^{c'}t^{b'})
 \Big]\ .
\end{split}
\end{align}
At large $N$ there is only one leading contribution, that with colour contractions given by \eqref{eq:NN}.  The corresponding amplitude is
\beq
A(1^{\phi_{12}}, 4^{\phi_{34}}, 3^{\psi_{234}}, 2^{\psi_1}) \,=\,  \frac{\b{13}\b{34}}{\b{14}\b{23}}\ . 
\eeq
Including the fermion polarisation spinors from the form factors \eqref{eq:form-factor-fermion-scalar} we get
\begin{align}
\begin{split}
-A(1^{\phi_{12}}, 4^{\phi_{34}}, 3^{\psi_{234}}, 2^{\psi_1}) \,\lambda^2_\beta  \tl^3_{\dot\beta}\, =  \, - \frac{(\ell_2\,\bar{\ell_1}\,\ell_3)_{\beta\dot{\beta}}}{2(\ell_1\cdot\ell_4)}\, \equiv   \,N_{\beta\dot{\beta}} \ ,
\end{split}
\end{align}
where we use the notation
\begin{align}
(\ell_i \bar{\ell}_j\ell_k)_{\alpha\dot{\alpha}}\,\equiv\, \lambda^i_{\alpha}[ij]\b{jk}\tl^k_{\dot\alpha}\,,\qquad (\bar{\ell}_i \ell_j \bar{\ell}_k)_{\dot{\alpha}\alpha}\,\equiv\, \tl^i_{\dot{\alpha}}\b{ij}[jk]\lambda^k_{\alpha}
\end{align}
and so on. The cut integral to consider is thus 
\beq
\label{888}
I_{\beta\dot{\beta}}\, \equiv\, \int\!\! d^4\ell_1  d^4\ell_3 \,
\delta^{(+)} (\ell_1^2) \,  
\delta^{(+)} (\ell_3^2) \,  
\delta^{(+)}\left( (L-\ell_1)^2\right) \,  
\delta^{(+)} \left((L+\ell_3)^2\right) \,  
\ \cdot \ N_{\beta\dot{\beta}}\ ,  
\eeq
where by Lorentz invariance $I_{\beta\dot{\beta}}$ must have the form 
\begin{align}
I_{\beta\dot{\beta}}\,=\,A\,L_{\beta\dot{\beta}}\ .
\end{align}
A simple PV reduction (shown in Appendix \ref{app:PV}, see \eqref{eq:A-Atilde}) determines that the UV-divergent part of the coefficient $A$ is  equal to $A_{\rm UV} =1/2$.


For the {\bf S} case, we get the single leading contribution to be
\begin{align}
\begin{split}
- A(1^{\phi_{12}}, 4^{\psi_{234}}, 3^{\phi_{34}}, 2^{\psi_1})\,\lambda^2_\beta  \tl^4_{\dot\beta} \,&=\,  - \frac{(\ell_2\,\bar{\ell_1}\,\ell_4)_{\beta\dot{\beta}}}{2(\ell_2\cdot\ell_3)}\, \equiv \,\tilde{N}_{\beta\dot{\beta}} \ .
\end{split}
\end{align}
The relevant integral is now  
\beqa
\tilde{I}_{\beta\dot{\beta}}&\equiv&
\label{999}
\int\!\! d^4\ell_1  d^4\ell_3 \,
\delta^{(+)} (\ell_1^2) \,  
\delta^{(+)} (\ell_3^2) \,  
\delta^{(+)} \left((L-\ell_1)^2\right) \,  
\delta^{(+)} \left((L+\ell_3)^2\right) \,  
\ \cdot \ \tilde{N}_{\beta\dot{\beta}}
\nonumber \\ 
&=&
\tilde{A}\,L_{\beta\dot{\beta}}
\ , 
\eeqa
where a PV reduction shows that   $\tilde{A}=-1/2$. Note that in arriving at this result we have discarded finite integrals, which do not contribute to the anomalous dimensions (more precisely, in all calculations the only other finite integral appearing is the kite, depicted in Figure \ref{finiteintegral}).

\begin{figure}[h]
\centering
\includegraphics[width=0.3\linewidth]{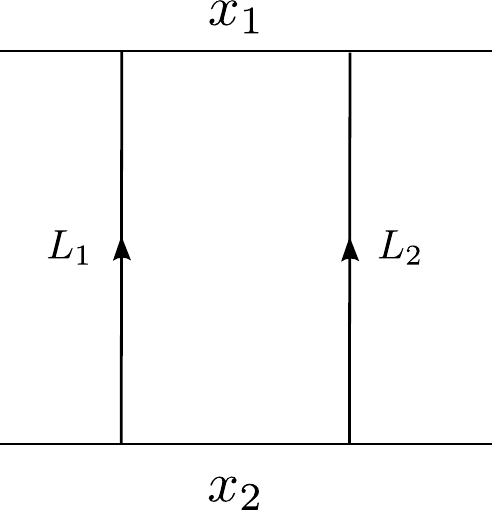}
\caption{\it
Tree-level planar contractions of two nearest neighbour fields of \eqref{eq:two-point-function}.
}
\label{fig:tree-level-contraction}
\end{figure}

Summarising, the scalar-fermion $\to$ scalar-fermion case  gives $\pm 1/2 \, L_{\beta\dot{\beta}}$ times a double-bubble integral, for the {\bf U}/{\bf S} case, respectively. 
This has  to be compared to the tree-level expression shown in \fref{fig:tree-level-contraction}, which is given by (using $L_2\equiv L-L_1$)
\begin{align}
\label{eq:fermion-scalar-tree}
\begin{split}
I^{\rm tree}_{\beta\dot{\beta}}\,&\equiv\, 
\int\frac{d^DL_1}{(2\pi)^D}\frac{L_{1\,\beta\dot{\beta}}}{L^2_1(L-L_1)^2} \ = \ {1\over 2} \, L_{\beta\dot{\beta}}\, 
{\rm Bub} (L^2)  \ .
\end{split}
\end{align}
Thus for the two-scalar two-fermion case we get: 
\beq
\uno: 1 \, , \qquad 
\mathbb{P}: -1 \ ,
\eeq
and the corresponding contribution to the spin-chain Hamiltonian is%
\footnote{Here we also reinstate powers of $g_{\rm YM}^2$ from the tree-level amplitudes, of $N$, arising from colour contractions, and a factor of $1/(8\pi^2)$ arising from the UV singularity \eqref{eq:double-bubble-divergence} of the double-bubble integral \eqref{eq:double-bubble}.}
\beq
\label{1f1s}
{\lambda \over 8 \pi^2}\left(  \left\{\begin{smallmatrix}A\,\beta\\\cr A\,\beta\end{smallmatrix}\right\} \, - \,
 \left\{\begin{smallmatrix}A\,\beta\\\cr \beta\,  A\end{smallmatrix}\right\} \right)
 \ , 
\eeq
in agreement with the corresponding terms in  \eqref{eq:dilsu23}.

\subsection*{Two-fermion $\to$ two-fermion}

In this case we consider the four-point correlator 
\begin{align}
\label{eq:ops-fermion-fermion}
\big\la (\psi_{1\,\alpha}^a \psi^b_{1\,\beta})(x_1)
(\psi^c_{234\,\dot{\alpha}}\psi^d_{234\, \dot\beta})(x_2)\big\ra  \ . 
\end{align}
The form factors of $\O(x_1)$ are given by
\begin{align}
\label{eq:form-factor-fermions}
\begin{split}
F_{\psi_{1\,\alpha}^a \psi^b_{1\,\beta}}(\ell_1^{\psi_{1\,\alpha}^{a'}},\ell_2^{\psi_{1\,\beta}^{b'}};L)\,&\equiv\,
\int\!\!d^4x \  e^{iL\cdot x}\ 
 \big\langle \, 0\, |(\psi_{1\,\alpha}^a \psi^b_{1\,\beta})(x_1)|\, \psi_{1}^{a'}(\ell_1),\psi_{1}^{b'}(\ell_2)\, \big\rangle\\
\,&=\, (2 \pi)^4 \delta^{(4)}\big(L-\ell_1-\ell_2\big)\,\cdot {1\over 2}  \big(\lambda^1_\alpha\lambda^2_\beta\, \delta^{aa'}\delta^{bb'}\, -\,  \lambda^1_\beta\lambda^2_\alpha\, \delta^{ab'}\delta^{ba'}\big)\, ,
\end{split}
\end{align}
and similarly for the form factor of $\bar\O(x_2)$. Note the factor of $1/2$ appearing because of the presence of two identical particles in the state. 
Contracting the two form factors with the four planar permutations of the full amplitude,
we get
\begin{align}
\label{eq:contraction-fermions}
\begin{split}
-\frac{1}{4}\big(&\lambda^1_\alpha\lambda^2_\beta\delta^{aa'}\delta^{bb'}- \lambda^1_\beta\lambda^2_\alpha\delta^{ab'}\delta^{ba'}\big)\big(\tl^3_{\dot\alpha}\tl^4_{\dot\beta}\delta^{cc'}\delta^{dd'}- \tl^3_{\dot\beta}\tl^4_{\dot\alpha}\delta^{cd'}\delta^{dc'}\big)\\
\times \Big[&A(1^{\psi_1},2^{\psi_1},3^{\psi_{234}},4^{\psi_{234}})\,\Tr(t^{a'}t^{b'}t^{c'}t^{d'}) - A(1^{\psi_1},2^{\psi_1},4^{\psi_{234}},3^{\psi_{234}})\,\Tr(t^{a'}t^{b'}t^{d'}t^{c'})\\
 + &A(1^{\psi_1},3^{\psi_{234}},4^{\psi_{234}},2^{\psi_{1}})\,\Tr(t^{a'}t^{c'}t^{d'}t^{b'}) - A(1^{\psi_1},4^{\psi_{234}},3^{\psi_{234}},2^{\psi_{1}})\,\Tr(t^{a'}t^{d'}t^{c'}t^{b'})
 \Big]\ .
\end{split}
\end{align}
In the large-$N$ limit we only need to keep the following terms out of those in \eqref{eq:contraction-fermions}:
\begin{align}
\label{eq:contraction-fermions-2}
\begin{split}
-{1\over 4} \bigg[ A(1^{\psi_1},2^{\psi_1},3^{\psi_{234}},4^{\psi_{234}})\,\lambda^1_\beta\lambda^2_\alpha \tl^3_{\dot\beta}\tl^4_{\dot\alpha}+ A(1^{\psi_1},2^{\psi_1},4^{\psi_{234}},3^{\psi_{234}})\,\lambda^1_\beta\lambda^2_\alpha\tl^3_{\dot\alpha}\tl^4_{\dot\beta} \\ 
 -\, A(1^{\psi_1},3^{\psi_{234}},4^{\psi_{234}},2^{\psi_{1}})\,\lambda^1_\alpha\lambda^2_\beta \tl^3_{\dot\beta} \tl^4_{\dot\alpha} - A(1^{\psi_1},4^{\psi_{234}},3^{\psi_{234}},2^{\psi_{1}})\,\lambda^1_\alpha\lambda^2_\beta  \tl^3_{\dot\alpha} \tl^4_{\dot\beta}\bigg] \ , 
\end{split}
\end{align}
where the  relevant four-fermion amplitudes are 
\begin{align}
&A(1^{\psi_1},2^{\psi_1},3^{\psi_{234}},4^{\psi_{234}})\,
\,=\,-\,  \frac{\b{34}^2}{\b{23}\b{41}}\ , \nonumber \\
&A(1^{\psi_1},2^{\psi_1},4^{\psi_{234}},3^{\psi_{234}})\,
\,=\,- \, \frac{\b{34}^2}{\b{24}\b{31}}\, , \nonumber \\
&A(1^{\psi_1},3^{\psi_{234}},4^{\psi_{234}},2^{\psi_{1}}) \,=\, \frac{\b{34}^2}{\b{13}\b{42}}\, , \nonumber \\
&A(1^{\psi_1},4^{\psi_{234}},3^{\psi_{234}},2^{\psi_{1}})\,=\,  \frac{\b{34}^2}{\b{14}\b{32}}\ .
\label{ffa}
\end{align}
Using \eqref{ffa}, we can rewrite   \eqref{eq:contraction-fermions-2} as
\begin{align}
\frac{1}{4}\left[\frac{(\ell_2\bar{\ell_1})_{\alpha\beta}(\bar{\ell_4}\ell_3)_{\dot{\alpha} \dot{\beta}}+(\ell_1\bar{\ell_2})_{\alpha\beta}(\bar{\ell_3}\ell_4)_{\dot{\alpha} \dot{\beta}}}{2(\ell_2\cdot\ell_3)}\,+\, \ell_1\leftrightarrow \ell_2\right]\ .
\end{align}
The term with $\ell_1\leftrightarrow \ell_2$ is simply a relabelling of the integration variables, and we conclude  that the one-loop integrand  is given by
\begin{align}
{1\over 2} \left[\frac{(\ell_2\bar{\ell_1})_{\alpha\beta}(\bar{\ell_4}\ell_3)_{\dot{\alpha} \dot{\beta}}+(\ell_1\bar{\ell_2})_{\alpha\beta}(\bar{\ell_3}\ell_4)_{\dot{\alpha} \dot{\beta}}}{2(\ell_2\cdot\ell_3)}\right]\equiv 
N_{\alpha\beta\dot{\alpha}\dot{\beta}}\ .
\end{align}
Thus we have to consider the cut-integral 
\beq
\label{111}
I_{\alpha\beta\dot{\alpha}\dot{\beta}} \equiv \int\!\! d^4\ell_1  d^4\ell_3 \,
\delta^{(+)}(\ell_1^2) \,  
\delta^{(+)}(\ell_3^2) \,  
\delta^{(+)} \left((L-\ell_1)^2\right) \,  
\delta^{(+)} \left((L+\ell_3)^2\right) \,  
\ \cdot \ N_{\alpha\beta\dot{\alpha}\dot{\beta}}\ . 
\eeq
It depends on only one scale $L$, hence it has the form 
\begin{align}
\label{eq:PV-2-fermions}
\,I_{\alpha\beta\dot{\alpha}\dot{\beta}}\,&=\,A\,L^2\epsilon_{\alpha\beta} \epsilon_{\dot{\alpha} \dot{\beta}}\,+\,B\,(L_{\alpha\dot{\alpha}}L_{\beta\dot{\beta}} + L_{\alpha\dot{\beta}}L_{\beta\dot{\alpha}})\ .
\end{align}
Contracting \eqref{111} and \eqref{eq:PV-2-fermions}
with $\epsilon^{\alpha\beta}\epsilon^{\dot{\alpha} \dot{\beta}}$ and $(\bar{L}^{\dot\alpha\alpha}\bar{L}^{\dot{\beta}\beta} + \bar{L}^{\dot{\beta}\alpha}\bar{L}^{\dot{\alpha}\beta})$ we can solve for the coefficients $A$ and $B$. 
The result for the corresponding UV-divergent parts is (as computed in \eqref{eq:two-fermions-one-loop})
\beq
\label{eq:coeffs-4-fermions}
A_{\rm UV}\, =\, 0\ , \qquad  B_{\rm UV}\, =\, 1/6\ . 
\eeq
At this point we lift the four cut propagators to full propagators, so
that the cut double bubble becomes a full double-bubble integral. The
conclusion is then that the UV-divergent part of the integral
representing the two-fermion $\to$ two-fermion process is a
double bubble with coefficient
\begin{align}
\frac{1}{6}\,(L_{\alpha\dot{\alpha}}L_{\beta\dot{\beta}} + L_{\alpha\dot{\beta}}L_{\beta\dot{\alpha}})\ . 
\end{align}
This result has to be compared with the planar contractions at  tree level, shown in \fref{fig:tree-level-contraction},
\begin{align}
\label{eq:4-fermion-tree}
\begin{split}
I^{\rm tree}_{\alpha\beta\dot{\alpha}\dot{\beta}}\,&\equiv\,
\int\frac{d^dL_1}{(2\pi)^d}\frac{L_{1\,\alpha\dot{\beta}}(L-L_1)_{\beta\dot{\alpha}}}{L^2_1(L-L_1)^2}\ .
\end{split}
\end{align}
Here $L_1$ and $L_2$ are the momenta of each fermion and $L=L_1+L_2$.
 After a similar PV reduction of the $L_1$ integration in \eqref{eq:4-fermion-tree}, also found explicitly in \eqref{eq:A-UV-tree} and \eqref{eq:B-UV-tree}, we find that $I^{\rm tree}_{\alpha\beta\dot{\alpha}\dot{\beta}}$  is given by a scalar (single) bubble with coefficient
\begin{align}
\label{eq:fermion-tree-bubble-coeff}
{1\over 4} \left[  -  L^2 \, \epsilon_{\alpha\beta} \epsilon_{\dot{\alpha}\dot{\beta}}\, + \, 
{1\over 3} \big( L_{\alpha\dot{\alpha}}L_{\beta\dot{\beta}}+ L_{\beta\dot{\alpha}}L_{\alpha\dot{\beta}}\big)\right]  \ . 
\end{align}
This is the ``identity" or $\left\{\begin{smallmatrix}\alpha\,\beta\\ \cr \alpha\,\beta\end{smallmatrix}\right\}
$. 
The permutation is obtained by swapping $\dot\alpha$ and $\dot\beta$, or $\left\{\begin{smallmatrix}\alpha\,\beta\\ \cr \beta\,\alpha\end{smallmatrix}\right\}
$.  Thus, we can write: 
\beqa
\left\{\begin{smallmatrix}\alpha\,\beta\\ \cr \alpha\,\beta\end{smallmatrix}\right\}
: \quad &&{1\over 4} \left[\, -L^2 \, \epsilon_{\alpha\beta} \epsilon_{\dot{\alpha}\dot{\beta}}\, + \, 
{1\over 3} \big( L_{\alpha\dot{\alpha}}L_{\beta\dot{\beta}}+ L_{\beta\dot{\alpha}}L_{\alpha\dot{\beta}}\big)\right]\, , 
\\ 
\left\{\begin{smallmatrix}\alpha\,\beta\\ \cr \beta\,\alpha\end{smallmatrix}\right\}
: \quad &&{1\over 4} \left[   \, L^2 \, \epsilon_{\alpha\beta} \epsilon_{\dot{\alpha}\dot{\beta}}\, + \, 
{1\over 3} \big( L_{\alpha\dot{\alpha}}L_{\beta\dot{\beta}}+ L_{\beta\dot{\alpha}}L_{\alpha\dot{\beta}}\big)\right]\, .
\eeqa
In this language, the tree-level contraction is represented as  
\beq
\left\{\begin{smallmatrix}\alpha\,\beta\\ \cr \alpha\,\beta\end{smallmatrix}\right\}
\ .
\eeq
Hence, reinstating powers of the 't Hooft coupling,  we obtain  that the term in the spin-chain Hamiltonian corresponding to the two-fermion $\to$ two-fermion process is
\beq
\label{2f}
{\lambda \over 8 \pi^2} \left( \left\{\begin{smallmatrix}\alpha\,\beta\\ \cr \alpha\,\beta\end{smallmatrix}\right\}
 \ + \ \left\{\begin{smallmatrix}\alpha\,\beta\\ \cr \beta\,\alpha\end{smallmatrix}\right\} \right)
  \ , 
\eeq
in agreement with the corresponding terms in \eqref{eq:dilsu23}. In conclusion, putting together the purely scalar result of \sref{sec:unitarity-SO(6)}, \eqref{eq:pss}, as well as the results  \eqref{1f1s} and \eqref{2f} for the two-fermion two-scalar and four-fermion cases, we have confirmed the complete expression \eqref{eq:dilsu23} for the spin-chain Hamiltonian in the $SU(2|3)$ sector.

\chapter{On-shell diagrams: planar and non-planar}
\label{ch:onshelldiagrams}

\section{Introduction and review of planar case}

In this chapter we go back to the study of on-shell scattering amplitudes, but as opposed to the vast majority of the $\N=4$ SYM literature, we focus on non-planar corrections.
Although there has been important progress in the study of non-planar amplitudes in $\mathcal{N}=4$ SYM \cite{Bern:1997nh,Bern:2007hh,Bern:2010tq,Carrasco:2011mn,Bern:2012uc,Badger:2015lda}, they are far less understood than amplitudes in the planar sector. The purpose of the work presented here is to study non-planar on-shell diagrams.

The methods presented in Chapter \ref{ch:Introduction} already suggested that momentum space is not the best way to represent amplitudes if one wants to make use of the large amount of symmetries that underlie them. There are many ways one can study scattering amplitudes which are not in ordinary momentum space. Here we will explore a dual formulation for planar amplitudes proposed in \cite{ArkaniHamed:2009dn,ArkaniHamed:2009vw,Kaplan:2009mh,Mason:2009qx,ArkaniHamed:2009dg}, where all N$^{k-2}$MHV loop leading singularities arise as residues of an integral over the \emph{Grassmannian} $Gr_{k,n}$~---~the space of $k$-dimensional planes in $\mathbb{C}^n$. Underlying this description is the idea that to all loops one only needs to consider on-shell data, and loop integration variables lie inside the Grassmannian space. This idea is made manifest with the concept of on-shell diagrams~---~graphs formed by nodes which are three-particle amplitudes conneted by edges which are on-shell momenta.

On-shell diagrams arise naturally planar $\N=4$ SYM because the all-loop integrand satisfy the all-loop BCFW recursion relation \cite{ArkaniHamed:2010kv}. Each BCFW term can in turn be represented as a planar on-shell diagram. Currently there exists a canonical definition of the planar integrand (see Figure \ref{fig:planar-integrand-canonical}) but there is no well-defined notion of loop integrands for the amplitudes beyond the planar limit due to the lack of canonical variables. Non-planar on-shell diagrams are, however, still worth studying since, to say the least, they provide a description for computing non-planar leading singularities of loop amplitudes. Leading singularities are important information which can be used to construct the full loop amplitudes; it is in fact believed, and supported by many non-trivial examples, that for special theories such as $\mathcal{N}=4$ SYM and $\mathcal{N}=8$ supergravity, the full loop amplitudes can be completely determined by the knowledge of their leading singularities \cite{Cachazo:2008vp}. More ambitiously, one could envision that a Grassmannian formulation of non-planar $\mathcal{N}=4$ SYM exists and, if so, it can perhaps be phrased in terms of non-planar on-shell diagrams. Moreover, as we shall see, on-shell diagrams are the mathematical objects that naturally provide the logarithmic singularities alluded to in \cite{Arkani-Hamed:2014via,Bern:2014kca}.

Before we begin the discussion on the Grassmannian formulation, it is useful to get acquainted with this space.
An element of $Gr_{k,n}$ is the span of $k$ vectors with $n$ complex components each, thus it can be parametrised by organising the components of these vectors as rows of of a $k\times n$ matrix $C$,
\begin{equation}
\label{eq:C}
C=\begin{pmatrix}
c_{11} & c_{12} & \ldots & c_{1n} \\
c_{21} &\ddots & & c_{2n}\\
\vdots & & \ddots & \vdots \\
c_{k1} & c_{k2} & \ldots & c_{kn}
\end{pmatrix}.
\end{equation}
Since any linear combination of the $k$ vectors span the same plane, the parametrisation above should be considered modulo a $GL(k)$ action. The dimension of $Gr_{k,n}$ is thus
$\text{dim}\left(Gr_{k,n}\right)=k(n-k).$

A suitable set of coordinates in $Gr_{k,n}$ are the $SL(k)$ invariant maximal minors of $C$, called \pl coordinates. There are two common notations for these determinants,
\begin{align}
\label{eq:Plucker-coords}
(i_1 i_2 \cdots i_k)\,=\,\Delta_{i_1,i_2,\dots,i_k}\,\equiv\,\left|\begin{array}{cccc}
c_{1i_1} & c_{1i_2} & \ldots & c_{1i_k} \\
c_{2i_1} &\ddots & & c_{2i_k}\\
\vdots & & \ddots & \vdots \\
c_{ki_1} & c_{ki_2} & \ldots & c_{ki_k}
\end{array}\right|\ .
\end{align}
\pl coordinates are not all independent because they satisfy the \emph{\pl relations}:
\begin{align}
\label{eq:Plucker}
\begin{split}
(b_1\cdots b_{k-1} a_1)&(a_2\cdots a_{k+1})-(b_1\cdots b_{k-1} a_2)(a_1\cdots a_{k+1})\\
+\,\dots\,  +\,&(-1)^{k}(b_1\cdots b_{k-1} a_{k+1})(a_1\cdots a_{k})\,=\,0\ .
\end{split}
\end{align} 

Due to the natural ordering of planar amplitudes (see Figure \ref{fig:amp-disk}), the Grassmannian formulation in this case can be simplified to a description in terms of the \emph{positive Grassmannian} $Gr^+_{k,n}$ which is a subspace of $Gr_{k,n}$. Here positivity means that $Gr^+_{k,n}$ can be parametrised by matrices $C$ that admit a parametrisation of its entries such that all ordered minors $\Delta_{i_1,i_2,\dots,i_k},\,i_1<i_2<\dots<i_k$ are positive\footnote{Strictly speaking, we are considering the \emph{totally non-negative Grassmannian} as the \pl coordinates can also be zero.}. From a mathematical point of view, the positive Grassmannian was extensively studied by Postnikov in \cite{2006math09764P} and appears in other Physical contexts apart from scattering amplitudes. In the Grassmannian formulation, the study of the singularities of the $S$-matrix~---~which is the fundamental guiding principle of the $S$-matrix theory~---~ boils down to the study of cells and boundaries of $Gr^+_{k,n}$. A cell in $Gr^+_{k,n}$ is characterised by which minors are non-zero and which are zero. On-shell diagrams provide a bridge between cells in $Gr^+_{k,n}$ and terms that enter the BCFW expansion of amplitudes, as will become clear later on.

To construct the Grassmannian formulation of scattering amplitudes, we start by encoding the external kinematic data $\{\lambda_{\alpha}^i,\tl_{\dot\alpha}^i,\eta^{iA}\}$ for the $i=1,\dots, n$ scattering particles as columns of the following matrices:
\begin{equation}
\Lambda=\begin{pmatrix}
\lambda^1_1 & \lambda^2_1 & \ldots & \lambda^n_1 \\
\lambda^1_2 & \lambda^2_2 & \ldots & \lambda^n_2
\end{pmatrix},\quad \widetilde{\Lambda}=\begin{pmatrix}
\tl^1_1 & \tl^2_1 & \ldots & \tl^n_1 \\
\tl^1_2 & \tl^2_2 & \ldots & \tl^n_2 
\end{pmatrix},\quad \eta=\begin{pmatrix}
\eta^{11} & \eta^{21} & \ldots & \eta^{n1} \\
\vdots  & \vdots & \ddots & \vdots \\
\eta^{14} & \eta^{24} & \ldots & \eta^{n4} 
\end{pmatrix}\ .
\end{equation}
Both $\Lambda$ and $\widetilde\Lambda$ span a bosonic two-plane in $\mathbb{C}^n$ while $\eta$ spans a fermionic four-plane in $\mathbb{C}^n$. Notice that The action of $SL(2)$ on $\Lambda,\,\widetilde{\Lambda}$ are Lorentz transformation and $SL(4)$ on $\eta$ is an $R$-symmetry transformation. As a consequence of momentum (super-momentum) conservation, the planes $\Lambda$ and $\widetilde{\Lambda}$ ($\Lambda$ and $\eta$) are orthogonal, that is,
\begin{align}
\sum\limits_{i=1}^n\lambda_\alpha^i\tl^i_{\dot\alpha}\,=\,0\quad\Rightarrow \quad\Lambda\cdot\widetilde{\Lambda}^T\,=\,\boldsymbol{0}_{2\times 2}\, ,\qquad
\sum\limits_{i=1}^n\lambda_\alpha^i\eta^{iA}\,=\,0\quad\Rightarrow \quad\Lambda\cdot\eta^T\,=\,\boldsymbol{0}_{2\times 4}\ .
\end{align}
The geometrical idea behind the Grassmannian formulation is to integrate over $k$-planes in $\mathbb{C}^n$, denoted by $C$, which satisfy
\begin{itemize}
\item The two-plane $\widetilde{\Lambda}$ and the four-plane $\eta$ are orthogonal to $C$, that is $C\cdot \widetilde{\Lambda}^T=0,\, C\cdot \eta=0$.
\item The two-plane $\Lambda$ is contained in $C$ or, in other words, $\Lambda$ is orthogonal to the orthogonal complement of $C$ which is an $(n-k)\times n$ matrix $C^\perp$. Thus we impose $C^{\perp}\cdot\Lambda^T=0$.
\end{itemize}
Using this, N$^{k-2}$MHV leading singularities with $n$ external states in planar $\N=4$ SYM is given by the following contour integral \cite{ArkaniHamed:2009dn},
\begin{equation}
\label{eq:planarG}
\mathcal{L}_{k,n}=\int\limits_{\Gamma_{k,n}} \frac{d^{k\times n} C}{\text{Vol}(GL(k))}\, \frac{\delta^{(2k)}\big(C\cdot \widetilde{\Lambda}^T\big)\,\delta^{(2(n-k))}\left(C^\perp\cdot\Lambda^T\right)\,\delta^{(0|4k)}\left(C\cdot \eta^T\right)}{(1\cdots k)(2\cdots k+1)\cdots(n\cdots k-1)}\ ,
\end{equation}
where $\Gamma_{k,n}$ stands for the integration contour, namely a prescription for which particular combination of $k\times k$ consecutive minors of the matrix $C$ must be set to zero in order to compute the residues. 
It is interesting to notice that the number of bosonic constraints in \eqref{eq:planarG} is always $2n-4$ (the $-4$ corresponds to the momentum conservation delta-functions which can always be factored out of the integral). For the MHV case, this is precisely the number of degrees of freedom of $Gr_{2,n}$ and thus there is no need for a contour. In fact, one can always choose two columns of $C$ to coincide to the $\Lambda$ plane. This gives rise to another Grassmannian formula in terms of momentum twistors\footnote{Momentum twistor variables, introduced in \cite{Hodges:2009hk}, are intrinsically associated to the region momentum variables $x_i$ defined in \eqref{eq:dual-variables}, which are in turn well defined only for planar amplitudes. Given that this chapter is devoted to non-planar amplitudes, it is more convenient to study the Grassmannian formula in terms of spinor-helicity variables.} where N$^{k-2}$MHV amplitudes are a residue over an integral over $Gr_{k-2,n}$ instead of $Gr_{k,n}$, with a prefactor corresponding to an MHV superamplitude \cite{Mason:2009qx}.

An interesting observation is that for $k=0,1$ there are more constraints than integration variables in \eqref{eq:planarG}, indeed it is not possible that a zero- or one-plane contains the two-plane $\Lambda$ (analogously for $k=n-1,n$, $C^\perp$ is a zero- or one-plane and cannot contain $\widetilde{\Lambda}$), hence it is immediate to see that these amplitudes are zero. A special case is when $n=3$. For the three-particle special kinematics already studied in \eqref{eq:3-pt-constraints}, momentum conservation admits a non-trivial solution for $k=1$ and $k=2$ when all momenta are collinear. For the MHV case, all $[ij]=0$ and thus $\tl^1\propto\tl^2\propto\tl^3$, so the $\widetilde{\Lambda}$ plane is actually a line. The same is true for the $\MHVb$ case, except now all $\b{ij}=0$ and $\lambda^1\propto\lambda^2\propto\lambda^3$, so $\Lambda$ is a line. These amplitudes are precisely the building blocks of the \emph{on-shell diagrams} which are bicoloured graphs whose nodes are MHV (black) and $\MHVb$ (white) superamplitudes and whose edges are on-shell momenta. These are shown in Table \ref{tab:BW}.
\begin{table}[h]
\centering
\begin{tabular}{cl}
$\black$ & $\quad\qquad\qquad\,=\, \dfrac{\delta^{(4)}\left(\lambda^a \tl^a+\lambda^b \tl^b +\lambda^c \tl^c \right) \delta^{(0|8)}\left(\lambda^a \eta^a+\lambda^b \eta^b +\lambda^c \eta^c\right)}{\la ab\ra \la bc\ra\la ca\ra} 
\ $\\[5pt]
$\white$ & $\quad\qquad\qquad\,=\,\dfrac{\delta^{(4)}\left(\lambda^a \tl^a+\lambda^b \tl^b +\lambda^c \tl^c \right) \delta^{(0|4)}\left(\eta^a[bc]+\eta^b[ca]+\eta^c[ab]\right)}{[ab][bc][ca]} $
\end{tabular}
\caption{\it Fundamental nodes of on-shell diagrams are MHV and $\MHVb$ trivalent superamplitudes.}
\label{tab:BW}
\end{table}

\begin{figure}[h]
\centering
\includegraphics[width=0.7\linewidth]{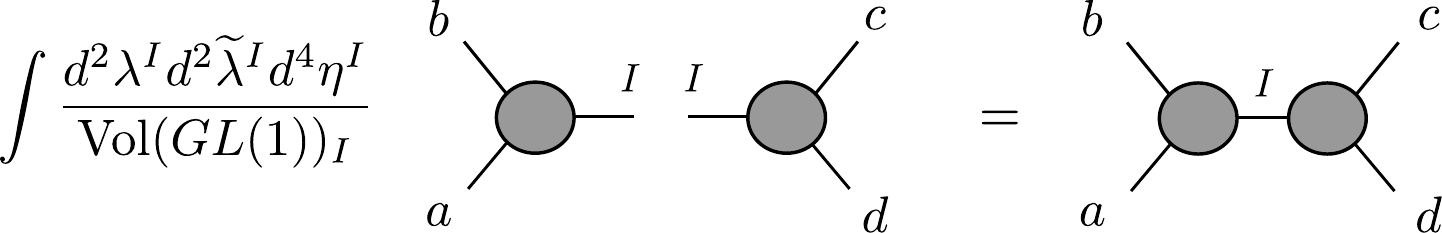}
\caption{\it Fundamental nodes can be merged upon integration over the one particle on-shell phase space of the shared edge. Grey nodes in the figure can be either black or white.}
\label{fig:glue}
\end{figure}

A generic on-shell diagram is obtained by gluing the fundamental nodes via the integration of the one-particle Lorentz invariant phase space of the particle with on-shell momentum $\lambda^I\tl^I$ shared by the two nodes (this is called the Nair measure \cite{Nair:1988bq}),
\begin{equation}
\label{eq:1-particle-LIPS}
\frac{d^2\lambda^I d^2\widetilde{\lambda}^I d^4 \eta^I}{\text{Vol}(GL(1))_I}=\left(\la\lambda^I d\lambda^I \ra d^2\tl^I-[\tl^Id\tl^I]d^2\lambda^I\right)d^4\eta^I \ ,
\end{equation}
where the integration over $\eta^I$ amounts to summing over all possible helicites of the intermediate particle. This is shown in \fref{fig:glue}.

\begin{figure}[h]
\centering
\includegraphics[width=0.35\linewidth]{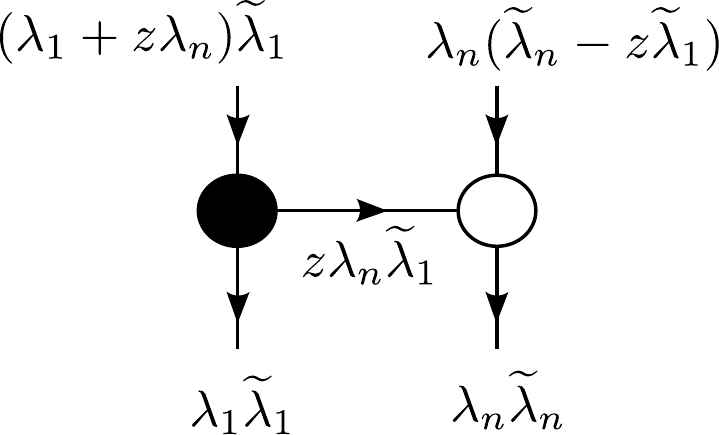}
\caption{\it BCFW-bridge. The fact that the $\widetilde{\lambda}$ ($\lambda$) variables are proportional around a black (white) node fixes the momentum on the bridge to be proportional to $\lambda_n\widetilde{\lambda}_1$. Momentum conservation implements a BCFW shift \eqref{eq:BCFW_shift}.}
\label{fig:BCFW-bridge}
\end{figure}

By gluing nodes one can generate arbitrary on-shell diagrams. Although from a mathematical point of view they are interesting objects in their own right (as will hopefully become clear soon), for planar $\N=4$ SYM a very precise combination of planar on-shell diagrams corresponds to tree amplitudes and loop integrands; that given by the all-loop BCFW recursion relation \cite{ArkaniHamed:2010kv}. The BCFW shift \eqref{eq:BCFW_shift} in the on-shell diagram language is simply the structure shown in \fref{fig:BCFW-bridge}.

\begin{figure}[h]
\centering
\includegraphics[scale=.8]{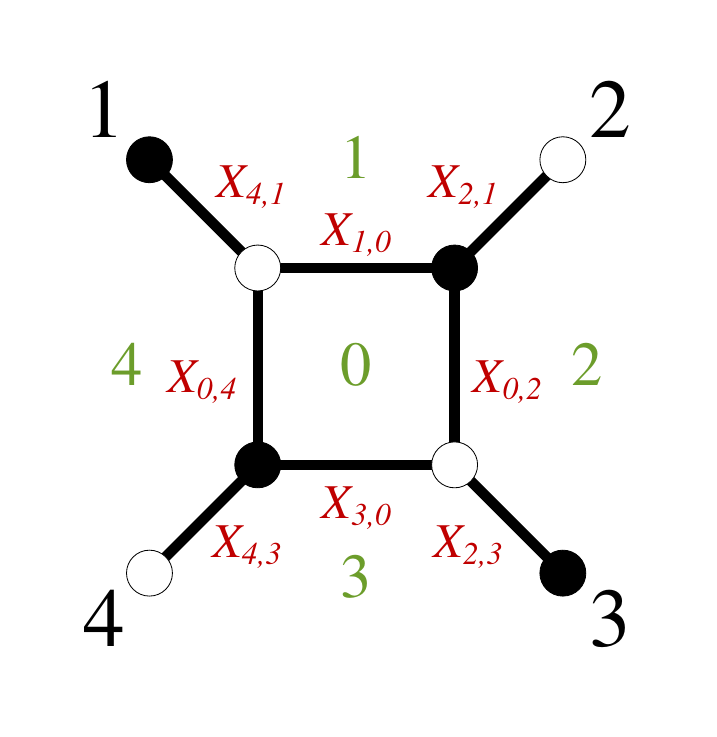}
\caption{\it On-shell diagram for the tree-level four-point MHV amplitude $A^{\text{MHV}}_4$. The number of degrees of freedom is $d=4$. Faces are labeled in green, external nodes in black and edge weights in red.}
\label{fig:squarenoarrows}
\end{figure}

\noindent One can then deduce that the four point MHV tree level amplitude is simply given by a box shown in \fref{fig:squarenoarrows}.

As vertices are glued together, the Grassmannians $Gr_{1,3}$ and $Gr_{2,3}$ associated to the nodes give rise to a larger Grassmannian $Gr_{k,n}$, where $k=2n_B+n_W-n_I$ for a trivalent diagram with $n$ external edges, $n_B$ black nodes, $n_W$ white nodes and $n_I$ internal edges. 


The number of degrees of freedom $d$ of a general on-shell diagram is obtained by associating weights $X_e$ to each edge and subtracting the $GL(1)$ gauge redundancy associated to every internal node (recall that $\text{dim}(Gr_{1,3})=\text{dim}(Gr_{2,3})=2$). This means that for a diagram with $E$ edges and $V$ internal vertices, we have 
\begin{equation}
\label{eq:d-general}
d\,=\,E-V\ .
\end{equation}
This expression is completely general. For a planar on-shell diagram with $F$ faces, this is equal to 
\begin{equation}
\label{eq:dof-planar}
d_{\rm planar}=F-1\ .
\end{equation}
As a consequence, all edge weights can be expressed in terms of $F-1$ independent ones. A more efficient parametrisation of an on-shell diagram is in terms of face variables $f_i,\,i=1,\dots,F$, which are subject to the constraint $\prod_{i=1}^{F}f_i=1$. They are given by the product of all \emph{oriented edge weights} around a face (closed or open) and, for concreteness, the face boundaries can be taken to be oriented clockwise. In what follows, we will adopt the convention in which oriented edge weights go from white to black nodes. As a result, some edge weights will appear in the numerator or denominator of the previous expressions depending on whether their orientation coincides or opposes that of the corresponding path, respectively. An example of the map between face and edge variables for the planar diagram of  \fref{fig:squarenoarrows} is
\begin{align}
\label{eq:edges-faces}
f_0\,=\,\frac{X_{1,0}X_{3,0}}{X_{0,4}X_{0,2}}\,,\; f_1\,=\,\frac{X_{2,1}X_{4,1}}{X_{1,0}}\,,\; f_2\,=\,\frac{X_{0,2}}{X_{2,3}X_{2,1}}\,,\; f_3\,=\,\frac{X_{4,3}X_{2,3}}{X_{3,0}}\,,\; f_4\,=\,\frac{X_{0,4}}{X_{4,1}X_{4,3}}\ .
\end{align}
Notice that $\prod_{i=0}^4f_i=1$. 

In \sref{sec:generalised-faces}, we will generalise face variables to non-planar diagrams and discuss how the counting of degrees of freedom \eqref{eq:dof-planar} is modified. 

To each on-shell diagram there is an associated differential form in the Grassmannian in terms of the edge weights of the graph,
\begin{align}
\label{eq:on-shell_form}
\begin{split}
&\left(\prod_{\text{int. nodes }V} \dfrac{1}{\text{Vol}(GL(1)_V)}\right) 
\left(\prod_{\text{edges }X_e} \dfrac{dX_e}{X_e}\right)\\
&\times \delta^{(2k)}\big(C(X)\cdot \widetilde{\Lambda}^T\big)\,\delta^{(2(n-k))}\big(C^\perp(X)\cdot\Lambda^T\big)\,\delta^{(0|4k)}\left(C(X)\cdot \eta\right) ,
\end{split}
\end{align}
where the first product is taken over all internal nodes and the entries of the matrix $C(X)\subset Gr_{k,n} $ is computed by studying paths in the graph that connect two external nodes\footnote{In the work presented here, following a standard approach in the combinatorics literature, we chose to include external nodes at the endpoints of legs of on-shell diagrams. We would like to emphasise that we are dealing with ordinary on-shell diagrams and that such external nodes have no physical significance. They can become useful bookkeeping devices when performing certain transformations of the diagram. For this reason we use the terms external nodes, edges or legs interchangeably.}. This matrix is called the \emph{boundary measurement} and will be discussed in more detail in \sref{sec:Bipartite}. We will refer to the form \eqref{eq:on-shell_form} excluding the delta-functions as the {\it on-shell form} $\Omega$ corresponding to a given on-shell diagram. The on-shell form associated to a $d$-dimensional planar on-shell diagram in terms of edge or face variables is of the ``$d\log$'' form \cite{ArkaniHamed:2012nw},
\begin{equation}
\label{eq:edges}
\Omega\,=\,\frac{dX_1}{X_1} \, \frac{dX_2}{X_2} \cdots \frac{dX_d}{X_d}\,=\,\frac{df_1}{f_1} \, \frac{df_2}{f_2} \cdots \frac{df_d}{f_d}.
\end{equation}
Note that the expression in terms of edge weights generalises straightforwardly to the non-planar case, whereas the $d\log$ form the $GL(1)_V$ invariant way needs to be modified.

\begin{figure}[h]
\centering
\includegraphics[width=0.9\linewidth]{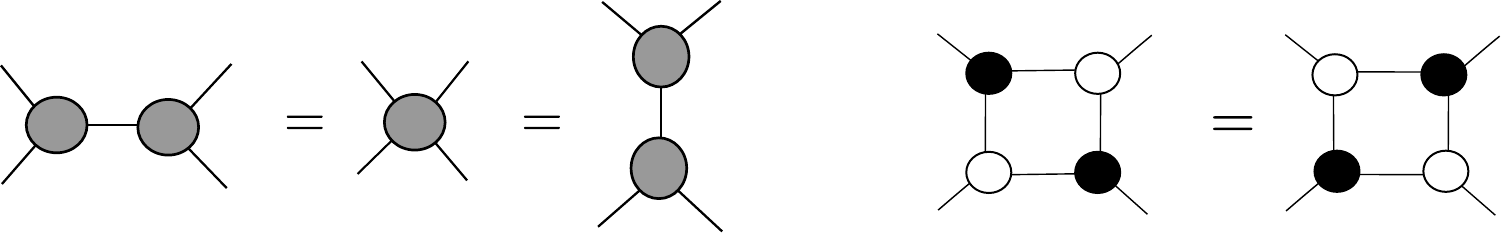}
\caption{\it (Left) Merger move: two connected internal nodes of the same colour are condensed and can also be expanded in a different kinematic channel. (Right) Square move.}
\label{fig:equivalence-moves}
\end{figure}

On-shell diagrams form {\it equivalence classes} and can be connected by {\it reductions}. Equivalent on-shell diagrams parametrise the same region of $Gr_{k,n}$ and are related by a sequence of the {\it equivalence moves} shown in \fref{fig:equivalence-moves}~---~\emph{merger} and \emph{square moves}. In the planar case, to say that two diagrams are equivalent is the same as to say that their boundary measurement $C(X)$ has the same set of non-vanishing minors, but this is no longer the case for non-planar graphs and we will come back to this question in \sref{section:complicatedcase}.

Any on-shell diagram can be made bipartite by using the operations of \fref{fig:equivalence-moves}. In the following we will thus focus almost exclusively on bipartite graphs.\footnote{For this reason, we will use the terms on-shell diagram, diagram, bipartite graph and graph interchangeably.} Mergers can be used in both directions, to either increase or decrease the valency of nodes.

\begin{figure}[h]
\centering
\includegraphics[width=10cm]{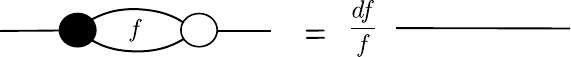}
\caption{\it Bubble reduction.}
\label{fig:bubble-reduction}
\end{figure}

In addition to equivalence moves, there is an operation that reduces the number of faces in the graph~---~the \emph{bubble reduction}, shown in \fref{fig:bubble-reduction}. In terms of the on-shell form, the variable associated to the deleted bubble factors out as a plain $d\log$ (that is, it does not appear in the boundary measurement).

\noindent Bubble reduction reduces the number of degrees of freedom in the diagram by one while preserving the associated region of the Grassmannian. A graph is said to be \emph{reduced} if it is impossible to remove an edge while preserving the cell in $Gr_{k,n}$ it parametrises. For planar diagrams it means that one cannot delete an edge without keeping the same set of non-vanishing \pl coordinates. For non-planar graphs this condition is too restrictive and it is possible to obtain more general reductions, for instance generating identities between \pl coordinates. This question will be revisited in \sref{section:complicatedcase}.


When the dimension of a reduced graph coincides with the dimension of $Gr_{k,n}$, i.e. $d=k(n-k)$, the on-shell form is said to be top-dimensional and \eqref{eq:edges} becomes equivalent to \eqref{eq:planarG} after including the delta-functions and the integral symbol \cite{ArkaniHamed:2012nw}. If the dimension of the graph is larger than the dimension of $Gr_{k,n}$, it indicates that there are some variables which are redundant and the graph may be reduced into a graph of dimension $d \leq k (n-k)$. If the dimension of the graph is smaller than the dimension of $Gr_{k,n}$, \eqref{eq:edges} arises as certain \textit{residue} of \eqref{eq:planarG}; the residue is taken around the vanishing of those minors which disappear once those graphical degrees of freedom have been removed. The special case is when a reduced graph has dimension $d=n-4$ which is the number of bosonic delta-functions in \eqref{eq:on-shell_form}. In this case the value of the integrated on-shell form is a rational function of the kinematics which corresponds to a leading singularity.

The singularity structure of \eqref{eq:on-shell_form} is inherited by amplitudes in planar $\N=4$ SYM, for instance, the MHV loop integrand has only logarithmic singularities and no poles at infinity. In some cases it is possible to find the same $d\log$ structure in momentum space, leading to pure integrals discussed in \sref{sec:Trans-func-symbols}. Recently, building on this observation, it has been conjectured that non-planar amplitudes share the same property \cite{Arkani-Hamed:2014via}. Further evidence supporting this conjecture was provided in \cite{Bern:2014kca} and led to the conjecture of the existence of an amplituhedron-like structure in the non-planar sector too \cite{Bern:2015ple}. This provides another physical motivation to study non-planar on-shell diagrams.

The study of non-planar on-shell diagrams recently began to be explored in \cite{Arkani-Hamed:2014bca},\footnote{See also \cite{Du:2014jwa,Chen:2014ara,Chen:2015qna,Chen:2015bnt} for relevant work.} primarily in the case of MHV leading singularities. This chapter is based on  \cite{Franco:2015rma} where we studied in detail general non-planar on-shell diagrams in $\mathcal{N}=4$ SYM.

Above we reviewed only a few features of planar on-shell diagrams. For a detailed presentation, we refer the reader to the original work \cite{ArkaniHamed:2012nw}. The remaining parts of this chapter are organised as follows. Before studying non-planar on-shell diagrams in full generality, we discuss in \sref{section:non-adjacentBCFW} a concrete scenario in which non-planar on-shell diagrams appear and are relevant, namely the computation of tree-level amplitudes using non-adjacent BCFW shifts. In \sref{sec:Bipartite} we review some concepts which are common in the study of bipartite graphs that will be used in the study of non-planar diagrams. \sref{sec:generalised-faces} introduces canonical variables for non-planar graphs generalising the planar face variables. Among other things, these variables amount for the most efficient way of packaging the degrees of freedom of a graph and automatically make the $d\log$ structure of the on-shell form manifest. We also discuss a systematic procedure for determining these canonical variables, based on the embedding of on-shell diagrams into bordered Riemann surfaces. Physical results are, of course, independent of the choice of embedding. 

On-shell diagrams are mapped into the Grassmannian via the boundary measurement. In \cite{Franco:2015rma} we proposed a boundary measurement for completely general on-shell diagrams. So far, the boundary measurement was only known for graphs admitting a genus-zero embedding \cite{2009arXiv0901.0020G,Franco:2013nwa}. Needless to say, the boundary measurement is an essential ingredient for developing a comprehensive theory of non-planar on-shell diagrams and the associated region of $Gr_{k,n}$. 

While going from an on-shell diagram to the corresponding on-shell form in terms of face variables is straightforward, it is however much more challenging to directly obtain its expression in terms of minors. In \sref{sec:computeform}, we generalise the prescription introduced in \cite{Arkani-Hamed:2014bca} beyond the MHV case, which allows us to directly write the on-shell form of reduced diagrams as a function of minors starting from the graph. This prescription bypasses the need to compute the boundary measurement. As a consistency check, we compare the results of this method with those obtained using the boundary measurement, finding agreement. An interesting new feature of non-planar on-shell diagrams we uncover is the possibility of a new kind of pole in the on-shell form, not given by the vanishing of a \pl coordinate. 

\section{Non-planar on-shell diagrams and non-adjacent BCFW shifts} \label{section:non-adjacentBCFW}

Before embarking into a fully general investigation of non-planar on-shell diagrams in the coming sections, we would like to collect a few thoughts about a concrete scenario in which non-planar on-shell diagrams appear and are important, namely the computation of tree-level amplitudes in $\mathcal{N}=4$ SYM via non-adjacent BCFW shifts \cite{Britto:2005fq}.

\begin{figure}[h]
\centering
\includegraphics[width=14cm]{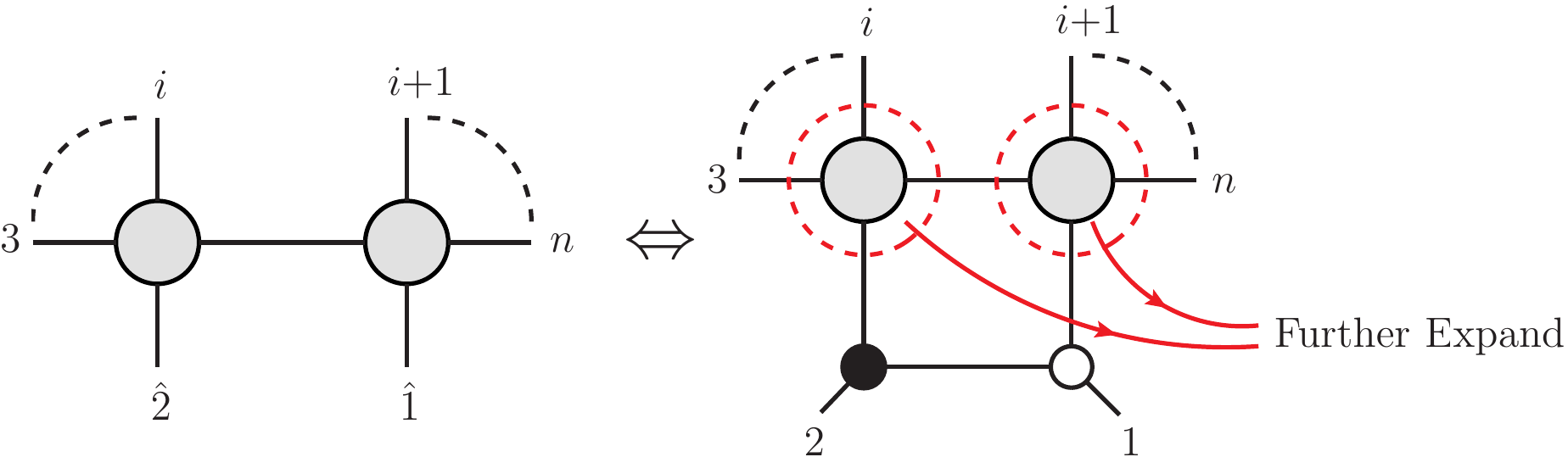}
\caption{\it A one-to-one correspondence between a BCFW diagram with an adjacent shift and a leading singularity of a two-mass-hard box. The tree-level amplitudes in the two massive corners can be further expanded into two-mass-hard boxes until reaching an on-shell diagram representation of the BCFW diagram.}
\label{fig:adjacentBCFW}
\end{figure}

It is a well known fact that there is a one-to-one correspondence between the quadruple cut of a planar two-mass-hard box integral\footnote{A two-mass-hard box integral is a box integral with the two massive momenta entering two adjacent corners, as opposed to the two-mass-easy box integral where the massive corners are diagonally opposite to each other. The two-mass-easy box integral appears in \fref{fig:scalarints}.} and a BCFW diagram with adjacent shifts \cite{Roiban:2004ix}, as shown in \fref{fig:adjacentBCFW}. In fact, this is how the BCFW recursion relations for tree-level amplitudes were originally derived in \cite{Britto:2004ap}. As emphasised in the figure, one can further recursively express the tree-level amplitudes entering the two massive corners of the box in terms of two-mass-hard boxes, obtaining a representation of the BCFW diagram with adjacent BCFW shifts in terms of on-shell diagrams.

\begin{figure}[h]
\centering
\includegraphics[width=13cm]{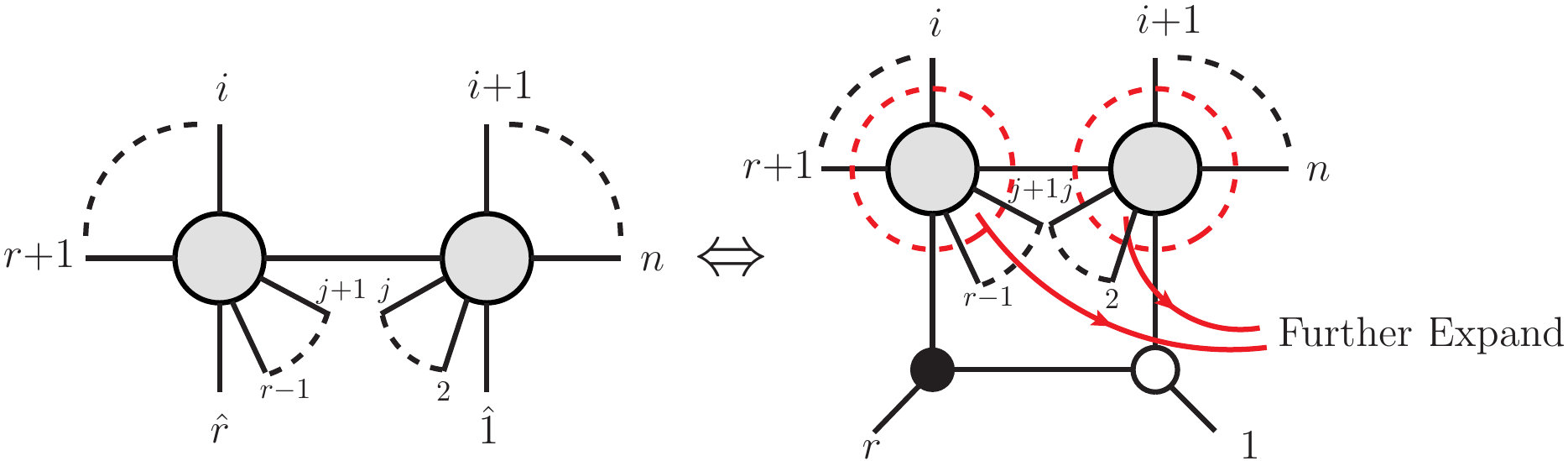}
\caption{\it A one-to-one correspondence between a BCFW diagram with non-adjacent shifts and a non-planar two-mass-hard box. The tree-level amplitudes at the two massive corners can be further expanded into either non-planar or planar two-mass-hard boxes until reaching an on-shell diagram representation of the BCFW diagram.}
\label{fig:nonadjacentBCFW}
\end{figure}

Since tree-level amplitudes can also be expressed in terms of BCFW diagrams with non-adjacent shifts, it is natural to wonder whether there is a corresponding on-shell diagram representation. Indeed, such a representation exists and the resulting objects are precisely non-planar on-shell diagrams. Similarly to what happens for BCFW diagrams with adjacent shifts, there is a one-to-one correspondence between a BCFW diagram with non-adjacent shifts and a non-planar two-mass-hard box, as shown in \fref{fig:nonadjacentBCFW}.\footnote{This type of non-planar diagrams can always be ``planarised" by means of the Kleiss-Kuijf
relations~\cite{Kleiss:1988ne} which are satisfied by the tree-level amplitudes in the two massive corners. This allows one to bring outside all the external legs that are originally inside the loop, giving rise to a planar two-mass box.} Once again, the tree-level amplitudes in the two massive corners can be further expanded into two-mass-hard boxes, either planar or non-planar. Doing this recursively, we can express any BCFW diagram with non-adjacent shifts in terms of non-planar on-shell diagrams.

\begin{figure}[h]
\centering
\includegraphics[width=14cm]{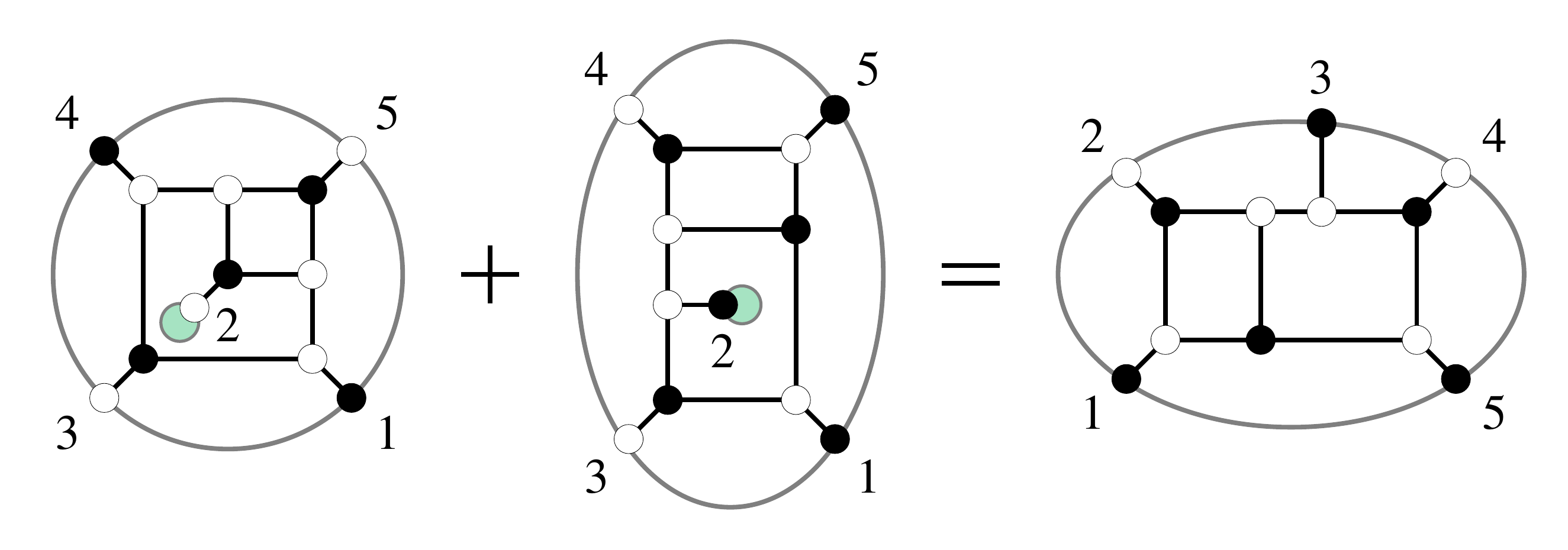}
\caption{\it Two ways to determine the tree-level five-point MHV amplitude from recursion relations. (Left) Non-planar on-shell diagrams obtained by a non-adjacent BCFW-shift on legs 1 and 3. (Right) Planar on-shell diagram obtained by the adjacent BCFW-shift on legs 4 and 5.}
\label{fig:fivept}
\end{figure}

It is possible to represent a given amplitude in terms of different on-shell diagrams obtained via different BCFW shifts. This procedure thus generates interesting identities between on-shell diagrams. We present an example of such an identity in \fref{fig:fivept}, where we provide two alternative expressions for the tree-level five-point MHV amplitude $\mathcal{A}^{\rm MHV}_5$. One of the expressions involves two non-planar diagrams and the other one involves a single planar diagram. Furthermore, it is known that there are additional relations between BCFW diagrams with non-adjacent shifts due to the so-called bonus relations \cite{ArkaniHamed:2008gz,Spradlin:2008bu,Feng:2010my}; it would be interesting to explore their application to non-planar on-shell diagrams. Finally, it would be interesting to investigate how general the construction of non-planar on-shell diagrams in terms of non-adjacent BCFW shifts can be.

\subsection{Bipartite graph technology
 and the boundary measurement}
\label{sec:Bipartite}

The main aim of this section is to explain how to obtain the boundary measurement. To  this end, it is useful to start by discussing a few concepts that are suitable for the analysis of bipartite graphs, both planar and non-planar. 

A {\it perfect matching} $p$ is a subset of the edges in the graph such that every internal node is the endpoint of exactly one edge in $p$ and external nodes belong to one or no edge in $p$. Given a bipartite graph, there exists an efficient procedure for obtaining its perfect matchings based on generalised Kasteleyn matrices, certain adjacency matrices of the graph \cite{Franco:2012mm}. 

\begin{figure}[h]
\centering
\includegraphics[width=0.35\linewidth]{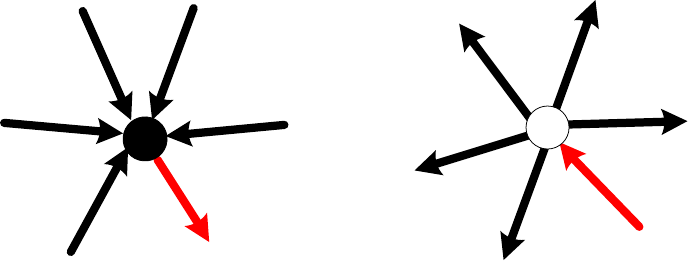}
\caption{\it Perfect orientation: each black node has a single outgoing arrow and each white node has a single incoming arrow.}
\label{fig:perfect-matching}
\end{figure}

The next step is to assign orientations to edges in order to produce a {\it perfect orientation}; an orientation such that each white vertex has a single incoming arrow and each black vertex has a single outgoing arrow, as shown in \fref{fig:perfect-matching}. Perfect orientations are in one-to-one correspondence with perfect matchings: the single edge with a distinctive orientation at each internal node is precisely the corresponding edge contained in the perfect matching \cite{2006math09764P,Franco:2012mm}. 

\begin{figure}[h]
\centering
\includegraphics[width=0.7\linewidth]{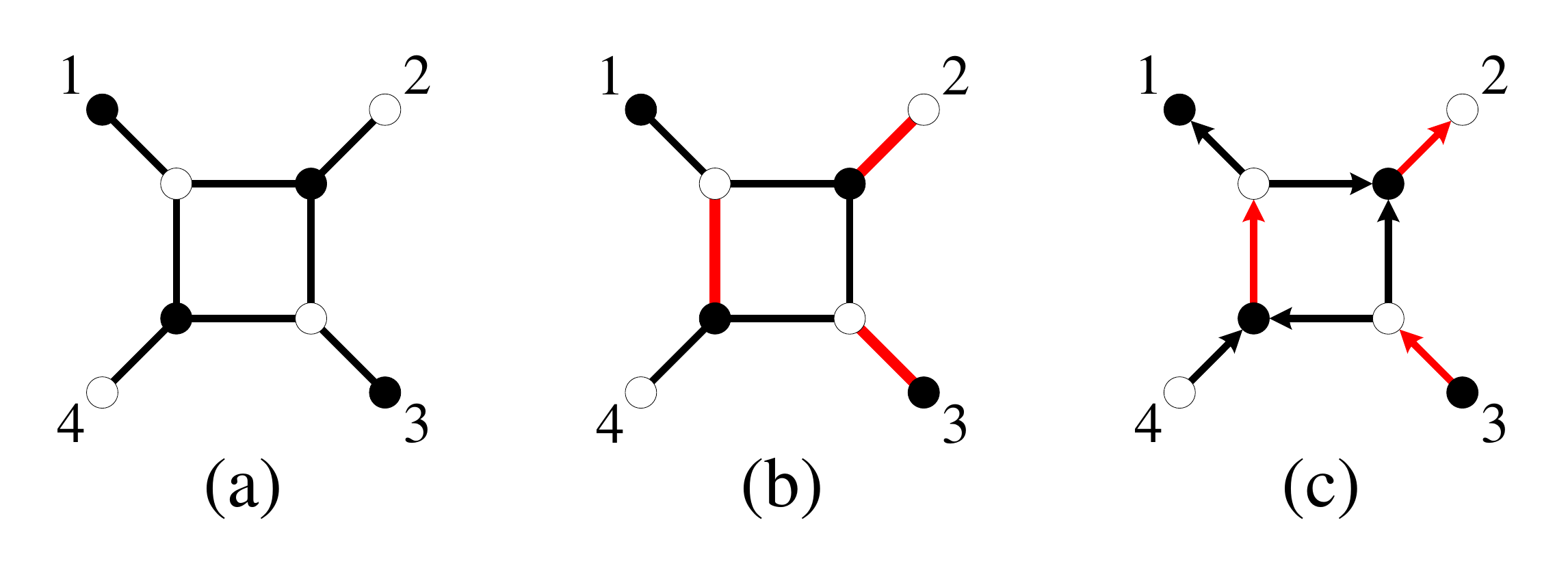}
\caption{\it (a) The graph for $A^{\text{MHV}}_4$, (b) a choice of a possible perfect matching is shown in red and (c) the perfect orientation associated to it. Here 3 and 4 are the sources while 1 and 2 are the sinks.}
\label{pm_and_perfect_orientation}
\end{figure}

Given a perfect orientation, external nodes are divided into sources and sinks, as shown in the example of \fref{pm_and_perfect_orientation}. We will now explain how bipartite graphs parametrise $Gr_{k,n}$. In this map, $k$ is the number of sources and $n$ is the total number of external nodes in any perfect orientation. This provides us with an alternative way for deriving $k$ for general graphs. 

We now have all the necessary ingredients for constructing the {\it boundary measurement}, which maps edge weights of the on-shell diagram to a $k\times n$ matrix $C$ in $Gr_{k,n}$ \cite{2006math09764P}. More rigorously, the boundary measurement is constructed in terms of oriented edge weights. The entries of the matrix $C$ are given by
\begin{equation} \label{eq:genericentries}
C_{ij}(X)=\sum_{\Gamma \in \{i \rightsquigarrow j\}}(-1)^{s_\Gamma}\prod_{e\,\in\, \Gamma}X_e\ ,
\end{equation}
where $i$ runs over the sources, $j$ runs over all external nodes and $\Gamma$ is an oriented path from $i$ to $j$. For two sources $i_1$ and $i_2$, this definition results in $C_{i_1 i_2}=\delta_{i_1 i_2}$. Here $X_e$ indicates oriented edge weights taken along the perfect orientation and $(-1)^{s_\Gamma}$ is a crucial sign depending on the details of each path. This sign prescription is discussed in detail in \cite{Franco:2015rma}, where a generalisation from graphs with genus zero embedding \cite{2009arXiv0901.0020G,Franco:2013nwa} to any graph was presented. This sign prescription ensures positivity in the planar case and allows a classification in terms of matroid polytopes in the non-planar case. In the following we will mostly focus on the on-shell form and recommend \cite{Franco:2015rma} for the details regarding the general boundary measurement together with many additional examples.

\begin{figure}[h]
\centering
\includegraphics[scale=.8]{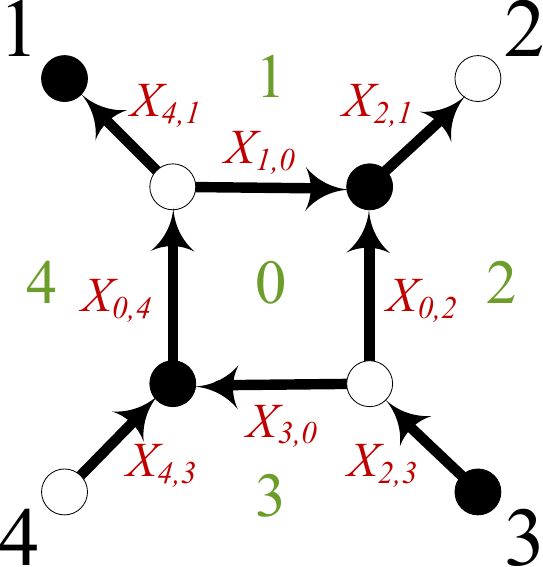}
\caption{\it On-shell diagram for the tree-level four-point MHV amplitude $A^{\text{MHV}}_4$ with a perfect orientation. Particles $1$ and $2$ are sinks while particles $3$ and $4$ are sources.}
\label{G24_edges_and_faces}
\end{figure}

In order to illustrate these ideas, let us consider the simple example shown in \fref{G24_edges_and_faces}, which is the same diagram of \fref{fig:squarenoarrows} endowed with a perfect orientation such that particles $3$ and $4$ are sources and particles $1$ and $2$ are sinks.

\noindent The boundary measurement for this graph is:
\begin{align}
\begin{split}
&C(X)\,=\,\left(\begin{array}{c|cccc} & \ 1 & 2 & 3 & 4 \\ \hline
\Gape[15pt][10pt]
3 \ \ & \dfrac{X_{3,0} X_{4,1}}{X_{2,3} X_{0,4}} & \dfrac{X_{0,2} }{X_{2,3} X_{2,1}} + \dfrac{X_{3,0}X_{1,0}}{X_{2,3} X_{0,4} X_{2,1}} & \ 1 & \ 0 \\ [10pt]
4 \ \ & \ \dfrac{X_{4,3} X_{4,1} }{X_{0,4}} \ & \ \dfrac{X_{4,3} X_{1,0} }{ X_{0,4}  X_{2,1} }  \ & \ 0 & \ 1 \end{array}\right) \\[10pt]
\Rightarrow\qquad & C(f)\,=\,\left(\begin{array}{c|cccc} & \ 1 & 2 & 3 & 4 \\ \hline
3 \ \ & f_0 f_1 f_2  & f_2+f_0 f_2 & \ 1 & \ 0 \\  [.2cm]
4 \ \ & \ f_0 f_1 f_2 f_3 \ & \ f_0 f_2 f_3 \ & \ 0 & \ 1 \end{array}\right)
\end{split}
\end{align}
As explained above, using the $GL(1)$ gauge redundancies associated to the the internal nodes, the boundary measurement in terms of edge variables can be expressed in terms of $d=4$ independent parameters.




\section{Generalised face variables}

\label{sec:generalised-faces}


In this section we begin by introducing canonical variables capturing the degrees of freedom of arbitrary graphs.  Although many of these ideas have already appeared in the literature in various forms \cite{Franco:2012wv,Franco:2013nwa}, their presentation as a set of tools for dealing with non-planar on-shell diagrams is new. 

We term these variables as \emph{generalised face variables} because they have the nice property of being invariant under the $GL(1)$ gauge redundancies associated to all internal nodes.

\subsection{Embedding into a Riemann surface} 
\label{section_embedding}

A useful auxiliary step for identifying generalised face variables is embedding the on-shell diagram into a bordered Riemann surface. While only the connectivity of an on-shell diagram matters, we would like to emphasise that considering such an embedding is very convenient. Given a graph, the choice of embedding is not unique. However we will later see that, as expected, physical results are independent of it.

It is interesting to notice that a choice of embedding is already implicit in the usual discussion of planar diagrams. Indeed, face variables are not an intrinsic property of planar graphs, but arise when imagining them to be embedded on a disk. Similarly, the discussion of zig-zag paths, which are tightly related to the concept of permutations, also depends on assuming planar graphs are embedded on a disk \cite{ArkaniHamed:2012nw}. In fact, as we will see in explicit examples, other embeddings are possible, they lead to different variables, but the final answers remain the same.

In the planar level, graphs embedded on a disk are accompanied by a trace of the gauge group generators following the order of the external legs around the border of the disk, as shown before in \fref{fig:amp-disk}. For non-planar graphs, one can similarly show that gluing the structure constants inherent to the trivalent nodes and using the $U(N)$ completeness relation $\sum_{a=1}^{N^2}(t^a)_{i}^{\phantom{i}j}(t^a)_{k}^{\phantom{k}l}=\delta_{i}^{\phantom{i}l}\delta_{k}^{\phantom{k}j}$, one obtains multitrace contributions, with each trace corresponding to the external legs ending on each boundary of the bordered Riemann surface, in the clockwise direction.

In the coming sections, we will present several explicit examples of graph embeddings and their applications.
\subsection{Canonical variables for non-planar diagrams: generalised faces} 
\label{section_generalised_faces}

Generalising the result for planar graphs, the boundary measurement for generic on-shell diagrams can be constructed in terms of oriented paths in an underlying perfect orientation. Physical answers are independent of the particular choice of perfect orientation as they correspond simply to different $GL(k)$ gauge fixings. It is convenient to describe such paths in terms of a basis, and this can be done by constructing the generalised face variables introduced in this section.  Here we will briefly review the ideas introduced in \cite{Franco:2012wv}. The first step, as discussed in \sref{section_embedding}, is to embed the graph into a bordered Riemann surface. Once this is done, we can associate to the the diagram $F$ faces, $B$ boundaries and a genus $g$. These ingredients are sufficient to construct the basis as follows:

\begin{itemize}
\item {\bf Faces:} A variable $f_i$, $i=1,\ldots, F$, is introduced for every path going clockwise around a face, either internal or external. Face variables satisfy
\begin{equation}
\prod_{i=1}^F f_i= 1\ .
\end{equation}
Hence, one of the face variables can always be expressed in terms of the others. 

\item {\bf Cuts between boundaries:} For $B > 1$, it is necessary to introduce $B-1$ paths, which we call $b_a$, $a=1,\ldots, B-1$, stretching between different boundary components. The particular choice of these $B-1$ paths, i.e.\ how we chose the pairs of boundaries to be connected by them, is unimportant. We will often refer to them as cuts.\footnote{These cuts have nothing to do with the familiar notion of cutting propagators. We hope the reader is not confused by our choice of terminology.}

\item {\bf Fundamental cycles:} For genus $g$ we need to consider $\alpha_m$ and $\beta_m$ pairs of variables, $m=1,\ldots g$, associated to the fundamental cycles in the underlying Riemann surface.

\end{itemize}
The paths $b_a$, $\alpha_\mu$ and $\beta_\mu$ are expressed as products of oriented edge weights in the same way as for $f_i$.\footnote{It is important to note that the definition of these variables, which correspond to oriented paths, {\it  does not} require an underlying perfect orientation. In fact, the orientation of edges in these paths typically does not agree with the one in any perfect orientation.} Furthermore, they are not unique and can be deformed.

These variables contain all of the degrees of freedom $d$ of a general on-shell diagram, which is simply determined by the generalisation of \eqref{eq:dof-planar},
\begin{equation}
d_{\text{general}}\,=\,F+B+2g-2 \ .
\label{d_general}
\end{equation}
There is a simple way of understanding the origin of this expression. Consider an embedding of the diagram with Euler characteristic $\chi$ and such that the diagram gives rise to $F$ faces. Since $\chi=F-E+V$, then the general expression \eqref{eq:d-general} precisely coincides with
\begin{equation}
d=F-\chi \ ,
\end{equation}
which in turn agrees with \eqref{d_general}.

\subsubsection{The $\boldsymbol{d}$log form} 

An important feature of on-shell diagrams is the $d\log$ on-shell form. This form arises automatically when using generalised face variables, without the need for solving for the $GL(1)$ redundancies associated to internal nodes when using edge variables.\footnote{The expression of the on-shell form in terms of edge variables \eqref{eq:edges} remains valid for non-planar diagrams.}
For arbitrary diagrams, the planar $d\log$ form \eqref{eq:on-shell_form} generalises to 
\begin{equation}
\label{general_integrand_face_variables}
\Omega\,=\,\frac{dX_1}{X_1} \, \frac{dX_2}{X_2} \cdots \frac{dX_d}{X_d}\,=\, \prod_{i=1}^{F-1} \dfrac{d f_i}{f_i} \ \prod_{a=1}^{B-1} \dfrac{d b_a}{b_a} \ \prod_{m=1}^{g} \dfrac{d\alpha_m}{\alpha_m} \ \dfrac{d\beta_m}{\beta_m}\ .
\end{equation}
The general form in \eqref{general_integrand_face_variables} is an embedding-independent statement, since ultimately it is only the connectivity of the graph which is of importance. 

Appendix \ref{section_simple_example} illustrates embedding independence in a very simple example: a box diagram embedded on a disk and on an annulus. By flipping an external leg, we lose the internal face but give rise to an additional boundary, which in turn produces a new cut. The independent set of generalised face variables would then go from $\{ f_1, f_2, f_3, f_4 \}$ to  $\{ f_1, f_2, f_3, b_1 \}$. The on-shell form, in both sets of variables, becomes

\begin{equation}
\dfrac{d f_1}{f_1}\dfrac{d f_2}{f_2}\dfrac{d f_3}{f_3}\dfrac{d f_4}{f_4} = \dfrac{d f_1}{f_1}\dfrac{d f_2}{f_2}\dfrac{d f_3}{f_3}\dfrac{d b_1}{b_1} \; .
\end{equation}

\subsection{A genus-one, $B=2$ example}
 \label{genus_1_example_generalised_faces}

\begin{figure}[h]
\centering
\includegraphics[scale=0.65]{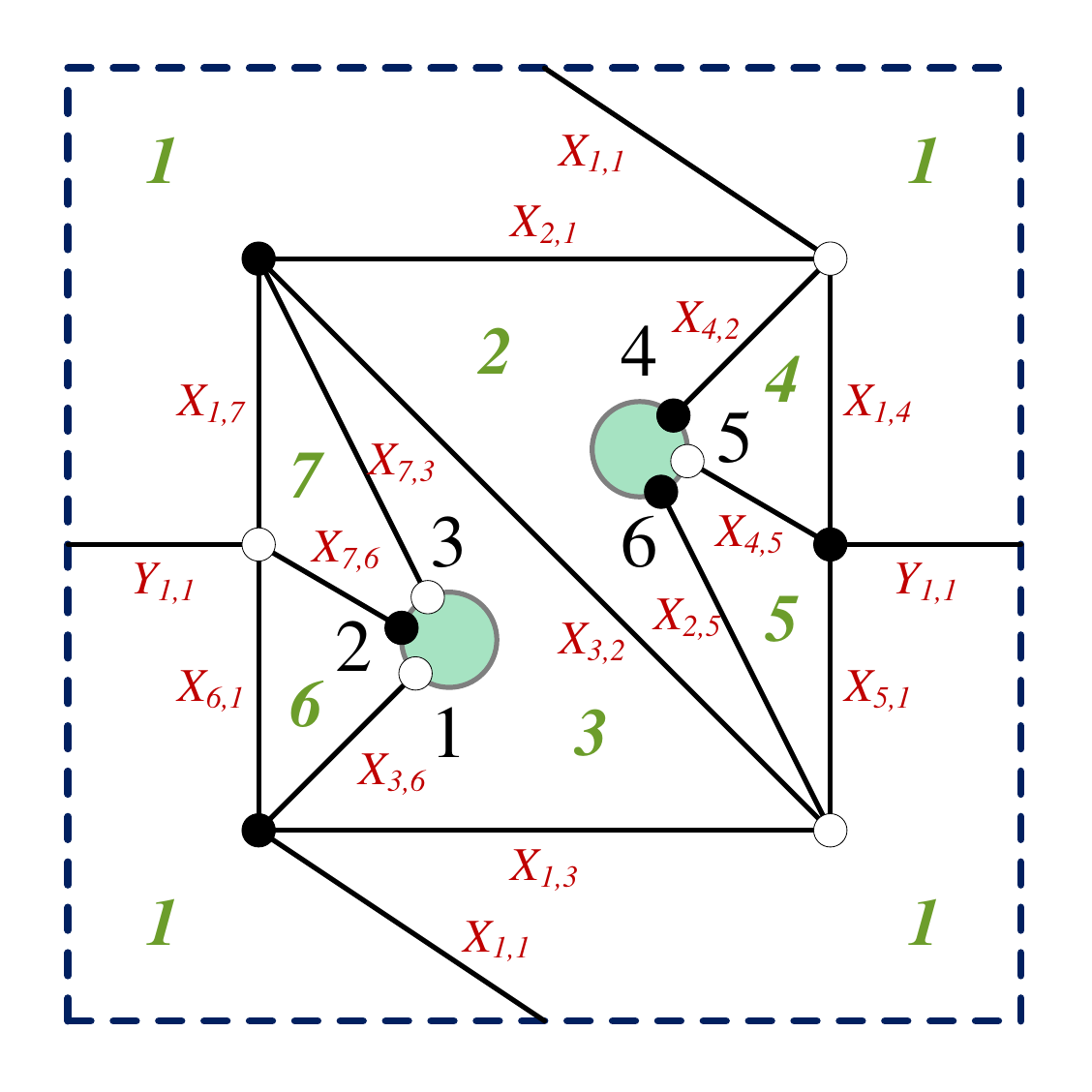}
\caption{\it A reduced on-shell diagram embedded into a torus with two boundaries. This graph cannot be embedded on any surface with $g=0$. Faces are labeled in green, external nodes in black and edge weights in red.}
\label{fig:nonplanarisabletorus1}
\end{figure}

In order to understand how generalised face variables work, it is instructive to study an explicit example. Let us consider the on-shell diagram embedded on a torus with two boundaries shown in \fref{fig:nonplanarisabletorus1}. This diagram does not admit any $g=0$ embedding. Moreover, it is \textit{reduced}, as can be verified using the tools of \cite{Franco:2015rma}.

\begin{figure}[h]
\centering
\includegraphics[width=15cm]{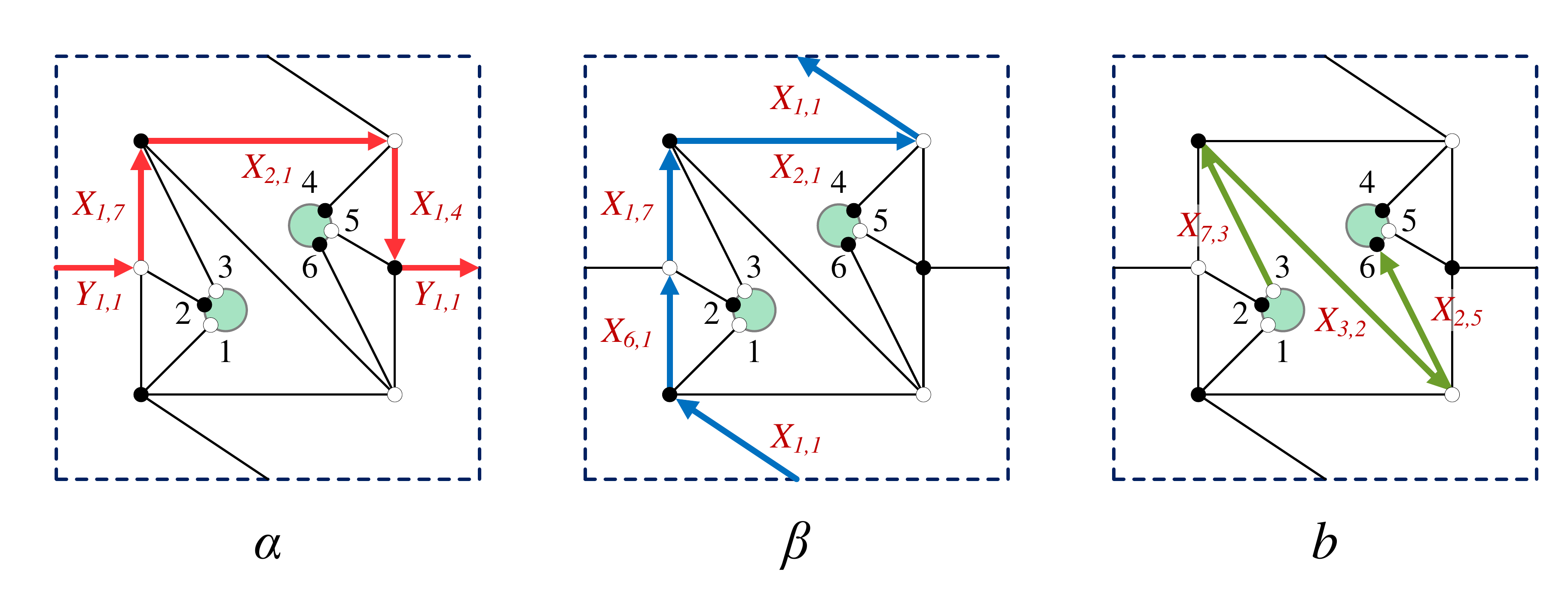}
\caption{\it Possible choices of the $\alpha$, $\beta$ and $b$ variables.}
\label{alphabetab1}
\end{figure}

This diagram is particularly interesting, since it exhibits the two new types of variables we introduced: cuts and fundamental cycles. Since the diagram is embedded into a torus, there is a pair of variables $\alpha$ and $\beta$ corresponding to its fundamental cycles. In addition, there is a cut $b$ connecting the two boundaries. \fref{alphabetab1} shows a possible set of these variables. As we mentioned earlier, the choice of these paths is not unique. In terms of edges, they are given by
 
 \begin{equation}
 \alpha =  \frac{X_{1,7} X_{1,4}}{ Y_{1,1}X_{2,1}} \ \ \ \ \ \ \ \ \beta =  \frac{X_{1,1} X_{1,7} }{ X_{6,1} X_{2,1}} \ \ \ \ \ \ \ \ b =  \frac{X_{7,3} X_{2,5}}{ X_{3,2}} 
 \label{alpha_beta_b_genus_1}
 \end{equation}

 In addition, the ordinary faces are
 \begin{equation}
 \begin{array}{cclccclcccl}
 f_1 & = & \dfrac{X_{2,1} X_{5,1} X_{6,1}}{X_{1,3} X_{1,4} X_{1,7}} & \quad \quad \quad \quad &  f_2 & = & \dfrac{X_{3,2}X_{4,2}}{X_{2,5}X_{2,1}}  & \quad \quad \quad \quad &  f_3 & = & \dfrac{X_{7,3}X_{1,3}}{X_{3,2}X_{3,6}} \\ [.5cm]
f_4 & = & \dfrac{X_{1,4}}{X_{4,2}X_{4,5}}  & & f_5 & = & \dfrac{X_{4,5}X_{2,5}}{X_{5,1}} & &  f_6 & = & \dfrac{X_{3,6}}{X_{6,1}X_{7,6}}  \\ [.5cm]
& & & & f_7 & = & \dfrac{X_{1,7}}{X_{7,6}X_{7,3}} & & & &  
 \end{array}
 \label{faces_genus_1}
 \end{equation}
 The faces satisfy $\prod_{i=1}^7 f_i=1$ so, without loss of generality, we can discard $f_7$. Interestingly, this example also serves to illustrate some non-trivial feature. Face $f_1$ overlaps with itself over two edges, $X_{1,1}$ and $Y_{1,1}$. This implies that when we circle $f_1$ completely in the clockwise orientation, we transverse each of these edges twice, each time in opposite directions. As a result, the contributions of both edges to $f_1$ cancel out.
 
It is possible to gauge fix the $GL(1)$ redundancies of the 6 internal nodes by setting to 1 one edge for each of them. One consistent way of picking these edges corresponds to setting
\begin{equation}
X_{7,6}=X_{3,6}=X_{4,5}=X_{4,2}=X_{1,3}=X_{1,7}=1\ .
\label{gauge_fixing_genus_1}
\end{equation}
The remaining edges are
\begin{equation}
X_{1,1}, \, X_{1,4}, \, X_{2,1}, \, X_{2,5}, \, X_{3,2}, \, X_{5,1}, \, X_{6,1}, \, X_{7,3}, \, Y_{1,1}\  .
\label{free_edges_genus1}
\end{equation}
We thus conclude that this on-shell diagram has $d=9$ degrees of freedom. Following \sref{sec:generalised-faces}, this counting of course agrees with the one based on generalised face variables; we have: 7 faces (6 of which are independent), an $\alpha$ and a $\beta$ cycle from being on a torus and $B-1=1$ cut. 

After this gauge fixing, the independent generalised face variables become
\spacing{1}
\begin{equation}
 \begin{array}{cclccclcccl}
 f_1 & = & \dfrac{X_{2,1} X_{5,1} X_{6,1}}{X_{1,4}} & \quad \quad \quad \quad &  f_2 & = & \dfrac{X_{3,2}}{X_{2,5}X_{2,1}}  & \quad \quad \quad \quad &  f_3 & = & \dfrac{X_{7,3}}{X_{3,2}} \\ [.5cm]
f_4 & = & X_{1,4}  & & f_5 & = & \dfrac{X_{2,5}}{X_{5,1}} & &  f_6 & = & \dfrac{1}{X_{6,1}}
\\ [.5cm]
\alpha & =  & \dfrac{X_{1,4}}{Y_{1,1}X_{2,1}} & & \beta & = & \dfrac{X_{1,1}}{X_{6,1} X_{2,1}} & & b & = & \dfrac{X_{7,3} X_{2,5}}{X_{3,2}} 
 \end{array}
 \label{generalised_faces_gauge_fixed_genus1}
 \end{equation}
 If desired, this map can be inverted, obtaining
 \begin{align}
 \begin{array}{cclccclcccl}
X_{1,1}&=&\dfrac{\beta f_1 f_3 f_4 f_5}{b} & \quad \quad \quad \quad & X_{1,4}&=&f_4& \quad \quad \quad \quad & X_{2,1}&=&\dfrac{f_1 f_3 f_4 f_5 f_6}{b} \\ [.25cm]
X_{2,5}&=&\dfrac{b}{f_3} & & X_{3,2}&=&f_1 f_2 f_4 f_5 f_6  & & X_{5,1}&=&\dfrac{b}{f_3 f_5} \\ [.25cm]
X_{6,1}&=&\dfrac{1}{f_6} & & X_{7,3}&=&f_1 f_2 f_3 f_4 f_5 f_6 & & Y_{1,1}&=&\dfrac{b}{\alpha f_1 f_3 f_5 f_6}
 \end{array}\nonumber\\[5pt]
\end{align}
\spacing{1.5}
Let us now translate the boundary measurement from the edge variables in \eqref{free_edges_genus1} to generalised face variables. It becomes
\begin{equation}
\begin{array}{ccl}
\Omega &= &\dfrac{d X_{1,1}}{X_{1,1}} \dfrac{d X_{1,4}}{X_{1,4}} \dfrac{d X_{2,1}}{X_{2,1}} \dfrac{d X_{2,5}}{X_{2,5}} \dfrac{d X_{3,2}}{X_{3,2}} \dfrac{d X_{5,1}}{X_{5,1}} \dfrac{d X_{6,1}}{X_{6,1}} \dfrac{d X_{7,3}}{X_{7,3}} \dfrac{d Y_{1,1}}{Y_{1,1}} \\ [.5cm]
 &= &\dfrac{f_1^2 f_2 f_4^4 f_5}{\alpha^2 f_3} \times \dfrac{\alpha}{b \beta f_1^3 f_2^2 f_4^5 f_5^2 f_6} \times d f_1\; d f_2\; d f_3\; d f_4\; d f_5\; d f_6\; d \alpha\; d \beta\; d b  \\ [.5cm]
 &= &\dfrac{d f_1}{f_1} \, \dfrac{d f_2}{f_2} \, \dfrac{d f_3}{f_3} \, \dfrac{d f_4}{f_4} \, \dfrac{d f_5}{f_5} \, \dfrac{d f_6}{f_6} \, \dfrac{d \alpha}{\alpha} \, \dfrac{d \beta}{\beta} \, \dfrac{d b}{b}
\end{array}
\end{equation}
where, in the middle line, the first factor comes from the Jacobian of the variable transformation and the second factor comes from the product of edge variables. We see that the on-shell form takes the general form in \eqref{general_integrand_face_variables}. In other words, generalised variables can be used to directly write the on-shell form in a $d\log$ form without having to work through the $GL(1)$ gauge fixing that is necessary for arriving at \eqref{free_edges_genus1}.

\begin{figure}[h]
\centering
\includegraphics[width=5cm]{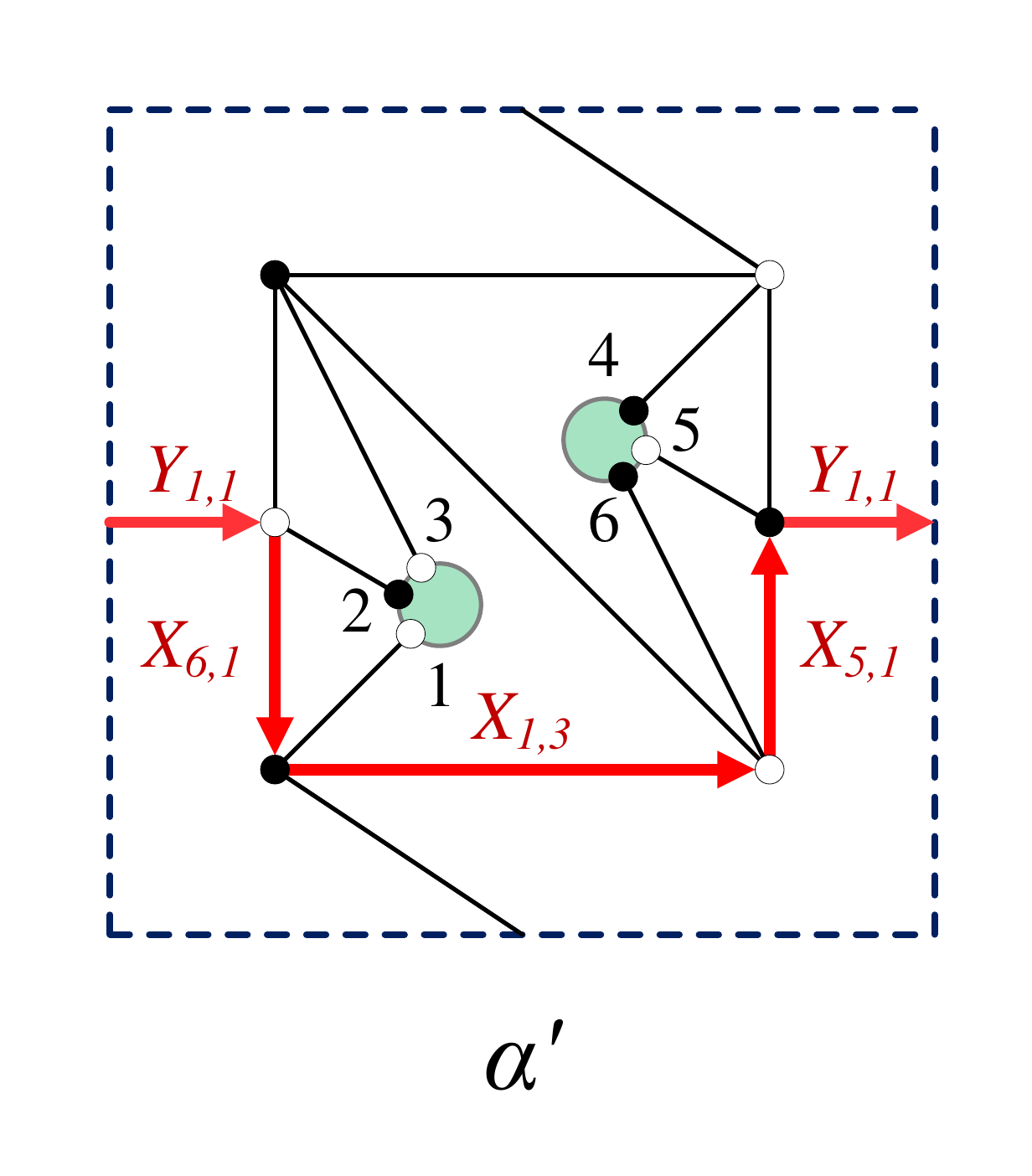}
\caption{\it An alternative choice for one of the fundamental cycles of the torus. The Jacobian of the change of variables is such that the on-shell form preserves its $d\log$ in terms of generalised face variables.}
\label{alphaprime}
\end{figure}

It is also easy to verify that the $d\log$ form of the on-shell form is independent of the explicit choice of generalised face variables. For example, we could trade $\alpha$ for another path $\alpha'$ also wrapping the torus along the horizontal direction, such as the one shown in \fref{alphaprime}. Once again, the Jacobian of the change of variables is such that the $d\log$ form is preserved.

\section{The non-planar on-shell form} \label{sec:computeform}

We shall now study the differential form associated to each non-planar on-shell diagram. As we have already seen in \sref{sec:generalised-faces} there are multiple ways of expressing it:
\begin{enumerate}
\item Using edge variables as in \eqref{eq:edges}, which straightforwardly extends to non-planar graphs. This has the advantage of manifestly displaying the $d\log$ form of the on-shell form and being independent of embedding. A slight disadvantage is that it depends on the choice of $GL(1)$ gauge at every internal node, which needs to be taken into account to identify $d$ independent edges.
\item Using generalised face variables as in \eqref{general_integrand_face_variables}. This approach has the advantage of both displaying the $d\log$ form as well as being independent of the choice of $GL(1)_V$. The determination of generalised face variables naturally involves an embedding of the diagram.
\item In terms of minors of $C$ as in the planar integral \eqref{eq:planarG}, which is only possible for reduced graphs. For generic diagrams it takes the form
\begin{equation}
\label{eq:F}
\Omega\,=\, \frac{d^{k\times n} C}{\text{Vol}(GL(k))}\,\frac{1}{(1 \cdots k)(2 \cdots k+1) \cdots (n \cdots k-1)} \times \mathcal{F}\ ,
\end{equation}
where the non-trivial factor $\mathcal{F}$ accounts for the non-planarity of the on-shell diagram. While this representation hides the $d\log$ form and has a $GL(k)$ redundancy, it has the advantage having a more direct connection to the geometry of $Gr_{k,n}$, naturally expressed in terms of \pl coordinates.
\end{enumerate}
In this section we will be primarily concerned with the third point. In particular, the on-shell forms obtained in this section correspond to having non-trivial factors $\mathcal{F}$ in \eqref{eq:F}. While the discussion in the previous sections applies to general on-shell diagrams, here we focus on reduced ones. This is physically motivated by being interested in leading singularities, which are represented by reduced diagrams. Formally, it is also required by a dimensionality argument; in order to express the on-shell form in terms of minors, its rank needs to match the number of independent \pl coordinates, implying the diagram must be reduced.

\subsection{From generalised face variables to minors} \label{sec:integrandfromface}

A possible way of obtaining the on-shell form in term of minors of $C$ is to use generalised face variables and the boundary measurement. More explicitly, starting with the form in \eqref{general_integrand_face_variables}, we can use the boundary measurement to obtain the map between \pl coordinates and generalised face variables. Solving for the generalised face variables will then yield the desired expression:
\begin{equation}
\prod_{i=1}^{F-1}\frac{df_i}{f_i}\prod_{j=1}^{B-1}\frac{db_j}{b_j}\prod_{m=1}^g\frac{d\alpha_m}{\alpha_m}\frac{d\beta_m}{\beta_m}\,=\,|\mathcal{J}|\, d^{d} C \prod_{i,j,m}\frac{1}{f_i(\Delta) b_j(\Delta)\alpha_m(\Delta)\beta_m(\Delta)}\ , 
\end{equation}
where $\Delta$ is the relevant set of \pl coordinates, and $\mathcal{J}$ is the Jacobian for the transformation between entries in the Grassmannian and generalised face variables.\footnote{It is possible to do a similar thing starting from the on-shell form in terms of edge weights and using the boundary measurement to connect it to \pl coordinates. The advantage of using generalised face variables is that they automatically produce the starting point \eqref{general_integrand_face_variables}.} 

\begin{figure}[h]
\centering
\includegraphics[scale=0.6]{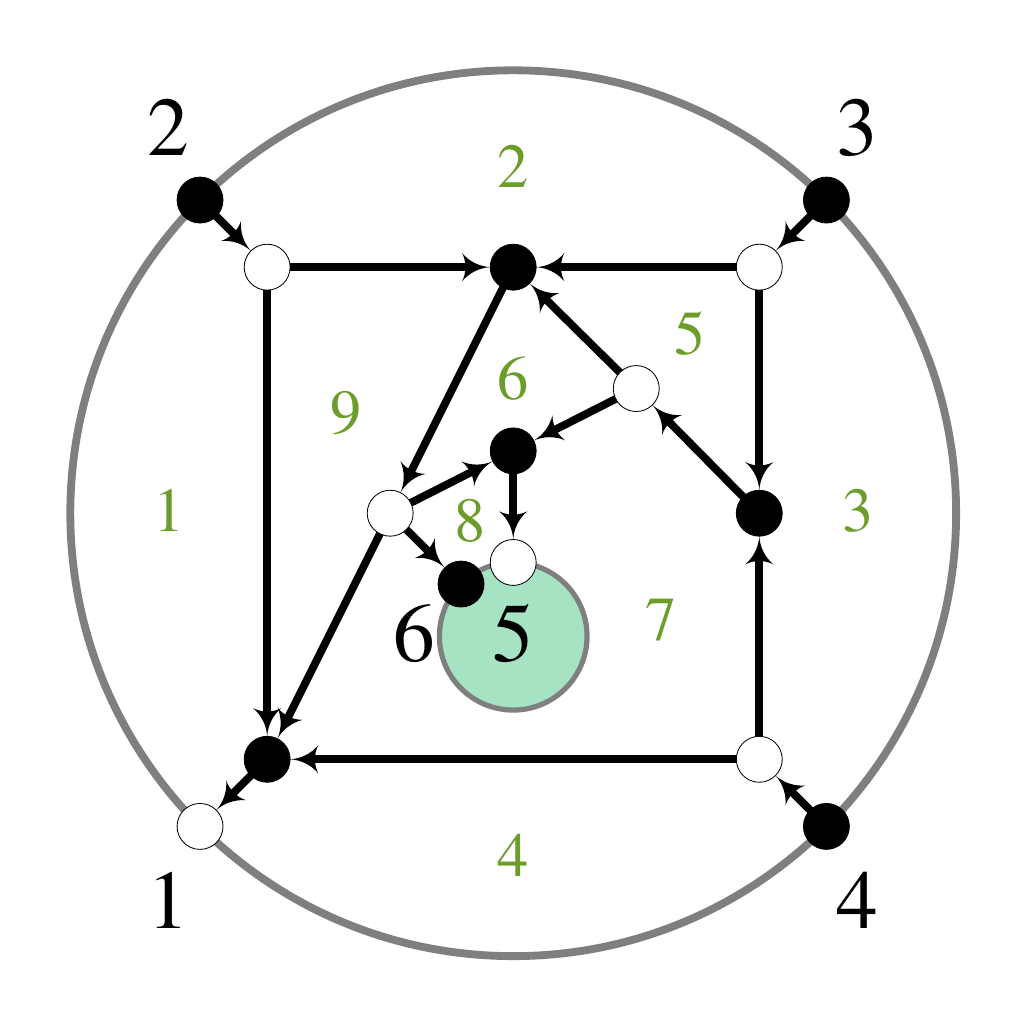}
\vspace{-15pt}\caption{\it A top-dimensional on-shell diagram in $Gr_{3,6}$ embedded on an annulus. The selected perfect orientation has source set $\{2,3,4\}$.}
\label{fig:NMHVex1}
\end{figure}

We shall now illustrate how this works in practice in a top-dimensional example in $Gr_{3,6}$ with two boundaries, shown in \fref{fig:NMHVex1}.

This example has 9 independent generalised face variables: 8 independent $f_i$ variables and one $b_j$. In terms of oriented edge weights, the generalised face variables are given by
\begin{equation}
\begin{array}{rlrlrlrl}
f_1=& \dfrac{X_{9,1}}{X_{1,2}X_{1,4}}\,, &    f_2= & \dfrac{X_{5,2}X_{1,2}}{X_{2,3}X_{2,9}}\,, &   f_3= & \dfrac{X_{7,3}X_{2,3}}{X_{3,4}X_{3,5}}\,, &  f_4= & \dfrac{X_{1,4}X_{3,4}}{X_{4,7}}\,,\\[12pt] 
f_5= & \dfrac{X_{6,5}X_{3,5}}{X_{5,7}X_{5,2}}\,, & 
f_6=&\dfrac{X_{7,6}X_{9,6}}{X_{6,8}X_{6,5}}\,, &    f_7= & \dfrac{Y_{8,7}X_{5,7}X_{4,7}X_{8,7}}{X_{7,6}X_{7,3}X_{7,9}}\,,  & f_8= & \dfrac{X_{6,8}}{X_{8,7}Y_{8,7}}\,, \\[12pt]
& &  f_9=& \dfrac{X_{2,9}X_{7,9}}{X_{9,6}X_{9,1}}\,, &  b_1=& \dfrac{X_{1,4}X_{8,7}}{X_{7,9}}\ .  && 
\end{array}
\end{equation} 
Eliminating $f_4$ using $\prod_{i=1}^9 f_i=1$ we obtain the on-shell form
\begin{equation}
\Omega = \frac{db_1}{b_1} \prod_{i \neq 4}^{9}\frac{df_i}{f_i}\  .
\end{equation}
Using the boundary measurement defined in \cite{Franco:2015rma}, we obtain the following matrix
\spacing{1.2}
{\footnotesize
\begin{equation}
C\,=\, \left(
\begin{array}{c|cccccc}
& 1 & \ \ 2 \ \ & \ \ 3 \ \ & \ \ 4 \ \ & 5 & 6 \\ \hline \Gape[7pt][0pt]
2 \ \ & f_1 (1 + f_9) &  \ \ 1 \ \ &  \ \ 0 \ \ & \ \ 0 \ \ &  b_1 f_1 f_8 f_9 &  b_1 f_1 f_9\\[10pt]
3 \ \ &  -f_1 f_2 (1 + f_5) f_9 &  
  \ \ 0 \ \ & \ \ 1 \ \ & \ \ 0 \ \ &  -b_1 f_1 f_2 (1 + f_5 + f_5 f_6) f_8 f_9 &  -b_1 f_1 f_2 (1 + 
     f_5) f_9\\[10pt]
4 \ \ &  f_1 f_2 f_3 f_5 (1 + f_6 f_7 f_8) f_9 & \ \ 0 \ \ & \ \ 0 \ \ & \ \ 1 \ \  &  
  b_1 f_1 f_2 f_3 f_5 (1 + f_6) f_8 f_9 &  b_1 f_1 f_2 f_3 f_5 f_9
\end{array} \right) .
\label{eq:CforG36}
\end{equation}}
\spacing{1.5}
The variable transformation from generalised face variables to elements of the above matrix, i.e.\ to $ d^9 C$, carries a Jacobian, which can also be expressed in terms of the generalised face variables. 

Using \eqref{eq:CforG36} we can express the \pl coordinates in terms of generalised face variables. Solving for the generalised face variables, we obtain the following differential form:
\begin{equation}
\label{eq:example1}
\Omega\,=\, \prod_{i\neq 4}^9 \frac{df_i}{f_i}\frac{db_1}{b_1}\,=\,d^9 C \dfrac{(246)^2}{(234)(345)(456)(612)(124)(146)(236)(256)}\ .
\end{equation}
An important remark is that the resulting expression in terms of minors is independent of the chosen embedding. The simple example in Appendix \ref{section_simple_example} illustrates this point.

\subsection{A combinatorial method} 

\label{section_combinatorial_on_shell}

In this section we present an alternative systematic procedure for computing the non-planar on-shell form in terms of \pl coordinates for any MHV degree $k$, which allows us to construct it without the need to compute the boundary measurement. This is a generalisation of the method developed in \cite{Arkani-Hamed:2014bca} for general non-planar MHV leading singularities. We begin by quickly reviewing the procedure in \cite{Arkani-Hamed:2014bca}, and then propose its generalisation to any $k$ followed by examples in \sref{sec:example1} and a proof in \sref{sec:proof}. As a consistency check, all results in this section have also been obtained using the method of \sref{sec:integrandfromface} using generalised face variables.

\subsubsection{MHV leading singularities}
\label{sec: MHV LS}

\begin{figure}[h]
\centering
\includegraphics[width=0.25\linewidth]{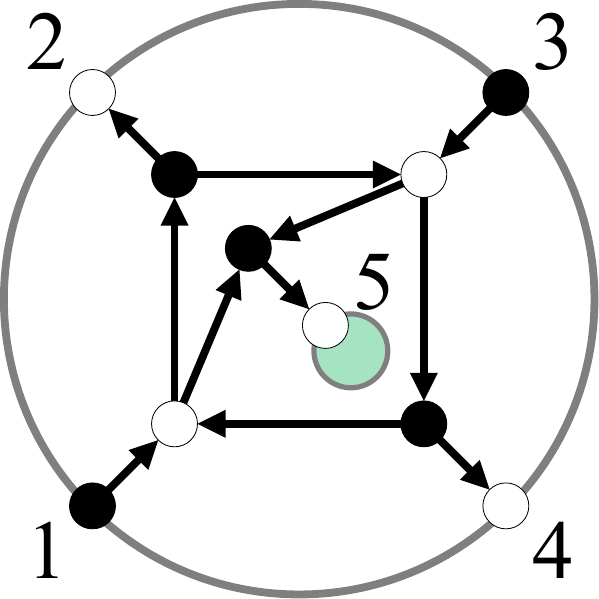}
\caption{\it A five-point MHV on-shell diagram with two boundaries.}
\label{fig:non-planar-MHV}
\end{figure}

A general method for obtaining non-planar MHV leading singularities was recently introduced in \cite{Arkani-Hamed:2014bca}. We now review this method with a simple example, shown in \fref{fig:non-planar-MHV}.

A general feature of MHV leading singularities is that every internal black vertex can be associated to a set of three external legs~---~those that are connected to the black node either directly or through a sequence of edges and internal white nodes. The previous sentence applies to non-necessarily bipartite on-shell diagrams. As explained earlier, every on-shell diagram can be turn into a bipartite one. We will continue focusing on bipartite diagrams, for which it is clear that there can only be at most one internal white node connecting an internal black node to an external leg.

The procedure of \cite{Arkani-Hamed:2014bca} for obtaining MHV non-planar leading singularities is as follows:
\begin{enumerate}
\item For each internal black node, find the three external legs associated to it. Then construct a $n_B \times 3$ matrix $T$, where each row contains the labels of the three external nodes associated to each black node (the order of the rows in $T$ does not matter). For the example in \fref{fig:non-planar-MHV}, $T$ is given by
\begin{align}
T\,=\,\begin{pmatrix}
\ 1 \ & \ 2 \ & \ 3 \ \ \\
\ 1 \ & \ 3 \ & \ 5 \ \ \\
\ 1 \ & \ 3 \ & \ 4 \ \
\end{pmatrix}\ .
\end{align}
\item Next, construct an $n_B \times n$ matrix $M$ in the following manner. For each row $\{i,j,k\}$ in $T$ populate the corresponding row in $M$ by inserting $(i\,j)$ at position $k$, $(j\,k)$ at position $i$, $(k\,i)$ at $j$, and zero for the remaining entries. For our example, we get
\begin{align}
M\,=\,\begin{pmatrix}
\ (23) & (31) & (12) & 0 & 0 \\
\ (35) & 0 & (51) &  0 & (13) \\
\ (34) & 0 & (41) & (13) & 0
\end{pmatrix}\ .
\end{align}
\item Delete two arbitrary columns $a$ and $b$ from the matrix $M$, to obtain the square matrix $\widehat{M}_{a,b}$ of size $n_B \times (n-2) = n_B \times n_B$. Compute next $\det(\widehat{M}_{a,b})/(ab)$. This quantity is independent of the choice of $a$ and $b$ \cite{Arkani-Hamed:2014bca}, as will become clear for any $k$ in \sref{sec:proof}. For the case at hand $\det (\widehat{M}_{a,b}/(ab))=-(13)^2$.
\item Finally, the on-shell form corresponding to a diagram for which
\begin{equation}
\label{eq:Tmatrix}
T=\begin{pmatrix}
i^{(1)}_1 & i^{(1)}_2 & i^{(1)}_{3}\\
i^{(2)}_1 & i^{(2)}_2 & i^{(2)}_{3}\\
\vdots & \vdots & \vdots\\
i^{(n_B)}_1 & i^{(n_B)}_2 &  i^{(n_B)}_{3}
\end{pmatrix}
\end{equation}
is given by
\begin{align}
\label{eq:LS2}
\Omega\,=\,\frac{d^{2\times n}C}{\text{Vol}(GL(2))} \left(\frac{\det(\widehat{M}_{a,b}) }{(a\,b)}\right)^2\frac{1}{\text{PT}^{(1)}\text{PT}^{(2)}\cdots\text{PT}^{(n_B)}} ,
\end{align}
where we denote by $\text{PT}^{(i)}$ the Parke-Taylor-like product corresponding to each row $i$ of $T$; for instance in \eqref{eq:Tmatrix}, $ \text{PT}^{(1)}\, =\, (i^{(1)}_1 i^{(1)}_2) (i^{(1)}_2i^{(1)}_{3})(i^{(1)}_{3} i^{(1)}_{1})$.
For the example in \fref{fig:non-planar-MHV}, the differential form obtained from the above procedure is
\begin{equation}
\Omega\,= \frac{d^{2\times 5}C}{\text{Vol}(GL(2))}\, \frac{ (13)^4 }{ (12)(23)(31)(13)(35)(51) (13)(34)(41) } \ .
\end{equation}
\end{enumerate}

The original rules of \cite{Arkani-Hamed:2014bca} are formulated in terms of spinor brackets $\langle i\,j \rangle$. However, recall that for MHV leading singularities the dimension of the on-shell diagram is $d=2n-4=\text{dim}(Gr_{2,n})$, thus each $\b{i\,j}$ is equivalent to $(i\,j)$ on the support of the kinematic constraints of \eqref{eq:planarG}.  Writing the rules in terms of minors hints at an appropriate generalisation to N$^{k-2}$MHV diagrams, for which the minors are $k \times k$. This is the subject we investigate next.

\subsubsection{Generalisation to N$^{k-2}$MHV on-shell diagrams} \label{sec:rules}

Here we propose a generalisation of the procedure shown above to $k>2$. The Subsequent section \sref{sec:example1} illustrates its inner workings with some non-trivial examples and a proof is presented in \sref{sec:proof}. 

MHV leading singularities only require us to take into account on-shell diagrams with trivalent black vertices, but for $k>2$ we will need to consider more general bipartite graphs. The complications arising when $k>2$ are twofold:

\smallskip
\begin{itemize}
\item In order to have $k \times k$ minors we need a matrix $T$ with $k+1$ columns. For $k>2$ it is possible that some internal black nodes do not connect to $k+1$ external legs in the way described for $k=2$.
\item The number of black nodes may exceed $(n-k)$, forcing $\widehat{M}$ to have more rows than columns, thus preventing us from taking its determinant.
\end{itemize}
\smallskip

\begin{figure}[h]
\centering
\includegraphics[scale=0.5]{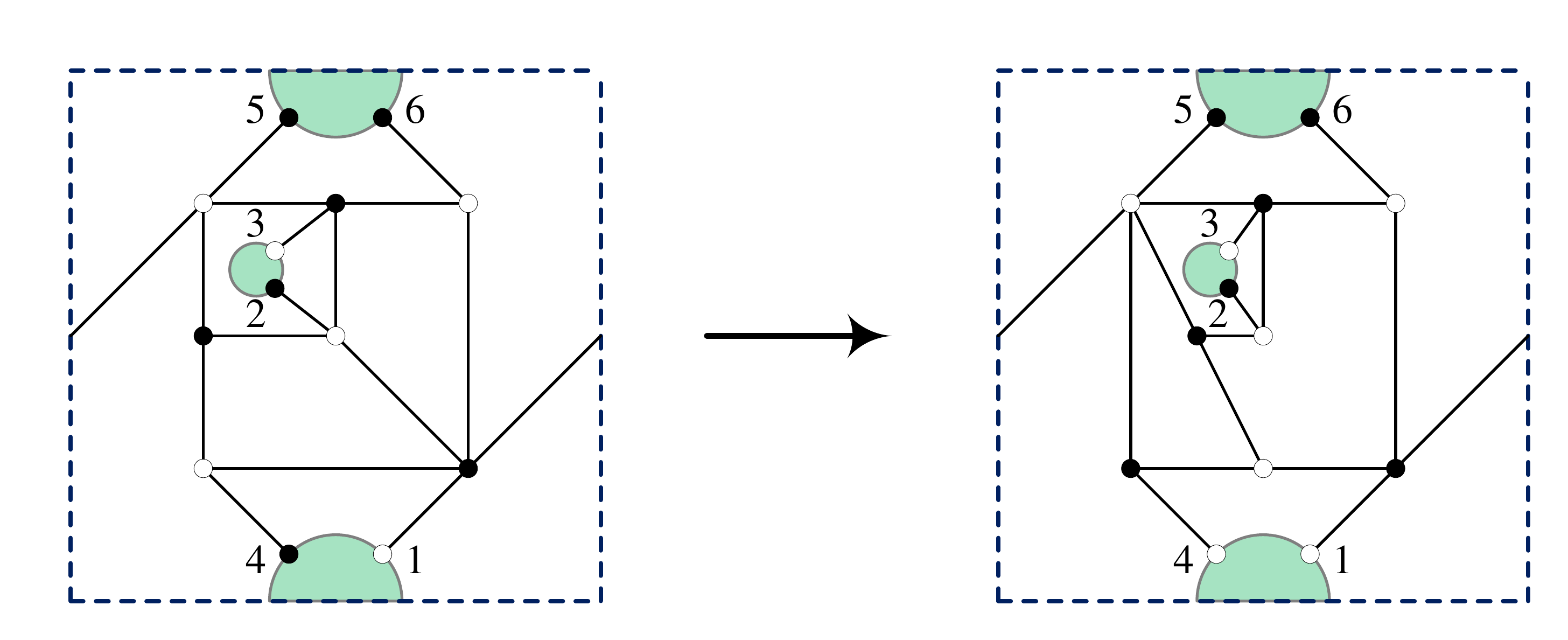}
\caption{\it On the left, an NMHV diagram where the black node attached to external node 1 has valency $v>k+1$. This is resolved by performing a square move, leading to the diagram on the right, where all nodes have $v \leq k+1$.}
\label{fig:largevalency}
\end{figure}

The first point is related to the valency $v$ of internal black nodes. There are two possible reasons why internal black nodes might fail to connect to $k+1$ external ones. The first one is that the valency of the node is $v>k+1$. Generally, performing a square move changes the valency of nodes in a diagram. In what follows we will assume that it is always possible to perform a series of equivalence moves to turn a diagram into one where every black node has $v\leq k+1$. An example of this procedure is given in \fref{fig:largevalency}.

If, on the other hand, the valency of an internal black node is $v<k+1$, we assign the first entries of the corresponding row in $T$ to the external nodes to which the black node connects to, i.e. $\{i_1,\dots, i_v\}$, 
 and leave the remaining $k+1-v$ entries free, which we denote by $\{*_{v+1}, \dots *_{k+1}\}$. The $(k+1)$-tuple associated to the given black node is then (the order of the labels is irrelevant):
\begin{equation}
\label{eq:*}
\{ i_1, \ldots, i_v, *_{v+1}, \dots *_{k+1}\}\ .
\end{equation}
We then fill these additional entries with external labels, chosen arbitrarily from the set of nodes that do not already appear in the row, i.e.\ $*_j \notin \{i_1, \ldots, i_v\}$. The final result is independent of the choice of $*_j$, as will be shown in \sref{sec:star}.


The second complication listed above, regarding the total number of black nodes, typically arises when the diagram has internal white nodes which are completely surrounded by black nodes. Notice that for bipartite graphs, this is always the case, unless when the internal white nodes are directly connected to some external leg. In the examples we have studied, it appears that\footnote{The use of this expression is inspired by Cachazo's talk \cite{freddy_talk}. We stress that $\alpha$ has nothing to do with the generalised face variable of graphs embedded on higher genus surfaces.}
\begin{equation}
\label{eq:alpha}
n_B\,=\,n-k+\alpha\ ,
\end{equation}
where $\alpha$ is the number of such white nodes in the diagram. This issue is resolved by adding an auxiliary external leg to every internal white node contributing to $\alpha$.\footnote{It is interesting to notice that, when thinking in terms of an embedding, this operation can generate new boundary components. In addition, if applied to a reducible graph it can turn it into a reduced one.} 
Once the form has been obtained, through the generalisation of the steps in \sref{sec: MHV LS} which we will outline shortly, we integrate over the extra variables $C_{ij}$, $j=n+1, \ldots, n+\alpha$ around $C_{ij}=0$. We will see this done in detail in several examples.

In summary, the procedure to obtain the differential form for general N$^{k-2}$MHV on-shell diagrams is as follows:
\begin{enumerate}
\item If any internal black node is connected to more than $k+1$ external nodes either directly or through a succession of edges and internal white nodes, perform a series of equivalence moves until all internal black nodes only connect to $k+1$ or fewer external nodes. Also, if $n_B>n-k$, add auxiliary external legs to the internal white nodes which are totally surrounded by internal black nodes, until $n_B=n-k$.
\item Construct the $n_B \times (k+1)$ matrix $T$ where each row corresponds to an internal black node. Every time there is an internal black node that connects to fewer than $k+1$ external nodes, choose the remaining entries freely as described above.
\item Construct the $n_B \times n$ matrix $M$ in the same way as for the MHV case. For each row $\{i_1,\dots, i_j,\dots,i_{k+1}\}$ in $T$ populate the same row in $M$ as follows. At each position $i_j$, insert the minor $(-1)^{j-1}(i_1 \cdots \hat{i}_{j}\cdots i_{k+1})$ obtained by removing $i_j$ and all other entries are zero.
\label{item3-rules}
\item Remove $k$ columns from $M$, chosen arbitrarily, to form $\widehat{M}_{a_1,\dots,a_{k}}$. Then compute the ratio $(-1)^{\sum\limits_{i=1}^{k} a_i}\det(\widehat{M}_{a_1,\dots,a_{k}})/(a_1\cdots a_{k})$.  We emphasise that this quantity is independent of the choice of $\{a_1,\dots, a_{k}\}$; as will be shown in \sref{sec:proof} different choices of ${a_1,\dots,a_{k}}$ simply correspond to different $GL(k)$ gauge choices.


\item The on-shell form corresponding to a diagram for which 
\begin{equation}
\label{eq:Tmatrix2}
T=\begin{pmatrix}
i^{(1)}_1 & i^{(1)}_2 & \cdots & i^{(1)}_{k+1}\\
i^{(2)}_1 & i^{(2)}_2 & \cdots & i^{(2)}_{k+1}\\
\vdots & & & \vdots\\
i^{(n_B)}_1 & i^{(n_B)}_2 & \cdots & i^{(n_B)}_{k+1}\\
\end{pmatrix}
\end{equation}
is given by
\begin{align}
\label{eq:LS}
\Omega\,=\,\frac{d^{k\times n}C}{\text{Vol}(GL(k))} \left(\frac{(-1)^{\sum\limits_{i=1}^k a_i}\det(\widehat{M}_{a_1,\dots,a_{k}}) }{(a_1\cdots a_{k})}\right)^k\frac{1}{\text{PT}^{(1)}\text{PT}^{(2)}\cdots\text{PT}^{(n_B)}}\ ,
\end{align}
where we denote by $\text{PT}^{(i)}$ the Parke-Taylor-like product corresponding to each row $i$ of $T$, for instance in \eqref{eq:Tmatrix2}, $ \text{PT}^{(1)}\, =\, (i^{(1)}_1\cdots i^{(1)}_k) (i^{(1)}_2\cdots i^{(1)}_{k+1}) \cdots (i^{(1)}_{k+1}\cdots i^{(1)}_{k-1})$. If there was no need for introducing auxiliary external legs, i.e. $\alpha=0$ in \eqref{eq:alpha}, this is the final answer.
\item In the presence of auxiliary legs, we now need to integrate over the extra variables $C_{ij}$, $j=n+1, \ldots, n+\alpha$ around $C_{ij}=0$. Below we present various examples in which this is done.
\end{enumerate}

An interesting observation is that for every row in $T$ where we have undetermined entries $\{i_1, \ldots ,i_v,*_{v+1}, \ldots ,*_{k+1} \}$, any minor involving the columns $\{i_1, \dots, i_v\}$ vanishes. This will be proven below in \sref{sec:star}.

\subsection{Examples}
\label{section_method_examples}

We now illustrate the rules introduced in the previous section with a few  explicit examples. Additional examples can be found in Appendices \ref{app:highergenus} and \ref{app:NNMHV}.

\subsubsection{NMHV with low valency} 
\label{sec:example1}

\begin{figure}[h]
\centering
\includegraphics[width=5cm]{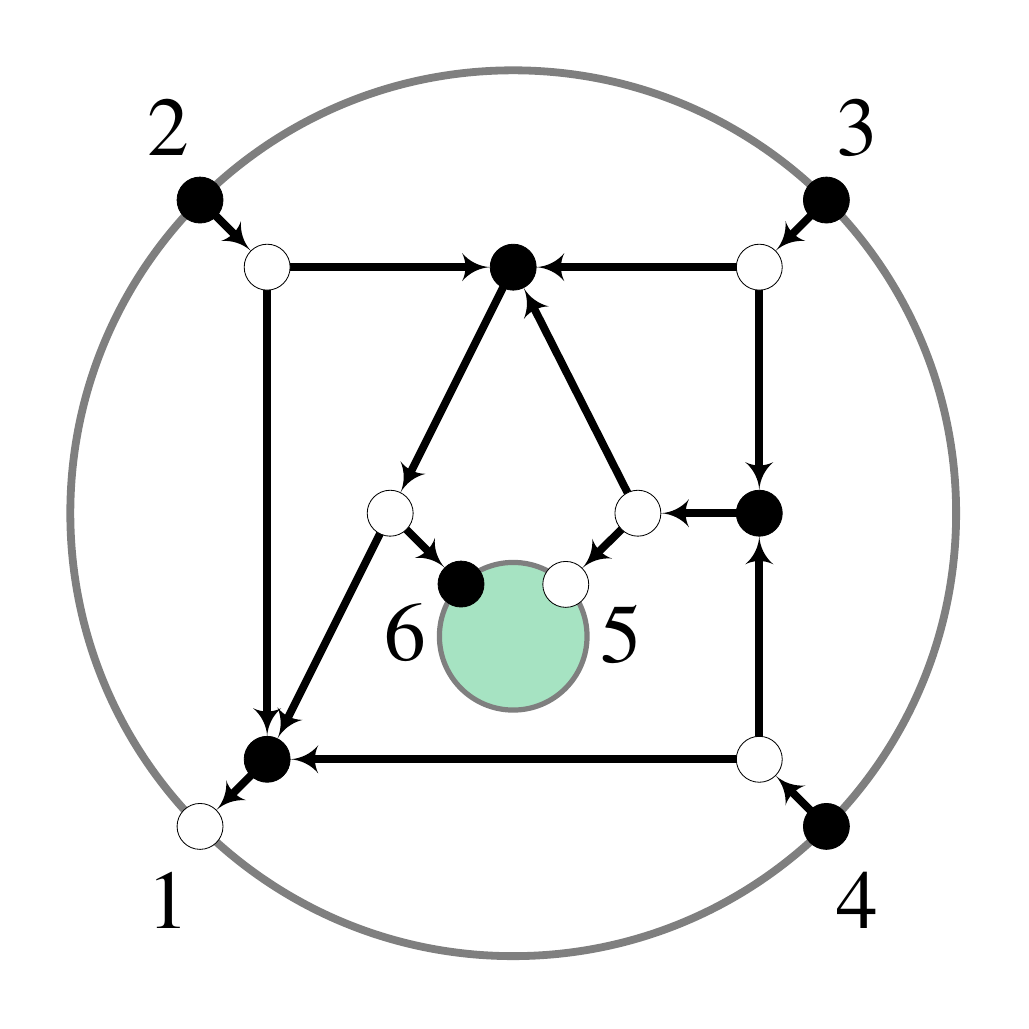}
\caption{\it NMHV leading singularity with $(345)=0$.}
\label{fig:NMHV1}
\end{figure}

The first example illustrates how to deal with cases where a black node has valency $v<k+1$ and as a result we need to introduce $*$ into the matrix $T$. The diagram we study is the NMHV leading singularity shown in \fref{fig:NMHV1}. We will also show that this diagram is decomposable into a sum of Parke-Taylor factors through the use of Kleiss-Kuijf relations \cite{Kleiss:1988ne}, thus independently confirming the answer.

\noindent Since $n_B=n-k$ and all internal black nodes connect to a maximum of $k+1=4$ external nodes, no manipulations of the diagram are required. The $T$ matrix is given by
\begin{align}
\label{eq:T-star}
T\,=\,\begin{pmatrix}
1 & 2 & 6 & 4 \\
2 & 3 & 5 & 6 \\
5 & 3 & 4 & *
\end{pmatrix} ,
\end{align}
where we may choose $*=1,$ $2$ or $6$. The final answer is independent of this choice, and in the following we choose $*=2$. From the bottom row we can also immediately read off that the minor $(345)=0$. We now construct the matrix $M$, 
\spacing{1.2}
\begin{align}
T\,=\,\begin{pmatrix}
1 & 2 & 6 & 4 \\
2 & 3 & 5 & 6 \\
5 & 3 & 4 & 2
\end{pmatrix} \; \rightarrow \;
M\,=\,\begin{pmatrix}
(264) & -(164) & 0 & -(126) & 0 & (124) \\
0 & (356) & -(256) & 0 & (236) & -(235) \\
0 & -(534) & -(542) & (532) & (342) & 0
\end{pmatrix} .
\end{align}
\spacing{1.5}
Deleting columns $2$, $3$, and $4$ we get
\spacing{1.2}
\begin{align}
\widehat{M}_{2,3,4}\,&=\,\begin{pmatrix}
(264) &  0 & (124) \\
0 &  (236) & -(235) \\
0 & (342) & 0
\end{pmatrix} \quad \Rightarrow \quad \frac{\det \widehat{M}_{2,3,4}}{(234)}=-(264)(235) .
\end{align}
\spacing{1.5}
Thus, the on-shell form corresponding to the leading singularity in \fref{fig:NMHV1} is given by
\begin{align}
\label{eq:LS1}
\Omega\,=\,\left. \frac{d^{3\times 6}C}{\text{Vol}(GL(3))} \frac{(264)^2(235)}{(126)(641)(412)(356)(562)(623)(342)(425)(345)}\right|_{(345)=0} .
\end{align}
For this particular example, \eqref{eq:LS1} can be explicitly confirmed to be correct; this leading singularity can be written in terms of planar leading singularities with the help of the Kleiss-Kuijf relations \cite{Kleiss:1988ne} on the four-point nodes present in the diagram in \fref{fig:NMHV1}. Explicitly, using \pl relations \eqref{eq:Plucker} at the pole $(345)=0$ one may rewrite the ratio in \eqref{eq:LS1} as
\begin{align}
\label{eq:L6S1planar}
\begin{split}
 & \left. \frac{(264)^2(235)}{(126)(641)(412)(356)(562)(623)(342)(425)(345)}\right|_{(345)=0} \\
=\, & I(1,6,2,3,5,4) +I(1,6,2,5,3,4) + I(1,2,6,3,5,4)+ I(1,2,6,5,3,4)\ ,
\end{split}
\end{align}
where $I(i_1, i_2, i_3, i_4, i_5, i_6)$ stands for the planar integrand in the Grassmannian formula \eqref{eq:planarG}, with ordering indicated by their arguments:
\begin{equation}
I(i_1, i_2, i_3, i_4, i_5, i_6) = {1 \over (i_1 i_2 i_3)(i_2 i_3 i_4)(i_3 i_4 i_5)(i_4 i_5 i_6)(i_5 i_6 i_1)(i_6 i_1 i_2)}\ .
\end{equation}

It was shown in \cite{Arkani-Hamed:2014bca} that every MHV non-planar leading singularity can be re-expressed as a sum of Parke-Taylor factors with coefficients $+1$. This is not a general feature of N$^{k-2}$MHV leading singularities, as will become clear with the last example of this section.

\subsubsection{NMHV with too many black nodes} \label{sec:example2}

\begin{figure}[h]
\centering
\includegraphics[scale=0.5]{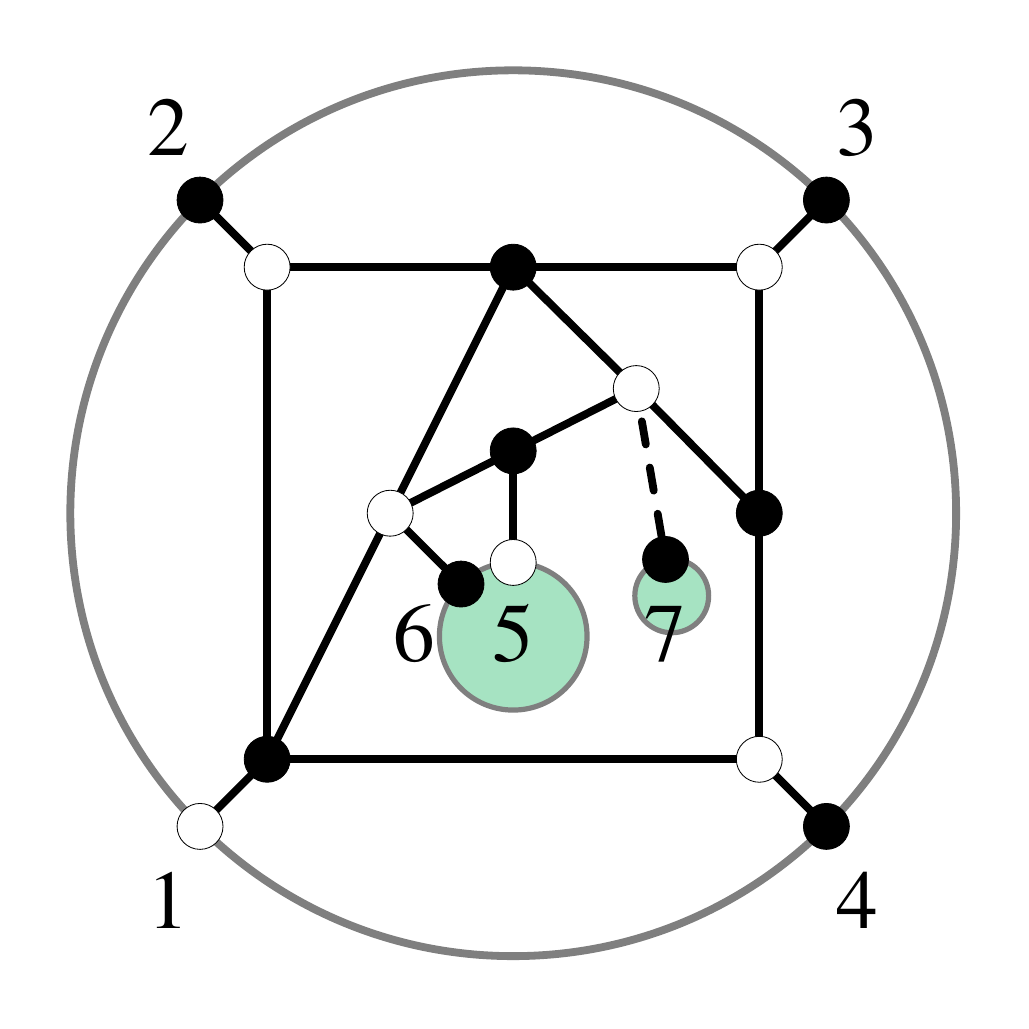}
\caption{\it NMHV leading singularity with $n_B>n-k$. This requires the introduction of an auxiliary leg, indicated by a dashed line and numbered 7.}
\label{fig:NMHV2}
\end{figure}

Let us now consider diagrams with $n_B > n-k$. An example of this type is provided in \fref{fig:NMHV2}, which is obtained by adding a BCFW bridge to legs 5 and 6 in \fref{fig:NMHV1}, thus lifting it to a top-cell. Hence, the two examples must agree on the pole $(345)=0$, which provides us with an additional check of the validity of the procedure in \sref{sec:rules}.

This example has $\alpha=1$ in \eqref{eq:alpha}. Following \sref{sec:rules}, we must introduce an auxiliary leg as shown in \fref{fig:NMHV2}. This new diagram yields the $T$ matrix
\spacing{1.2}
\begin{align}
T\,=\,\begin{pmatrix}
\ 1 \ \ & \ 2 \ \ & \ 6 \ \ & \ 4 \ \ \\
2 & 3 & 7 & 6 \\
7 & 3 & 4 & * \\
5 & 6 & 7 & * \\
\end{pmatrix}\quad\xrightarrow{\text{Choice of }*}\quad T=\begin{pmatrix}
\ 1 \ \ & \ 2 \ \ & \ 6 \ \ & \ 4 \ \ \\
2 & 3 & 7 & 6 \\
7 & 3 & 4 & 2 \\
5 & 6 & 7 & 2 \\
\end{pmatrix}\ .
\end{align}
\spacing{1.5}
\noindent Notice how from the last two rows of $T$ we learn that $(734)=(567)=0$. This gives the following matrix $M$:
\spacing{1.2}
\begin{align}
M\,=\,\begin{pmatrix}
\label{eq:M7}
(264) & -(164) & 0 & -(126) & 0 & (124) & 0 \\
0 & (376) & -(276) & 0 & 0 & -(237) & (236) \\
0 & -(734) & -(742) & (732) & 0 & 0 & (342) \\
0 & -(567) & 0 & 0 & (672) & -(572) & (562) 
\end{pmatrix} ,
\end{align}
\spacing{1.5}
\noindent which results in the on-shell form
\begin{align}
\Omega\, =\,\frac{d^{3\times 7}C}{\text{Vol}(GL(3))} {(264)^2 \over (126)(641)(412)(623)(234)(256)} \times \left.I\right|_7\ ,
\end{align}
where we separated the dependence on the auxiliary external node $7$ on the factor $\left.I\right|_7$. On the poles $(347)=(567)=0$, $\left.I\right|_7$ can be recast as
\begin{align}
\left. I\right|_7\,=\, {(256)  \over (456)(347)(567)(725)}\ .
\end{align}

The final step is to remove the effect of the auxiliary edge. This is done by taking a generic element of the ``extended'' Grassmannian $Gr_{k,n+1}$ and integrating the extra variables $C_{i7}$ around $C_{i7}=0$. To do so, we write a generic $3\times 7$ matrix $C$ and compute the residues of $\left. I\right|_7$ around $C_{i7}=0$ for $i=1,2,3$. We finally obtain
\begin{align}
\Omega\, =\,\frac{d^{3\times 6}C}{\text{Vol}(GL(3))}\dfrac{(246)^2}{(234)(345)(456)(612)(124)(146)(236)(256)}\ .
\end{align}
As expected, this result agrees with the leading singularity \eqref{eq:LS1} on the support of $(345)=0$.

With the two previous examples, we have illustrated the full set of tools required to use the method of \sref{section_combinatorial_on_shell}. As an additional demonstration of the power of this procedure, in Appendix \ref{app:highergenus} we compute the on-shell form of an NMHV graph embedded on a genus-one surface and on Appendix \ref{app:NNMHV} we compute a highly non-trivial N$^2$MHV example that requires the addition of two auxiliary edges.

\subsubsection{NMHV with a new type of poles} \label{section:complicatedcase}

\begin{figure}[h]
\centering
\includegraphics[scale=0.75]{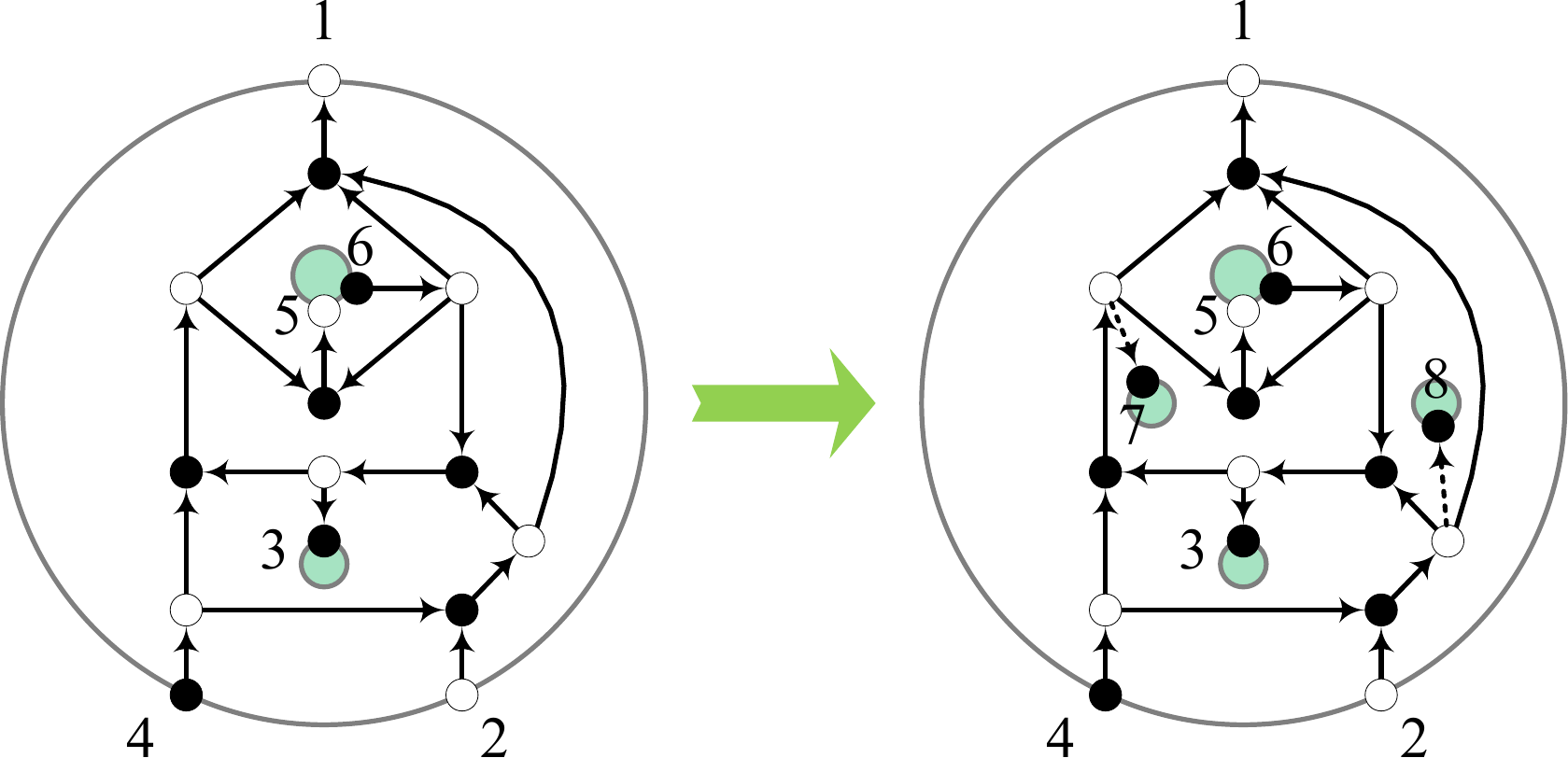}
\caption{\it (Left) An NMHV top-dimensional diagram in $Gr_{3,6}$. (Right) this diagram requires the addition of two auxiliary legs, here shown with dashed arrows and terminating on the external nodes 7 and 8. This example has a non-standard singularity when $(124)(346)(365)-(456)(234)(136)=0$.} 
\label{fig:ProcedureWeirdPole}
\end{figure}

Having learned how to find the on-shell form in terms of minors for arbitrary graphs, we now compute a top-dimensional example in $Gr_{3,6}$ that displays a novel feature: a differential form with a singularity which is not of the form $(ijk)=0$. This fact is intrinsically non-planar and ultimately prohibits the diagram from being able to be written as a sum of planar terms. The on-shell diagram we study is shown in \fref{fig:ProcedureWeirdPole}. 

\noindent The $T$ matrix is
\spacing{1.2}
{\small
\begin{align}
T\,=\,\begin{pmatrix}
\ 1 \ \ & \ 8 \ \ & \ 6 \ \ & \ 7 \ \ \\
5 & 6 & 7 & * \\
6 & 8 & 3 & * \\
8 & 2 & 4 & * \\
7 & 3 & 4 & * \\
\end{pmatrix}\quad\xrightarrow{\text{Choice of }*}\quad T=\begin{pmatrix}
\ 1 \ \ & \ 8 \ \ & \ 6 \ \ & \ 7 \ \ \\
5 & 6 & 7 & 2 \\
6 & 8 & 3 & 2 \\
8 & 2 & 4 & 6 \\
7 & 3 & 4 & 2 \\
\end{pmatrix}\ ,
\end{align}
}
\spacing{1.5}
\noindent from which we can immediately read off that
\begin{align}
\label{eq:poles-7-8}
(347)=(567)=(368)=(248)=0 . 
\end{align}

\noindent From $T$, we construct the matrix $M$
\spacing{1.2}
{\small
\begin{align}
M\,=\,\begin{pmatrix}
\label{eq:M8}
(867) & 0 & 0 & 0 & 0 & (187) & -(186) & -(167) \\
0 & -(567) & 0 & 0 & (672) & -(572) & (562) & 0 \\
0 & -(683) & (682) & 0 & 0 & (832) & 0 & -(632) \\
0 & -(846) & 0 & (826) & 0 & -(824) & 0 & (246) \\
0 & -(734) & -(742) & (732) & 0 & 0 & (342) & 0
\end{pmatrix} .
\end{align}
}
\spacing{1.5}
\noindent The resulting on-shell form can be simplified on the poles \eqref{eq:poles-7-8} to 
\begin{align}
\label{eq:I78}
\Omega\,&=\,\frac{d^{3\times 8}C}{\text{Vol}(GL(3))}  {(346)^2(356) \over (234)(345)(456)(561)(136)(236)} \times \left.I\right|_{7,8}\ ,
\end{align}
where $\left.I\right|_{7,8}$ encodes all the dependence on the extra legs $7$ and $8$,
\begin{equation}
\left.I\right|_{7,8}\, =\, {1 \over (781)(567)(368)(248)(347)}\ .
\end{equation}
As in the previous examples, we now compute the residues of $\left.I\right|_{7,8}$ around $C_{i7}=C_{i8}=0$ for $i=1,2,3$ and obtain
\begin{align}
\left.I\right|_{7,8}\rightarrow {1 \over (124)(346)(365)-(456)(234)(136) }\ .
\end{align}
Thus we find that the on-shell form of the six-point diagram in \fref{fig:ProcedureWeirdPole} is given by
\begin{align}
\Omega\,=\,\frac{d^{3\times 6}C}{\text{Vol}(GL(3))} {(346)^2(356) \over (234)(345)(456)(561)(136)(236) \left( (124)(346)(365)-(456)(234)(136) \right)}\ .
\label{Omega_complicated_case}
\end{align}

\noindent The appearance of the factor
\begin{equation}
\label{eq:strangepole}
(124)(346)(365)-(456)(234)(136)
\end{equation}
in the denominator through this process is rather non-trivial and shows that this diagram, unlike the NMHV leading singularity \eqref{eq:LS1}, cannot be written as a linear combination of planar diagrams. This example thus provides concrete evidence for a behavior already announced in \cite{Arkani-Hamed:2014bca}, that starting from $k=3$ and $n=6$ not all leading singularities can be expressed as linear combinations of planar ones.

This diagram was further studied in \cite{Franco:2015rma} using matching and matroid polytopes. In this perspective, the appearance of this pole becomes clear as it is possible to identify an edge which when removed does not set any \pl coordinates to zero but instead relates \pl coordinates to each other, i.e.\ it imposes the relation $(124)(346)(365)-(456)(234)(136)=0$. The leading singularity that arises through the removal of this edge is also computed in \cite{Franco:2015rma}. 

\begin{figure}[h]
\centering
\includegraphics[width=0.8\linewidth]{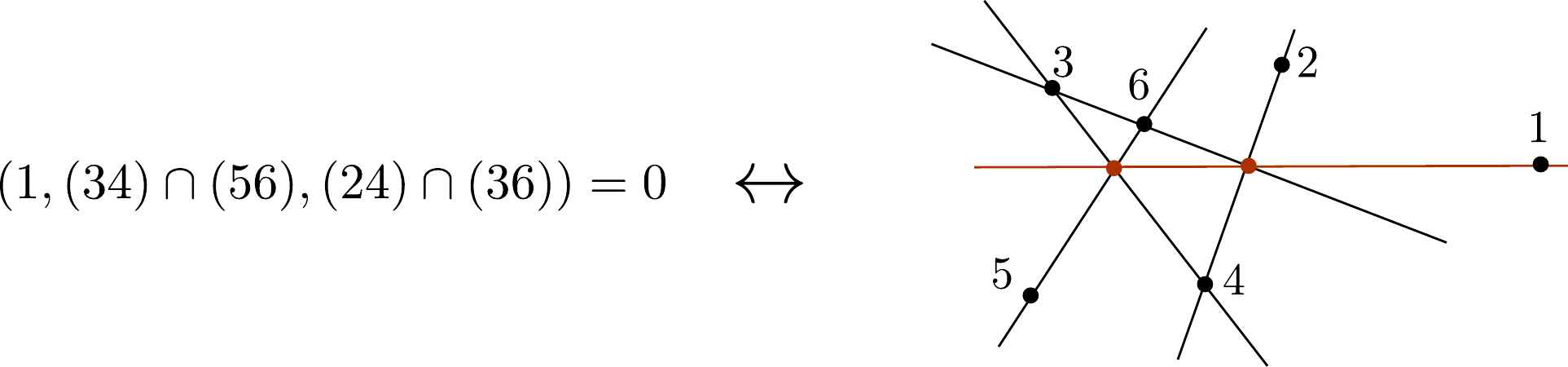}
\caption{\it Configuration of points in $\mathbb{P}^2$ corresponding to the singularity $(124)(346)(365)-(456)(234)(136)=0$.}
\label{fig:strange-pole}
\end{figure}

We now expose the geometry of this singularity. Each column $\vec{c}_i$ of $C$ can be thought of as a point in $\mathbb{P}^2$. A usual pole of the form $(ijk)=0$ means that the three points $\vec{c}_i,\,\vec{c}_j$ and $\vec{c}_k$ are on the same line. In contrast with this simple configuration, denoting by $(ij)$ the line defined by points $\vec{c}_i$ and $\vec{c}_j$, the relation \eqref{eq:strangepole} between minors can be rewritten in a more illuminating way,
\begin{equation}
(124)(346)(365)-(456)(234)(136)\,=\,(1,(34)\cap(56),(24)\cap(36))\ .
\end{equation}
 where $(ij)\cap(kl)$ stands for the point of intersection between the lines $(ij)$ and $(kl)$. The geometrical configuration of points in $\mathbb{P}^2$ is shown in \fref{fig:strange-pole}.

\subsection{Proof of the combinatorial method}
\label{sec:proof}

In this section we present a proof of the method proposed in \sref{sec:rules} for constructing the on-shell form of N$^{k-2}$MHV in terms of \pl coordinates. We consider the class of on-shell diagrams with $n_B=n-k$ and hence without white nodes surrounded by black nodes ($\alpha=0$ in \eqref{eq:alpha}). Since the addition of an auxiliary edge on diagrams for which $n_B>n-k$ leads to a graph with $\alpha=0$ we argue that the proof is valid for these cases as well.

\subsubsection{Top forms with $n_B=n-k$}

Let us consider first on-shell diagrams that are top forms in $Gr_{k,n}$ and with $\alpha=0$, i.e.\ $n_B=n-k$. This means that the matrix $T$ has no arbitrary entries $*_i$, so every black node has valency $k+1$ and is associated to a local Grassmannian $Gr_{k,k+1}$. We denote elements of the Grassmannian $Gr_{k,k+1}$ by $\widetilde{C}$ to distinguish them from the elements of the Grassmannian $Gr_{k,n}$ associated with the complete graph. The proof in this section follows the same logic used for MHV leading singularities in \cite{Arkani-Hamed:2014bca}. In the following we will discuss also the case for which $T$ has arbitrary entries.

We start by studying the contribution of each black node to the on-shell form of the full diagram \eqref{eq:F}. For an internal black node associated with the subset $\{i_1,\dots, i_k\} $ of external particles, the corresponding constraint $\delta^{(2)}(\widetilde{C}^{\perp}\cdot\lambda)$ provides a linear relation satisfied by the set $\{\lambda^{i_1},\dots, \lambda^{i_k}\} $ connected to the node. $Gr_{k,k+1}$ has $k$ degrees of freedom, which can be parametrised by the entries of the $1\times (k+1)$ matrix $\widetilde{C}^{\perp}$ modulo $GL(1)$,
\begin{equation}
\label{eq:c-perp-black}
\widetilde{C}^{\perp}\,=\, \big(\alpha_{i_1} \; \cdots \; \alpha_{i_{k+1}}\big) \ .
\end{equation}
Then, we associate the following form to every internal black node 
\begin{align}
\label{eq:Cperp}
\begin{split}
\{i_1,\dots, i_{k+1}\}\quad \leftrightarrow\;\quad & \frac{1}{\text{Vol}(GL(1))}\prod_{j=1}^{k+1} \frac{d\alpha_{i_j}}{\alpha_{i_j}} \delta^{(2)}\Big(\sum_{j=1}^{k+1} \alpha_{i_j} \lambda^{i_j}\Big) \ .
\end{split}
\end{align}
Recalling that the matrices $\widetilde{C}$ and $\widetilde{C}^{\perp}$ associated to the local $Gr_{k,k+1}$ are complementary matrices, we may equivalently write
\begin{equation}
\label{eq:relation-minors}
\alpha_{i_j}\,=\, (i_j)\Big|_{\widetilde{C}^{\perp}}\, =\, (-1)^{j-1} (i_1\cdots \hat{i}_j\cdots i_{k+1})\Big|_{\widetilde{C}} \ ,
\end{equation}
where $(i_j)\Big|_{\widetilde{C}^{\perp}}$ is a $1\times 1$ minor of $\widetilde{C}^{\perp}$ and $(i_1\cdots \hat{i}_j\cdots i_{k+1})\Big|_{\widetilde{C}}$ is a $k\times k$ minor of $\widetilde{C}$ obtained by deleting the column $i_j$. Using this, \eqref{eq:Cperp} may be recast as
\begin{align}
\label{eq:PTs}
\{i_1,\dots, i_{k+1}\}\; \leftrightarrow\;  \frac{d^{k\times (k+1)}\widetilde{C}}{\text{Vol}(GL(k))} \frac{\delta^{(2)}\Big(\sum_{j=1}^{k+1} (-1)^{j-1}(i_1\cdots \hat{i}_j\cdots i_{k+1}) \lambda_{i_j}\Big)}{(i_1\cdots i_k)(i_2\cdots i_{k+1})\cdots(i_{k+1}\cdots i_{k-1})}\ .
\end{align}
It is clear that the product of $k\times k$ minors in the denominators of the above expression gives rise to the Parke-Taylor-like factors introduced in \eqref{eq:LS}.

The next step is to consider the complete diagram instead of each internal black node separately. We write the matrix $C \in Gr_{k,n}$ as
\begin{equation}
C\,=\,\big(
\vec{c}_1 \; \cdots \; \vec{c}_n \big)\ ,
\end{equation}
where $\vec{c}_i$ are $k$-vectors.
At this point, we recall that the matrix $M$ introduced on item \ref{item3-rules} of \sref{sec:rules} provides a representative of the $(n-k)\times n$ matrix $C^{\perp}$ since
\begin{equation}
\label{eq:Cramer}
\vec{c}_{i_1}(i_2\cdots i_{k+1})- \vec{c}_{i_2}(i_1\cdots i_{k+1})  + \dots + (-1)^{k}\, \vec{c}_{i_{k+1}}\,(i_1\cdots i_{k})\,=\,0\quad \Rightarrow\quad  M\cdot C^{\rm T} \,=\,0\ ,
\end{equation}
where at this point we identified
\begin{equation}
\label{eq:minor-identification}
(i_1\cdots i_k)\Big|_{\widetilde{C}}\,=\, (i_1\cdots i_k)\Big|_{C}\ .
\end{equation}
The next step is to relate $C^{\perp}$ to $M$. In order to do so, we gauge fix the $GL(k)$ redundancy in $C$ by writing each column as a linear combination of $k$ columns $\{\vec{c}_{a_1},\dots,\vec{c}_{a_k}\}$. This fixes columns $a_1,\dots,a_k$ to the identity matrix. Denoting the matrix gauge fixed this way by $C_{a_1,\dots,a_k}^{\rm gf}$, the corresponding constraint $\delta^{(2k)}(C\cdot\widetilde{\lambda})$ acquires a Jacobian factor of $\dfrac{1}{(a_1\cdots a_k)^k}$.  This gauge fixing in $C$ induces a gauge fixing in $C^{\perp}$ for which all columns except $a_1,\dots,a_k$ are gauge fixed to the identity matrix, which we denote by $C_{a_1,\dots,a_k}^{\perp \rm gf}$.
Relating $C_{a_1,\dots,a_k}^{\perp \rm gf}$ to $M$ amounts to multiplying $M$ by $\widehat{M}_{a_1,\dots,a_k}^{-1}$, the inverse of $\widehat{M}_{a_1,\dots,a_k}$ defined in item \ref{item3-rules} of \sref{sec:rules}.
Thus, we finally arrive at the result
\begin{align}
\label{eq:jacobianM}
\begin{split}
\frac{\delta^{(2k)}(C\cdot\widetilde{\lambda})\,\delta^{(2(n-k))}(C^{\perp} \cdot\lambda)}{\text{Vol}(GL(k))}\,=\, &\left(\frac{(-1)^{\sum\limits_{i=1}^k a_i}\det(\widehat{M}_{a_1,\dots,a_k})}{(a_1\cdots a_k)}\right)^k  \\
& \times\,\delta^{(2k)}(C_{a_1\cdots a_k}^{\rm gf}\cdot\widetilde{\lambda})\,\delta^{(2(n-k))}(C_{a_1\cdots a_k}^{\perp \rm gf} \cdot\lambda)\ .
\end{split}
\end{align}
Combining \eqref{eq:jacobianM} with the Parke-Taylor denominators of \eqref{eq:PTs} we obtain precisely \eqref{eq:LS}, upon omitting the delta-functions.

\subsubsection{Diagrams with $*$}
\label{sec:star}
We now discuss diagrams for which one or more black nodes have valency $v<k+1$ and thus the matrix $T$ has undetermined entries. This situation corresponds to the case where the diagram is not a top-dimensional form, as will become clear soon.

\begin{figure}[h]
\centering
\includegraphics[width=0.8\linewidth]{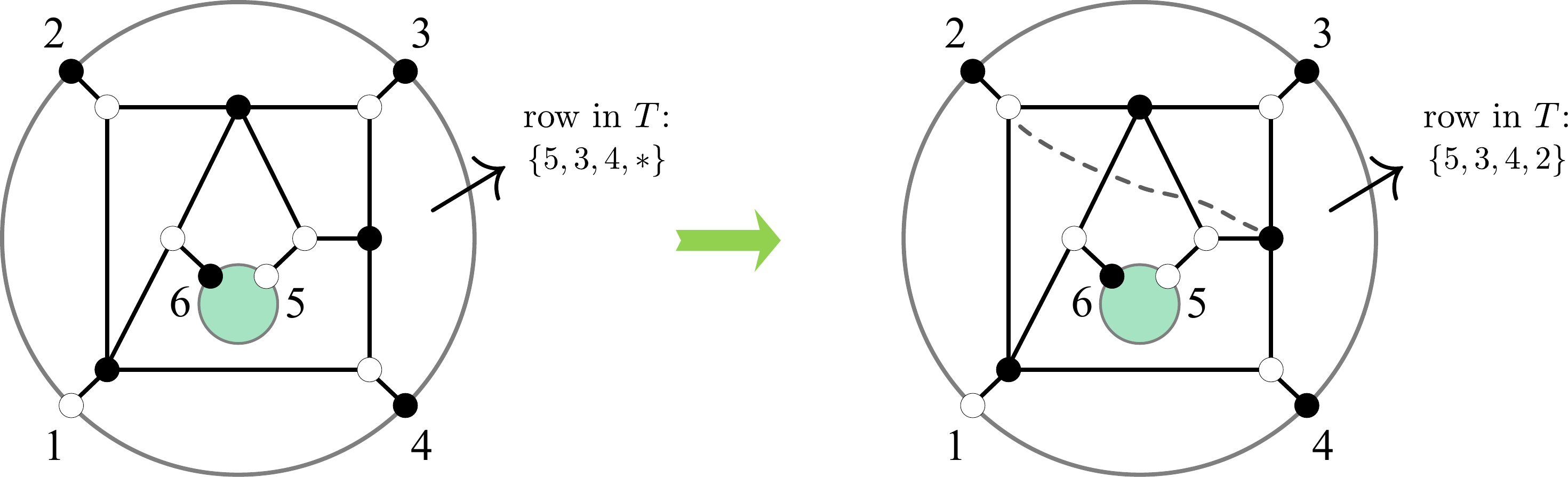}
\caption{\it Addition of an auxiliary edge to a black node with valency $v<k+1$. The grey line fixes the arbitrary entry in the matrix $T$ \eqref{eq:T-star} to be $*=2$. We do not show the new embedding surface since it is not relevant for the computation of the on-shell form and the additional edge is to be deleted in the following step.}
\label{fig:NMHV1extraleg}
\end{figure}

A black node of valency $v$ is associated to the Grassmannian $Gr_{v-1,v}$. Consider for instance a black node for which the corresponding row in $T$ is \begin{equation}
\{i_1,\dots, i_v,*_{v+1},\dots, *_{k+1} \}\ .
\end{equation}
 The first step is to add auxiliary degrees of freedom until the diagram is lifted to a top-cell. This is done by adding extra edges to the black nodes until all of them have valency $k+1$. As a result the analogue of the matrix \eqref{eq:c-perp-black} is
\begin{equation}
\widetilde{C}^{\perp}\,=\, \big( \alpha_{i_1} \; \cdots \; \alpha_{i_v} \; \alpha_{*_{v+1}} \; \cdots \; \alpha_{*_{k+1}} \big)\ .
\end{equation}
The auxiliary edges may connect the black node with any other white node of the graph which are not already connected to it (otherwise the graph would become reducible but not a top-cell). The entries $*_i$ now become labels present in the graph. There are several possible ways to lift the diagram to a top-dimensional cell in $Gr_{k,n}$. Consider for example the diagram from \fref{fig:NMHV1extraleg}, where an auxiliary leg sets the unfixed entry $*=2$, however, one could similarly add a leg in a way such that $*=6$ or $*=1$.
The proof now proceeds as if there were no undetermined entries and in the end we remove the auxiliary degrees of freedom by taking residues around $\alpha_{*_v+1},\dots,\alpha_{*_{k+1}}=0$. Notice that this implies that the complementary minors of $C$ vanish, in analogy with \eqref{eq:relation-minors},
\begin{align}
\alpha_{*_i}\,=\, (*_i)\Big|_{\widetilde{C}^{\perp}}\,=\,0\quad\Rightarrow\quad (i_1\cdots \hat{*}_i\cdots *_{k+1})\Big|_{\widetilde{C}}\,=\,0\ .
\end{align}
Taking the residue around all $\alpha_{*_i}=0$ imposes that the columns $\vec{c}_1,\dots,\vec{c}_v$ are linearly dependent vectors after the identification \eqref{eq:minor-identification}.

The independence of the choice of the labels $*_i$ (or in other words how the lift to a top-cell is made) can also be seen in a simple way. Take for instance the example on the left of \fref{fig:NMHV1extraleg} that has a row in $T$ given by $\{5,3,4,*\}$. Since $k=3$, we can choose three linearly independent vectors to form a basis, thus a general redefinition of the column $\vec{c}_*$ of $C$ can be written, for instance, as
\begin{align}
\vec{c}_*\,\rightarrow x\, \vec{c}_* + y\, \vec{c}_3 + z\, \vec{c}_4\ .
\end{align}
Note that we could not choose $\vec{c}_3,\,\vec{c}_4$ and $\vec{c}_5$ to form a basis since $(345)=0$.
The dependence of the general formula \eqref{eq:LS} on $\vec{c}_*$ is through the minors
\begin{align}
(*53)\,\rightarrow\, x\,(*53) + z\,(453)\,,\quad (34*)\,\rightarrow\, x\,(34*)\,,\quad (4*5)\,\rightarrow\, x\,(4*5) + y\,(453)\ .
\end{align}
Since $(345)=0$ every minor involving $\vec{c}_*$ simply gets rescaled. Is it clear that under such a scaling \eqref{eq:LS} transforms as $x^k/x^k$ which guarantees that it is independent of the choice of $\vec{c}_*$.
This completes the proof of the procedure of \sref{section_combinatorial_on_shell} to any on-shell diagram.

\bigskip

We conclude this chapter by stating that the original work \cite{Franco:2015rma} contains more than what was covered here, a short summary of our additional findings are:
\begin{itemize}
\item A characterisation of non-planar diagrams based on the generalised \emph{matching} and \emph{matroid polytopes} \cite{Franco:2012mm,Franco:2012wv,Amariti:2013ija,Franco:2013nwa,Franco:2014nca}. In this classification, each perfect matching is mapped to a point in the matching polytope, whereas perfect matchings that give rise to the same source set (recall that perfect matchings and orientations are in one-to-one correspondence, see \fref{pm_and_perfect_orientation}) lead to a point in the matroid polytope. In this way the polytopes provide a characterisation of equivalence classes of non-planar on-shell diagrams and, moreover, the question of reducibility in the non-planar case can be phrased in terms of polytopes too. We have seen here that the full $Gr_{k,n}$ is far more complex than the $Gr^+_{k,n}$ associated to planar diagrams, for instance, in \eqref{Omega_complicated_case} we found that the the boundary of a cell in $Gr_{3,6}$ can be associated to a relation between minors which is beyond the \pl relations, as opposed to the planar case where all boundaries are of the form $\Delta_i=0$. This example was further studied in \cite{Franco:2015rma}, where this boundary structure was seen to emerge from the matching and matroid polytopes.
\item A generalisation of the boundary measurement~---~previously defined for graphs admitting a genus-zero embedding with arbitrarily many boundaries \cite{2006math09764P,2009arXiv0901.0020G,Franco:2013nwa}~---~to graphs embedded on surfaces of any genus. This boundary measurement required a refined sign prescription that allowed for a consistent characterisation using the matroid polytope, namely a $k\times k$ \pl coordinate $\Delta_{i_1,i_2,\cdots ,i_k}$ is expressed as linear combinations of perfect matchings with the corresponding source set $\{i_1,i_2,\cdots ,i_k\}$.
\end{itemize}

\chapter{Conclusions}
\label{ch:conclusions}

This thesis consisted of three main parts. In the first two we presented applications of on-shell methods to particular off-shell quantities in $\N=4$ SYM~---~form factors and the dilatation operator~---~and in the third part we presented a generalisation of the on-shell diagram formulation beyond the planar limit. In this final chapter we present a short summary of our main findings, concluding remarks and an outlook of possible future research.

\medskip

In Chapter \ref{ch:formfactors} we investigated supersymmetric form
factors of an infinite class of half-BPS operators which we called $\T_k$~---~whose totally bosonic component is $\Tr(\phi^k)$~---~up to two loops. At tree level, the BCFW construction using non-adjacent shifts produced a boundary contribution which then led to a recursion relation involving MHV form factors of $\T_k$ and $\T_{k-1}$. We conjectured a solution for all MHV form factors of $\T_k$ for arbitrary number of external legs $n$. As a consistency check, we observed that the solution satisfies cyclicity for some values of $k$ and $n$, however a general proof is still lacking. It would be interesting to investigate whether form factors with higher MHV degree, or perhaps different operators, satisfy similar recursion relations.

At one loop, the universal IR structure of form factors determines the part proportional to the tree level result, and we computed the extra finite contributions in the MHV case using quadruple cuts. This way we obtained all MHV super form factors of $\T_k$ at one loop and found that they are formed by one-mass triangles and finite box functions.

Following the one-loop computation, we restricted ourselves to minimal form factors of $\T_k$ (i.e. with $k$ external states) and studied them at two loops using generalised unitarity. After constructing the integrand in this way, we arrived at a basis of integral functions which were available in the literature \cite{Gehrmann:1999as,Gehrmann:2000zt}, however, the results for the integrated expressions were complicated, containing various multiple polylogarithms. Using this result to define a finite two-loop remainder function, we observed that its symbol was considerably simpler. For $\T_3$ we were able to integrate the symbol to obtain a compact remainder function of uniform transcendentality four and containing only classical polylogarithms. For higher $k$ we decomposed the symbol into building blocks depending only on three variables each, and identified the part of the symbol that could not be integrated to classical polylogarithms. Doing so, we obtained an analytic expression for all remainder functions for the minimal form factors of $\T_k$.

Beyond the BPS case, loop form factors have since been studied in \cite{Wilhelm:2014qua,Nandan:2014oga,Loebbert:2015ova}. In particular, in \cite{Loebbert:2015ova} the authors found that the leading transcendentality part of the two-loop remainder function of non-protected operators in the $SU(2)$ sector is universal and corresponds to our BPS result. It would be interesting to study non-minimal form factors of non-protected operators and to perhaps find more connections with QCD results, in the same spirit as the relations found between non-minimal form factors of $\T_2$ and Higgs plus multi-gluon amplitudes  \cite{Gehrmann:2011aa,Brandhuber:2012vm}.

\medskip

In Chapter \ref{ch:dilatation} we obtained the one-loop dilatation operator in the $SO(6)$ and $SU(2|3)$ sectors by applying on-shell methods to the the two-point correlation functions $\b{\O(x)\bar{O}(y)}$ in each sector. Firstly, inspired by \cite{Koster:2014fva}, we studied the dilatation operator in the $SO(6)$ sector using MHV diagrams. This computation was subsequently simplified by directly applying generalised unitarity to the calculation of the two-point functions, which allowed for a simple treatment of fermions in the $SU(2|3)$ sector too. 

It would be interesting to apply MHV diagrams to the calculation of the dilatation operator in other sectors of $\N=4$ SYM, also containing fermions and derivatives.  Applications to different Yang-Mills theories with less supersymmetry can also be considered, given the validity of the MHV diagram method beyond $\N=4$ SYM. In the unitarity-based approach, the use of gluon amplitudes remains a future direction of research, and we expect these to be relevant for the study of the $SL(2)$ sector as well as for single-trace operators made of field strengths in QCD \cite{Ferretti:2004ba}.

An obvious goal  is the extension of our calculation to higher loops. This has proved difficult for amplitudes using MHV diagrams, but addressing the calculation of just the UV-divergent part of the two-point correlation function may simplify this task  enormously. At one loop the complete dilatation operator is known \cite{Beisert:2003jj}, while direct perturbative calculations at higher loops~---~without the assumption of integrability~---~have been performed only up to two   \cite{Eden:2005bt,Belitsky:2005bu,Georgiou:2011xj}, three  
\cite{Beisert:2003ys, Eden:2005ta,Sieg:2010tz} and four loops \cite{Beisert:2007hz} in particular sectors. A simplified route to such a calculation would be greatly desirable, and would provide further verification of this crucial assumption. The expected structure remains that of \eqref{eq:double-bubble}, with the double-bubble integral replaced by  more complicated (but still single-scale) loop integrals. 

It is important to point out other works being carried out in the same spirit of connecting on-shell methods with the dilatation operator. Firstly, in \cite{Koster:2014fva},  twistor-space  MHV diagrams were used to find the dilatation operator in the $SO(6)$ sector at one loop directly from two-point correlators, leading to the position-space form of the correlator as found by \cite{Minahan:2002ve}. In \cite{Wilhelm:2014qua} the complete one-loop dilatation operator was obtained by calculating form factors for generic single-trace operators using generalised unitarity, making interesting contact with earlier work of  
\cite{Zwiebel:2011bx}. In particular, the integral form for the dilatation operator in \cite{Zwiebel:2011bx} is mapped to a phase-space integral, which appears naturally in a unitarity-based approach. The calculation of two-loop form factors using unitarity was also employed to obtain  the two-loop anomalous dimension of the Konishi operator in  \cite{Nandan:2014oga}.  A comprehensive summary of these methods appear in \cite{Wilhelm:2016izi}. 

In comparing the two main lines of approach, using form factors or the two-point correlators, one notices the following main points. In order to extract $L$-loop anomalous dimensions from form factors, an $L$-loop calculation is required, while for the two-point correlators in momentum space in principle $2 L$-loop integrals can appear. However, form factors also have (universal) infrared divergences which need to be disentangled from the UV divergences, and with increasing loop order one obtains integrals with an increasing number of scales. 
In the case of two-point correlators, one has the advantage of only having to consider single-scale integrals, albeit at higher-loop order in momentum space, and one never encounters infrared divergences. 


Finally, our result hints at a link between the Yangian symmetry of amplitudes in $\N=4$ SYM \cite{Drummond:2009fd} and integrability of the dilatation operator  of the theory 
\cite{Minahan:2002ve,Beisert:2003jj,Beisert:2003tq,Bena:2003wd, Beisert:2003yb,Beisert:2004ry}. This point was later explored in \cite{Brandhuber:2015dta}, where the commutation relations between Yangian generators and the dilatation operator of \cite{Dolan:2003uh} were rederived using the realisation of the Yangian on tree level scattering amplitudes.

\medskip

In Chapter \ref{ch:onshelldiagrams} we studied a generalisation of on-shell diagrams in $\N=4$ SYM beyond the planar limit. In our approach, we considered the embedding of on-shell diagrams on Riemann surfaces with boundaries. This embedding allowed us to define a generalisation of the efficient face-variable parametrisation of a cell in $Gr_{k,n}$ associated to the on-shell diagram. Following this, we developed a combinatorial method to determine the on-shell form in terms of $k\times k$ minors. This method is a generalisation of the one presented in \cite{Arkani-Hamed:2014bca} for MHV leading singularities and its main advantage is that it allows the determination of the on-shell form without the need to compute the boundary measurement for each individual diagram.

The natural goal of this program is to achieve a level of understanding of non-planar on-shell diagrams similar to the existing one for the planar case, and in particular if and how they determine a notion of a non-planar integrand.  It is also interesting to investigate whether there are non-planar counterparts for some of the objects which followed on-shell diagrams in planar $\mathcal{N}=4$ SYM, such as deformed on-shell diagrams \cite{Ferro:2012xw,Ferro:2013dga,Beisert:2014qba,Kanning:2014maa,Broedel:2014pia}\footnote{Deformed amplitudes have been studied in \cite{Broedel:2014hca,Ferro:2014gca}.} and on-shell diagrams for theories with $\N<4$ SUSY \cite{ArkaniHamed:2012nw,Benincasa:2015zna}.
Another question to explore is whether there is a non-planar generalisation of the connection between scattering amplitudes in ABJM theory \cite{Aharony:2008ug} and the positive orthogonal Grassmannian \cite{Huang:2013owa,Huang:2014xza}. 

Finally, for planar amplitudes, on-shell diagrams are not the state of the art, in particular this program goes further and culminates in the complete geometrisation of scattering amplitudes in terms of the \emph{amplituhedron} \cite{Arkani-Hamed:2013jha,Arkani-Hamed:2013kca}, where tree amplitudes and the loop integrands are thought of as the volume of a polytope. A hint of an amplituhedron-like structure beyond the planar limit was recently found in \cite{Bern:2015ple}. Is is known that the planar Grassmannian formulation is a consequence of the Yangian symmetry of planar leading singularities, thus an exciting question is what fixes the form of the non-planar Grassmannian integral. In this regard, the recent work \cite{Frassek:2016wlg} precisely finds Yangian-like symmetries of non-planar on-shell forms.


\newpage
\chapter*{Acknowledgements}
\thispagestyle{empty}

It is with great pleasure that I express my gratitude to my supervisor Gabriele Travaglini for the presence, guidance and close collaboration in the past years. The work done throughout the PhD would not have been possible without my second supervisor Bill Spence and my additional collaborators Congkao Wen, Andreas Brandhuber, Donovan Young, Daniele Galloni and Sebasti\'{a}n Franco.

My deepest thank you to all members of Queen Mary University of London, including the fellow PhD students of many generations, Felix Rudolph, James McGrane, Zac Kenton, Ed Hughes, Paolo Mattioli, Martyna Kostaci\'{n}ska, Emanuele Moscato, Rodolfo Panerai, Joel Berkeley, David Garner, Sam Playle, Dimitrios Korres, \"{O}mer G\"{u}rdo\v{g}an, Rob Mooney, Jurgis Pasukonis, Edvard Musaev and especially Joe Hayling for his support, companionship, and for proofreading parts of this thesis. Thank you also to the Queen Mary friends Sophia Goldberg, Asmi Barot, Serena Maugeri, Manting Qiu, Viraj Sanghai, Marco Bianchi and Raquel Ribeiro. Together they made the workplace one of friendship and joy.

In addition, I would like to thank the faculty members Sanjaye Ramgoolam, David Berman, Costis Papageorgakis, Rodolfo Russo, Steve Thomas, Brian Wecht and Richard Nelson, as well as the staff member Sarah Cowls, Jessica Henry, Lucie Bone, Jazmina Moura, Karen Wilkinson, Predrag Micakovic, Terry Arter and John Sullivan for ensuring that everything ran smoothly.

Momentarily leaving the United Kingdom, I would like to show deep gratitude to my BSc supervisors Sandra Vianna, Marcelo Leite and especially Bruno Cunha from Universidade Federal de Pernambuco, Brazil, for their invaluable guidance, enthusiasm and teachings in the early days.  Thank you also to my MSc adviser Freddy Cachazo from Perimeter Institute, Canada, who introduced me to scattering amplitudes, the subject of this thesis. Thank you to all members of the African Institute of Mathematical Sciences in South Africa, in particular the good friends Emile Chimusa and Mekdes Awalew. Special thanks to the old friends Rafael de Lima, Rebeca Holanda and Jorge Rehn from the undergraduate days and Grisha Sizov, Kate Hughes, Robert Schuhmann, Yvonne Geyer, Pavel Chvykov, Eduardo Casali, Dalimil Maza\v{c} and Zuzka Mas\'{a}rov\'{a} whom I met at Perimeter.

I would like to thank Humboldt University of Berlin for the hospitality in the last few months, and where this thesis was mostly written, and my co-workers  Laura Koster, Sourav Sarkar, Christian Marboe, David Meidinger, Gregor Richter, Yumi Ko, Christoph Sieg, Stijn van Tongeren, Burkhard Eden, Johannes Br\"{o}del, Matthias Staudacher, Jan Plefka, Valentina Forini, Vladimir Smirnov, Thomas Klose, Dhritiman Nandan, Pedro Liendo, Gang Yang, Matteo Rosso, Florian Loebbert, Edoardo Vescovi,  Hagen M\"{u}nkler, Dennis M\"{u}ller, Annie Spiering, Wadim Wormsbecher, Josua Faller, Sylvia Richter, Matthias Wilhelm, Rouven Frassek and Michael Borinsky. Special thanks to Leo Zipellius for proofreading this thesis.

My deepest thank you to my parents Acy and Fl\'{a}vio Penante, my sister Diana Penante and the rest of my family and friends in Brazil whom I miss very much. Although our contact have mostly been virtual in the past few years, their love, support and help in times of need were crucial.

Back to the United Kingdom, I would like to thank the members of the London Buddhist Centre to whom I owe much of my piece of mind, including T\^{a}nia Azevedo, Kate Hayler, Cait Crosse, Catrine Skeppar, Dorota Mu\l czy\'{n}ska, Lucy Norris, Miranda Brennan, Lydia Parussol, Mia Kos, Marsha Saunders, Livia \v{C}a\v{c}kan\'{a}, Tara Allitt, and the teachers Annie Gogarty, Sugati, Lilamani, Singhamanas, Subhadramati, Maitreyabandhu and Jnanavaca. Special thanks also to the Berliner counterpart Buddhistisches Tor Berlin, in particular Lalitaratna.

My research at Humboldt University of Berlin is supported by the People Programme (Marie Curie Actions) of the European Union, Grant Agreement No. 317089 (GATIS).

\appendix

\chapter{Spinor conventions}
\label{app:spinor-conventions}

In this thesis we have extensively used the spinor-helicity variables introduced in \sref{sec:SH_Formalism}. The purpose of this appendix is to show the conventions we used to manipulate these variables.\\
 It is usual to use the following vectors of Pauli matrices:
\begin{equation}
(\sigma^{\mu})_{\alpha \dot\alpha}\,=\,(\uno,\vec{\sigma})_{\alpha\dot\alpha}\,,\quad 
(\bar{\sigma}^{\mu})^{\dot\alpha \alpha}\,=\,(\uno,-\vec{\sigma})^{\dot\alpha \alpha}\ ,
\end{equation}
where the Pauli matrices are
\spacing{1.2}
\begin{equation}
\sigma^0=\begin{pmatrix}
1 & 0 \\
0 & 1
\end{pmatrix},\quad \sigma^1=\begin{pmatrix}
0 & 1 \\
1 & 0
\end{pmatrix},\quad \sigma^2=\begin{pmatrix}
0 & - i \\
i & 0
\end{pmatrix},\quad \sigma^3=\begin{pmatrix}
1 & 0 \\
0 & - 1
\end{pmatrix}\ .
\end{equation}
\spacing{1.5}
\noindent In this notation, the $SU(2)$ invariant tensors and their inverse are
\spacing{1.2}
\begin{align}
\eps_{\alpha\beta}=\eps_{\dot\alpha\dot\beta}\,=\,i\sigma^2\,=\,\begin{pmatrix}
0 & 1 \\
-1 & 0
\end{pmatrix}\ ,\qquad \eps^{\alpha\beta}=\eps^{\dot\alpha\dot\beta}\,=\,-i\sigma^2\,=\,\begin{pmatrix}
0 & -1 \\
1 & 0
\end{pmatrix}\ .
\end{align}
\spacing{1.5}
\noindent Thus
\begin{equation}
\epsilon_{\alpha\beta}\epsilon^{\beta\gamma}\,=\,\delta^\gamma_\alpha\,,\quad \epsilon_{\dot\alpha\dot\beta}\epsilon^{\dot\beta\dot\gamma}\,=\,\delta^{\dot\gamma}_{\dot\alpha}\ .
\end{equation}
According to \eqref{eq:P-spinor}, the on-shell momentum of a particle labeled by $i$ is defined in terms of spinors as
\begin{align}
p^i_{\alpha\dot{\alpha}}\,=\,p^i_\mu\sigma^\mu_{\alpha\dot{\alpha}}\,=\,\lambda^i_\alpha\tl^i_{\dot\alpha}\,,\quad \bar{p}^{i\dot{\alpha}\alpha}\,=\,p^i_\mu\bar{\sigma}^{\mu\dot{\alpha}\alpha}\,=\,\lambda^{i\alpha}\tl^{i\dot\alpha}\ .
\end{align}
Spinor indices are raised and lowered according to
\begin{align}
\begin{split}
\lambda_\alpha\,=\,\eps_{\alpha\beta}\lambda^\beta\,,\quad \lambda^\alpha\,=\,\epsilon^{\alpha\beta}\lambda_\beta \ ,\\
\tl_{\dot\alpha}\,=\,\eps_{\dot\alpha\dot\beta}\tl^{\dot\beta}\,,\quad \tl^{\dot\alpha}\,=\,\eps^{\dot\alpha\dot\beta}\tl_{\dot\beta}\ ,
\end{split}
\end{align}
and the $\sigma/\bar{\sigma}$-matrices are related via
\begin{align}
(\bar{\sigma}^\mu)^{\dot\alpha \alpha}\,=\,\epsilon^{\alpha\beta}\sigma^{\mu}_{\beta\dot\beta}\eps^{\dot\alpha\dot\beta}\,,\quad \sigma^{\mu}_{\alpha\beta}\,=\, \epsilon_{\dot\alpha\dot\beta}(\bar{\sigma}^\mu)^{\dot\beta \beta}\eps_{\alpha\beta}\ .
\end{align}
The spinor brackets \eqref{eq:spinor-brackets} are given by
\begin{align}
\begin{split}
 \braket{ij}&\,\equiv\,\braket{\lambda^{i}\lambda^{j}}\,=\, \lambda^{i\alpha}\lambda^j_{\alpha}\,=\,\epsilon_{\alpha\beta}\lambda^{i\alpha}\lambda^{j\beta}, \qquad\, \braket{ij}\,=\,-\braket{ji},\\
 [ij]&\,\equiv\,[\widetilde{\lambda}_{i}\,\widetilde{\lambda}_{j}]\,=\,\tl^i_{\dot\alpha}\tl^{j\dot\alpha}\,=\,\epsilon^{\dot{\alpha}\dot{\beta}} \widetilde{\lambda}^i_{\dot{\alpha}}\tl^j_{\dot{\beta}}, \quad\quad\quad [ij]\,=\,-\,[ji]\ .
 \end{split}
\end{align}
Note that in our conventions
\begin{equation}
\epsilon_{\alpha\beta}\lambda^{i \alpha}\lambda^{j\beta}\,=\,
\lambda^{i\alpha}\lambda_{j\alpha}\,=\,-\lambda^i_{\alpha}\lambda^{j\alpha}\ ,
\end{equation}
and thus
\begin{equation}
2(p^i\cdot p^j)\,=\,(\bar{p}^{i})^{\dot\alpha \alpha}p^j_{\alpha\dot\alpha}\,=\,\lambda^{i\alpha} \tl^{i \dot\alpha}\lambda^j_{\alpha}\tl^j_{\dot\alpha}\,=\,\b{ij}[ji]\ .
\end{equation}
Throughout the computations, we systematically use the following expansion for the trace of four momenta:
\begin{align}
\label{eq:trace-expand}
\begin{split}
&\b{ab}[bc]\b{cd}[da]\,=\,\text{Tr}_+(a \, b \,c \,  d) = \text{Tr}(\tfrac{1}{2}(1+\gamma^5)\slashed{a} \, \slashed{b} \, \slashed{c} \,  \slashed{d})\\
=\,& 2 \left((a\cdot b)(c \cdot d) +  (b\cdot c)(a \cdot d) - (a\cdot c)(b \cdot d) - i \epsilon^{\mu\nu\rho\sigma}a_\mu b_\nu c_\rho d_\sigma\right).
\end{split}
\end{align}

\chapter{Integrals}
\section{One-loop scalar integrals}
\label{app:scalar-integrals}

\begin{figure}[htb]
\centering
\includegraphics[width=\linewidth]{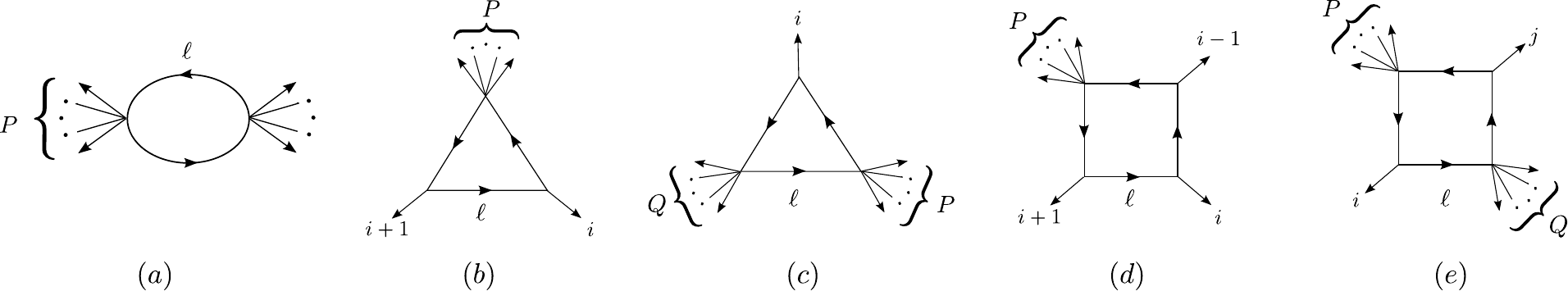}
\caption{\it Scalar integrals which appear in the calculation of the dilatation operator in Chapter \ref{ch:dilatation} and loop-level form factors of half-BPS operators in Chapter \ref{ch:formfactors}~---~$(a)$ bubble integral, $(b)$ one-mass triangle, $(c)$ two-mass triangle, $(d)$ one-mass box and $(e)$ two-mass-easy box.}
\label{fig:scalarints}
\end{figure}

In this appendix we give the explicit expressions for the integral functions used throughout this thesis. We consider them in the context of dimensional regularisation, so $d=4-2\eps$. For the definition of the various momentum assignments we refer to Figure \ref{fig:scalarints} and we use the conventions of \cite{Bern:1994cg}.\\[12pt]

\noindent The bubble integral is defined as
\begin{align}
I_2(P^2)\,\equiv\,-i(4\pi)^{2-\eps}\int\! \frac{d^{4-2\eps} \ell}{(2\pi)^{4-2\eps}}\,\frac{1}{\ell^2 (\ell-P)^2}\,=\, \frac{r_{\Gamma}}{\eps(1-2\eps)} \, (-P^2)^{-\eps}\ ,
\end{align}
where 
\begin{equation}
\label{eq:rgamma}
 r_\Gamma \, \equiv\,  
 \dfrac{\Gamma(1+\epsilon)\Gamma^2(1-\epsilon)}{\Gamma(1-2\epsilon)} \ . 
\end{equation}
\noindent The one-mass and two-mass triangle integrals are given by
\begin{align}
\label{eq:1massTri}
I_{3;i}^{1\rm m}(P^2)\,\equiv\,i(4\pi)^{2-\eps}\int\! \frac{d^{4-2\eps} \ell}{(2\pi)^{4-2\eps}}\,\frac{1}{\ell^2 (\ell-p_i)^2(\ell+p_{i+1})^2} &=\frac{r_\Gamma}{\epsilon ^2} (- P^2)^{-1-\epsilon}\ ,\\[10pt]
\label{eq:2massTri}
I_{3;i}^{2\rm m}(P^2,Q^2)\,\equiv\,i(4\pi)^{2-\eps}\int\! \frac{d^{4-2\eps} \ell}{(2\pi)^{4-2\eps}}\,\frac{1}{\ell^2 (\ell-P)^2(\ell+Q)^2}&=\frac{r_\Gamma}{\epsilon ^2} \frac{(- P^2)^{-\epsilon}-(- Q^2)^{-\epsilon}}{P^2-Q^2} \ .
\end{align} 
The one-mass box is given by
\begin{align}
\begin{split}
I^{\rm 1m}_{4;i}(s,t,P^2)  \,\equiv&\,-i(4\pi)^{2-\eps}\int\! \frac{d^{4-2\eps} \ell}{(2\pi)^{4-2\eps}}\,\frac{1}{\ell^2 (\ell-p_i)^2(\ell+p_{i+1})^2(\ell-p_i-p_{i-1})^2}\\
=&\, -\frac{2r_{\Gamma}}{st}\Big\{-\frac{1}{\eps^2}\big[(-s)^{-\eps}+(-t)^{-\eps} -(-P^2)^{-\eps}\big]\\ 
&\,+ \text{Li}_2\Big(1-\frac{P^2}{s}\Big) + \text{Li}_2\Big(1-\frac{P^2}{t}\Big) +\frac{1}{2}\log^2\Big(\frac{s}{t}\Big) + \frac{\pi^2}{6}\Big\} \ ,
\end{split}
\end{align}
where $s\equiv (p_i+p_{i-1})^2$ and $t\equiv(p_i+p_{i+1})^2 $.
For the two-mass-easy box integral,
\begin{equation}
I^{\rm 2me}_{4;i,j}(s,t,P^2,Q^2)  \,\equiv\,-i(4\pi)^{2-\eps}\int\! \frac{d^{4-2\eps} \ell}{(2\pi)^{4-2\eps}}\,\frac{1}{\ell^2 (\ell+p_i)^2(\ell+p_{i}+P)^2(\ell-Q)^2}\ ,
\end{equation}
it is more useful to define the \emph{box function} $F^{\rm 2me}$, which is related to $I^{\rm 2me}$ according to \cite{Bern:1994zx}
\begin{equation}
F^{\rm 2me}_{4;i,j}= -\frac{1}{2r_\Gamma}(P^2Q^2-st)  I^{\rm 2me}_{4;i,j}\ .
\end{equation}
\noindent The two-mass easy box function is given by
\begin{align}
\label{eq:2me}
\begin{split}
F^{\rm 2me}_{4;i,j}(s,t,P^2,Q^2)\,& = 
- \frac{1}{\epsilon^2}  \Big[ 
(-s)^{-\epsilon}  +(-t)^{-\epsilon}-(-P^2)^{-\epsilon}-(-Q^2)^{-\epsilon} \Big]  \cr
&+  {\rm Fin^{2me}}
(s,t,P^2,Q^2) 
\ ,
\end{split}
\end{align}
where 
 $s\equiv (p_i+P)^2$ and  $t\equiv (p_i+Q)^2$ and ${\rm Fin^{2me}}$ stands for finite terms. The  finite part of the  two-mass-easy box function, in the form of  \cite{Duplancic:2000sk,Brandhuber:2004yw}, is 
\begin{equation}
\label{2mebst}
  {\rm  Fin^{2me}} (s,t,P^2, Q^2) \, = \, 
 \Li(1-aP^2)\, + \, \Li(1-aQ^2)  \, -\,  \Li(1-as)
\,  -\,  \Li(1-at)
\ , 
\end{equation}
where 
\begin{equation}
\label{adef} 
a \ = \
\frac{P^2+Q^2-s-t}{P^2Q^2-st} \ . 
\end{equation}
An analytic proof of the equivalence of \eqref{2mebst} and the form given in 
\cite{Bern:1993kr}  can be found in \cite{Brandhuber:2004yw}. 

\section{Tensor integrals and Passarino-Veltman reduction}
\label{app:PV}

In Chapter \ref{ch:dilatation} there appear integrals which are not scalar but instead have a numerator which depend on the loop integration variable. In this section we perform what is called a \emph{Passarino-Veltman} reduction \cite{Passarino:1978jh}, which makes use of the Lorentz invariance of the integrated result to expand tensor integrals as a linear combination of scalar integrals, like the ones presented above in \sref{app:scalar-integrals}.

The first example is the reduction of a linear bubble into a scalar bubble integral, used in \eqref{eq:Linear-Bubble}. Due to Lorentz invariance we can write the following ansatz for the linear bubble integral
\begin{align}
\label{eq:PV-linear-bubble}
\int\!{d^d K \over (2 \pi)^d} {K^\mu \over K^2 (K \pm L)^2} \ = \ A \, L^\mu 
\, \int\!{d^d K \over (2 \pi)^d} {1 \over K^2 (K \pm L)^2}  \ .
\end{align}
Contracting both sides of \eqref{eq:PV-linear-bubble} with $L_\mu$ we get the following relation between the integrands,
\begin{align}
L_\mu K^{\mu}\,=\, A L^2\ .
\end{align}
The next step is to expand the scalar product into full propagators,
\begin{equation}
L_\mu K^\mu\,=\,\pm \frac{1}{2}\big[(L\pm K)^2 -L^2 - K^2\big]\ .
\end{equation}
The factors $(L\pm K)^2$ and $K^2$ cancel a propagator of \eqref{eq:PV-linear-bubble} and thus lead to tadpole integrals which are zero in dimensional regularisation. The only term that survives is $L^2$, which gives the result quoted in \eqref{eq:Linear-Bubble}, namely $A=\mp 1/2$.
\medskip

The next set of reductions are the ones used in equations \eqref{888} and \eqref{999} that feature in the fermion-scalar terms of the one-loop dilatation operator $\Gamma^{SU(2|3)}$. We repeat them here for convenience
\begin{align}
I_{\beta\dot{\beta}} & \equiv \int\!\! d^4\ell_1  d^4\ell_3 \,
\delta^{(+)} (\ell_1^2) \,  
\delta^{(+)} (\ell_3^2) \,  
\delta^{(+)}\left( (L-\ell_1)^2\right) \,  
\delta^{(+)} \left((L+\ell_3)^2\right) \,  
\ \cdot \ N_{\beta\dot{\beta}}\ ,  \\
\tilde{I}_{\beta\dot{\beta}} & \equiv 
\int\!\! d^4\ell_1  d^4\ell_3 \,
\delta^{(+)} (\ell_1^2) \,  
\delta^{(+)} (\ell_3^2) \,  
\delta^{(+)} \left((L-\ell_1)^2\right) \,  
\delta^{(+)} \left((L+\ell_3)^2\right) \,  
\ \cdot \ \tilde{N}_{\beta\dot{\beta}}
\ , 
\end{align}
where the integrands are
\begin{align}
\begin{split}
  \,N_{\beta\dot{\beta}}\, &=  \, - \frac{(\ell_2\,\bar{\ell_1}\,\ell_3)_{\beta\dot{\beta}}}{2(\ell_1\cdot\ell_4)}\,=\,-\frac{\lambda^2_\beta\, [21]\b{13}\tl^3_{\dot{\beta}}}{2(\ell_1\cdot\ell_4)}\ ,\\
\tilde{N}_{\beta\dot{\beta}}\,& =\,  - \frac{(\ell_2\,\bar{\ell_1}\,\ell_4)_{\beta\dot{\beta}}}{2(\ell_2\cdot\ell_3)}\,=\,-\frac{\lambda^2_\beta\, [21]\b{14}\tl^4_{\dot{\beta}}}{2(\ell_2\cdot\ell_3)} \ . 
\end{split}
\end{align}
We are only interested in the UV-divergent part of $I_{\beta\dot{\beta}}$ and $\tilde{I}_{\beta\dot{\beta}}$. Using Lorentz invariance we can write
\begin{align}
\label{eq:I1-PV}
I_{\beta\dot{\beta}}\Big|_{\rm UV}\,=\, A L_{\beta\dot{\beta}}\times \text{DB}(L^2)\Big|_{\rm UV}\ ,\qquad  \tilde{I}_{\beta\dot{\beta}}\Big|_{\rm UV}\,=\,
\tilde{A}\,L_{\beta\dot{\beta}}\times \text{DB}(L^2)\Big|_{\rm UV}\ , 
\end{align}
where $\text{DB}(L^2)$ stands for the double-bubble integral of \fref{fig:double-bubble}. After Fourier transforming to position space, the UV-divergent part of the double-bubble integral is given by \eqref{eq:double-bubble-divergence}.

In order to find $A$ and $\tilde{A}$ we will discard terms which lead to the kite integral of \fref{finiteintegral} as it is not UV-divergent. We use $L=\ell_1+\ell_2=-(\ell_3+\ell_4)$ and the cut conditions $\ell_1^2=\ell_2^2=\ell_3^2=\ell_4^2=0$.

\noindent Contracting both sides of \eqref{eq:I1-PV} with $\bar{L}^{\dot{\beta}\beta}$ we get
\begin{align}
\label{eq:scalar-fermion-PV}
\begin{split}
2AL^2\,&=\,-\frac{\Tr_+(\ell_2\,\ell_1 \,\ell_3\,L)}{s_{\ell_1\ell_4}}\,=\,-\frac{\Tr_+(\ell_2\,\ell_1 \,\ell_3\,\ell_1)}{s_{\ell_1\ell_4}}\ ,\\
2\tilde{A}L^2\,&=\,-\frac{\Tr_+(\ell_2\,\ell_1 \,\ell_4\,L)}{s_{\ell_2\ell_3}}\,=\,-\frac{\Tr_+(\ell_2\,\ell_1 \,\ell_4\,\ell_1)}{s_{\ell_2\ell_3}}\ .
\end{split}
\end{align}
According to \eqref{eq:trace-expand}, the traces can be expanded as
\begin{align}
\begin{split}
\Tr_+(\ell_2\,\ell_1 \,\ell_3\,\ell_1)\,&=\,s_{\ell_1\ell_2}s_{\ell_1\ell_3}\,=\, L^2 (-s_{\ell_1\ell_2}-s_{\ell_1\ell_4})\,=\,-L^4-L^2 s_{\ell_1\ell_4}\ ,\\
\Tr_+(\ell_2\,\ell_1 \,\ell_4\,\ell_1)\,&=\,s_{\ell_1\ell_2}s_{\ell_1\ell_4}\,=\,L^2 s_{\ell_2\ell_3}\ .
\end{split}
\end{align}
The $L^4$ term gives rise to a kite integral, whereas the $-L^2 s_{\ell_1\ell_4}$ and $L^2 s_{\ell_2\ell_3}$ terms cancel the additional propagator of \eqref{eq:scalar-fermion-PV}, leading to a double bubble.
Thus we find 
\begin{align}
\label{eq:A-Atilde}
\begin{split}
I_{\beta\dot{\beta}}\Big|_{\rm UV-divergent}\,&=\,-\frac{1}{2} L_{\beta\dot{\beta}}\times \text{DB}(L^2)\Big|_{\rm UV}\quad\Rightarrow\quad A_{\rm UV}\,=\,-\frac{1}{2}\ ,\\
\tilde{I}_{\beta\dot{\beta}}\Big|_{\rm UV-divergent}\,&=\,\frac{1}{2} L_{\beta\dot{\beta}}\times \text{DB}(L^2)\Big|_{\rm UV}\quad\Rightarrow\quad \tilde{A}_{\rm UV}\,=\,\frac{1}{2}\ .
\end{split}
\end{align}
Next we compute the PV reduction of the integral \eqref{111} appearing in the four-fermion component of $\Gamma^{SU(2|3)}$.
\beq
\label{eq:int-4-fermions-PV}
I_{\alpha\beta\dot{\alpha}\dot{\beta}} \equiv \int\!\! d^4\ell_1  d^4\ell_3 \,
\delta^{(+)}(\ell_1^2) \,  
\delta^{(+)}(\ell_3^2) \,  
\delta^{(+)} \left((L-\ell_1)^2\right) \,  
\delta^{(+)} \left((L+\ell_3)^2\right) \,  
\ \cdot \ N_{\alpha\beta\dot{\alpha}\dot{\beta}}\ ,
\eeq
where
\begin{align}
\begin{split}
N_{\alpha\beta\dot{\alpha}\dot{\beta}}\,&\equiv\,{1\over 2} \left[\frac{(\ell_2\bar{\ell_1})_{\alpha\beta}(\bar{\ell_4}\ell_3)_{\dot{\alpha} \dot{\beta}}+(\ell_1\bar{\ell_2})_{\alpha\beta}(\bar{\ell_3}\ell_4)_{\dot{\alpha} \dot{\beta}}}{s_{\ell_2\ell_3}}\right]\\
\,&=\,\frac{1}{2}\left[\frac{\lambda^2_{\alpha}[21]\lambda^1_{\beta} \tl^4_{\dot\alpha}\b{43}\tl^3_{\dot{\beta}}+\lambda^1_{\alpha}[12]\lambda^2_{\beta} \tl^3_{\dot\alpha}\b{34}\tl^4_{\dot{\beta}}}{s_{\ell_2\ell_3}}\right]\ .
\end{split}
\end{align}
It depends on only one scale $L$, hence it has the form 
\begin{align}
\label{eq:PV-4-fermions}
\,I_{\alpha\beta\dot{\alpha}\dot{\beta}}\Big|_{\rm UV}\,&=\,\big[A\,L^2\epsilon_{\alpha\beta} \epsilon_{\dot{\alpha} \dot{\beta}}\,+\,B\,(L_{\alpha\dot{\alpha}}L_{\beta\dot{\beta}} + L_{\alpha\dot{\beta}}L_{\beta\dot{\alpha}})\big]\text{DB}(L^2)\Big|_{\rm UV}\ .
\end{align}
Contracting \eqref{eq:int-4-fermions-PV} and \eqref{eq:PV-4-fermions} with $\epsilon^{\alpha\beta}\epsilon^{\dot\alpha\dot\beta}$ and $(\bar{L}^{\dot\alpha \alpha}\bar{L}^{\dot\beta \beta}+\bar{L}^{\dot\alpha \beta}\bar{L}^{\dot\beta \alpha})$ and using the rules of Appendix \ref{app:spinor-conventions} we get
\begin{align}
\label{eq:two-fermions-one-loop}
\begin{split}
\eps^{\alpha\beta}\eps^{\dot\alpha\dot\beta}\,N_{\alpha\beta\dot{\alpha}\dot{\beta}} \,&=\,-\frac{L^4}{s_{\ell_2\ell_3}}\,\xrightarrow{\text{UV-divergent} }\,0\,=\,4AL^2\,\Rightarrow A_{\rm UV}\,=\,0\ ,\\
(\bar{L}^{\dot\alpha \alpha}\bar{L}^{\dot\beta \beta}+\bar{L}^{\dot\alpha \beta}\bar{L}^{\dot\beta \alpha})\,N_{\alpha\beta\dot{\alpha}\dot{\beta}}\,&=\, 2L^4 + \frac{L^6}{s_{\ell_2\ell_3}}\,\xrightarrow{\text{UV-divergent} }\,2L^4\,=\,12BL^2\, \Rightarrow B_{\rm UV}=\frac{1}{6}\ ,
\end{split}
\end{align}
which is the result of \eqref{eq:coeffs-4-fermions}.

The last PV reduction is that of the tree-level contraction of the correlator with two fermions, whose momentum assignment is shown in \fref{fig:tree-level-contraction}. The integral is a tensor single bubble \eqref{eq:4-fermion-tree},
\begin{align}
\label{eq:tensor-single-bubble}
\begin{split}
I^{\rm tree}_{\alpha\beta\dot{\alpha}\dot{\beta}}\,&\equiv\,
\int\frac{d^dL_1}{(2\pi)^d}\frac{L_{1\,\alpha\dot{\beta}}(L-L_1)_{\beta\dot{\alpha}}}{L^2_1(L-L_1)^2}\quad \Rightarrow \quad N^{\rm tree}_{\alpha\beta\dot{\alpha}\dot{\beta}}\,\equiv\,L_{1\,\alpha\dot{\beta}}(L-L_1)_{\beta\dot{\alpha}}  \ .
\end{split}
\end{align}
In complete analogy with \eqref{eq:PV-4-fermions} we write
\begin{align}
\label{eq:PV-4-fermions-tree}
I^{\rm tree}_{\alpha\beta\dot{\alpha}\dot{\beta}}\Big|_{\rm UV}\,&=\,\big[A\,L^2\epsilon_{\alpha\beta} \epsilon_{\dot{\alpha} \dot{\beta}}\,+\,B\,(L_{\alpha\dot{\alpha}}L_{\beta\dot{\beta}} + L_{\alpha\dot{\beta}}L_{\beta\dot{\alpha}})\big]\text{Bub}(L^2)\Big|_{\rm UV}\ ,
\end{align}
where $\text{Bub}(L^2)$ is the scalar single-bubble integral \eqref{eq:scalar-single-bubble}.
Contracting \eqref{eq:tensor-single-bubble} and \eqref{eq:PV-4-fermions} with $\epsilon^{\alpha\beta}\epsilon^{\dot\alpha\dot\beta}$ we get the value of $A$,
\begin{align}
\label{eq:A-UV-tree}
\begin{split}
\eps^{\alpha\beta}\eps^{\dot\alpha\dot\beta}\,N^{\rm tree}_{\alpha\beta\dot{\alpha}\dot{\beta}} \,&=\,-L^2\,=\,4AL^2\quad\Rightarrow\quad A_{\rm UV}\,=\,-\frac{1}{4}\ .
\end{split}
\end{align}
Now contracting with $(\bar{L}^{\dot\alpha \alpha}\bar{L}^{\dot\beta \beta}+\bar{L}^{\dot\alpha \beta}\bar{L}^{\dot\beta \alpha})$ we get
\begin{align}
\begin{split}
\label{eq:B-term-scalar}
(\bar{L}^{\dot\alpha \alpha}\bar{L}^{\dot\beta \beta}+\bar{L}^{\dot\alpha \beta}\bar{L}^{\dot\beta \alpha})\,N^{\rm tree}_{\alpha\beta\dot{\alpha}\dot{\beta}}\,&=\, 2L^4- \Tr_+(L\,L_1\,L\,L_1)\,=\,2L^4-[4(L\cdot L_1)^2 -L^2L_1^2]
\end{split}
\end{align}
We can rewrite the scalar product as
\begin{equation}
4(L\cdot L_1)^2\,=\,[-(L-L_1)^2+L^2+L_1^2]^2\ ,
\end{equation}
Notice that terms with $(L-L_1)^2$ and $L_1^2$ will delete propagators of \eqref{eq:tensor-single-bubble}, so the only term that contributes is $L^4$.
Plugging this back in \eqref{eq:B-term-scalar} we find the value of $B$,
\begin{align}
\label{eq:B-UV-tree}
\begin{split}
(\bar{L}^{\dot\alpha \alpha}\bar{L}^{\dot\beta \beta}+\bar{L}^{\dot\alpha \beta}\bar{L}^{\dot\beta \alpha})\,N^{\rm tree}_{\alpha\beta\dot{\alpha}\dot{\beta}}\,&=\, L^4 \,=\, 12 B L^4  \quad\Rightarrow\quad B_{\rm UV}\,=\,\frac{1}{12}\ ,
\end{split}
\end{align}
which is the result of \eqref{eq:fermion-tree-bubble-coeff}.

\chapter{Form factors}

\section{Cyclicity of $\F^{\rm MHV}_{4,n}$} 
\label{app:cyclicity}

In this appendix we prove the cyclicity of the form factor $\F^{\rm MHV}_{4,n}$. This is given in \eqref{eq:O4}, but for convenience  we repeat its expression here:
\begin{align}
\F^{\text{MHV}}_{4,n}\ = \ \F^{\text{MHV}}_{2,n} \sum\limits_{1\leq i\leq j}^{n-3} \sum\limits_{j< k\leq l}^{n-2} (2-\delta_{ij})(2-\delta_{kl}) \frac{\b{n\,i}\b{j\,k}\b{l\,n-1}}{\b{n-1\,n}}(\eta^{-, i}\cdot \eta^{-,j})(\eta^{-,k}\cdot \eta^{-,l})\ .
\end{align}
The procedure we will follow consists in  eliminating $\eta^{-, 1}$ using supermomentum conservation in the $Q^{-}$ direction,  
and showing that the result one obtains in this way is the same as the original expression but with all relevant indices shifted by one unit. 
After substituting in the solution for $\eta^{-,1}$ from supermomentum conservation,  
we consider contributions to  terms of different structure in the various  $\eta^{-}$'s separately. 
In what follows we will list all possible structures and their corresponding coefficients:

\begin{itemize}

\item $(\eta^{-,i} \cdot \eta^{-,j})(\eta^{-,n-1})^2$ : 
\begin{align}
(2-\delta_{ij})\frac{\b{n\,1}\b{1\,i}\b{j\,n\!-\!1}}{\b{n\!-\!1\,n}} \frac{\b{n\!-\!1 \, n}^2 }{\b{n \, 1 }^2} \ = \  (2-\delta_{ij}) \frac{\b{1\, i}\b{j\,n\!-\!1}\b{n\!-\!1\,n}}{\b{n\,1}} \, .
\end{align}

\item $(\eta^{-,i})^2(\eta^{-,k})^2, \quad {\rm with} \quad i<k$: 
\begin{align}
\begin{split}
&\frac{\b{n\,i}\b{i\,k}\b{k\,n\!-\!1}}{\b{n\!-\!1\,n}}
+
\frac{\b{n\,1}\b{1\,i}\b{i\,n\!-\!1}}{\b{n\!-\!1\,n}}\frac{\b{k \, n}^2 }{\b{n \, 1}^2}
+
\frac{\b{n\,1}\b{1\,k}\b{k\,n\!-\!1}}{\b{n\!-\!1\,n}}\frac{\b{i \, n}^2 }{\b{n \, 1}^2}\\
-&2 \frac{\b{i\,n}\b{k\,n}}{\b{n \, 1}^2} \frac{\b{n\,1}\b{1\,i}\b{k\,n\!-\!1}}{\b{n\!-\!1 \, n}}
+
2\frac{\b{i\,n}}{\b{n \, 1}} \frac{\b{n\,1}\b{i\,k}\b{k\,n\!-\!1}}{\b{n\!-\!1 \, n}}
 \ = \  \frac{\b{1 \, i} \b{i \, k} \b{k \, n } }{\b{n \, 1 }}\, .
\end{split}
\end{align}

\item $(\eta^{-,i} \cdot \eta^{-,j})(\eta^{-,k})^2, \quad {\rm with} \quad i< j<k$ :
\begin{align}
\begin{split}
&2\frac{\b{n\,i}\b{j\,k}\b{k\,n\!-\!1}}{\b{n\!-\!1\,n}}
+
2\frac{\b{n\,1}\b{1\,i}\b{j\,n\!-\!1}}{\b{n\!-\!1\,n}}\frac{\b{k \, n}^2 }{\b{n \, 1}^2}
+
2\frac{\b{n\,1}\b{1\,k}\b{k\,n\!-\!1}}{\b{n\!-\!1\,n}}\frac{\b{i \, n} \b{j\, n} }{\b{n \, 1}^2}\\
-&2 \frac{\b{j\,n}\b{k\,n}}{\b{n \, 1}^2} \frac{\b{n\,1}\b{1\,i}\b{k\,n\!-\!1}}{\b{n\!-\!1 \, n}}
-2 \frac{\b{i\,n}\b{k\,n}}{\b{n \, 1}^2} \frac{\b{n\,1}\b{1\,j}\b{k\,n\!-\!1}}{\b{n\!-\!1 \, n}}\\
-&
2\frac{\b{k\,n}}{\b{n \, 1}} \frac{\b{n\,1}\b{i\,j}\b{k\,n\!-\!1}}{\b{n\!-\!1 \, n}} 
+
2\frac{\b{j\,n}}{\b{n \, 1}} \frac{\b{n\,1}\b{i\,k}\b{k\,n\!-\!1}}{\b{n\!-\!1 \, n}}
+
2\frac{\b{i\,n}}{\b{n \, 1}} \frac{\b{n\,1}\b{j\,k}\b{k\,n\!-\!1}}{\b{n\!-\!1 \, n}}\\
 \ = \  &2 \frac{\b{1 \, i} \b{j \, k} \b{k \, n } }{\b{n \, 1 }} \, .
\end{split}
\end{align}

\item $(\eta^{-,i} \cdot \eta^{-,j})(\eta^{-,k})^2, \quad {\rm with} \quad k<i< j $ :
\begin{align}
\begin{split}
&2\frac{\b{n\,k}\b{k\,i}\b{j\,n\!-\!1}}{\b{n\!-\!1\,n}}
+
2\frac{\b{n\,1}\b{1\,i}\b{j\,n\!-\!1}}{\b{n\!-\!1\,n}}\frac{\b{k \, n}^2 }{\b{n \, 1}^2}
+
2\frac{\b{n\,1}\b{1\,k}\b{k\,n\!-\!1}}{\b{n\!-\!1\,n}}\frac{\b{i \, n} \b{j\, n} }{\b{n \, 1}^2}\\
-&2 \frac{\b{k\,n}\b{j\,n}}{\b{n \, 1}^2} \frac{\b{n\,1}\b{1\,k}\b{i\,n\!-\!1}}{\b{n\!-\!1 \, n}}
-2 \frac{\b{k\,n}\b{i\,n}}{\b{n \, 1}^2} \frac{\b{n\,1}\b{1\,k}\b{j\,n\!-\!1}}{\b{n\!-\!1 \, n}}
\\
+&
4\frac{\b{k\,n}}{\b{n \, 1}} \frac{\b{n\,1}\b{k\,i}\b{j\,n\!-\!1}}{\b{n\!-\!1 \, n}} 
 \ = \ 2 \frac{\b{1 \, k} \b{k \, i} \b{j \, n } }{\b{n \, 1 }} \, .
\end{split}
\end{align}

\item $(\eta^{-,i} \cdot \eta^{-,j})(\eta^{-,k})^2, \quad {\rm with} \quad i<k< j$ : 
\begin{align}
\begin{split}
&
2\frac{\b{n\,1}\b{1\,i}\b{j\,n\!-\!1}}{\b{n\!-\!1\,n}}\frac{\b{k \, n}^2 }{\b{n \, 1}^2}
+
2\frac{\b{n\,1}\b{1\,k}\b{k\,n\!-\!1}}{\b{n\!-\!1\,n}}\frac{\b{i \, n} \b{j\, n} }{\b{n \, 1}^2}\\
-&2 \frac{\b{k\,n}\b{j\,n}}{\b{n \, 1}^2} \frac{\b{n\,1}\b{1\,i}\b{k\,n\!-\!1}}{\b{n\!-\!1 \, n}}
-2 \frac{\b{k\,n}\b{i\,n}}{\b{n \, 1}^2} \frac{\b{n\,1}\b{1\,k}\b{j\,n\!-\!1}}{\b{n\!-\!1 \, n}}
\\
- &
2\frac{\b{k\,n}}{\b{n \, 1}} \frac{\b{n\,1}\b{i\,k}\b{j\,n\!-\!1}}{\b{n\!-\!1 \, n}}+
2\frac{\b{j\,n}}{\b{n \, 1}} \frac{\b{n\,1}\b{i\,k}\b{k\,n\!-\!1}}{\b{n\!-\!1 \, n}} 
 \ = \ 0  \, .
\end{split}
\end{align}

\item $(\eta^{-,i} \cdot \eta^{-,j})(\eta^{-,k} \cdot \eta^{-,l}), \quad {\rm with} \quad i<j<k<l$ :
\begin{align}
\begin{split}
&
4\frac{\b{n\,i}\b{j\,k}\b{l\,n\!-\!1}}{\b{n\!-\!1\,n}}
+
4\frac{\b{n\,1}\b{1\,i}\b{j\,n\!-\!1}}{\b{n\!-\!1\,n}}\frac{\b{k \, n} \b{l\, n} }{\b{n \, 1}^2}
+
4\frac{\b{n\,1}\b{1\,k}\b{l\,n\!-\!1}}{\b{n\!-\!1\,n}}\frac{\b{i \, n} \b{j\, n} }{\b{n \, 1}^2}\\
+& 4 \frac{\b{j\,n}}{\b{n \, 1}} \frac{\b{n\,1}\b{i\,k}\b{l\,n\!-\!1}}{\b{n\!-\!1 \, n}}
+4 \frac{\b{i\,n}}{\b{n \, 1}} \frac{\b{n\,1}\b{j\,k}\b{l\,n\!-\!1}}{\b{n\!-\!1 \, n}} 
-4 \left[ \frac{\b{k\,n}}{\b{n \, 1}} \frac{\b{n\,1}\b{i\,j}\b{l\,n\!-\!1}}{\b{n\!-\!1 \, n}}\right. \\
+&
\left.\frac{\b{i\,n}\b{k\,n}}{\b{n \, 1}^2 } \frac{\b{n\,1}\b{1\,j}\b{l\,n\!-\!1}}{\b{n\!-\!1 \, n}} 
+
\frac{\b{j\,n}\b{l\,n}}{\b{n \, 1}^2 } \frac{\b{n\,1}\b{1\,i}\b{k\,n\!-\!1}}{\b{n\!-\!1 \, n}} \right]
 \ = \  
 4\frac{\b{1\,i}\b{j\,k}\b{l\,n}}{\b{n \, 1}} \, .
\end{split}
\end{align}

\item $(\eta^{-,i} \cdot \eta^{-,j})(\eta^{-,k} \cdot \eta^{-,l}), \quad {\rm with} \quad i<k<j<l$ :
\begin{align}
\begin{split}
&
4\frac{\b{n\,1}\b{1\,i}\b{j\,n\!-\!1}}{\b{n\!-\!1\,n}}\frac{\b{k \, n} \b{l\, n} }{\b{n \, 1}^2}
+
4\frac{\b{n\,1}\b{1\,k}\b{l\,n\!-\!1}}{\b{n\!-\!1\,n}}\frac{\b{i \, n} \b{j\, n} }{\b{n \, 1}^2}\\
+ & 4 \frac{\b{j\,n}}{\b{n \, 1}} \frac{\b{n\,1}\b{i\,k}\b{l\,n\!-\!1}}{\b{n\!-\!1 \, n}} 
-4 \left[ \frac{\b{l\,n}}{\b{n \, 1}} \frac{\b{n\,1}\b{i\,k}\b{j\,n\!-\!1}}{\b{n\!-\!1 \, n}}\right. \\
+ &
\frac{\b{k\,n}\b{j\,n}}{\b{n \, 1}^2 } \frac{\b{n\,1}\b{1\,i}\b{l\,n\!-\!1}}{\b{n\!-\!1 \, n}} 
+
\left.\frac{\b{i\,n}\b{l\,n}}{\b{n \, 1}^2 } \frac{\b{n\,1}\b{1\,k}\b{j\,n\!-\!1}}{\b{n\!-\!1 \, n}} \right]
 \ = \ 
0\, .
\end{split}
\end{align}

\end{itemize}
Thus we have shown that all terms $(\eta^{-,i} \cdot \eta^{-,j})(\eta^{-,k} \cdot \eta^{-,l})$ with the right ordering, namely when $i \leq j < k \leq l$, have the correct coefficients, whereas when $i,j,k,l$ are in a wrong ordering the corresponding coefficients vanish. This completes the proof of the cyclicity of $\F_{4,n}$.

\section{Explicit computation of $ F^{(1)}_{3}(1^{\phi_{12}},2^{\phi_{12}},3^{\phi_{12}},4^+;q)$}
\label{app:1loopComponent}

In this section we compute a particular component of a four-point form factor of $\O_3$ at one loop, namely $ F^{(1)}_{3}(1^{\phi_{12}},2^{\phi_{12}},3^{\phi_{12}},4^+;q)$. We show that, after many cancellations between the IR divergent parts of one-mass triangles and box-functions, it matches the structure \eqref{eq:all-1loop}. In order to do so, we will compute the discontinuity across all kinematic channels and at the end lift the cut integrals off shell. We show every step of the computation in detail as it might be useful for a reader who is learning how to do them for the first time. However, we recommend the more experienced reader to skip to the summary of the cuts shown \sref{sec:summary-cuts}.

\subsection*{$\boldsymbol{(q-p_1)^2}$-channel}

\begin{figure}[htb]
\centering
\includegraphics[scale=0.8]{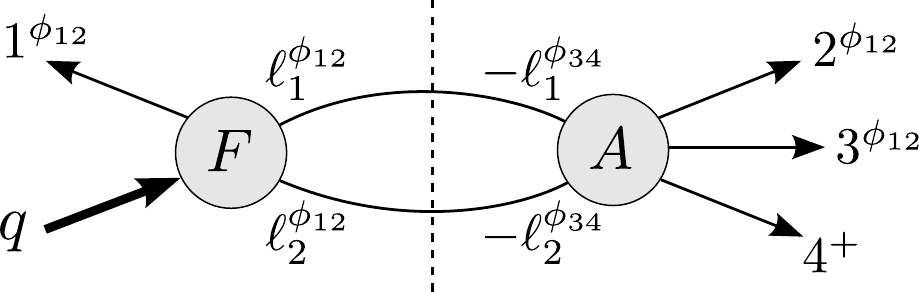}
\caption{\it Cut along the $(q-p_1)^2 $-channel for $F^{(1)}_{3}(1^{\phi_{12}},2^{\phi_{12}},3^{\phi_{12}},4^+;q)$.}
\label{fig:cut4-1}
\end{figure}

\begin{figure}[htb]
\centering
\includegraphics[width=0.8\linewidth]{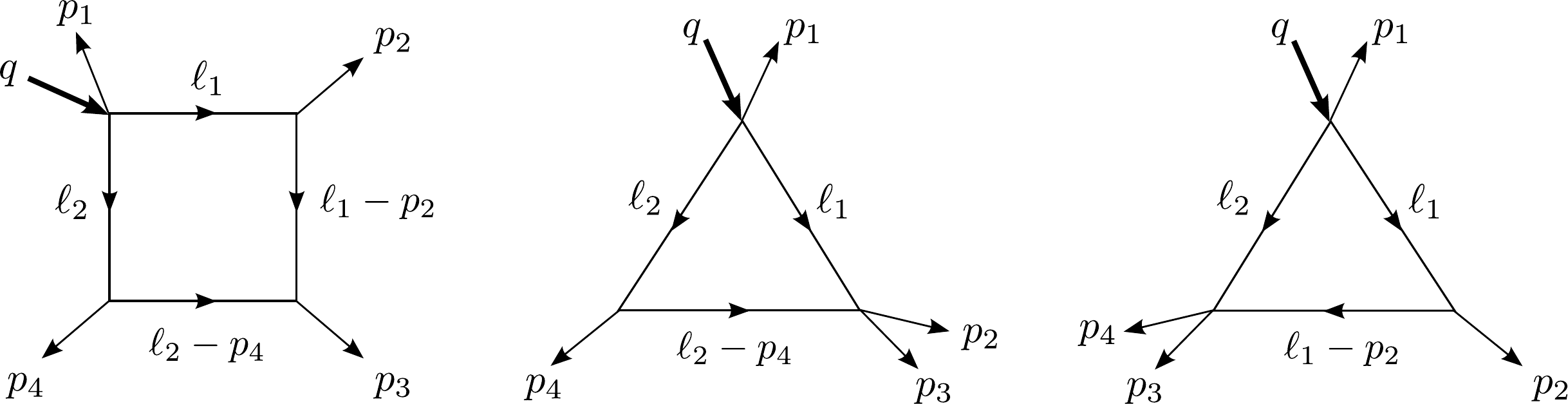}
\caption{\it One-loop result for the $(q-p_1)^2$-channel of $F^{(1)}_{3}(1^{\phi_{12}},2^{\phi_{12}},3^{\phi_{12}},4^+;q)$. }
\label{fig:q-p1}
\end{figure}

We start by inspecting the $(q-p_1)^2$-channel, where there is only one contribution given by
\begin{equation}
\int d\text{LIPS}(\ell_1,\ell_2;P)F^{(0)}_{3}(1^{\phi_{12}},\ell_1^{\phi_{12}},\ell_2^{\phi_{12}};q) A^{\text{MHV}}(2^{\phi_{12}},3^{\phi_{12}},4^+,-\ell_2^{\phi_{34}},-\ell_1^{\phi_{34}}), \quad P=q-p_1\ .
\end{equation}
This is shown in \fref{fig:cut4-1}, where the helicities are assigned assuming all particles outgoing.
\\
We will look at the integrand and use the fact that we are on the cut, so
\begin{equation}
\ell_1+\ell_2=q-p_1,\quad\ell_1^2=\ell_2^2=0\ .
\end{equation} 
Plugging in the tree level expressions for the form factor and amplitude, and factoring out $ F^{(0)}=\dfrac{\b{31}}{\b{34}\b{41}}$, we get
\begin{align}
\label{eq:boson1loop1}
\frac{1}{4} F^{(0)} \frac{\b{23}\b{41}[\ell_1\,2][4\,\ell_2]\b{\ell_2\,\ell_1}}{ \b{31}(p_2\cdot \ell_1)(p_4\cdot \ell_2)} = -\frac{1}{4}\left(\dfrac{\b{23}\b{14}}{\b{24}\b{13}}\right)    \frac{\text{Tr}_+(\ell_1\, \ell_2 \,  p_4 \,  p_2)}{(p_2\cdot \ell_1)(p_4\cdot \ell_2)}\ ,
\end{align}
where we used \eqref{eq:trace-expand}. On the cut, the trace can be written as 
\begin{align}
\begin{split}
\label{eq:trace4}
\text{Tr}_+(\ell_1 \, \ell_2 \, p_4 \, p_2)&=\text{Tr}_+(\ell_1 \, P \, p_4 \, p_2)\\
=  (\ell_1\cdot P) s_{24} + 2(\ell_1\cdot p_2)&(p_4\cdot P) - 2(\ell_1\cdot p_4)(p_2\cdot P) \ .
\end{split}
\end{align}
Noting that $(\ell_1\cdot P)=(\ell_1\cdot\ell_2)=\tfrac{1}{2} P^2$ and writing the last term in \eqref{eq:trace4} as
\begin{equation}
- 2(\ell_1\cdot p_4)(p_2\cdot P)=2 (\ell_2\cdot p_4)(p_2\cdot P)  - 2  (P \cdot p_4)(p_2\cdot P)\ ,
\end{equation}
we can recast \eqref{eq:boson1loop1} as
\begin{align}
\label{eq:1-loop-4-cut1}
\begin{split}
\left.F^{(1)}\right|_{(q-p_1)^2\text{-cut}}&= \\
\frac{1}{2} F^{(0)}&\left(\frac{\b{41}\b{23}}{\b{13}\b{24}}\right) \left(\frac{s_{23}s_{34}}{(\ell_1-p_2)^2 (\ell_2-p_4)^2} + \frac{(q-p_1)^2-s_{23} }{(\ell_2-p_4)^2 } + \frac{(q-p_1)^2-s_{34} }{(\ell_1-p_2)^2 } \right)\ ,
\end{split}
\end{align}
where we rewrote the numerators using
\begin{align}
\begin{split}
\frac{1}{2}s_{24} P^2-2 (p_2\cdot P) (p_4\cdot P)\,=\,-\frac{1}{2}s_{23}s_{34}&\ ,\\
 2(p_4\cdot P)\,=\,s_{24}+s_{34}=(q-p_1)^2-s_{23}&\ ,\\
 2(p_2\cdot P)\,=\,s_{23}+s_{24}=(q-p_1)^2-s_{34}&\ .
 \end{split}
\end{align}
We can immediately recognise \eqref{eq:1-loop-4-cut1} as a sum of a one-mass box with massive corners $q-p_1$ and two two-mass triangles with massless corners $p_4$ and $p_2$ respectively, as shown in \fref{fig:q-p1}.

Notice that the overall factor of \eqref{eq:1-loop-4-cut1} is not the plain tree-level form factor, but there is also a cross ratio $\dfrac{\b{41}\b{23}}{\b{13}\b{24}}$. If we had computed a form factor with different helicity configuration, say we take $F^{(1)}_{3}(1^{\phi_{12}},2^+,3^{\phi_{12}},4^{\phi_{12}};q)$ then we would find that the result is \textit{almost} the same as the present one, except that the overall cross ratio is different. To show this we need to look at the factorisation
\begin{equation}
\label{eq:boson1loop2}
\int d\text{LIPS}(\ell_1,\ell_2;P)F^{(0)}_{3}(1^{\phi_{12}},\ell_1^{\phi_{12}},\ell_2^{\phi_{12}};q) A^{\text{MHV}}(2^+,3^{\phi_{12}},4^{\phi_{12}},-\ell_2^{\phi_{34}},-\ell_1^{\phi_{34}})\ .
\end{equation}
This time $F^{(0)}=\dfrac{\b{23}}{\b{12}\b{23}}$, so factoring this out we have that the integrand of \eqref{eq:boson1loop2} is
\begin{align}
\frac{1}{4}F^{(0)} \left(\dfrac{\b{34}\b{12}}{\b{13}\b{24}}\right)    \frac{\text{Tr}_+(\ell_2 \, \ell_1 \, p_2 \, p_4)}{(p_2\cdot \ell_1)(p_4\cdot \ell_2)}\ .
\end{align}
Since the contractions we get from $\text{Tr}_+(P \,\ell_1 \, p_2 \, p_4)$ are identical to the ones we had before, that is,  $\text{Tr}_+(\ell_1 \, P  \, p_4  \, p_2)$, we know that the integrand we obtain is the same, but we have a different overall cross ratio in comparison with \eqref{eq:1-loop-4-cut1}.

Lastly, one can check that the form factor $F^{(1)}_{3}(1^{\phi_{12}},2^{\phi_{12}},3^+,4^{\phi_{12}};q)$ comes yet with another cross ratio, in this case just ``1''. In conclusion, the tree level expression cannot be factored out in the one-loop correction of super form factor $\F^{\text{MHV}}_{3,n}$ as happens for the form factor of the chiral part of the stress tensor, $\T_2$ \cite{Brandhuber:2011tv}. This is not unexpected since for $\T_2$ the tree level formula has a much simpler numerator, namely just super-momentum conservation. In our case, however, although we cannot pull out the tree-level super form factor $\F_{3,4}^{(0)}$ from $\F_{3,4}^{(1)}$, this can be done without too much effort provided we choose a particular component, as we will see explicitly in \sref{sec:summary-cuts}.

Let us focus on the helicity configuration we started with, $\{1^{\phi_{12}},2^{\phi_{12}},3^{\phi_{12}},4^+\}$, and obtain the full one-loop result by examining all kinematic channels.

\subsection*{$\boldsymbol{(q-p_2)^2}$-channel}
In the $(q-p_2)^2$ there is only one factorisation
\begin{align}
&F^{(0)}_{3}(2^{\phi_{12}}, \ell_1^{\phi_{12}},\ell_2^{\phi_{12}};q) A^{\text{MHV}}(3^{\phi_{12}},4^+,1^{\phi_{12}},-\ell_2^{\phi_{34}},-\ell_1^{\phi_{34}})\ ,
\end{align}
which is given by
\begin{align}
\begin{split}
&F^{(0)}\frac{\b{34}\b{41}}{\b{31}}\frac{\b{\ell_1\,\ell_2}^2\b{13}^2}{\b{\ell_2\,\ell_1}\b{\ell_1\,3}\b{34}\b{41}\b{1\ell_2}}\,=\, \frac{1}{4} F^{(0)}\frac{\b{\ell_1\,\ell_2}\b{13}[\ell_1\,3][1\ell_2]}{(\ell_1\cdot p_3)(p_1\cdot\ell_2)}\\
&=\,\frac{1}{4} F^{(0)}\frac{\Tr_+(p_1 \, p_3  \, \ell_1  \, \ell_2 )}{(\ell_1\cdot p_3)(p_1\cdot\ell_2)}\,=\,\frac{1}{4} F^{(0)}\frac{\Tr_+(p_1 \, p_3  \, \ell_1  \, P)}{(\ell_1\cdot p_3)(p_1\cdot\ell_2)},\qquad P\,=\, q-p_2\ .
\end{split}
\end{align}
The trace is given by
\begin{align}
\Tr_+(p_1 \, p_3  \, \ell_1  \, P) &\,=\, s_{13}(\ell_1\cdot P) + 2 (\ell_1\cdot p_3) (p_1\cdot P) -2 (p_1 \cdot \ell_1)(p_3\cdot P) \ .
\end{align}
Using $(\ell_1\cdot P)=\tfrac{1}{2}P^2$ for the first term and $(p_1\cdot \ell_1)= (p_1\cdot P)-(p_1\cdot \ell_2)$ on the last we get
\begin{align}
\label{eq:result-q-p2-cut}
\left.F^{(1)}\right|_{(q-p_2)^2\text{-cut}}\,=\,-\frac{1}{2}F^{(0)} \left(\frac{s_{41}s_{340}}{(\ell_1-p_3)^2(\ell_2-p_1)^2}+\frac{(q-p_2)^2-s_{34}}{(\ell_2-p_1)^2}+\frac{(q-p_2)^2-s_{41}}{(\ell_1-p_3)^2}\right)\ ,
\end{align}
where we simplified the numerator using
\begin{align}
\begin{split}
\frac{1}{2}s_{13}P^2-2(p_1\cdot P)(p_3\cdot P)=-\frac{1}{2}s_{41}s_{34}&\ ,\\
2(p_1\cdot P)= s_{13}+s_{41}=(q-p_2)^2-s_{34}&\ ,\\
2(p_3\cdot P) = s_{13}+s_{34}=(q-p_2)^2-s_{41}&\ .
\end{split}
\end{align}

\subsection*{$\boldsymbol{(q-p_3)^2}$-channel}
In the $(q-p_3)^2$ there is only one factorisation
\begin{align}
&F^{(0)}_{3}(3^{\phi_{12}}, \ell_1^{\phi_{12}},\ell_2^{\phi_{12}};q) A^{\text{MHV}}(4^+,1^{\phi_{12}},2^{\phi_{12}},-\ell_2^{\phi_{34}},-\ell_1^{\phi_{34}})\ ,
\end{align}
which is given by
\begin{align}
\begin{split}
&F^{(0)}\frac{\b{34}\b{41}}{\b{31}} \frac{\b{\ell_2\,\ell_1}\b{12}}{\b{41}\b{2\,\ell_2} \b{\ell_1\,4}}=  \frac{1}{4} F^{(0)}\frac{\b{34}\b{12} \b{\ell_2\,\ell_1}[\ell_1\,4]\b{42}[2\ell_2]}{\b{31}\b{42}(p_4\cdot \ell_1)(p_2\cdot\ell_2)}\\
=&\frac{1}{4} F^{(0)}\left(\frac{\b{12}\b{34}}{\b{13}\b{24}}\right) \frac{\Tr_+(p_4 \, p_2  \, \ell_2  \, \ell_1 )}{(p_4\cdot \ell_1)(p_2\cdot\ell_2)} \ .
\end{split}
\end{align}
As before, we use that on the cut $P\equiv q-p_3=\ell_1+\ell_2 $ to rewrite the trace as
\begin{align}
\Tr_+(p_4 \, p_2  \, \ell_2  \, P ) &\,=\, s_{24}(\ell_2 \cdot P) + 2 (p_2\cdot \ell_2) (p_4\cdot P) -2 (p_4 \cdot \ell_2)(p_2\cdot P)\ .
\end{align}
Using $(\ell_2\cdot P)=\tfrac{1}{2}P^2$ for the first term and $(p_4\cdot \ell_2)= (p_4\cdot P)-(p_4\cdot \ell_1)$ on the last we get
{\small
\begin{align}
\label{eq:result-q-p3-cut}
\left.F^{(1)}\right|_{(q-p_3)^2\text{-cut}}=-\frac{1}{2}F^{(0)}\left(\frac{\b{12}\b{34}}{\b{13}\b{24}}\right) \left(\frac{s_{41}s_{12}}{(\ell_1-p_4)^2(\ell_2-p_2)^2}+\frac{(q-p_3)^2-s_{12}}{(\ell_1-p_4)^2}+\frac{(q-p_3)^2-s_{41}}{(\ell_2-p_2)^2}\right),
\end{align}}
where we simplified the numerator using
\begin{align}
\begin{split}
\frac{1}{2}s_{24}P^2-2(p_4\cdot P)(p_2\cdot P)=-\frac{1}{2}s_{41}s_{12}&\ ,\\
2(p_4\cdot P)= s_{41}+s_{24}=(q-p_3)^2-s_{12}&\ ,\\
2(p_2\cdot P) = s_{12}+s_{24}=(q-p_3)^2-s_{41}&\ .
\end{split}
\end{align}

\subsection*{$\boldsymbol{(q-p_4)^2}$-channel}

The cut across the $(q-p_4)^2$-channel is identically zero as there is no consistent helicity assignments for the internal momenta.

\subsection*{$\boldsymbol{(q-p_1-p_2)^2}$-channel}

\begin{figure}[htb]
\centering
\includegraphics[width=0.85\linewidth]{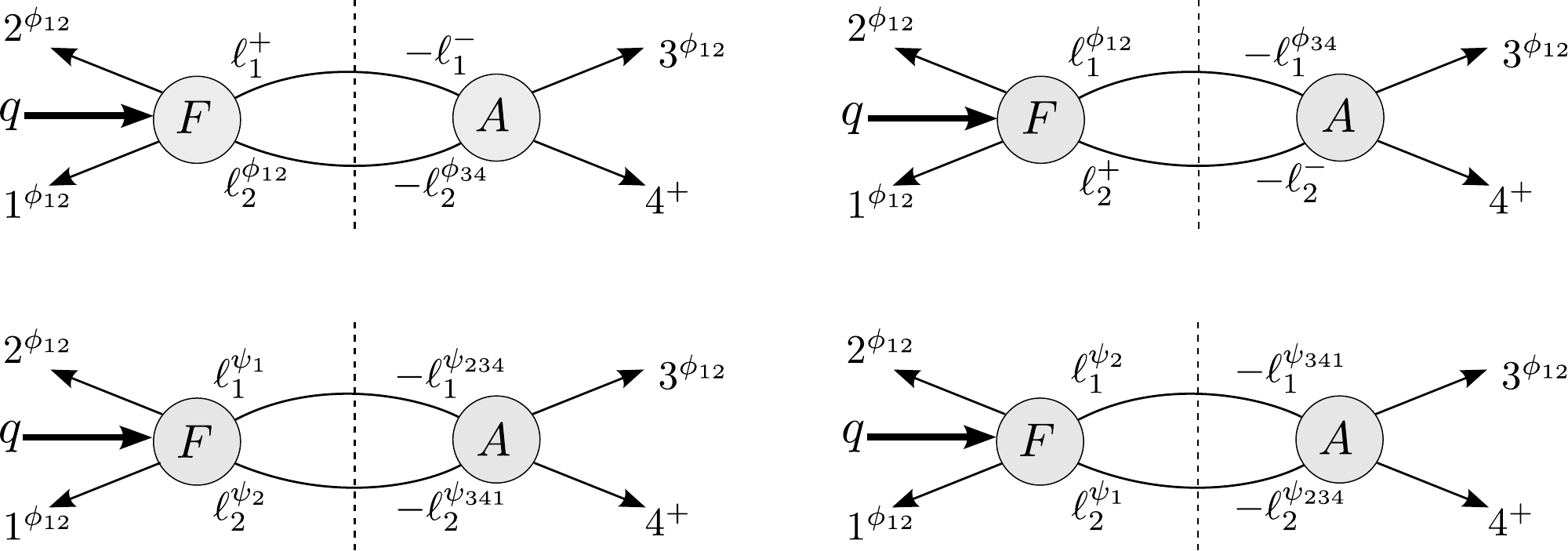}
\caption{\it Four helicity configurations present in the cut across the $(q-p_1-p_2)^2$ cut of $ F^{(1)}_{3}(1^{\phi_{12}},2^{\phi_{12}},3^{\phi_{12}},4^+;q)$.}
\label{fig:cut-4-2}
\end{figure}

\begin{figure}[htb]
\centering
\includegraphics[width=0.8\linewidth]{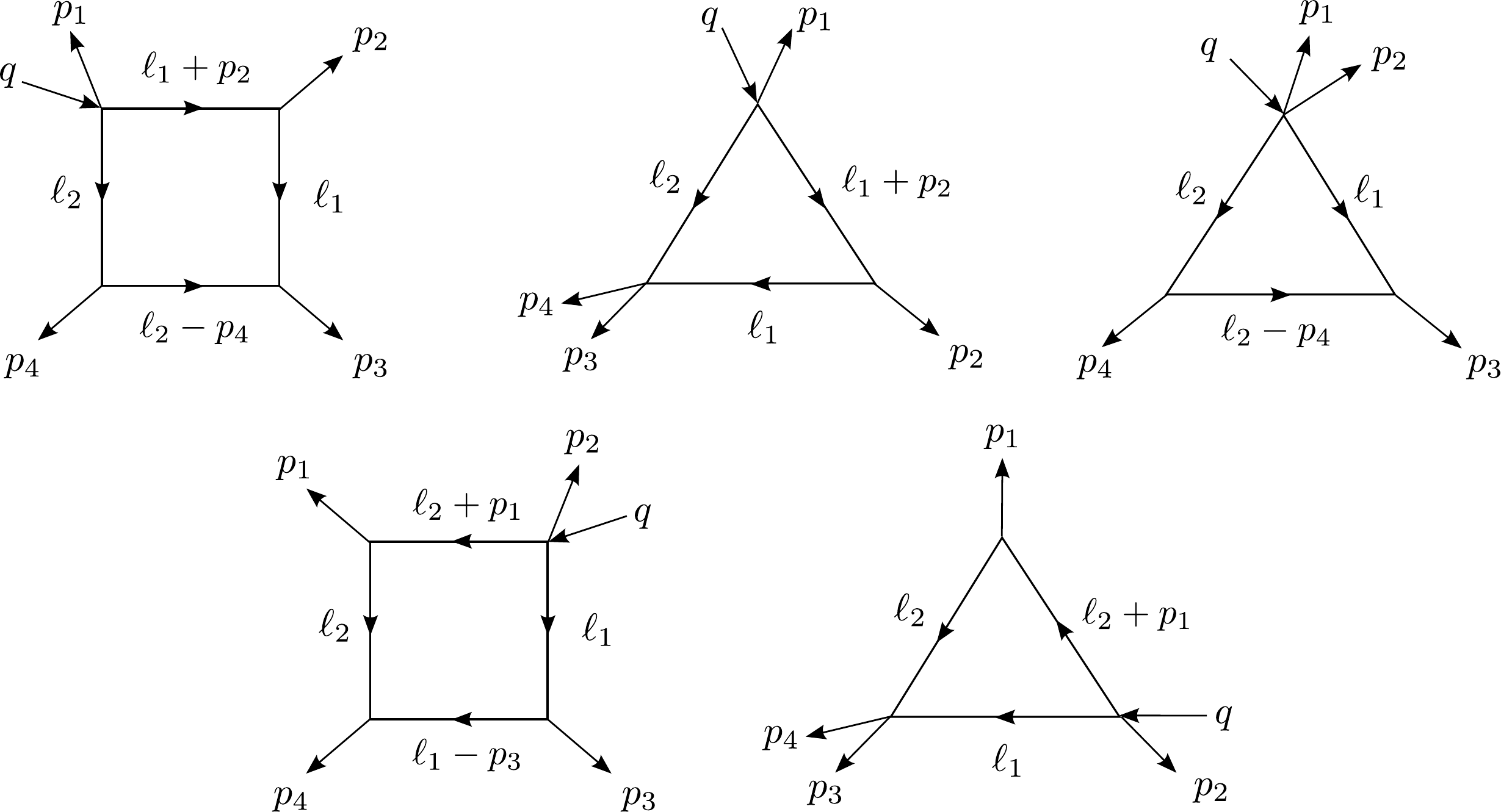}
\caption{\it One-loop result for the $(q-p_1-p_2)^2$-channel of $F^{(1)}_{3}(1^{\phi_{12}},2^{\phi_{12}},3^{\phi_{12}},4^+;q)$. }
\label{fig:q-p1-p2}
\end{figure}

Here we compute the discontinuity of $ F^{(1)}_{3}(1^{\phi_{12}},2^{\phi_{12}},3^{\phi_{12}},4^+;q)$ across the $(q-p_1-p_2)^2$ cut. There are four cases one must consider:
\begin{align}
\label{eq:1loop4-b1}
&F^{(0)}_{3}(1^{\phi_{12}}, 2^{\phi_{12}}, \ell_1^{+},\ell_2^{\phi_{12}};q) A^{\text{MHV}}(3^{\phi_{12}},4^+,-\ell_2^{\phi_{34}},-\ell_1^{-})\ ,\\
\label{eq:1loop4-b2}
&F^{(0)}_{3}(1^{\phi_{12}}, 2^{\phi_{12}}, \ell_1^{\phi_{12}},\ell_2^{+};q) A^{\text{MHV}}(3^{\phi_{12}},4^+,-\ell_2^{-},-\ell_1^{\phi_{34}})\ ,\\
\label{eq:1loop4-b3}
&F^{(0)}_{3}(1^{\phi_{12}}, 2^{\phi_{12}}, \ell_1^{\psi_{1}},\ell_2^{\psi_{2}};q) A^{\text{MHV}}(3^{\phi_{12}},4^+,-\ell_2^{\psi_{341}},-\ell_1^{\psi_{234}})\ ,\\
\label{eq:1loop4-b4}
&F^{(0)}_{3}(1^{\phi_{12}}, 2^{\phi_{12}}, \ell_1^{\psi_{2}},\ell_2^{\psi_{1}};q) A^{\text{MHV}}(3^{\phi_{12}},4^+,-\ell_2^{\psi_{234}},-\ell_1^{\psi_{341}})\ ,
\end{align}
where above we ommited the integration over the phase space:
\begin{equation}
\int d\text{LIPS}(\ell_1,\ell_2;P), \qquad P=q-p_1-p_2 \ .
\end{equation}
The four cases above are shown on \fref{fig:cut-4-2}.
\\
We start with \eqref{eq:1loop4-b1}. Keeping in mind $ F^{(0)}=\dfrac{\b{31}}{\b{34}\b{41}} $, we plug in the tree level expressions to  get:
\begin{align}
\label{eq:1-1loop4}
\frac{\b{2\,\ell_2}}{\b{2\,\ell_1}\b{\ell_1\,\ell_2}} \frac{\b{\ell_1\,3}\b{\ell_2\,\ell_1}}{\b{34}\b{4\,\ell_2}}\, =\, 
F^{(0)} \frac{\b{41} \b{2\,\ell_2} \b{3\,\ell_1}}{\b{31}\b{2\,\ell_1}\b{4\,\ell_2}}\ .
\end{align}
Analogously, for \eqref{eq:1loop4-b2} we obtain
\begin{align}
\label{eq:2-1loop4}
\frac{\b{\ell_1\,1}}{\b{\ell_1\,\ell_2}\b{\ell_2\,1}} \frac{\b{3\,\ell_2}^2\b{\ell_2\,\ell_1}}{\b{34}\b{4\,\ell_2}\b{\ell_1\,3}}\, = \,
F^{(0)} \frac{\b{41} \b{3\,\ell_2}^2 \b{1\,\ell_1}}{\b{31}\b{\ell_2\,1}\b{4\,\ell_2}\b{\ell_1\,3}}\ .
\end{align}
For \eqref{eq:1loop4-b3} and \eqref{eq:1loop4-b4}, let us first investigate the tree level form factors with two scalars and two fermions: $F_{3}(1^{\phi_{12}},2^{\phi_{12}},3^{\psi_1},4^{\psi_2};q)$ and $F_{3}(1^{\phi_{12}},2^{\phi_{12}},3^{\psi_2},4^{\psi_1};q)$.
Starting from the supersymmetric expression, which can be written as
\begin{equation}
\label{eq:2fermions}
-\prod_{a=1}^2\delta^{(2)}(\lambda^1\eta^{a,1}+\lambda^2\eta^{a,2}+ \lambda^3\eta^{a,3}+\lambda^4\eta^{a,4}) \left(\frac{\eta_1^1\eta_1^2}{\b{23}\b{34}\b{42}}+\frac{\eta_3^1\eta_3^2}{\b{12}\b{24}\b{41}}\right)\ .
\end{equation}
We are interested in the coefficients of $ -(\eta^{1,1} \eta^{1,2} \eta^{1,3} \;\, \eta^{2,1} \eta^{2,2} \eta^{2,4}) $ and $ \eta^{1,1} \eta^{1,2} \eta^{1,4}\;\, \eta^{2,1} \eta^{2,2} \eta^{2,3} $, so we can neglect the second term in the sum and set $\eta_1\rightarrow 0$ in the $\delta$-functions. Now we can rewrite \eqref{eq:2fermions} as:
\begin{align}
-\frac{\prod_{a=1}^2\delta(\eta^{a,2}\b{23}+\eta^{a,4}\b{43}) \delta(\eta^{a,2}\b{24}+\eta^{a,3}\b{34})}{\b{34}^2} \frac{\eta_1^1\eta_1^2}{\b{23}\b{34}\b{42}},
\end{align}
so the term proportional to $\eta^{1,1}\eta^{1,2}\;\,\eta^{2,1}\eta^{2,2}$ is given by 
$
\dfrac{-\eta^{1,3}\eta^{2,4}+\eta^{2,3}\eta^{1,4}}{\b{34}}$, thus we have the result for the form factor with fermions:
\begin{align}
F_{3}(1^{\phi_{12}},2^{\phi_{12}},3^{\psi_1},4^{\psi_2};q)=\frac{1}{\b{34}}=- F_{3}(1^{\phi_{12}},2^{\phi_{12}},3^{\psi_2},4^{\psi_1};q)\ .
\end{align}
Now we use this result to compute \eqref{eq:1loop4-b3} and \eqref{eq:1loop4-b4}, which give identical results (a minus sign from the form factor is compensated by a form factor coming from the amplitude):
\begin{align}
\label{eq:fermion-a}
\frac{1}{\b{\ell_2 \,\ell_1}}\left(-\frac{\b{3\,\ell_2}\b{\ell_2\,\ell_1}}{\b{34}\b{4\,\ell_2}}\right)&=- F^{(0)} \frac{\b{41}\b{3\,\ell_2}}{\b{31}\b{4\,\ell_2}}\ .
\end{align}
We can combine the two diagrams with fermions with the first two diagrams which do not involve fermions. Consider first  \eqref{eq:1-1loop4} summed with 
\begin{align}
\begin{split}
\label{eq:diagwfermion1}
F^{(0)}\frac{\b{41}}{\b{31}\b{4\,\ell_2}} &\left(\frac{\b{2\,\ell_2} \b{3\,\ell_1}-\b{2\,\ell_1}\b{3\,\ell_2}}{\b{2\,\ell_1}} \right)= F^{(0)}\frac{\b{41}\b{23}\b{\ell_2\,\ell_1}}{\b{31}\b{4\,\ell_2}\b{2\,\ell_1}}\\
= \frac{1}{4}&F^{(0)}\left(\frac{\b{41}\b{23}}{\b{31}\b{42}}\right)\frac{\Tr_+(\ell_2 \, \ell_1 \, p_2  \, p_4)}{(p_2\cdot \ell_1)(p_4\cdot \ell_2)}\ .
\end{split}
\end{align}
In the cut, $\ell_1+\ell_2=p_3+p_4$, so
\begin{align}
\begin{split}
&\Tr_+(\ell_2 \, \ell_1  \, p_2  \, p_4)=\Tr_+(p_3 \, \ell_1 \, p_2  \, p_4)\\
=\,&  s_{24} (p_3\cdot\ell_1)+  s_{34} (p_2\cdot\ell_1) -  s_{23}(\ell_1\cdot p_4) \\
=\,& ( s_{24}+s_{23}) (p_3\cdot\ell_1)+  s_{34} (p_2\cdot\ell_1) - \tfrac{1}{2}s_{23}s_{34}\ .
\end{split}
\end{align}
Here we used $ 2(\ell_1\cdot p_4) =-2 (\ell_1\cdot p_3) + s_{34} $ which holds on the cut.
Noticing that $ (p_3 \cdot \ell_1)= (p_4 \cdot \ell_2)$,  \eqref{eq:diagwfermion1} becomes
\begin{align}
\label{eq:q-p1-p2-1}
\begin{split}
&\frac{1}{2} F^{(0)}\left(\frac{\b{41}\b{23}}{\b{13}\b{24}}\right) \left(  \frac{  s_{24}+s_{23}}{(\ell_1+p_2)^2}-\frac{  s_{34}}{(\ell_2-p_4)^2}+\frac{ s_{23}s_{34}}{(\ell_1+p_2)^2(\ell_2-p_4)^2}\right)\\
=&\, \frac{1}{2} F^{(0)}\left(\frac{\b{41}\b{23}}{\b{13}\b{24}}\right) \left( \frac{ s_{23}s_{34}}{(\ell_1+p_2)^2(\ell_2-p_4)^2} + \frac{ (q-p_1)^2-s_{34}}{(\ell_1+p_2)^2}-\frac{ s_{34}}{(\ell_2-p_4)^2}\right).
\end{split}
\end{align}
Again we used $s_{23}+s_{24}=(q-p_1)^2$.
The first term is a one-mass box with massive corner $p_1-q$, the second term is a two-mass triangle with massive corners $p_3+p_4 $  and $p_1-q$ and lastly the third term is a one-mass triangle with massive corner $p_1+p_2-q$ (see \fref{fig:q-p1-p2}).

Now let us look at \eqref{eq:2-1loop4} summed with the other diagram with fermions:
\begin{align}
\begin{split}
&F^{(0)} \frac{\b{41}\b{3\,\ell_2}}{\b{31}\b{4\,\ell_2}}\left(\frac{\b{1\,\ell_1}\b{3\,\ell_2}-\b{\ell_2\,1}\b{\ell_1\,3}}{\b{\ell_2\,1}\b{\ell_1\,3}}\right)= F^{(0)} \frac{\b{41}\b{3\,\ell_2}\b{\ell_2\,\ell_1}}{\b{\ell_2\,1}\b{\ell_1\,3}\b{4\,\ell_2}}  \\
=\,  \frac{1}{4}&F^{(0)} \frac{\b{41}\b{3\,\ell_2}\b{\ell_2\,\ell_1}[\ell_2\,1][4\,\ell_2]}{\b{\ell_1\,3} (\ell_2\cdot p_1) (\ell_2\cdot p_4)}  =  \frac{1}{4}F^{(0)} \frac{\b{41}\b{\ell_2\,\ell_1}[\ell_2\,1][4\,\ell_1]}{ (\ell_2\cdot p_1) (\ell_2\cdot p_4)} \\
= \, \frac{1}{4} & F^{(0)}\frac{\Tr_+(p_4 \, p_1 \, \ell_2  \, \ell_1 )}{(\ell_2\cdot p_1)(\ell_2\cdot p_4)}  = \frac{1}{4}  F^{(0)}\frac{\Tr_+(p_4 \, p_1 \, \ell_2  \, p_3 )}{(\ell_2\cdot p_1) (\ell_2\cdot p_4)}\ .
\end{split}
\label{eq:diagwfermion2}
\end{align}\\[5pt]
On the second line we used that $\b{3\,\ell_2}[4\,\ell_2]=-\b{3\,\ell_1}[4\,\ell_1]$. The trace gives
\begin{align}
\begin{split}
\Tr_+(p_4 \, p_1 \, \ell_2  \, p_3 )\,=\,& s_{41}(\ell_2\cdot p_3)+s_{34}(\ell_2\cdot p_1) -s_{13}(\ell_2\cdot p_4)\\
=\,& \tfrac{1}{2} s_{41}s_{34}-(s_{41}+s_{13})(\ell_2\cdot p_4)+s_{34}(\ell_2\cdot p_1)\ .
\end{split}
\end{align}
Once again we used that, on this cut,
$ 2(\ell_2\cdot p_3) = -2(\ell_2\cdot p_4) + s_{34} $.
Thus \eqref{eq:diagwfermion2} is
\begin{align}
\label{eq:q-p1-p2-2}
-\frac{1}{2} F^{(0)} \left(\frac{ s_{34} s_{41}}{(\ell_2+p_1)^2(\ell_2-p_4)^2}+\frac{ s_{34}}{(\ell_2-p_4)^2}+\frac{ (q-p_2)^2 - s_{34} }{(\ell_2+p_1)^2} \right).
\end{align}
The first term gives a  a one-mass box with massive corner $p_2-q$, the second term gives a one-mass triangle with massive corner $p_1+p_2-q$ and finally the last term is a two mass triangle with massive corners $p_2-q$ and $p_3+p_4$.


Putting together \eqref{eq:q-p1-p2-1} and \eqref{eq:q-p1-p2-2} we get the result for the cut across the $(q-p_1-p_2)^2$-channel. There is only one function present in both expressions, which is the one-mass triangle with massive corner $q-p_1-p_2$. The final expression is
\begin{align}
\label{eq:result-q-p1-p2}
\begin{split}
\left.F^{(1)}\right|_{(q-p_1-p_2)^2\text{-cut}}= &\frac{1}{2} F^{(0)} \left\{\left(\frac{\b{41}\b{23}}{\b{13}\b{24}}\right)\right.  \left( \frac{ s_{23}s_{34}}{(\ell_1+p_2)^2(\ell_2-p_4)^2} + \frac{ (q-p_1)^2-s_{34}}{(\ell_1+p_2)^2}\right)\\
-& \left(\frac{\b{12}\b{34}}{\b{13}\b{24}}\right)\left. \frac{ s_{34}}{(\ell_2-p_4)^2}
-  \frac{ s_{34} s_{41}}{(\ell_2+p_1)^2(\ell_2-p_4)^2} - \frac{ (q-p_2)^2 - s_{34} }{(\ell_2+p_1)^2} \right\}.
\end{split}
\end{align}
The functions appearing in \eqref{eq:result-q-p1-p2} are shown in order in \fref{fig:q-p1-p2}.
\\[12pt]
Comparing the results of the $(q-p_1)^2$ and $(q-p_1-p_2)^2$ cuts, \eqref{eq:1-loop-4-cut1} and \eqref{eq:result-q-p1-p2}, we see that indeed the functions which are detectable on both channels appear with the same coefficient.

\subsection*{$\boldsymbol{(q-p_2-p_3)^2}$-channel}

Here we compute the discontinuity of $ F^{(1)}_{3}(1^{\phi_{12}},2^{\phi_{12}},3^{\phi_{12}},4^+;q)$ across the $(q-p_2-p_3)^2$ cut. There are four cases one must consider:
\begin{align}
\label{eq:1loop4-c1}
&F^{(0)}_{3}(2^{\phi_{12}}, 3^{\phi_{12}}, \ell_1^{+},\ell_2^{\phi_{12}};q) A^{\text{MHV}}(4^+,1^{\phi_{12}},-\ell_2^{\phi_{34}},-\ell_1^{-})\ ,\\
\label{eq:1loop4-c2}
&F^{(0)}_{3}(2^{\phi_{12}}, 3^{\phi_{12}}, \ell_1^{\phi_{12}},\ell_2^{+};q) A^{\text{MHV}}(4^+,1^{\phi_{12}},-\ell_2^{-},-\ell_1^{\phi_{34}})\ ,\\
\label{eq:1loop4-c3}
&F^{(0)}_{3}(2^{\phi_{12}}, 3^{\phi_{12}}, \ell_1^{\psi_{1}},\ell_2^{\psi_{2}};q) A^{\text{MHV}}(4^+,1^{\phi_{12}},-\ell_2^{\psi_{341}},-\ell_1^{\psi_{234}})\ ,\\
\label{eq:1loop4-c4}
&F^{(0)}_{3}(2^{\phi_{12}}, 3^{\phi_{12}}, \ell_1^{\psi_{2}},\ell_2^{\psi_{1}};q) A^{\text{MHV}}(4^+,1^{\phi_{12}},-\ell_2^{\psi_{234}},-\ell_1^{\psi_{341}})\ .
\end{align}
We start with \eqref{eq:1loop4-c1}. Keeping in mind that $ F^{(0)}=\dfrac{\b{31}}{\b{34}\b{41}} $, we plug in the tree level expressions to  get:
\begin{align}
\label{eq:1-1loop4c}
\frac{\b{3\,\ell_2}}{\b{3\,\ell_1}\b{\ell_1\,\ell_2}} \frac{\b{1\,\ell_1}^2\b{\ell_2\,\ell_1}}{\b{\ell_1\,4}\b{41}\b{1\,\ell_2}}\,=\, 
-F^{(0)} \frac{\b{34} \b{3\,\ell_2} \b{1\,\ell_1}^2}{\b{3\,\ell_1} \b{\ell_1\,4} \b{1\,\ell_2}\b{31}}\ .
\end{align}
Analogously, for \eqref{eq:1loop4-c2} we obtain
\begin{align}
\label{eq:2-1loop4c}
\frac{\b{\ell_1\,2}}{\b{\ell_1\,\ell_2}\b{\ell_2\,2}} \frac{\b{1\,\ell_2}\b{\ell_2\,\ell_1}}{\b{41}\b{\ell_1\,4}}\,=\, 
-F^{(0)} \frac{\b{34} \b{1\,\ell_2} \b{\ell_1\,2}}{\b{31}\b{\ell_2\,2}\b{\ell_1\,4}}\ .
\end{align}
The two factorisations with fermions, \eqref{eq:1loop4-c3} and \eqref{eq:1loop4-c4}, give the same result as happened in the previous section
\begin{align}
\label{eq:fermion-c}
\frac{1}{\b{\ell_2 \,\ell_1}}\left(-\frac{\b{1\,\ell_1}\b{\ell_1\,\ell_2}}{\b{\ell_1\,4}\b{41}}\right)&\,=\, F^{(0)} \frac{\b{1\,\ell_1}\b{34}}{\b{\ell_1\,4}\b{31}}\ .
\end{align}
We can combine the two diagrams with fermions with the first two diagrams which do not involve fermions. Consider first  \eqref{eq:1-1loop4c} summed with \eqref{eq:fermion-c}:
\begin{align}
\begin{split}
F^{(0)} \frac{\b{1\,\ell_1}\b{34}}{\b{\ell_1\,4}\b{31}}&\left(\frac{-\b{1\,\ell_1}\b{3\,\ell_2}+\b{1\,\ell_2}\b{3\,\ell_1}}{\b{1\,\ell_2}\b{3\,\ell_1}}\right)\,=\,-F^{(0)} \frac{\b{1\,\ell_1}\b{34}\b{\ell_2\,\ell_1}}{\b{\ell_1\,4}\b{1\,\ell_2}\b{3\,\ell_1}}\\
\,=\, -\frac{1}{4} &F^{(0)} \frac{\b{1\,\ell_1}\b{34}\b{\ell_2\,\ell_1}[\ell_1\,4][3\,\ell_1]}{(\ell_1\cdot p_4) (\ell_1\cdot p_3) \b{1\,\ell_2}}\ .
\end{split}
\end{align}
Using that on the cut $ \b{\ell_2\,\ell_1}[\ell_1\,4] = \b{\ell_2\,1}[14] $ we get
\begin{align}
\frac{1}{4} &F^{(0)} \frac{\b{1\,\ell_1}\b{34} [14] [3\,\ell_1]}{(\ell_1\cdot p_4) (\ell_1\cdot p_3) } \,=\, \frac{1}{4} F^{(0)} \frac{\Tr_+(p_1 \, \ell_1  \, p_3  \, p_4 )}{(\ell_1\cdot p_4) (\ell_1\cdot p_3) }\ .
\end{align}
The trace is
\begin{align}
\Tr_+(p_1 \, \ell_1  \, p_3  \, p_4 )\, =\, s_{41}(\ell_1\cdot p_3) + s_{34}(\ell_1\cdot p_1) - s_{13} (\ell_1\cdot p_4) \, .
\end{align}
Using that $(\ell_1\cdot p_1)= -(\ell_1\cdot p_4) + \frac{1}{2}s_{41}$, we get
\begin{align}
\label{eq:q-p1-p4-a}
\begin{split}
-\frac{1}{2}F^{(0)} \left( \frac{s_{34}s_{41}}{(\ell_1-p_4)^2(\ell_1+p_3)^2} + \frac{s_{41}}{(\ell_1-p_4)^2} + \frac{(q-p_2)^2-s_{41}}{(\ell_1+p_3)^2}  \right)\ .
\end{split}
\end{align}
The first term is a one-mass box with massive corner $(q-p_2)$, the second term is a one-mass triangle with massive corner $(q-p_2-p_3)$ and finally the last term is a two-mass triangle with massive corners $(q-p_2)$ and $-(p_4+p_1)$.

\noindent Now we sum \eqref{eq:2-1loop4c} summed with \eqref{eq:fermion-c} to get the second half of the answer:
\begin{align}
\begin{split}
&F^{(0)} \frac{\b{34}}{\b{\ell_1\,4}\b{31}}\left(\frac{-\b{1\,\ell_2}\b{\ell_1\,2}+\b{1\,\ell_1}\b{\ell_2\,2}}{\b{\ell_2\,2}}\right)=F^{(0)} \frac{\b{12}\b{34} \b{\ell_2\,\ell_1}}{\b{\ell_1\,4}\b{31}\b{\ell_2\,2}}\\
= -\frac{1}{4} &F^{(0)} \left(\frac{\b{12}\b{34}}{\b{13}\b{24}}\right) \frac{\b{\ell_2\,\ell_1} \b{24} [\ell_1\,4] [\ell_2\,2]}{(\ell_1\cdot p_4) (\ell_2\cdot p_2) } = -\frac{1}{4} F^{(0)} \left(\frac{\b{12}\b{34}}{\b{13}\b{24}}\right) \frac{\Tr_+(\ell_2 \,\ell_1  \, p_4  \, p_2)}{(\ell_1\cdot p_4) (\ell_2\cdot p_2) }\ .
\end{split}
\label{eq:q-p4-p2-2}
\end{align}
On the cut the trace gives
\begin{align}
\Tr_+(\ell_2 \, p_1  \, p_4  \, p_2)\,=\, s_{41}(\ell_2\cdot p_2) + s_{24}(\ell_2\cdot p_1) - s_{12} (\ell_2 \cdot p_4) \ .
\end{align}
We use that $(\ell_2\cdot p_1)=(\ell_1\cdot p_4) $ and $(\ell_2 \cdot p_4)=-(\ell_1 \cdot p_4)+\frac{1}{2}s_{41}$, so \eqref{eq:q-p4-p2-2} becomes
\begin{align}
\label{eq:q-p1-p4-b}
-\frac{1}{2} F^{(0)} \left(\frac{\b{12}\b{34}}{\b{13}\b{24}}\right) \left( \frac{s_{12}s_{41}}{(\ell_1-p_4)^2(\ell_2+p_2)^2} -\frac{s_{41}}{(\ell_1-p_4)^2}+\frac{(q-p_3)^2-s_{41}}{(\ell_2+p_2)^2}\right)\ .
\end{align}
Now we put together \eqref{eq:q-p1-p4-a} and \eqref{eq:q-p1-p4-b} to get the result for the cut
\begin{align}
\label{eq:q-p2-p3-cut}
\begin{split}
&\left.F^{(1)}\right|_{(q-p_2-p_3)^2\text{-cut}}=-\frac{1}{2}F^{(0)} \left\{\left( \frac{s_{34}s_{41}}{(\ell_1-p_4)^2(\ell_1+p_3)^2}\right. + \frac{(q-p_2)^2-s_{41}}{(\ell_1+p_3)^2}  \right)\\
&+ \left(\frac{\b{12}\b{34}}{\b{13}\b{24}}\right) \left( \frac{s_{12}s_{41}}{(\ell_1-p_4)^2(\ell_2+p_2)^2} +\frac{(q-p_3)^2-s_{41}}{(\ell_2+p_2)^2} \right)-\left. \left(\frac{\b{41}\b{23}}{\b{13}\b{24}}\right)\frac{s_{41}}{(\ell_1-p_4)^2}\right\}\ .
\end{split}
\end{align}

\subsection*{$\boldsymbol{(q-p_3-p_4)^2}$-channel}

We now look at the cut $ F^{(1)}_{3}(1^{\phi_{12}},2^{\phi_{12}},3^{\phi_{12}},4^+;q)$ across the $(q-p_3-p_4)^2$ cut. There is only one factorisation given by
\begin{align}
\label{eq:1loop4-q-p3-p4}
&F^{(0)}_{3}(3^{\phi_{12}},4^+, \ell_1^{\phi_{12}},\ell_2^{\phi_{12}};q) A^{\text{MHV}}(1^{\phi_{12}},2^{\phi_{12}},-\ell_2^{\phi_{34}},-\ell_1^{\phi_{34}})\ .
\end{align}
Plugging the tree level expressions we get
\begin{align}
F^{(0)}\frac{\b{34}\b{41}}{\b{31}} \frac{\b{3\,\ell_1}}{\b{34}\b{4\,\ell_1}} \frac{\b{12}\b{\ell_2\,\ell_1}}{\b{2\,\ell_2}\b{\ell_1\,1}}\, =\,\frac{1}{4} F^{(0)}  \frac{\b{41}\b{12}\b{3\,\ell_1}\b{\ell_2\,\ell_1}[2\,\ell_2][4\,\ell_1]}{\b{31}\b{\ell_1\,1}(p_2\cdot \ell_2)(p_4\cdot \ell_1)} \ .
\end{align}
We simplify the expression above by noting that on the cut $\b{\ell_2\,\ell_1}[2\,\ell_2]=\b{\ell_1\,1}[12]$ and also writing $\b{3\,\ell_1}\b{41} =\b{34} \b{\ell_1\,1} + \b{31}\b{4\,\ell_1} $, so we get
\begin{align}
\frac{1}{4} F^{(0)}  \frac{s_{12}\b{41}\b{3\,\ell_1}[4\,\ell_1]}{\b{31}(p_2\cdot \ell_2)(p_4\cdot \ell_1)}\, =\,\frac{1}{4}F^{(0)} \frac{s_{12}\b{34}[4\,\ell_1]\b{\ell_1\,1}}{\b{31}(p_2\cdot \ell_2)(p_4\cdot \ell_1)}+\frac{1}{2}F^{(0)} \frac{s_{12}  }{(\ell_2\cdot p_2)} \ .
\end{align}
The second term clearly gives a one-mas triangle, while we can manipulate the first term a little further:
\begin{align}
\frac{1}{4}F^{(0)} \frac{s_{12}\b{34}[4\,\ell_1]\b{\ell_1\,1}}{\b{31}(p_2\cdot \ell_2)(p_4\cdot \ell_1)}\, =\,\frac{1}{4}F^{(0)}\left(\frac{\b{12}\b{34}}{\b{24}\b{31}}\right) \frac{\Tr_+(p_2 \, p_4  \, \ell_1  \, p_1)}{(p_2\cdot \ell_2)(p_4\cdot \ell_1)}\ .
\end{align}
Using that on the cut $(p_2\cdot \ell_1)=-(p_1\cdot \ell_1)+\frac{1}{2}s_{12}$ and $(p_1\cdot \ell_1)=(p_2\cdot \ell_2) $, the trace gives
\begin{align}
\Tr_+(p_2 \, p_4  \, \ell_1  \, p_1)\, =\,(s_{14}+s_{24})(p_2\cdot\ell_2)+ s_{12}(p_4\cdot \ell_1)-\tfrac{1}{2}s_{14}s_{12}\ .
\end{align}
Now collecting everything we obtain the result for this cut
\begin{align}
\label{eq:q-p3-p4-cut}
\begin{split}
\left.F^{(1)}\right|_{(q-p_3-p_4)^2\text{-cut}}= &-\frac{1}{2} F^{(0)} \left(\frac{\b{12}\b{34}}{\b{13}\b{24}}\right)  \left( \frac{ s_{12}s_{41}}{(\ell_1+p_4)^2(\ell_2-p_2)^2} + \frac{ (q-p_3)^2-s_{12}}{(\ell_1+p_4)^2}- \frac{s_{12}}{(\ell_2-p_2)^2}\right)\\
&- F^{(0)} \frac{ s_{12}}{(\ell_2-p_2)^2} \ .
\end{split}
\end{align}

\subsection*{$\boldsymbol{(q-p_4-p_1)^2}$-channel}

Here we compute the discontinuity of $ F^{(1)}_{3}(1^{\phi_{12}},2^{\phi_{12}},3^{\phi_{12}},4^+;q)$ across the $(q-p_4-p_1)^2$ cut. There is only one factorisation:
\begin{align}
\label{eq:1loop4-q-p4-p1}
&F^{(0)}_{3}(4^+,1^{\phi_{12}}, \ell_1^{\phi_{12}},\ell_2^{\phi_{12}};q) A^{\text{MHV}}(2^{\phi_{12}},3^{\phi_{12}},-\ell_2^{\phi_{34}},-\ell_1^{\phi_{34}})\ .
\end{align}
This is given by
\begin{align}
F^{(0)}\frac{\b{34}\b{41}}{\b{31}} \frac{\b{\ell_2\,1}}{\b{\ell_2\,4}\b{41}}\frac{\b{\ell_2\,\ell_1}^2\b{23}^2}{\b{23}\b{3\,\ell_2}\b{\ell_2\,\ell_1}\b{\ell_1\,2}}= \frac{1}{4}F^{(0)} \frac{\b{\ell_2\,1} \b{34}\b{23} \b{\ell_2\,\ell_1}[\ell_1\,2][\ell_2\,4]}{\b{31}\b{3\,\ell_2}(p_4\cdot \ell_2)(p_2\cdot \ell_1)}\ .
\end{align}
We simplify the expression above by noting that on the cut $\b{\ell_2\,\ell_1}[\ell_1\,2]=[23]\b{3\,\ell_2}$ and also writing $\b{\ell_2\,1} \b{34}=\b{\ell_2\,3} \b{14}+\b{\ell_2\,4} \b{31}$, so we get
\begin{align}
\frac{1}{4}F^{(0)} \frac{s_{23}\b{\ell_2\,1} \b{34} [\ell_2\,4]}{\b{31}(p_4\cdot \ell_2)(p_2\cdot \ell_1)}= \frac{1}{4}F^{(0)} \frac{s_{23}\b{\ell_2\,3} \b{14} [\ell_2\,4]}{\b{31}(p_4\cdot \ell_2)(p_2\cdot \ell_1)}+\frac{1}{2}F^{(0)} \frac{s_{23}  }{(p_2\cdot \ell_1)}\ .
\end{align}
The second term gives a one-mas triangle. Let us explore the first term:
\begin{align}
\begin{split}
\frac{1}{4}F^{(0)} \frac{s_{23}\b{\ell_2\,3} \b{14} [\ell_2\,4]}{\b{31}(p_4\cdot \ell_2)(p_2\cdot \ell_1)}&\, =\, -\frac{1}{4}F^{(0)}\left(\frac{\b{23}\b{41}}{\b{24}\b{13}}\right) \frac{\b{\ell_2\,3} \b{42} [23] [\ell_2\,4]}{(p_4\cdot \ell_2)(p_2\cdot \ell_1)}\\
&\, =\, \frac{1}{4}F^{(0)}\left(\frac{\b{23}\b{41}}{\b{24}\b{13}}\right) \frac{\Tr_+(\ell_2 \, p_3 \, p_2  \, p_4)}{(p_4\cdot \ell_2)(p_2\cdot \ell_1)}\ .
\end{split}
\end{align}
Using that on the cut $(p_2\cdot \ell_2)=-(p_3\cdot \ell_2)+\frac{1}{2}s_{23}$ and $(p_3\cdot \ell_2)=(p_2\cdot \ell_1) $, the trace gives
\begin{align}
\begin{split}
\Tr_+(\ell_2 \, p_3 \, p_2  \, p_4)&\, =\,s_{24} (p_3\cdot \ell_2) + s_{23} (p_4\cdot \ell_2) - s_{34}(p_2\cdot \ell_2)\\
&= (s_{24} + s_{34}) (p_2 \cdot \ell_1) + s_{23} (p_4\cdot \ell_2)  -\tfrac{1}{2}s_{34}s_{23}\ .
\end{split}
\end{align}
Putting everything together we obtain the result for the cut:
\begin{align}
\label{eq:q-p4-p1-cut}
\begin{split}
\left.F^{(1)}\right|_{(q-p_4-p_1)^2\text{-cut}}= &\frac{1}{2} F^{(0)} \left(\frac{\b{41}\b{23}}{\b{13}\b{24}}\right)  \left( \frac{ s_{23}s_{34}}{(\ell_1-p_2)^2(\ell_2+p_4)^2} + \frac{ (q-p_1)^2-s_{23}}{(\ell_2+p_4)^2}-\frac{ s_{23}}{(\ell_1-p_2)^2}  \right)\\
&-F^{(0)} \frac{ s_{23}}{(\ell_1-p_2)^2}\ .
\end{split}
\end{align}
On the first line we have the functions which already appeared in other channels (and the coefficient are consistent): one-mass box with massive corner $q-p_1$ and two-mass triangle with massless corner $p_4$. On the second line we have a one mass triangle with massive corner $q-p_1-p_4$. There is no one-mass box with massive corner $q-p_4$ as there is no possible helicity assignment to the internal propagators.

\subsection{Summary of the cuts}
\label{sec:summary-cuts}
Here we collect the results of the cuts in all channels. One can check that the functions that are detectable in different channels consistently appear with the same coefficient.

{\small
\begin{align*}
\begin{split}
\left.F^{(1)}\right|_{(q-p_1)^2\text{-cut}}=& 
\frac{1}{2} F^{(0)} \left(\frac{\b{41}\b{23}}{\b{13}\b{24}}\right) \left(\frac{s_{23}s_{34}}{(\ell_1-p_2)^2 (\ell_2-p_4)^2} + \frac{(q-p_1)^2-s_{23} }{(\ell_2-p_4)^2 } + \frac{(q-p_1)^2-s_{34} }{(\ell_1-p_2)^2 } \right),\\[12pt]
\left.F^{(1)}\right|_{(q-p_2)^2\text{-cut}}=&-\frac{1}{2}F^{(0)} \left(\frac{s_{41}s_{34}}{(\ell_1-p_3)^2(\ell_2-p_1)^2}+\frac{(q-p_2)^2-s_{34}}{(\ell_2-p_1)^2}+\frac{(q-p_2)^2-s_{41}}{(\ell_1-p_3)^2}\right),
\\[12pt]
\left.F^{(1)}\right|_{(q-p_3)^2\text{-cut}}=&-\frac{1}{2}F^{(0)}\left(\frac{\b{12}\b{34}}{\b{13}\b{24}}\right) \left(\frac{s_{41}s_{12}}{(\ell_1-p_4)^2(\ell_2-p_2)^2}+\frac{(q-p_3)^2-s_{12}}{(\ell_1-p_4)^2}+\frac{(q-p_3)^2-s_{41}}{(\ell_2-p_2)^2}\right),\\[12pt]
\left.F^{(1)}\right|_{(q-p_4)^2\text{-cut}}=& 0,\\
\left.F^{(1)}\right|_{(q-p_1-p_2)^2\text{-cut}}= &\frac{1}{2} F^{(0)} \left\{\left(\frac{\b{41}\b{23}}{\b{13}\b{24}}\right)\right.  \left( \frac{ s_{23}s_{34}}{(\ell_1+p_2)^2(\ell_2-p_4)^2} + \frac{ (q-p_1)^2-s_{34}}{(\ell_1+p_2)^2}\right)\\
-& \left(\frac{\b{12}\b{34}}{\b{13}\b{24}}\right)\left. \frac{ s_{34}}{(\ell_2-p_4)^2} 
-  \frac{ s_{34} s_{41}}{(\ell_2+p_1)^2(\ell_2-p_4)^2} - \frac{ (q-p_2)^2 - s_{34} }{(\ell_2+p_1)^2} \right\},\\[12pt]
\left.F^{(1)}\right|_{(q-p_2-p_3)^2\text{-cut}}=&-\frac{1}{2}F^{(0)} \left\{\left( \frac{s_{34}s_{41}}{(\ell_1-p_4)^2(\ell_1+p_3)^2}\right. + \frac{(q-p_2)^2-s_{41}}{(\ell_1+p_3)^2}  \right)\\
+ \left(\frac{\b{12}\b{34}}{\b{13}\b{24}}\right)& \left( \frac{s_{12}s_{41}}{(\ell_1-p_4)^2(\ell_2+p_2)^2} +\frac{(q-p_3)^2-s_{41}}{(\ell_2+p_2)^2} \right)-\left. \left(\frac{\b{41}\b{23}}{\b{13}\b{24}}\right)\frac{s_{41}}{(\ell_1-p_4)^2}\right\},\\[12pt]
\left.F^{(1)}\right|_{(q-p_3-p_4)^2\text{-cut}}= &-\frac{1}{2} F^{(0)} \left(\frac{\b{12}\b{34}}{\b{13}\b{24}}\right)  \left( \frac{ s_{12}s_{41}}{(\ell_1+p_4)^2(\ell_2-p_2)^2} + \frac{ (q-p_3)^2-s_{12}}{(\ell_1+p_4)^2}- \frac{s_{12}}{(\ell_2-p_2)^2}\right)\\
&- F^{(0)} \frac{ s_{12}}{(\ell_2-p_2)^2}, \\[12pt]
\left.F^{(1)}\right|_{(q-p_4-p_1)^2\text{-cut}}= &\frac{1}{2} F^{(0)} \left(\frac{\b{41}\b{23}}{\b{13}\b{24}}\right)  \left( \frac{ s_{23}s_{34}}{(\ell_1-p_2)^2(\ell_2+p_4)^2} + \frac{ (q-p_1)^2-s_{23}}{(\ell_2+p_4)^2}-\frac{ s_{23}}{(\ell_1-p_2)^2}  \right)\\
&-F^{(0)} \frac{ s_{23}}{(\ell_1-p_2)^2}  \ .
\end{split}
\end{align*}}

\subsection{Summary of integrals with coefficients}
In this section we will combine all the cuts summarised in \sref{sec:summary-cuts} to obtain the result for $F^{(1)}_{3}(1^{\phi_{12}},2^{\phi_{12}},3^{\phi_{12}},4^+;q)$.

Firstly we examine the IR divergent terms coming of each scalar integral in multiples of $\dfrac{r_\Gamma}{2\epsilon^2}F^{(0)}$, where $r_{\Gamma}$ is defined in \eqref{eq:rgamma}. These are shown in Table \ref{tab:IRdiv} where we denote $q_i\equiv (q-p_i)^2$ and represent the cross ratios as
\begin{align}
\text{CR}_1\,=\, \frac{\b{41}\b{23}}{\b{13}\b{24}}\ , \qquad \text{CR}_2\,=\, \frac{\b{12}\b{34}}{\b{13}\b{24}}\ ,\qquad 
\text{CR}_2-\text{CR}_1\,=\,1\ .
\end{align}
\begin{table}[h]
\centering
{\footnotesize
\begin{tabular}{cclccl}
$(i)\mathord{\parbox[c]{1em}{\includegraphics[scale=0.5]{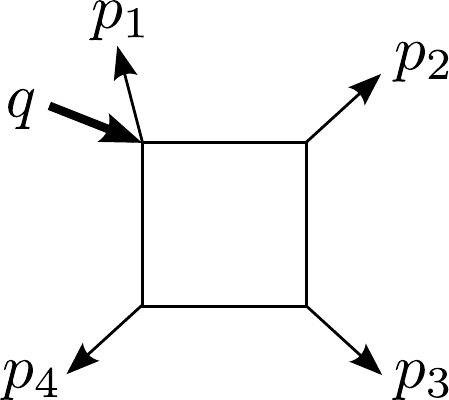}}}$  & \hspace*{2cm} & $ = 2 \,\text{CR}_1 (s_{23}^{-\epsilon}+s_{34}^{-\epsilon}-q_1^{-\epsilon})$ & $ (ii) \mathord{\parbox[c]{1em}{\includegraphics[scale=0.5]{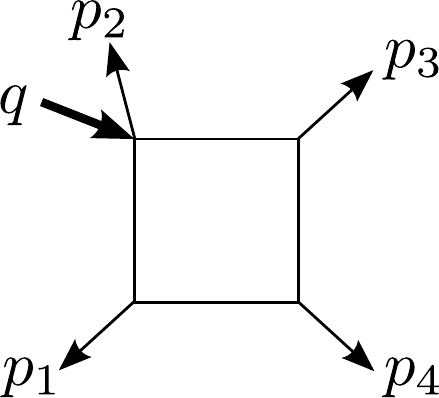}}} $ & \hspace*{2cm} &  $= -2 \, (s_{34}^{-\epsilon}+s_{41}^{-\epsilon}-q_2^{-\epsilon})$ \\
$ (iii)\mathord{\parbox[c]{1em}{\includegraphics[scale=0.5]{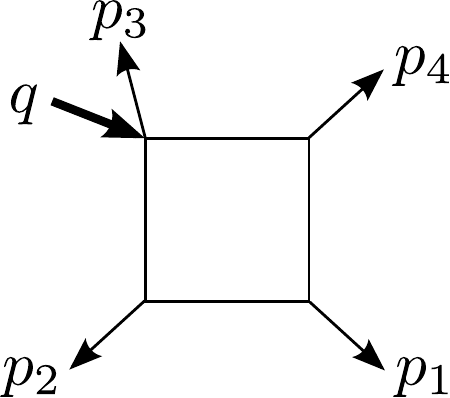}}}$  & \hspace*{2cm} & $= -2 \,\text{CR}_2 (s_{41}^{-\epsilon}+s_{12}^{-\epsilon}-q_3^{-\epsilon})$ & 
$ (iv) \mathord{\parbox[c]{1em}{\includegraphics[scale=0.5]{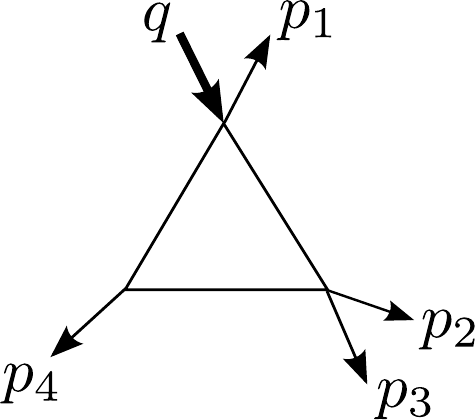}}} $ & 
\hspace*{2cm} & $ =  \text{CR}_1 (q_1^{-\epsilon}-s_{23}^{-\epsilon})$ \\
$ (v) \mathord{\parbox[c]{1em}{\includegraphics[scale=0.5]{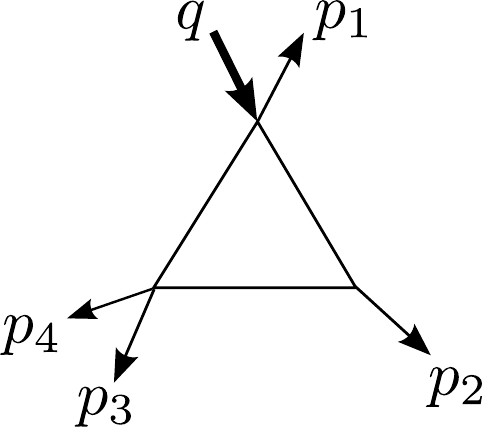}}}$ & \hspace*{2cm} & $= -\text{CR}_1 (s_{34}^{-\epsilon}-q_1^{-\epsilon})$ &
$ (vi)\mathord{\parbox[c]{1em}{\includegraphics[scale=0.5]{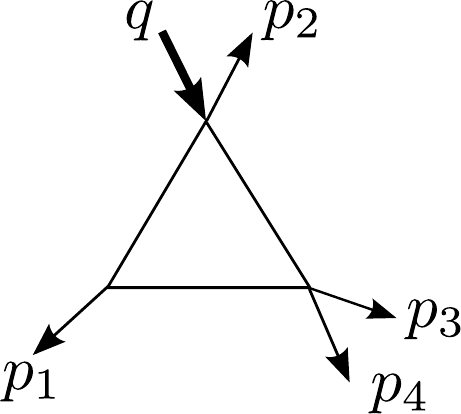}}}$ & 
\hspace*{2cm} & $= - (q_2^{-\epsilon}-s_{34}^{-\epsilon})$ \\
$ (vii)\mathord{\parbox[c]{1em}{\includegraphics[scale=0.5]{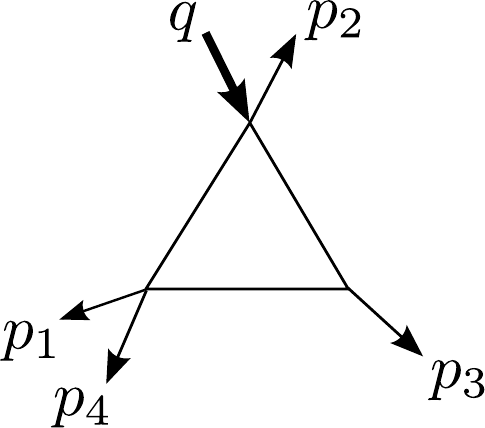}}}$ & \hspace*{2cm} & $= (s_{41}^{-\epsilon}-q_2^{-\epsilon})$
 & 
$ (viii)\mathord{\parbox[c]{1em}{\includegraphics[scale=0.5]{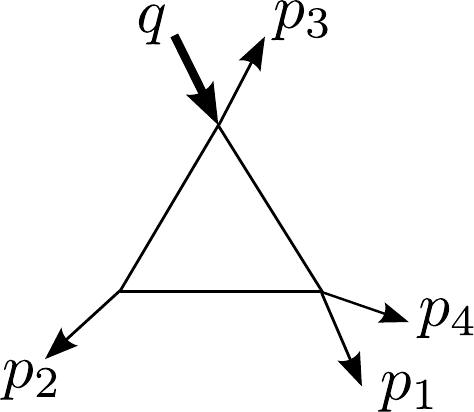}}}$ & 
\hspace*{2cm} & $= - \text{CR}_2 (q_3^{-\epsilon}-s_{41}^{-\epsilon})$
\\ $ (ix)\mathord{\parbox[c]{1em}{\includegraphics[scale=0.5]{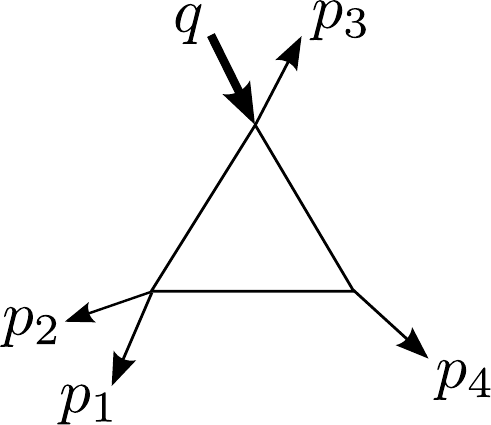}}}$ & \hspace*{2cm} & $= \text{CR}_2\, (s_{12}^{-\epsilon}-q_3^{-\epsilon})$
& $ (x)\mathord{\parbox[c]{1em}{\includegraphics[scale=0.5]{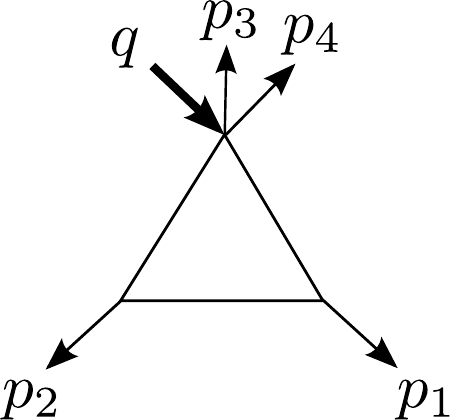}}}$ & 
\hspace*{2cm} & $=  (\text{CR}_2 - 2)\, s_{12}^{-\epsilon}$ \\
$ (xi)\mathord{\parbox[c]{1em}{\includegraphics[scale=0.5]{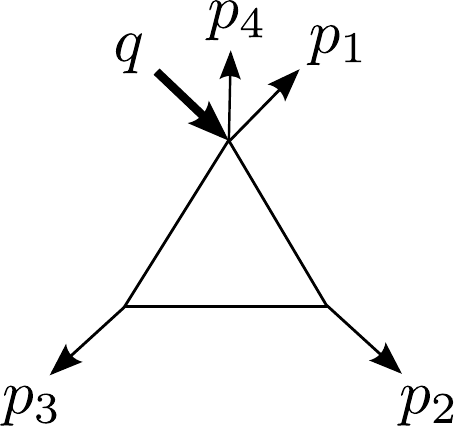}}}$ & \hspace*{2cm} & $= -(\text{CR}_1 + 2)\, s_{23}^{-\epsilon}$
 & $ (xii)\mathord{\parbox[c]{1em}{\includegraphics[scale=0.5]{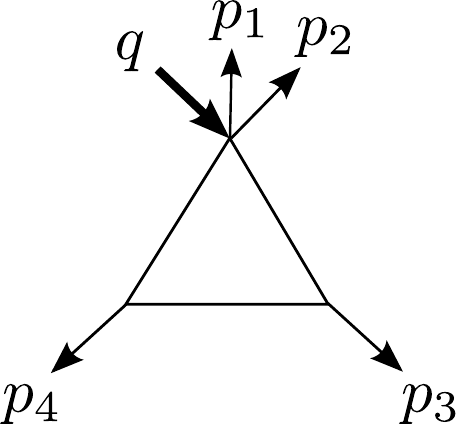}}} $ & 
\hspace*{2cm} & $= -(\text{CR}_1+1)\, s_{34}^{-\epsilon}$ \\
$ (xiii)\mathord{\parbox[c]{1em}{\includegraphics[scale=0.5]{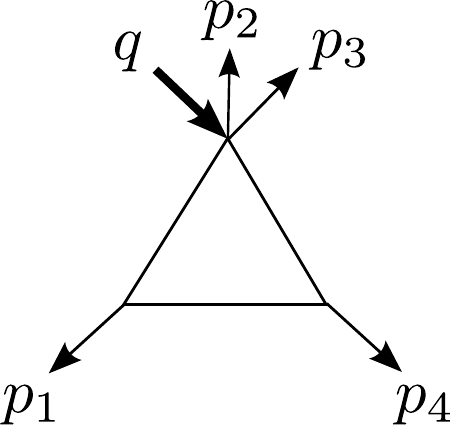}}} $ & \hspace*{2cm} & $= (\text{CR}_2-1)\, s_{41}^{-\epsilon}$
\end{tabular}}
\caption{\it IR divergent terms of $F^{(1)}_{3}(1^{\phi_{12}},2^{\phi_{12}},3^{\phi_{12}},4^+;q)$.}
\label{tab:IRdiv}
\end{table}
\pagebreak
\noindent The result of the IR divergent terms is the sum of all terms of Table \ref{tab:IRdiv}:
\begin{equation}
\label{eq:F-IR}
F^{(1)}_{3}(1^{\phi_{12}},2^{\phi_{12}},3^{\phi_{12}},4^+;q)\Big|_{\rm IR}\,=\,(i)+(ii)+\dots +(xiii)\ .
\end{equation}
Combining the terms depending on $q_1,\,q_2, q_3$ and no $q$ separately, we find that the all dependence on $q$ drops out,
\begin{align}
\begin{split}
q_1:\quad &(i)+(iv)+(v)\, = \,\text{CR}_1(s_{23}^{-\epsilon}+s_{34}^{-\epsilon})\,\equiv\,(xiv)\ , \\
q_2:\quad &(ii)+(vi)+(vii)\, =\, -(s_{34}^{-\epsilon}+s_{41}^{-\epsilon})\,\equiv\,(xv)\ ,\\
q_3:\quad &(iii)+(viii)+(ix)  \,=\, -\text{CR}_2(s_{41}^{-\epsilon}+s_{12}^{-\epsilon})\,\equiv\, (xvi)\ ,
\\
\text{No }q:\quad & (x)+(xi)+(xii)+(xiii)\,=\, \text{CR}_2 (s_{12}^{-\eps}+s_{41}^{-\eps})+\text{CR}_1(s_{23}^{-\eps}+s_{34}^{-\eps} )\\
&\qquad \qquad \qquad \qquad\qquad \qquad\quad  -(2 s_{12}^{-\eps}+2 s_{23}^{-\eps} + s_{34}^{-\eps} + s_{41}^{-\eps})\,\equiv \,(xvii) \ .
\end{split}
\end{align}
\noindent Then, \eqref{app:1loopComponent} becomes
\begin{align}
\begin{split}
F^{(1)}_{3}(1^{\phi_{12}},2^{\phi_{12}},3^{\phi_{12}},4^+;q)\Big|_{\rm IR}\,&=\, (xiv)+(xv)+(xvi)+(xii)\\
&=\, -2 (s_{12}^{-\epsilon}+s_{23}^{-\epsilon}+s_{34}^{-\epsilon}+s_{41}^{-\epsilon})\ .
\end{split}
\end{align}
So we conclude that the one-loop result for $ F^{(1)}_{3}(1^{\phi_{12}},2^{\phi_{12}},3^{\phi_{12}},4^+;q)$ is
\begin{align}
\label{eq:result-1loop-4-bosonic}
\begin{split}
-F^{(0)} \sum_{i=1}^4 s_{i\,i+1} I_{3;i}^{1m}(s_{i\,i+1}) + F^{(0)} \text{Fin}\left\{\left(\frac{\b{41}\b{23}}{\b{13}\b{24}}\right) I^{1m}_{4;3}(s_{23},s_{34},(q-p_1)^2)\right.\\ - \left. I^{1m}_{4;3}(s_{34},s_{41},(q-p_2)^2) -\left(\frac{\b{12}\b{34}}{\b{13}\b{24}}\right) I^{1m}_{4;1}(s_{41},s_{12},(q-p_3)^2) \right\}.
\end{split}
\end{align}
The part proportional to the tree level expression contains all information about the IR divergences and agrees with the expected result \eqref{eq:1-loop-MHV-all-k}. Indeed, the only physical IR divergences are soft and collinear and must be related to massless adjacent particles. For this reason, all dependence on $q$ dropped out.

\chapter{Non-planar on-shell diagrams}

\section{Embedding independence}

\label{section_simple_example}

\begin{figure}[h]
\begin{center}
\includegraphics[width=0.3\linewidth]{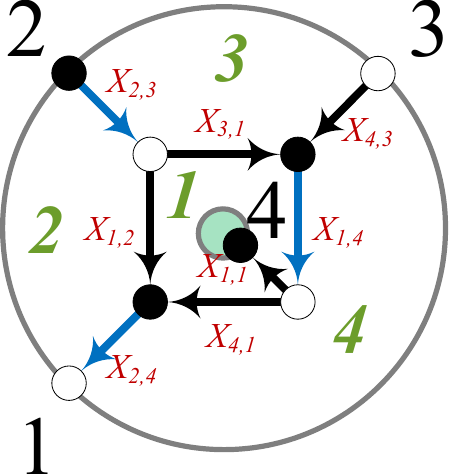}
\caption{\it An on-shell diagram on an annulus. This particular graph can be planarised by flipping the $X_{1,1}$ edge. Faces are labeled in green, external nodes in black and edges in red.}
\label{fig:sqbcrossed}
\end{center}
\end{figure}

Here we illustrate the independence on the embedding of the on-shell diagram with the simple example shown in \fref{fig:sqbcrossed}. It is clear that the non-planarity of this diagram is fake, since it can be embedded on a disk by flipping $X_{1,1}$. 

%
%

Here we have four face variables, three of which are independent, and one cut. In terms of oriented edge weights, they are given by
\begin{equation}
f_1=\frac{X_{3,1} X_{4,1}}{X_{1,2} X_{1,4}} \; , \quad f_2=\frac{X_{1,2}}{X_{2,3} X_{2,4}} \; , \quad f_3=\frac{X_{2,3} X_{4,3}}{X_{3,1}} \; , \quad b_1=\frac{X_{4,1}}{X_{1,1} X_{2,4}} \; .
\end{equation}

Let us consider the perfect orientation corresponding to the reference perfect matching $p_{\text{ref}} = X_{1,4} X_{2,3} X_{2,4} $, which has source set $\{2,3\}$. Using our prescription for the boundary measurement, we obtain the Grassmannian matrix
{\small
\begin{equation}
C = \left(
\begin{array}{c|cccc}
 & 1 & 2 & 3 & 4\\
 \hline
2 \ \ & \dfrac{X_{1,2}}{X_{2,3} X_{2,4}}+\dfrac{X_{3,1} X_{4,1}}{X_{1,4} X_{2,3} X_{2,4}} & 1 & 0 & \Gape[3pt][0pt]{-\dfrac{X_{1,1} X_{3,1}}{X_{1,4} X_{2,3}}} \\
3 \ \ &  -\dfrac{X_{4,1} X_{4,3}}{X_{1,4} X_{2,4}} & 0 & 1 & \dfrac{X_{1,1} X_{4,3}}{X_{1,4}} \\
\end{array}
\right) = \left(
\begin{array}{c|cccc}
 & 1 & 2 & 3 & 4 \\
 \hline
2 \ \ & f_1 f_2+f_2 & 1 & 0 & \Gape[3pt][0pt]{-\dfrac{f_1 f_2}{b_1}} \\
3 \ \ & -f_1 f_2 f_3 & 0 & 1 & \dfrac{f_1 f_2 f_3}{b_1} \\
\end{array}
\right) .
\end{equation}}

\noindent The on-shell form becomes
\begin{equation}
\Omega\,=\,\frac{df_1}{f_1} \frac{df_2}{f_2} \frac{df_3}{f_3} \frac{db_1}{b_1} .
\end{equation}
In terms of minors, it can be rewritten as
\begin{equation}
\Omega\,=\, {d^{2\times 4} C \over \text{Vol(GL}(2))} \frac{1 }{(12)(23)(34)(41)}\ ,
\end{equation}
which is simply the form for the planar embedding, i.e.\ the ordinary square box in \fref{G24_edges_and_faces}. This illustrates the independence of the on-shell form on the embedding and shows that the generalised face variables maintain a $d\log$ form regardless of its choice.

\section{On-shell form for a genus-one NMHV diagram} \label{app:highergenus}

\begin{figure}[h]
\begin{center}
\includegraphics[scale=0.45]{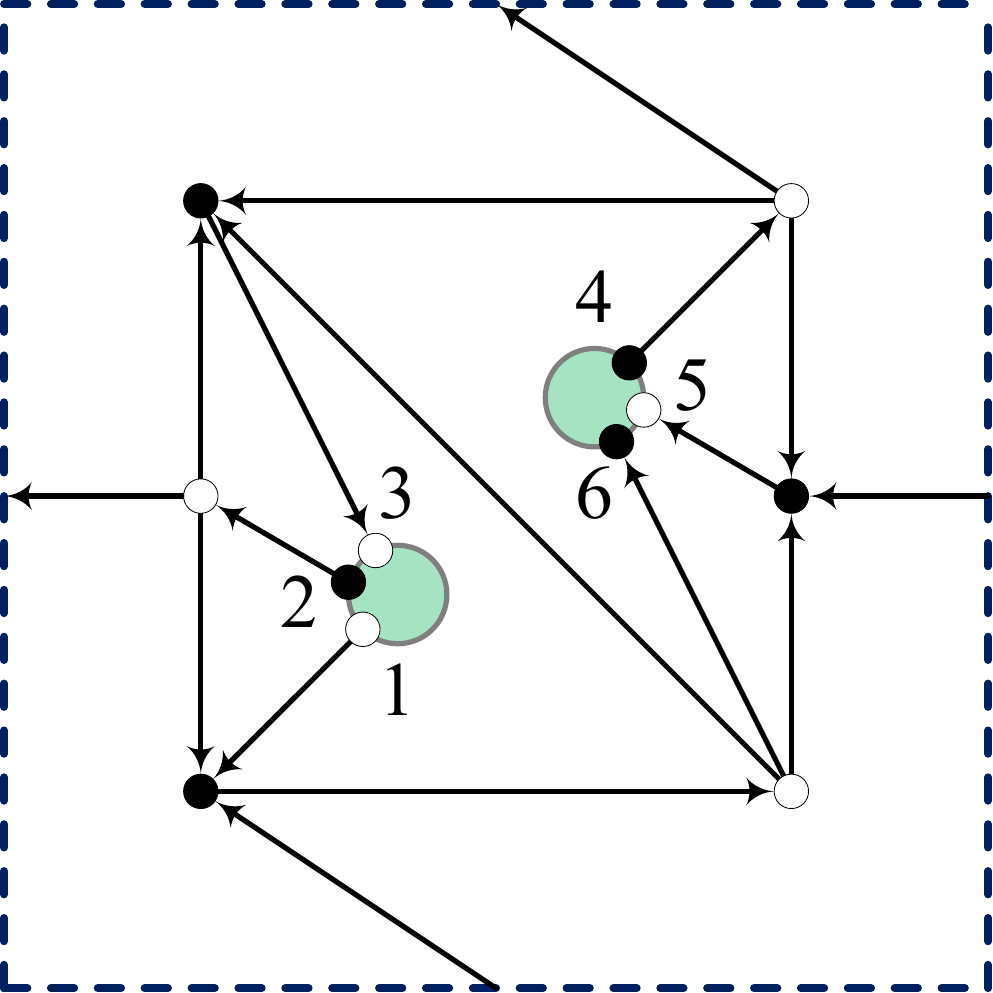}
\caption{\it An on-shell diagram embedded on a torus with two boundaries.}
\label{fig:nonplanarisableNolabel}
\end{center}
\end{figure}

To show that the method prescribed in \sref{sec:rules} works just as well for graphs with higher genus, we now consider a non-planarisable genus-one example shown in \fref{fig:nonplanarisableNolabel}.

\noindent Following the prescription in \sref{sec:rules}, we find the matrices $T$ and $M$ to be
\spacing{1.2}
\begin{align}
T\,=\,\begin{pmatrix}
1 & 6 & 4 & 2 \\
3 & 2 & 4 & 6 \\
5 & 4 & 2 & 6
\end{pmatrix}\, ,\quad M\,=\,\begin{pmatrix}
(642) & -(164) & 0 & (162) & 0 & -(142) \\
0 & -(346) & (246) & (326) & 0 & -(324) \\
0 & (546) & 0 & -(526) & (426) & -(542)
\end{pmatrix}\ .
\end{align}
\spacing{1.5}
\noindent It is easy to see that the simplest way to obtain the on-shell form is by deleting columns \{2,4,6\}, 
\spacing{1.2}
\begin{align}
\widehat{M}_{2,4,6}\,&=\,\begin{pmatrix}
(642) &  0 & 0 \\
0 &  (246) & 0 \\
0 & 0 & (426)
\end{pmatrix},\quad \frac{\det \widehat{M}_{2,4,6}}{(246)}=(246)^2\ ,
\end{align}
\spacing{1.5}
which gives the on-shell form
\begin{align}
\label{eq:int-genus1}
\Omega\,=\,\frac{d^{3\times 6}C}{\text{Vol(GL}(3))}\frac{(246)^3}{(164)(421)(216)(324)(463)(632)(542)(265)(654)}\ .
\end{align}
We have checked that this result coincides with the result obtained by using the boundary measurement as described in \sref{sec:integrandfromface}, giving further evidence to both methods as well as to the validity of the boundary measurement of \cite{Franco:2015rma}.

\section{N$^2$MHV example with two auxiliary edges}
\label{app:NNMHV}

\begin{figure}[h]
\begin{center}
\includegraphics[scale=0.45]{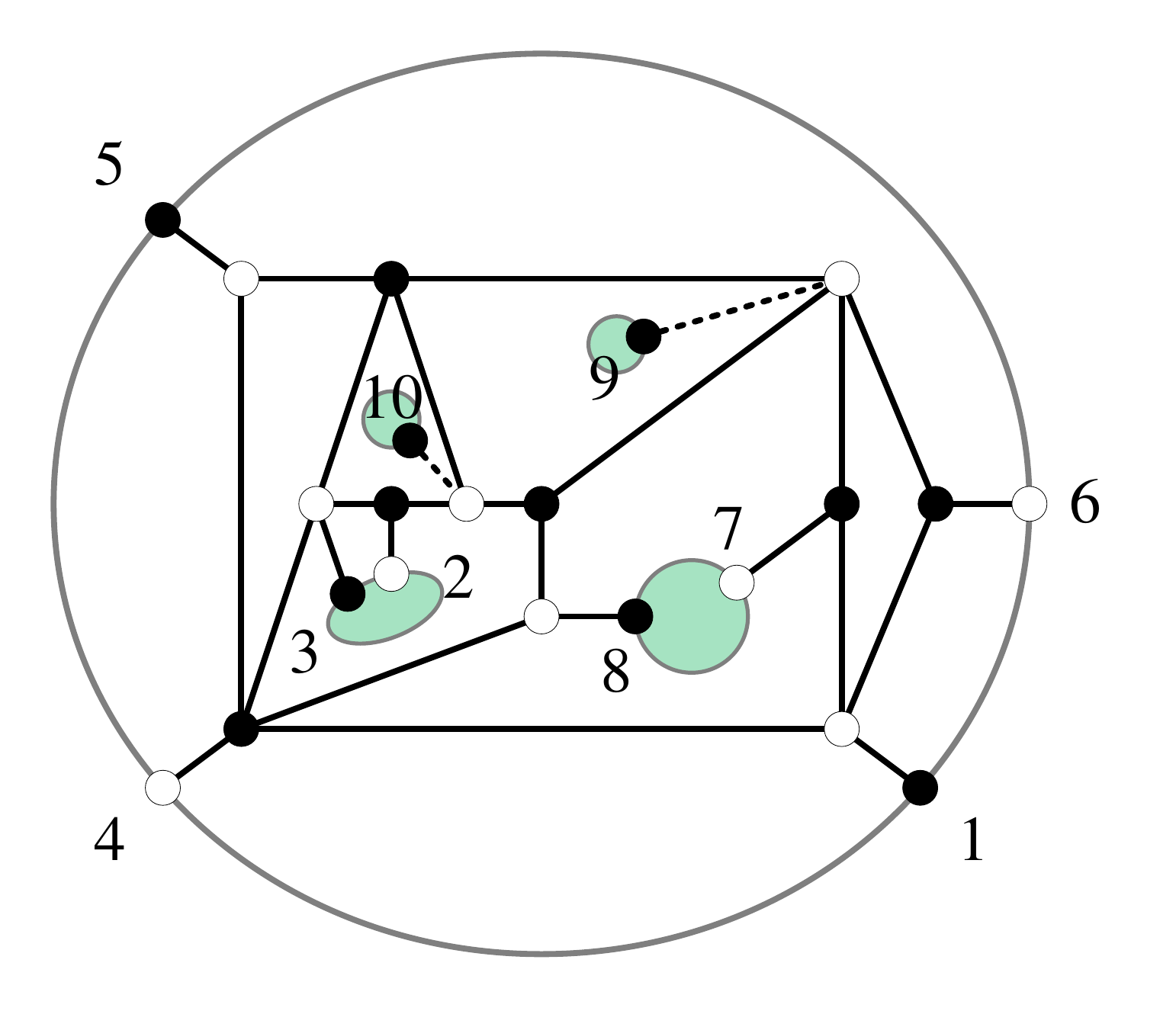}
\caption{\it An N$^2$MHV on-shell diagram for which $n_B=n-k+2$. In this case it is necessary to add two auxiliary external nodes, 9 and 10, for determining the on-shell form.}
\label{fig:NNMHV8}
\end{center}
\end{figure}

Let us consider the N$^2$MHV example in \fref{fig:NNMHV8}. The $T$ matrix is given by
\spacing{1.2}
{
\begin{align}
\label{eq:TNNMHV}
T\,=\,\begin{pmatrix}
\ 6 \ \ & \ 1 \ \ & \ 9 \ \ & \ * \ \ & \ * \ \ \\
1 & 7 & 9 & * & * \\
8 & 10 & 9 & * & *\\
10 & 3 & 5 & 9 & * \\
5 & 3 & 8 & 1 & 4 \\
2 & 3 & 10 & * & *
\end{pmatrix}\quad\xrightarrow{\text{Choice of }*}\quad T=\begin{pmatrix}
\ 6 \ \ & \ 1 \ \ & \ 9 \ \ & \ 3 \ \ & \ 8 \ \ \\
1 & 7 & 9 & 3 & 8 \\
8 & 10 & 9 & 1 & 3\\
10 & 3 & 5 & 9 & 1 \\
5 & 3 & 8 & 1 & 4 \\
2 & 3 & 10 & 1 & 8
\end{pmatrix}\ .
\end{align}
}
\spacing{1.5}
\noindent This leads to the following matrix $M$,
\spacing{1.2}
{\small
\begin{align}
M\,=\,\begin{pmatrix}
\label{eq:M82}
(9386) & 0 & (8619) & 0 & 0 & (1938) & 0 & (6193) & (3861) & 0\\
(7938) & 0 & (8179) & 0 & 0 & 0 & (9381) & (1793) & (3817) & 0 \\
(38109) & 0 & (81091) & 0 & 0 & 0 & 0 & (10913) & (13810) & (9138)\\
(10359) & 0 & (59110) & 0 & (91103) & 0 & 0 & 0 & (11035) & (3591) \\
(4538) & 0 & (8145) & (5381) & (3814) & 0 & 0 & (1453) & 0 & 0 \\
(82310) & (31018) & (10182) & 0 & 0 & 0 & 0 & (23101) & 0 & (1823) \\
\end{pmatrix}\ ,\nonumber\\[5pt]
\end{align}} 
where we eliminated the minus signs on the entries of $M$ by using the fact that an equivalent way to write \eqref{eq:Cramer} for even $k$ is $ \vec{c}_{i_1}(i_2\cdots i_{k+1})+\text{cyclic}(i_1,i_2,\dots, i_{k+1})=0$.
The result of the procedure in \sref{sec:rules} gives
{\small
\begin{align}
\begin{split}
\Omega\,&=\,\frac{d^{4\times 10}C}{\text{Vol(GL}(4))}\frac{(1358)^3(1389)^5(13810)^2(13910)^2}{(1238)(12310)(12810)(1345)(1348)(1359)(13510)(1368)(1369)(1378)(1379)}\\
&\times\frac{1}{(1458)(15910)(1689)(1789)(18910)(23810)(3458)(35910)(3689)(3789)(38910)}\ 
.
\end{split}\nonumber
\end{align}
}
This can be simplified using the fact that the points $\{1,6,7,9\}$ are collinear,  $\{8,9,10\}$ are collinear,  $\{2,3,10\}$ are collinear and  $\{3,5,9,10\}$ are coplanar, as can be read off from \eqref{eq:TNNMHV}. After these simplifications, the dependence on nodes $9$ and $10$ is encoded in the ratio
\begin{align}
\label{eq:ratioNNMHV}
\left.I\right|_{9,10}\,&=\,\frac{1}{(38910)(12310)(1369)(1689)(18910)(23810)}\ ,
\end{align}
which after the residues around $C_{i9}=C_{i10}=0$ for $i=1,\dots 4$ gives
\begin{align}
\left.I\right|_{9,10}\,&=\,\frac{1}{(1368)^2(1238)^2}\ .
\end{align}
Putting everything together, we obtain the following on-shell form
{\small
\begin{align}
\Omega\,=\,\frac{d^{4\times 8}C}{\text{Vol(GL}(4))}\frac{(1358)^3(1386)}{(7812)(1345)(1348)(1356)(1458)(1568)(1376)(6781)(2345)(3528)(3568)(3782)}\ .
\end{align}
}
This differential form has been independently confirmed using the boundary measurement procedure from \sref{sec:integrandfromface}.

\spacing{1}
\bibliographystyle{utphys}
\bibliography{thesis}
\end{document}